\documentclass[12pt,preprint]{aastex}
\usepackage{natbib}
\usepackage{bibentry}
\usepackage{amsmath}
\usepackage{epstopdf}
\usepackage{color}
\usepackage{hyperref}
\usepackage{exscale,relsize}
\usepackage{mathrsfs}

\newcommand{\MAP}{\textsl{WMAP}}
\newcommand{\WMAP}{\textsl{WMAP}}
\newcommand{\map}{{\MAP}}
\newcommand{\wmap}{{\WMAP}}

\newcommand{\Bennett}{{C. L. Bennett}}

\newcommand{\Dunkley}{{J. Dunkley}}
\newcommand{\Gold}{{B. Gold}}

\newcommand{\Halpern}{{M. Halpern}}
\newcommand{\Hill}{{R. S. Hill}}
\newcommand{\Hinshaw}{{G. Hinshaw}}
\newcommand{\Jarosik}{{N. Jarosik}}
\newcommand{\Kogut}{{A. Kogut}}
\newcommand{\Komatsu}{{E. Komatsu}}
\newcommand{\Smith}{{K. M. Smith}}
\newcommand{\Larson}{{D. Larson}}
\newcommand{\Limon}{{M. Limon}}
\newcommand{\Meyer}{{S. S. Meyer}}
\newcommand{\Nolta}{{M. R. Nolta}}
\newcommand{\Odegard}{{N. Odegard}}
\newcommand{\Page}{{L. Page}}

\newcommand{\Spergel}{{D. N. Spergel}}
\newcommand{\Tucker}{{G. S. Tucker}}

\newcommand{\Weiland}{{J. L. Weiland}}
\newcommand{\Wollack}{{E. Wollack}}
\newcommand{\Wright}{{E. L. Wright}}

\newcommand{\Adnet}{{ADNET Systems, Inc., 
            7515 Mission Dr., Suite A100 Lanham, Maryland 20706}}
\newcommand{\Brown}{{Dept. of Physics, Brown University, 
            182 Hope St., Providence, RI 02912-1843}}
\newcommand{\Cita}{{Canadian Institute for Theoretical Astrophysics, 
            60 St. George St, University of Toronto, 
	    Toronto, ON  Canada M5S 3H8}}
\newcommand{\Columbia}{{Columbia Astrophysics Laboratory, 
            550 W. 120th St., Mail Code 5247, New York, NY  10027-6902}}
\newcommand{\Goddard}{{Code 665, NASA/Goddard Space Flight Center, 
            Greenbelt, MD 20771}}
\newcommand{\Hopkins}{{Dept. of Physics \& Astronomy, 
            The Johns Hopkins University, 3400 N. Charles St., 
	    Baltimore, MD  21218-2686}}
\newcommand{\Minn}{{University of Minnesota, School of Physics \& Astronomy, 
            116 Church Street S.E., Minneapolis, MN 55455}}
\newcommand{\Oxford}{{Oxford Astrophysics, Denys Wilkinson Building, 
            Keble Road, Oxford, OX1 3RH, UK}}
\newcommand{\Perimeter}{{Perimeter Institute for Theoretical Physics, Waterloo, ON N2L 2Y5, Canada}}
\newcommand{\PrincetonPhysics}{{Dept. of Physics, Jadwin Hall, 
            Princeton University, Princeton, NJ 08544-0708}}
\newcommand{\PrincetonAstro}{{Dept. of Astrophysical Sciences, 
            Peyton Hall, Princeton University, Princeton, NJ 08544-1001}}
\newcommand{\UBC}{{Dept. of Physics and Astronomy, University of 
            British Columbia, Vancouver, BC  Canada V6T 1Z1}}
\newcommand{\UChicago}{{Depts. of Astrophysics and Physics, KICP and EFI, 
            University of Chicago, Chicago, IL 60637}}
\newcommand{\UTexas}{{Texas Cosmology Center and Dept. of Astronomy,  
            Univ. of Texas, Austin, 2511 Speedway, RLM 15.306, Austin, TX 78712}}
\newcommand{\UCLA}{{UCLA Physics \& Astronomy, PO Box 951547, 
            Los Angeles, CA 90095--1547}}
\newcommand{\MPA}{{Max-Planck-Institut f\"{u}r Astrophysik, Karl-Schwarzschild Str. 1, 
            85741 Garching, Germany}}
\newcommand{\IPMU}{{Kavli Institute for the Physics and Mathematics of the Universe, 
            Todai Institutes for Advanced Study, the University of Tokyo, Kashiwa, 
            Japan 277-8583 (Kavli IPMU, WPI)}}
\newcommand{\refeqn}[1]{(\ref{#1})}
\newcommand{\reffig}[1]{Figure~\ref{#1}}
\newcommand{\reftbl}[1]{Table~\ref{#1}}

\newcommand{\var}{{\rm var}}
\newcommand{\ukelvin}{\mbox{$\mu$\rm K}}
\newcommand{\ddeg}         {\mbox{${\rlap.}^\circ$}}
\newcommand{\mrad} {\mbox{\rm mrad} }

\newcommand{\lcdm}{\ensuremath{\Lambda\mathrm{CDM}}}

\renewcommand{\ell}{\ensuremath{l}}
\newcommand{\be}{\begin{equation}}
\newcommand{\ee}{\end{equation}}
\newcommand{\beq}{\begin{equation}}
\newcommand{\eeq}{\end{equation}}
\newcommand{\GHz}{GHz}
\newcommand{\gt}{>}
\newcommand{\degrees}{\ensuremath{^\circ}}

\def\ba{\begin{eqnarray}}
\def\ea{\end{eqnarray}}
\def\nn{\nonumber}
\newcommand{\barr}{\begin{array}}
\newcommand{\earr}{\end{array}}

\def\Npix{N_{\rm pix}}

\def\hC{{\widehat C}}

\def\hE{{\widehat{\mathcal E}}}

\def\Cov{\mbox{Cov}}
\def\Var{\mbox{Var}}
\def\ellmax{\ell_{\rm max}}

\def\N{{\mathcal N}}
\newcommand\lsim{\mathrel{\rlap{\lower4pt\hbox{\hskip1pt$\sim$}}
        \raise1pt\hbox{$<$}}}
\newcommand\gsim{\mathrel{\rlap{\lower4pt\hbox{\hskip1pt$\sim$}}
        \raise1pt\hbox{$>$}}}

\def\Tr{{\rm Tr}}

\def\fnlloc{f_{NL}^{\rm loc}}
\def\fnleq{f_{NL}^{\rm eq}} 
\def\fnlorth{f_{NL}^{\rm orth}}
\def\Mpl{M_{\rm Pl}}

\begin{document}
\title{Nine-Year Wilkinson Microwave Anisotropy Probe (\WMAP) Observations:
  Final Maps and Results}

\author{
{{\Bennett}}\altaffilmark{1}, 
{{\Larson}}\altaffilmark{1}, 
{{\Weiland}}\altaffilmark{1}, 
{{\Jarosik}}\altaffilmark{2},  
{{\Hinshaw}}\altaffilmark{3},
{{\Odegard}}\altaffilmark{4}, 
{{\Smith}}\altaffilmark{5,6}, 
{{\Hill}}\altaffilmark{4}, 
{{\Gold}}\altaffilmark{7},  
{{\Halpern}}\altaffilmark{3}, 
{{\Komatsu}}\altaffilmark{8,9,10}, 
{{\Nolta}}\altaffilmark{11}, 
{{\Page}}\altaffilmark{2}, 
{{\Spergel}}\altaffilmark{6,9},  
{{\Wollack}}\altaffilmark{12}, 
{{\Dunkley}}\altaffilmark{13}, 
{{\Kogut}}\altaffilmark{12}, 
{{\Limon}}\altaffilmark{14}, 
{{\Meyer}}\altaffilmark{15}, 
{{\Tucker}}\altaffilmark{16}, 
{{\Wright}}\altaffilmark{17}
}

\altaffiltext{1}{{\Hopkins}}
\altaffiltext{2}{{\PrincetonPhysics}}
\altaffiltext{3}{{\UBC}}
\altaffiltext{4}{{\Adnet}}
\altaffiltext{5}{{\Perimeter}}
\altaffiltext{6}{{\PrincetonAstro}}
\altaffiltext{7}{{\Minn}}
\altaffiltext{8}{{\MPA}}
\altaffiltext{9}{{\IPMU}}
\altaffiltext{10}{{\UTexas}}
\altaffiltext{11}{{\Cita}}
\altaffiltext{12}{{\Goddard}}
\altaffiltext{13}{{\Oxford}}
\altaffiltext{14}{{\Columbia}}
\altaffiltext{15}{{\UChicago}}
\altaffiltext{16}{{\Brown}}
\altaffiltext{17}{{\UCLA}}
\email{cbennett@jhu.edu}
\begin{abstract}
We present the final nine-year maps and basic results from the {\it Wilkinson Microwave Anisotropy Probe} (\WMAP) mission. The full nine-year analysis of the time-ordered data provides updated characterizations and calibrations of the experiment. We also provide new nine-year full sky  temperature maps that were processed to reduce the asymmetry of the effective beams. Temperature and polarization sky maps are examined to separate cosmic microwave background (CMB) anisotropy from foreground emission, and both types of signals are analyzed in detail. We provide new point source catalogs as well as new diffuse and point source foreground masks. An updated template-removal process is used for cosmological analysis; new foreground fits are performed, and new foreground-reduced CMB maps are presented. We now implement an optimal $C^{-1}$ weighting to compute the temperature angular power spectrum.  

The \wmap\ mission has resulted in a highly constrained $\Lambda$CDM cosmological model with precise and accurate parameters in agreement with a host of other cosmological measurements.  

When \wmap\ data are combined with finer scale CMB, baryon acoustic oscillation, and Hubble constant measurements, we find that Big Bang nucleosynthesis is well supported and there is no compelling evidence for a non-standard number of neutrino species (\ensuremath{N_{\rm eff} = 3.84\pm 0.40}).  The model fit also implies that the age of the universe is \ensuremath{t_0 = 13.772\pm 0.059\ \mbox{Gyr}}, and the fit Hubble constant is \ensuremath{H_0 = 69.32\pm 0.80}~km~s$^{-1}$~Mpc$^{-1}$.  Inflation is also supported: the fluctuations are adiabatic, with Gaussian random phases; the detection of a deviation of the scalar spectral index from unity, reported earlier by the \wmap\ team, now has high statistical significance (\ensuremath{n_s = 0.9608\pm 0.0080}); and the universe is close to flat/Euclidean (\ensuremath{\Omega_k = -0.0027^{+ 0.0039}_{- 0.0038}}).  

Overall, the \WMAP\ mission has resulted in a reduction of the cosmological parameter volume by a factor of 68,000 for the standard six-parameter $\Lambda$CDM model, based on CMB data alone.  For a model including tensors, the allowed seven-parameter volume has been reduced by a factor 117,000. Other cosmological observations are in accord with the CMB predictions, and the combined data reduces the cosmological parameter volume even further.  With no significant anomalies and an adequate goodness-of-fit, the inflationary flat $\Lambda$CDM model and its precise and accurate parameters rooted in \wmap\ data stands as the standard model of cosmology.
\end{abstract}


\section{Introduction}

Since its discovery in 1965, the cosmic microwave background (CMB) has played a central role in cosmology. The discovery of the CMB \citep{penzias/wilson:1965} confirmed a major prediction of the big bang theory and was difficult to reconcile with the steady state theory.  The precision measurement of the CMB spectrum by NASA's \textsl{Cosmic Background Explorer} (\textsl{COBE}) mission \citep{mather/etal:1990,mather/etal:1994} confirmed the predicted CMB blackbody spectrum, which results from thermal equilibrium between matter and radiation in the hot, dense early universe. The \textsl{COBE} detection of CMB anisotropy \citep{smoot/etal:1992, bennett/etal:1992b, kogut/etal:1992, wright/etal:1992} established the amplitude of the primordial scalar fluctuations and supported the case for the gravitational evolution of structure in the universe from primordial fluctuations. While \textsl{COBE} mapped the full sky anisotropy on angular scales $> 7^\circ$, greater than the horizon size at decoupling, \wmap\ mapped the full sky CMB anisotropy on both superhorizon and subhorizon angular scales. \wmap\ provided independent replication and confirmation of the COBE maps on angular scales $> 7^\circ$ as well as the determination of precision cosmological parameters from fits to the well-established physics of the observed sub-horizon acoustic oscillations.

This paper together with its companion paper on cosmological parameter determination  \citep{hinshaw/etal:prep} mark the nine-year and final official data release of the \textit{Wilkinson Microwave Anisotropy Probe} (\WMAP) mission. \WMAP\ was designed to make full sky maps of the CMB in five frequency bands straddling the spectral region where the CMB-to-foreground ratio is near its maximum.   

The overall \WMAP\ mission design was described by \cite{bennett/etal:2003}. The optical design was described by \cite{page/etal:2003} with the feeds and pre-flight beam patterns described by \cite{barnes/etal:2002}. The radiometer design and characterization was presented by \cite{jarosik/etal:2003}.

The \WMAP\ Science Team previously issued four major data releases, each with an accompanying set of publications.  The first-year results included a presentation of the full sky maps and basic results \citep{bennett/etal:2003b}, on-orbit radiometer characteristics \citep{jarosik/etal:2003b}, beam profiles and window functions \citep{page/etal:2003b}, Galactic emission contamination in the far-sidelobes of the beams \citep{barnes/etal:2003}, a description of data processing and systematic measurement errors \citep{hinshaw/etal:2003b}, an assessment of foreground emission \citep{bennett/etal:2003c}, tests of CMB Gaussianity \citep{komatsu/etal:2003}, the angular power spectrum \citep{hinshaw/etal:2003}, the temperature-polarization correlation \citep{kogut/etal:2003}, 
cosmological parameters \citep{spergel/etal:2003}, parameter estimation methodology \cite{verde/etal:2003}, implications for inflation \citep{peiris/etal:2003}, and an interpretation of the temperature-temperature and temperature-polarization cross-power spectrum peaks \citep{page/etal:2003c}.

The three-year \wmap\ results included full use of the polarization data and improvements to temperature data analysis.  The beam profile analysis, data processing changes, radiometer characterization, and systematic error limits were presented in \cite{jarosik/etal:2007}. An analysis of the temperature data carried through to the angular power spectrum was described by \cite{hinshaw/etal:2007}, and the corresponding polarization analysis was presented by \cite{page/etal:2007}. An analysis of the  polarization of the foregrounds was presented by \cite{kogut/etal:2007}. The cosmological implications of the three-year results were summarized by \cite{spergel/etal:2007}.

The five-year \wmap\ results included updates on data processing, sky maps, and the basic results \citep{hinshaw/etal:2009}, and updates on the beam maps and window functions \citep{hill/etal:2009}. The five-year results also included improvements to characterizing the Galactic foreground emission \citep{gold/etal:2009} and the point source catalog \cite{wright/etal:2009}. The angular power spectra \citep{nolta/etal:2009}, likelihoods and parameter estimates \citep{dunkley/etal:2009}, a discussion of the cosmological interpretation of these data \citep{komatsu/etal:2009}, and a Bayesian estimation of the CMB polarization maps \citep{dunkley/etal:2009} completed the five-year results.

The seven-year \WMAP\ results comprised sky maps, systematic errors, and basic results \citep{jarosik/etal:2011}, observations of planets and celestial calibration sources \citep{weiland/etal:2011}, Galactic foreground emission \citep{gold/etal:2011}, angular power spectra and cosmological parameters based only on \WMAP\ data \citep{larson/etal:2011}, cosmological interpretations based on a wider set of cosmological data \citep{komatsu/etal:2011}, and a discussion of the goodness of fit of the $\Lambda$CDM model and potential anomalies \citep{bennett/etal:2011}.

All of the \WMAP\ data releases have been accompanied by an up-to-date Explanatory Supplement, including this final nine-year release \citep{greason/etal:prep}. All \wmap\ data are public along with a large number of associated data products; they are made available by the Legacy Archive for Microwave Background Data Analysis (LAMBDA)\footnote{http://lambda.gsfc.nasa.gov/}.

Each \WMAP\ release improved cosmological constraints through three types of advances: (1) the addition of \WMAP\ data from extended observations; (2) improvements in the analysis of all of the \WMAP\ data included in the release, including more optimal analysis approaches and the use of additional seasons of data to arrive at improved experiment models (e.g., by trending); and (3) improvements in non-\WMAP\ cosmological measurements that are combined into the \WMAP\ team's combined likelihood analysis.

This paper is organized as follows.  The data processing changes from previous analyses are described in Section \ref{sec:processing}.  Beam patterns and window functions are discussed in Section \ref{sec:beamswindows}.  Temperature and polarization sky maps are presented in Section \ref{sec:mapmaking}. In Section \ref{sec:foregrounds} updated masks and an updated point source catalog are presented in addition to several different approaches to diffuse foreground evaluation, which are compared.  Angular power spectra are given in Section \ref{sec:powerspectra}.  An analysis of the model goodness-of-fit and a discussion of anomalies are in Section \ref{sec:goodnessoffit}. Cosmological implications are then presented in Section \ref{sec:cosmology}. Conclusions are given in Section \ref{sec:conclusions}.  The accompanying paper \citep{hinshaw/etal:prep} presents an in-depth analysis of cosmological parameter solutions from various combinations of data and models and offers cosmological conclusions.  
  
\section{Data Processing: Overview and Updates}  \label{sec:processing}

In this section we summarize changes in the \wmap\ data processing since the previous (seven-year) data release. 

\subsection{Time-Ordered Data}

\subsubsection{Data Archive Definition}

The full nine-year \wmap\ archive of nominal survey data covers 00:00:00 UT 2001 August 10 (day number 222) to 00:00:00 UT 2010 August 10 (day number 222).  
Individual year demarcations begin at 00:00:00 UT on day number 222 of a year and end at 23:59:59 UT on
day 221 of the following year.  In addition to processing improvements, the \wmap\ nine-year release
includes new data accumulated during mission years 8 and 9.  Flight operations during those final two years 
included five scheduled station-keeping maneuvers, a lunar shadow passage, and
special commanding procedures invoked within the last mission year to accommodate a compromised battery 
and transmitter.  
Overall, \wmap\ achieved a total mission observing efficiency of roughly 98.4\%.  The bulk of data excluded
from science analysis use are dominated by time intervals that do not exhibit sufficient thermal stability.

\subsubsection{Battery-Driven Thermal Effects}

The \wmap\ solar arrays were exposed to constant sunlight so the battery was trickle charged for almost a decade. This activated an internal battery design imperfection and caused battery voltage fluctuations in the final months of the mission \citep{greason/etal:prep}. The resulting thermal variations were beyond what had been experienced earlier in the mission. A detailed analysis of time-ordered data with sky signal subtracted showed no detectable dependence on thermal variations associated with battery events, and thus preservation of data was preferred to excision. Out of an abundance of caution, time sequences that contained some of the more egregious temperature excursions were flagged as suspect and omitted from use in the nine-year data processing even though there was no specific evidence of adverse effects.

\subsubsection{Pointing}

For each observation, sky pointings of individual \wmap\ feed horns are computed using boresight
vectors in spacecraft body coordinates coupled with the spacecraft
attitude solution provided by on-board star trackers.  After the first mission year,
it was discovered that
the apparent attitude computed by the trackers includes small errors induced by thermal 
flexure of the tracker mounting structure, as described by \citet{jarosik/etal:2007}. 
The amplitude of the flexure is time-dependent and driven by
spacecraft temperature gradients. The spacecraft temperature responds both to solar heating 
and internal power dissipation, and is monitored by thermistors mounted at different locations 
on the spacecraft \citep{greason/etal:prep}.

Telemetered spacecraft quaternions from the star trackers are corrected for this thermal effect at the
very beginning of ground processing, when the raw science archive is created.
Originally, we adopted a simple linear model, assuming a fixed angular rate of elevation change in units
of arcsec per unit temperature change.
As the mission progressed and additional data was used to improve the accumulated 
thermal profile history, the model has evolved to include angular
corrections both in elevation (the dominant term) and azimuth. The nine-year quaternion correction model 
updates the rate coefficients in both azimuth and elevation, and uses readings from two separate thermistors 
to characterize the spacecraft temperature gradients.  A more detailed description is provided by 
\citet{greason/etal:prep}.  The residual pointing error
after applying of the correction algorithm is computed using observations of Jupiter and Saturn. The upper limit of the estimated error is $10\arcsec$.

Beam boresight vectors have been updated based on the full nine-year archive.  The largest difference
between the seven-year and nine-year line-of-sight vectors is $3\arcsec$. 
Both the calibrated and uncalibrated \wmap\ archive data products include documentation of these line-of-sight vectors.

\subsubsection{Calibration}

Calibration of time-ordered data (TOD) from each \wmap\ radiometer
channel requires the derivation of time-dependent gains (responsivity, in units of counts~mK$^{-1}$) 
and baselines (in units of counts) that are used to convert raw differential data into temperature 
units. Algorithmic details and underlying concepts are set forth in \citet{hinshaw/etal:2007}.
\citet{jarosik/etal:2011} outline the calibration process as consisting of two general steps.
The first step determines baselines and preliminary gains on an hourly or daily basis via an iterative 
process that combines a sky-map estimation with a calibration solution that updates with each iteration.
Baselines and gains are computed by fitting sky-subtracted TOD to the dipole anisotropy induced 
by the motion of the \wmap\ spacecraft with respect to the CMB rest frame.  The second calibration step determines absolute gain 
and fits a parameterized gain model to the dipole gains derived in the first step.  

The form of the parameterized gain model is based on a physical understanding of radiometer performance,
and uses telemetered measures of instrument temperatures and the radio frequency (RF) biases.
The model provides a smooth characterization of the responsivity with time and allows higher
time resolution than provided by the dipole-fit gains. 
For the nine-year analysis, we augment the gain model by adding
a time-dependent linear trend term, $m{\Delta}t + c$, to the parameterized form presented in
\citet{jarosik/etal:2007}. Here ${\Delta}t$ is an elapsed mission time in days, and $m$, $c$ are additional
fit parameters. Physically, the linear trend can be thought of as a radiometer aging term.  Without the
addition of this term, model fits to the nine-year dipole gain measurements exhibited 
small systematic deviations from zero-mean residuals
for nine of the 40 \wmap\ channels. The four Ka1 channels were most affected; 
the inclusion of the gain model aging term prevents an induced total gain error of about 0.1\% in this band. 
Of the 40 \wmap\ radiometer channels, W323 alone has shown poor convergence in the iterative
procedure that determines dipole-fit gains.  Upon investigation we found that this problem is
peculiar to the iterative algorithm and not the data itself. The W323 calibration has not been
substantially affected in previous releases, but for the nine-year analysis the diverging mode was identified and we disallowed it in the gain model fit.
 
We continue to conservatively estimate an absolute calibration uncertainty of 0.2\% (1-sigma), based on end-to-end gain recovery 
simulations.  The overall change in calibration for the nine-year processing relative to the seven-year release
is -0.031,+0.048,-0.005,+0.041 and +0.025 \% for K-, Ka-, Q-, V- and W-bands
respectively;  a positive change indicates that features in the nine-year maps are slightly larger than those in
the equivalent seven-year maps (i.e., a slight decrease in nine-year absolute gain 
compared to seven-year).

\subsubsection{Transmission Imbalance Factors}

The transmission efficiencies of sky signals through the A-side and B-side optical systems into each \wmap\ radiometer
differ slightly from one another.  This deviation from ideal behavior is characterized in map-making 
and data analysis through the use of time-independent transmission imbalance factors.  
The method by which these factors are
determined from the \wmap\ data was described by \citet{jarosik/etal:2007}.  The determination
improves with additional data.
These factors have been updated for the nine-year analysis and are presented
in Table~\ref{tab:x_im}.  The nine-year values compare well against the previously
published seven-year values \citep{jarosik/etal:2011} within the quoted uncertainties.

\begin{deluxetable}{crc|crc}
\tablecaption{ Nine-year Fractional Transmission Imbalance \label{tab:x_im}}
\tablehead{ \colhead{Radiometer} & \colhead{$x_{\rm im}$} & \colhead{Uncertainty} & \colhead{Radiometer} & 
\colhead{$x_{\rm im}$} & \colhead{Uncertainty}}
\startdata 
       K11 &     -0.00067 &      0.00017 &
       K12 &      0.00536 &      0.00014 \\
      Ka11 &      0.00353 &      0.00014 &
      Ka12 &      0.00154 &      0.00008 \\
       Q11 &     -0.00013 &	 0.00046 &
       Q12 &      0.00414 &	 0.00025 \\
       Q21 &      0.00756 &	 0.00052 &
       Q22 &      0.00986 &	 0.00115 \\
       V11 &      0.00053 &	 0.00020 &
       V12 &      0.00250 &	 0.00057 \\
       V21 &      0.00352 &	 0.00033 &
       V22 &      0.00245 &	 0.00098 \\
       W11 &      0.01134 &	 0.00199 &
       W12 &      0.00173 &	 0.00036 \\
       W21 &      0.01017 &	 0.00216 &
       W22 &      0.01142 &	 0.00121 \\ 
       W31 &     -0.00122 &	 0.00062 &
       W32 &      0.00463 &	 0.00041 \\ 
       W41 &      0.02311 &	 0.00380 &
       W42 &      0.02054 &	 0.00202 \\
\enddata 
\tablecomments{The fractional transmission imbalance, $x_{\rm im}$, and its
uncertainty is determined from 
the nine-year observational data.  The fractional transmission imbalance is
defined as $x_{im} = (\epsilon_A - \epsilon_B)/(\epsilon_A + \epsilon_B)$, where
$\epsilon_{A}$ and $\epsilon_{B}$ are the input transmission coefficients for the A- and B-side
optics \citep{jarosik/etal:2003b}.  For an ideal differential radiometer, $x_{\rm im}=0$.}
\end{deluxetable}

\subsection{Map-Making}

\subsubsection{Standard Map-Making}

The standard \wmap\ map-making procedure is unchanged from the previous release and the resulting maps are used for the core cosmological analyses. Progress has been made on the algorithm for estimating the noise properties of the maps. The Stokes I noise levels ($\sigma_0$) are now more self-consistent between maps at angular resolution r9 and r10\footnote{The map resolution levels refer to the HEALPix pixelization scheme \citep{gorski/etal:2005} where r4, r5, r9, and r10 refer to $N_{\rm side}$ values of 16, 32, 512, and 1024, respectively.} than they had been previously.  Another difference from previous analyses is that this procedure now determines the noise in the polarized maps from the Stokes Q and U year-to-year differences while including a spurious (``S") map term, and a mean monopole is subtracted from each S map, as is done separately for Stokes I in the temperature map analysis. A detailed discussion is in Section \ref{sec:standard_mapmaking}.

Data are masked in the map-making process when one feed observes bright foregrounds (e.g., in the Galactic plane) while the corresponding differencing feed observes a far fainter sky. This masking prevents the contamination of faint pixels. Previous \wmap\ data analysis efforts used a single processing mask, based on the K-band temperature maps, to define which pixel-pairs to mask for all of the frequency bands. In the current processing we have changed to masking based on the brightness in each individual band. 

\subsubsection{Beam Pattern Determination}
The standard maps are used to subtract the background from Jupiter observations to create beam maps, as has been done in previous processing. We correct three seasons of Jupiter maps in the latter part of the mission for the proximity of Uranus and Neptune to Jupiter.  Two-dimensional profiles from the newly updated beam map data are now also used as inputs for the new beam-symmetrized map-making procedure, described below.

\subsubsection{Beam-Symmetrized Map-Making}

In addition to the standard map-making, a new map-making procedure, described in Section  \ref{sec:beamsymmaps}, effectively deconvolves the beam sidelobes to produce maps with the true sky signal convolved by symmetrized beams. As a result of this new procedure, the previously reported map power asymmetry, which we speculated was due to the asymmetric beams and not cosmology \citep{bennett/etal:2011} has indeed been mitigated in the new beam-symmetrized maps. 

In this paper we use the beam-symmetrized maps for diffuse foreground analysis
(Section~5.3), but not for estimating the angular power spectrum and
cosmological parameters.  This is because the deconvolution process introduces
correlations in the pixel noise on the beam scale and it is impractical to track
these correlations at the full pixel resolution.  Diffuse foreground analyses,
on the other hand, used maps smoothed to a $1^\circ$ scale.  Appendix B of
\citet{hinshaw/etal:2007} demonstrated that the cosmological power spectrum,
$C_\ell$, is insensitive to beam asymmetry at \WMAP's sensitivity level.  (It is
the 4-point bipolar power spectrum, not the 2-point angular power spectrum, that
is sensitive to beam asymmetry.)  Use of the beam symmetrized maps for
high-$\ell$ angular power spectrum estimation would invoke the need for high
resolution noise covariance matrices, along with far greater computational and
storage demands than are now feasible.  Given that dense r9 noise covariance
matrices are computationally undesirable and the cosmological power spectrum is
insensitive to beam asymmetry, we do not use beam-symmetrized maps for
cosmology.

\section{Beam Maps and Window Functions} \label{sec:beamswindows}

The \wmap\ full beams are considered as a combination of main beams and sidelobes.
These are treated separately in the data processing.
The sidelobe beam patterns were determined from early 
mission observations of the moon together with pre-flight ground-based measurements, as
described in \citet{barnes/etal:2003}. Potential contamination from sidelobe 
pickup was computed and removed from the calibrated time-ordered data prior to
map-making \citep{hinshaw/etal:2009}. In this section, we address the
main-beam response; treatment of the sidelobes remains unchanged from
the seven-year release.

\wmap\ beams are measured using observations of the planet Jupiter that occur
during the normal course of full-sky observing.  Two Jupiter observing
seasons of $\sim 50$ days each occur every $395-400$
days.  In the nine-year \wmap\ mission, a total of 17 seasons of Jupiter data were
obtained.  Time intervals for the four observing seasons occurring during
the last two mission years are presented in Table~\ref{tab:jup_seasons}; those for
seasons 1 - 13 are presented in Table~1 of \citet{weiland/etal:2011}. 

\begin{deluxetable}{cllccc}
  \tabletypesize{\scriptsize}
  \tablewidth{0pt}
  \tablecaption{ \wmap\ Jupiter Observing Seasons (2008-2010) \label{tab:jup_seasons}}
  \tablehead{
    \colhead{Season\tablenotemark{a}} &
    \colhead{Begin}  &
    \colhead{End}    &
    \colhead{Nearby Planet\tablenotemark{b}} &
    \colhead{Projected Separation\tablenotemark{c}}  &
    \colhead{\% excess\tablenotemark{d}} \\
    \colhead{}  &
    \colhead{}  &
    \colhead{}  &
    \colhead{}  &
    \colhead{}  
    }
\startdata
  14  & 2008 Aug 21  & 2008 Oct 06  & \nodata   &  \nodata                & \nodata       \\
  15  & 2009 May 17  & 2009 Jul 03  & Neptune &  0.4\arcdeg - 2.4\arcdeg  & 0.4 - 0.2   \\
  16  & 2009 Sep 26  & 2009 Nov 10  & Neptune &  3.8\arcdeg - 6.8\arcdeg  & 0.08 - 0.0  \\
  17  & 2010 Jun 24  & 2010 Aug 10  & Uranus  &  0.5\arcdeg - 3.1\arcdeg  & 0.9 - 0.4   \\
\enddata
  \tablenotetext{a}{An observing season is defined as a contiguous time interval during which an object is in the \wmap\ viewing swath. Observing seasons 1-13 are listed in \citet{weiland/etal:2011}}
  \tablenotetext{b}{Jupiter sky coordinates are in proximity to those of the planet listed.}
  \tablenotetext{c}{Seasonal range of projected separations between Jupiter's position and that of the other planet.}
  \tablenotetext{d}{Estimated excess integrated beam response, in \%, that would have been contributed to the Jupiter beam 
  by contaminating planet, if no correction had been applied.
  Provided as a range; the first number is for K-band, the last is for W-band; other frequencies are between these two values.}
\end{deluxetable}

The beams enter into CMB data analysis primarily through the 10 beam transfer
functions, $b_\ell$, which give the beam response in spherical harmonic space
for each differencing assembly (DA).  Beam response on the sphere is measured in a coordinate system
fixed to the \wmap\ spacecraft \citep{barnes/etal:2003}, 
and a computation of several steps is required to
generate $b_\ell$.  The nine-year beam analysis follows the process described
previously by \cite{hill/etal:2009} and \cite{jarosik/etal:2011}.

For a given DA, Jupiter is observed with only one feed at a time, so 
initially the A and B side beams are mapped separately.  After correction for
the static sky background, the data are coadded in a planar grid surrounding
each of the 20 A- and B-side boresights.  A physical optics code\footnote{DADRA: Y. Rahmat-Sahmi, W. Imbriale, \& V. Galindo-Israel 1995, YRS
Assocates, rahmat@ee.ucla.edu} is
used to compute beam models, which are optimized by $\chi^2$ minimization using
a modified conjugate gradient algorithm.  Two minor refinements were added to this
process for the nine-year analysis: first,  
a more rigorous treatment of the removal of the Galactic signal was adopted by
including the common-mode loss imbalance term; in practice this is a small effect since
strong Galactic signals are masked from use in the beam archive. Second, computation
of the interpolated beam model utilized an increase in secondary mirror samplings from
$200\times200$ to $235\times235$; this produced a smoother far-field tail for the W2 and W3 DAs.

Standard processing nominally rejects from analysis those Jupiter observations whose sky 
positions lie within a $7\arcdeg$ radius of other planets.  
Table~\ref{tab:jup_seasons} shows the seasonal range of projected sky separations between
Jupiter and planets that lie within the exclusion radius for the last three observing seasons.
Based on projected proximity to Uranus or Neptune, application of nominal exclusion criteria 
would have excised these three Jupiter seasons from use.  To preserve the ability
to characterize the beam response during the latter part of the mission, we chose instead to
correct the last three seasons of Jupiter data for excess contributions from Uranus and Neptune.
Excess response from these planets is computed and removed from each Jupiter observation assuming that the
response to Uranus and Neptune may be modeled using a symmetrized beam 
template with peak response inferred from \citet{weiland/etal:2011} .
An estimate of the magnitude of the correction is provided in 
the last column of Table~\ref{tab:jup_seasons},
provided as a percentage contribution in excess of the uncontaminated integrated Jupiter beam response
for each season.
Observations which occur when Jupiter's sky coordinates lie within the confines of a spatial ``Galaxy mask'' are also 
excluded from use in the analysis \citep{weiland/etal:2011}.
During observing season 14, the Galactic latitude of Jupiter is $\sim-18\arcdeg$, close enough to the Galactic
plane that some observations are rejected based on the masking criterion.  Masking is frequency
dependent: roughly 30\% of season 14 K-band observations are excluded, decreasing to 17\% for Ka, 13\% for Q and
less than 0.1\% for V- and W-bands.

For each DA, the Jupiter data for
sides A and B are combined with the best-fit models in a ``hybrid'' beam map,
which is used to construct the symmetrized radial beam profile, $b(\theta)$.  A
Legendre transform gives $b_\ell$.
The beam hybridization procedure is described in detail by
\cite{hill/etal:2009}.  Essentially, the process edits the Jupiter TOD by
replacing faint, noisy Jupiter samples with noise-free predicted values taken
from the 2-dimensional beam model.  This process is controlled by one parameter
for each DA, the threshold gain, $B_\mathrm{thresh}$:  all observed
beam samples with gain lower than $B_\mathrm{thresh}$ are replaced with their
counterpart model values.  This test is applied to the model samples,
rather than the observed ones, in order to avoid bias from observational noise.  
$B_\mathrm{thresh}$ is optimized statistically for each DA using a Monte Carlo
method, whereby uncertainty belonging to the beam model is traded against the
noise in the observed data points.  The figure of merit to be minimized is the
uncertainty of the resultant solid angle in the hybridized beam.  For this
purpose, the error in the model is assumed to be a $100\%$ uncertainty in
the overall scaling of the low-sensitivity ``tails,'' which is the only portion
of a beam model that is used in the hybrid.  For the nine-year data, $B_\mathrm{thresh}$
is set $1$ dB lower than for the seven-year data; values are 2, 3, 5, 6 and 9 dBi for
K- through W-bands, respectively.

\cite{hill/etal:2009} give the procedure for transforming the hybrid beam
profiles into beam transfer functions.  This computation also yields main beam
solid angles and estimates of the temperature of the Jupiter disk.
Beam-related quantities are summarized in Table~\ref{tab:beam_quantities}.
The last three columns list quantities that are valid for a point source with
spectral index $\alpha=-0.1$ (flux $F_\nu \propto \nu^\alpha$), typical of sources in the \wmap\ point source catalog.
They were determined as described in \cite{jarosik/etal:2011}, except a small
correction for bandpass drift was included in the calculation of
effective frequency for K-, Ka-, Q-, and V-bands as described in Appendix \ref{sec:bandcenter}.

The nine-year and seven-year $b_\ell$ are consistent with each other,
although the $b_\ell$ for W4 is about $0.6\%$ higher 
in the nine-year analysis than in the seven-year analysis for $\ell>100$, a shift
that is at the edge of the error band.   

The error bands for $b_\ell$ are computed using Monte Carlo simulations of the
beam map hybridization; details of the simulations follow the description provided in
\citet{ hill/etal:2009}.  As Jupiter observations have accumulated over the \wmap\ mission lifetime, the
contribution of the model tails to the hybrid beam has become less important. 
The nine-year hybrid beams are data dominated:  for each of the ten beams, less than $0.25\%$ of the 
integrated hybrid beam response is attributable to the model tails.

\begin{deluxetable}{cccccccc}
  \tabletypesize{\scriptsize}
  \tablewidth{0pt}
  \tablecolumns{5}
  \tablecaption{\wmap\ Nine-year Main Beam Parameters
    \label{tab:beam_quantities}}
  \tablehead{
    \colhead{}&
    \colhead{$\Omega_{\mathrm{9yr}}^S$\tablenotemark{a}}&
    \colhead{$\Delta(\Omega_{\mathrm{9yr}}^S)/\Omega^S$\tablenotemark{b}}&
    \colhead{$\frac{\Omega_{\mathrm{9yr}}^S}{\Omega_{\mathrm{7yr}}^S}-1$\tablenotemark{c}}&
    \colhead{$G_m$\tablenotemark{d}}&
    \colhead{$\nu_{\mathrm{eff}}^{\mathrm{ff}}$\tablenotemark{e}}&
    \colhead{$\Omega_{\mathrm{eff}}^{\mathrm{ff}}$\tablenotemark{f}}&
    \colhead{$\Gamma_{\mathrm{\mathrm{ff}}}$\tablenotemark{g}} \\
    \colhead{DA}&
    \colhead{(sr)}&
    \colhead{(\%)}&
    \colhead{(\%)}&
    \colhead{(dBi)}&
    \colhead{(\GHz)}&
    \colhead{(sr)}&
    \colhead{($\mu$K Jy$^{-1}$)}}
  \startdata
  \cutinhead{For 10 Maps}
  K1&  $2.469\times10^{-4}$&  0.5&   0.1&  47.07&22.69&$2.522\times10^{-4}$&  250.6 \\
  Ka1& $1.442\times10^{-4}$&  0.4&   0.0&  49.40&32.94&$1.465\times10^{-4}$&  204.9 \\
  Q1&  $8.815\times10^{-5}$&  0.5&  -0.2&  51.54&40.72&$8.934\times10^{-5}$&  219.7 \\
  Q2&  $9.113\times10^{-5}$&  0.5&  -0.1&  51.40&40.51&$9.234\times10^{-5}$&  214.8 \\
  V1&  $4.164\times10^{-5}$&  0.4&  -0.1&  54.80&60.09&$4.226\times10^{-5}$&  213.3 \\
  V2&  $4.236\times10^{-5}$&  0.4&   0.1&  54.72&60.96&$4.283\times10^{-5}$&  204.5 \\
  W1&  $2.038\times10^{-5}$&  0.4&  -0.2&  57.90&92.87&$2.040\times10^{-5}$&  185.0 \\
  W2&  $2.204\times10^{-5}$&  0.4&   0.2&  57.56&93.43&$2.203\times10^{-5}$&  169.2 \\
  W3&  $2.135\times10^{-5}$&  0.5&  -0.2&  57.70&92.44&$2.135\times10^{-5}$&  178.4 \\
  W4&  $1.994\times10^{-5}$&  0.5&  -0.6&  57.99&93.22&$1.997\times10^{-5}$&  187.6 \\
  \cutinhead{For 5 Maps}
  K&  $2.469\times10^{-4}$&   0.5&   0.1&  47.07&22.69&$2.522\times10^{-4}$&  250.6 \\
  Ka&  $1.442\times10^{-4}$&  0.4&  0.0&  49.40&32.94&$1.465\times10^{-4}$&  204.9 \\
  Q&  $8.964\times10^{-5}$&   0.5&  -0.2&  51.47&40.62&$9.084\times10^{-5}$&  217.2 \\
  V&  $4.200\times10^{-5}$&   0.4&   0.0&  54.76&60.52&$4.255\times10^{-5}$&  208.9 \\
  W&  $2.093\times10^{-5}$&   0.5&  -0.2&  57.78&92.99&$2.094\times10^{-5}$&  180.0 \\
  \enddata
  \tablenotetext{a}{Solid angle in azimuthally symmetrized beam.}
  \tablenotetext{b}{Relative error in $\Omega^S$.}
  \tablenotetext{c}{Relative change in $\Omega^S$ between nine-year and seven-year analyses.}
  \tablenotetext{d}{Forward gain $=$ maximum of gain relative to isotropic,
    defined as $4\pi/\Omega^S$.  Values of $G_m$ in Table 2 of Hill et al.
    (2009) were taken from the physical optics model, rather than
    computed from the solid angle in the table, and therefore are slightly different.}
  \tablenotetext{e}{The effective center frequency for a point source with flux
    spectral index $\alpha = -0.1$.  The estimated uncertainty, due to uncertainties 
    in the pre-flight passband response measurements, is 0.1\% for all DAs.}
  \tablenotetext{f}{The effective beam solid angle for a point source with flux
    spectral index $\alpha = -0.1$.  The uncertainties are estimated as 
    0.5, 0.4, 0.5, 0.4, and 0.5\% for K-, Ka-, Q-, V-, and W-band DAs, respectively.
    These include contributions from uncertainty in the beam solid angles,
    $\Delta(\Omega_{\mathrm{9yr}}^S)/\Omega^S$ (column 3), and uncertainty in
    the correction of pre-flight forward gain measurements for scattering
    described in \cite{jarosik/etal:2011}.  }
  \tablenotetext{g}{Conversion factor to obtain flux density from the peak \wmap\ antenna
    temperature, for a point source with flux spectral index $\alpha=-0.1$.
    Uncertainties in these factors are estimated as 0.6, 0.4, 0.5, 0.5 and 0.7\%
    for K-, Ka-, Q-, V- and W-band DAs respectively. These include contributions from
    uncertainty in the beam solid angles, $\Delta(\Omega_{\mathrm{9yr}}^S)/\Omega^S$ 
    (column 3), uncertainty in the pre-flight
    passband response measurements, and uncertainty in the correction of
    pre-flight forward gain measurements for scattering described in
    \cite{jarosik/etal:2011}.}
\end{deluxetable}
\section{Map-making}\label{sec:mapmaking}

\subsection{Standard Map Processing}\label{sec:standard_mapmaking}

\subsubsection{Individual Band Processing Masks} \label{sec:proc_mask}
The algorithm used to reconstruct sky maps from differential data masks selected observations to minimize artifacts
associated with regions of high foreground intensity.~\citep{jarosik/etal:2011}. Observations for which one of the 
telescope beams is in a region of high foreground intensity gradients while the other is in a low gradient region
are only applied to the pixel in the high foreground region as the map solutions are generated. This `asymmetric'
masking suppresses map reconstruction artifacts in the low foreground emission regions used for CMB analysis. These
artifacts arise from small variations in the power sampled by the telescope beams for different observations that
fall within the same map pixel. The variations result from a combination of the finite pixel size and beam ellipticity
that both couple to spatial intensity gradients. A processing mask is used to delineate the 
regions of high foreground intensity gradients.
Previous data releases used a common processing mask for all frequency
bands based on the K-band temperature maps, even though the foreground intensities vary greatly by band. The current
release uses different masks for each frequency band and therefore utilizes the data more efficiently.

 Masks for each frequency band are generated  using an algorithm that estimates the magnitude of processing artifacts in each r4 pixel
given the \wmap\ scan pattern, a candidate processing mask and the seven-year map of the sky temperature in that band. The magnitude of artifacts, $\xi$, 
in a resolution r4 pixel, $p_4$, is modeled as proportional to the mean magnitude of the temperature gradients within all the reference pixels used
in the observations contributing to the original pixel, 
\begin{eqnarray} 
\xi(p_4, n) \simeq \frac{\alpha}{N_{tot}(p_4, n)}\left[ \sum_{ p_A(i)=p_4}w_n(p_B(i))|\nabla T(p_B(i))| +  \sum_{ p_B(i)=p_4}w_n(p_A(i))|\nabla T(p_A(i))| \right],\\
N_{tot}(p_4, n) =   \sum_{ p_A(i)=p_4}w_n(p_B(i)) +  \sum_{ p_B(i)=p_4}w_n(p_A(i)).
\end{eqnarray}
Here $p_A(i)$ and $p_B(i)$ are the r4 pixel indices for the A and B side beams for TOD observation $i$, $w_n$ represents a candidate 
processing mask with $n$ pixels masked,  and the sums
are over observations for which the A-side beam and B-side beam point to pixel $p_4$. The proportionality constant $\alpha$ was 
evaluated as the amplitude of the response
for each telescope beam as it was rotated about its axis while viewing a uniform temperature gradient, yielding 
values from $0\ddeg032$ to $0\ddeg087$ for the different beams. The magnitude of the
temperature gradient in each r4 pixel is approximated as the standard deviation of the r9 pixels comprising each r4 pixel
\begin{eqnarray}
  |\nabla T(p_4)| &\simeq& \beta \cdot [\var(p_9 \in p_4) - \sigma^2(p_4)]^{1/2}\label{eqn:grad_eval}, \\
  \sigma^2(p_4) &=&  \frac{\displaystyle \sum_{p_9 \in p_4} \sigma_0^2/N_{obs}(p_9)}{\displaystyle \sum_{p9 \in p4}1},
\end{eqnarray}
where the last term in Equation~\refeqn{eqn:grad_eval} removes the bias introduced by the radiometer noise, $\sigma_0$ is the noise for
 one observation and $N_{obs}(p_9)$ is the number of observations in r9 pixel $p_9$. The constant $\beta \simeq 1.1\ {\rm deg}^{-1}$ for r4 pixels. 

\begin{figure}
\epsscale{0.70}
\plotone{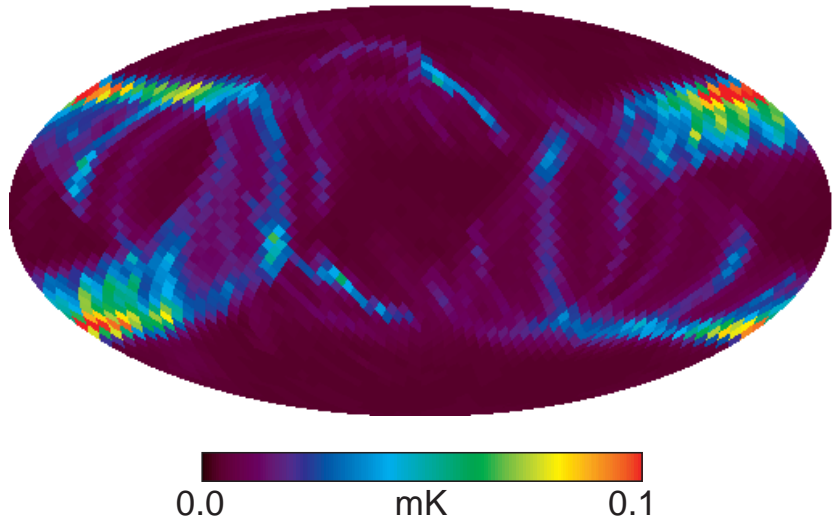}
\caption{
The estimated level of artifacts ($\xi$) that would have occurred in the Ka-band
map if no processing mask had been used.  Band-dependent processing masks were
used and tailored to minimize these artifacts when converting from time-ordered
to sky map data.  This map is in Galactic coordinates and the high intensity
regions arise from observations when one of the beams is near the Galactic
center and the processing mask is not used. (See Figure \ref{fig:masks} to
compare with the analysis sky cuts.) Since bright artifacts originate primarily
from beam crossings of bright Galactic plane regions, the nature of the unmasked
artifact pattern is similar for all DAs.  Although the patterns are similar for
all bands, the highest amplitude artifacts occur in K- and Ka- bands because
these have the brightest foregrounds.  To prevent significant artifacts,
processing masks are constructed for each band by growing the number of pixels
in the mask until $\xi$ is sufficiently reduced.  The estimated mean residual
level of artifacts ($\overline{\xi}$) is given in \reftbl{tab:masks_radii}.  We required
$\overline{\xi}< 5\,\ukelvin$ for all but K-band.  Construction of the K-band
mask is more complex (see text) yet still achieves $\overline{\xi} <
8\,\ukelvin$.
\vspace{1mm} \newline (A color version of this figure is available in the online journal.)}
\label{fig:RCBM}
\end{figure}

\begin{figure}
\epsscale{0.70}
\plotone{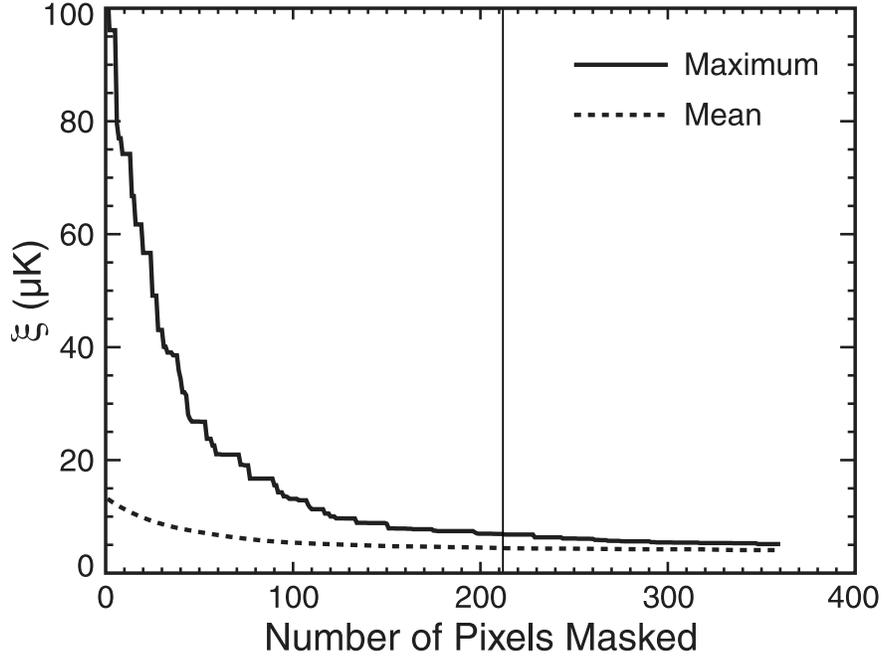}
\caption{Plots of the maximum and mean magnitude of the estimated map artifacts ($\xi$) for Ka-band versus the number of pixels masked by the 
processing mask. The vertical line indicates the adopted mask which is  the smallest mask for which $\max(\xi) < 1.15~\xi^{max}_{sat}$ as described in the text.}
\label{fig:MaxMean}
\end{figure}

Figure \ref{fig:RCBM} shows a map of $\xi(p_4, 0)$ for the Ka1 DA with no pixels masked in the candidate processing mask (n=0). The highest value areas in
this map correspond to regions that are   $\approx 140^\circ$ from the Galactic center corresponding to the spacing between the \wmap\ A-side and B-side 
telescope beams.
Processing masks for each frequency band are generated  iteratively starting from an empty mask, $n = 0$. The r4 pixel added to the 
candidate mask $w_n$ at each step is that which produces the greatest reduction in the mean value of $\xi(p_4,n)$ for the current value 
of n. The value of $\xi$ is then recalculated 
with the updated candidate mask, $w_{n+1}$,  and the process repeated. \reffig{fig:MaxMean} displays how the maximum and 
mean value of $\xi(p_4, n)$ vary as pixels are added to the mask. The mean and maximum values
decrease rapidly as $n$ increases and asymptotes to an approximately constant value for large $n$ . 
The value maximum values in the asymptotic region is calculated as
\begin{eqnarray}
\xi^{max}(n) &=& \max_{p_4} \xi(p_4, n),\\
\xi^{max}_{sat} &=& \overline{ \xi^{max}(n)},\ 180 \ge n \ge  360.
\end{eqnarray}  
These steps are executed for each DA
and masks for the Ka1-, Q-, V-, and W-band DAs are selected by choosing the smallest value of $n$ for which 
$\xi^{max}(n) < 1.15~\xi^{max}_{sat}$. This criterion was selected by requiring that $\overline{\xi} < 5\ukelvin$ for Ka-, Q-, V- and W-bands 
and that the resulting Q-band mask have approximately the same number of excluded pixels as the mask used in earlier data releases.   The mask created in this manner for the Ka1 DA
is the final processing mask. Masks for frequency bands with multiple DAs are formed by merging the individual
DA masks such that if a pixel was masked in either of the DA masks it is masked in the combined mask. 
The K-band processing mask requires special treatment due to the brightness of the foregrounds. Applying the criterion above yields a
very large sky mask that leaves many pixels with few or no observations causing convergence problems in the  conjugate gradient map solution.
The adopted K-band processing mask is the largest $w_n$ formed with K-band inputs for which the sky map solution converges for all years except year 2. 
Year 2 is particularly problematic due to the location of Jupiter. Achieving convergence requires
selection of the $w_{270}$ mask and reduction of the Jupiter exclusion radius to $2\ddeg5$. Even with these special considerations
the size of the processing mask is still substantially larger than used in previous data releases and should result in 
reduced artifacts. Table~\ref{tab:masks_radii} summarizes the mask sizes and planet exclusion radii for the nine-year maps.
\begin{deluxetable}{cccccccc}
\tabletypesize{\small}
\tablecaption{Map Generation Masking Parameters\label{tab:masks_radii}} 
\tablehead{&\colhead{masked pixels}&$\overline{\xi}$&\multicolumn{5}{c}{Planet Exclusion Radii (in $^\circ$)}\\
\colhead{Band} &\colhead{(of 3072 total)} &($\mu$K)& \colhead{Mars} & \colhead{Jupiter} & \colhead{Saturn} & \colhead{Uranus} & \colhead{Neptune}}
\startdata
K(yr $\ne$ 2)    & 312 & 7.12 & 2.0 & 3.0 & 2.0 & 2.0 & 2.0\\
K (yr $=$ 2)         & 270 & 7.59 & 2.0 & 2.5 & 2.0 & 2.0 & 2.0\\
Ka              & 212 & 4.46 & 1.5 & 2.5 & 1.5 & 1.5 & 1.5\\
Q               & 201 & 4.31 & 1.5 & 2.5 & 1.5 & 1.5 & 1.5\\
V               & 125 & 3.78 & 1.5 & 2.2 & 1.5 & 1.5 & 1.5\\
W               & 98  & 3.66 & 1.5 & 2.0 & 1.5 & 1.5 & 1.5
\enddata
\end{deluxetable}
     
\subsubsection{Summary of Standard Map-making} \label{sec:std_proc_summary}

The time-ordered-data (TOD), ${\bf d}$, for a differential radiometer sensitive only to a Stokes I signal may be written as
\begin{equation} \label{eqn:sky_signal_model}
{\bf d} = {\bf Mt} + {\bf n}.
\end{equation}
Here ${\bf M}$ is a sparse $N_{\rm t} \times N_{\rm p}$  matrix that contains information about the scan pattern and
transforms the input sky signal array, ${\bf t}$,  into a sequence of differential observations, {\bf d}.
The number of time-ordered observations is given by $N_{\rm t}$, the number of pixels in array {\bf t} is  $N_{\rm p}$,  and {\bf n} is an $N_{t}$ element array
representing the radiometer noise. For the standard map processing each  row of ${\bf M} $ contains two non-zero elements representing 
the weights given to the input map pixels nearest the A and B-side telescope lines-of-sight (LOS). 
The first step in generating a sky map is evaluation of the ``iteration zero'' map,  
\begin{equation}
{\bf \tilde{t}_0} = {\bf M}^{\rm T}_{\rm am} {\bf N}^{-1}{\bf d},
\end{equation} 
where ${\bf M_{\rm am}^{\rm T}}$ is the 
transpose of a masked version of the observation matrix, and ${\bf N}^{-1}$
is the inverse of the radiometer noise covariance matrix, 
\begin{equation}
  {\bf N}^{-1} = \langle{\bf nn}^{\rm T}\rangle^{-1},
\end{equation}
 with the angle brackets representing an average.
The masking contained in ${\bf M}_{\rm am}^{\rm T}$ prevents contamination of regions 
of the map with low foreground emission that can occur when one of the telescope beams is
 in a region of high foreground emission. (See Section \ref{sec:proc_mask}.) 
The reconstructed sky map, ${\bf \tilde{t}}$, is then calculated by solving 
\begin{equation} \label{eqn:basic_cg_soln}
{\bf \tilde{t}} = ({\bf M}_{\rm am}^{\rm T} {\bf N}^{-1}{\bf M})^{-1}~{\bf \tilde{t}_0}.
\end{equation}
The form of matrix ${\bf M}$ described above ignores the effects of the finite \wmap\ 
beam sizes since each observation is 
modeled using only the value of the input sky signal nearest the LOS direction. The actual radiometric data is an average of the 
input sky signal spatially weighted by the beam response. Each row of ${\bf M}$ should therefore contain additional non-zero elements describing the
signal contribution from the off-axis beam response. If the beam response was the same for the A and B side beams and azimuthally 
symmetric about the LOS, the observation matrix including the off-axis signal contributions, ${\bf M_{\rm s}}$, could be written in the form
\begin{equation} \label{eqn:sym_beam_M}
{\bf M_{\rm s}} = {\bf M C},
\end{equation}
where ${\bf C}$ is an $N_{\rm p} \times N_{\rm p}$ element matrix that performs a convolution by the symmetric beam pattern. 
Substituting ${\bf M_{\rm s}}$ for ${\bf M}$
in Equation~\refeqn{eqn:sky_signal_model} shows that in this limit the sky map reconstructed using Equation~\refeqn{eqn:basic_cg_soln} is the 
input map convolved by the symmetric beam pattern, ${\bf \tilde{t}}_{\rm c} = {\bf C t}$.

\begin{figure}
\epsscale{0.99}
\plotone{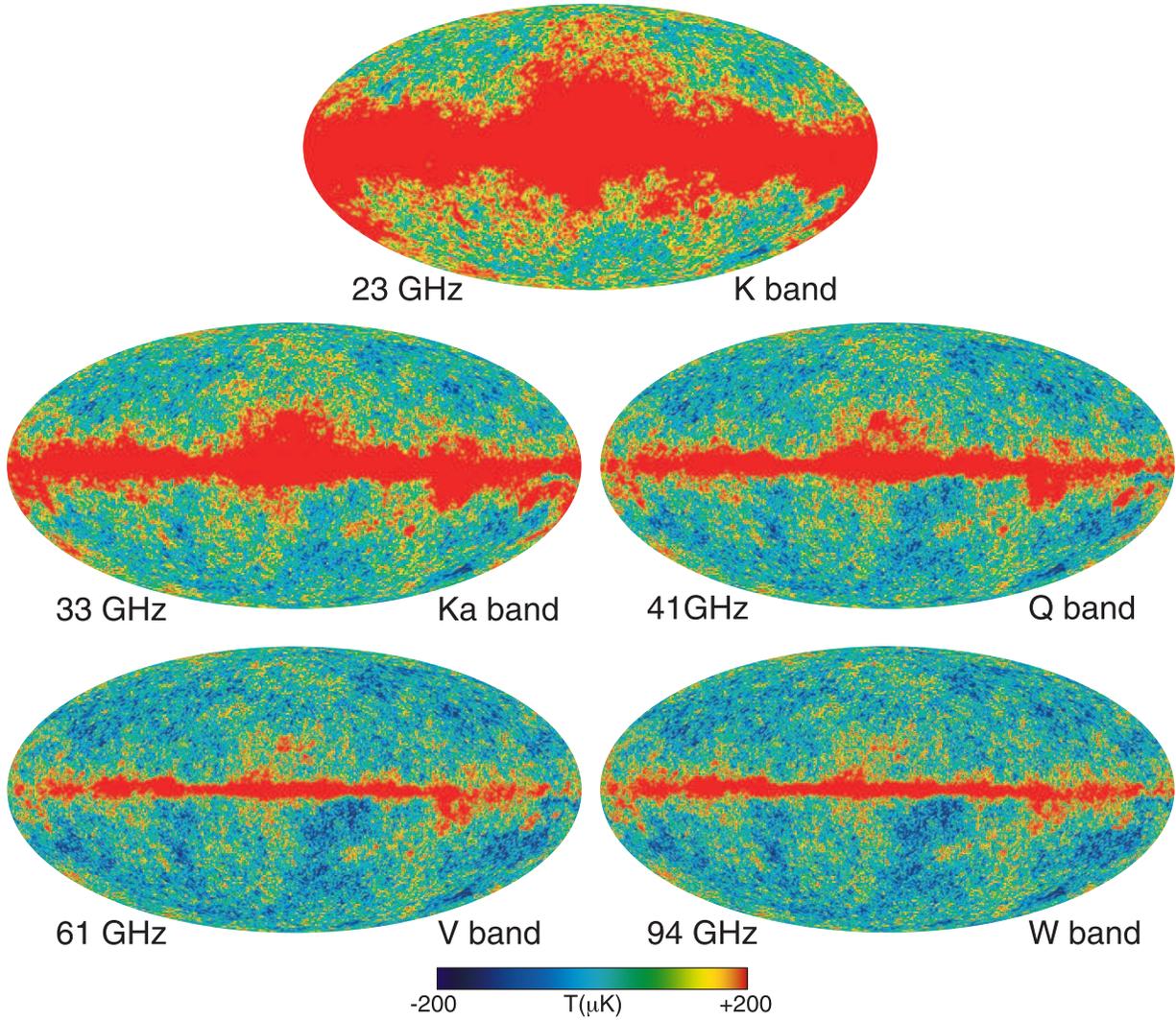}
\caption{Nine-year temperature sky maps in Galactic coordinates shown in a Mollweide
projection.  Maps have been slightly smoothed with a $0\ddeg2$ Gaussian beam.
\vspace{1mm} \newline (A color version of this figure is available in the online journal.)}
\label{fig:tmaps}
\end{figure}

\begin{figure}
\epsscale{0.99}
\plotone{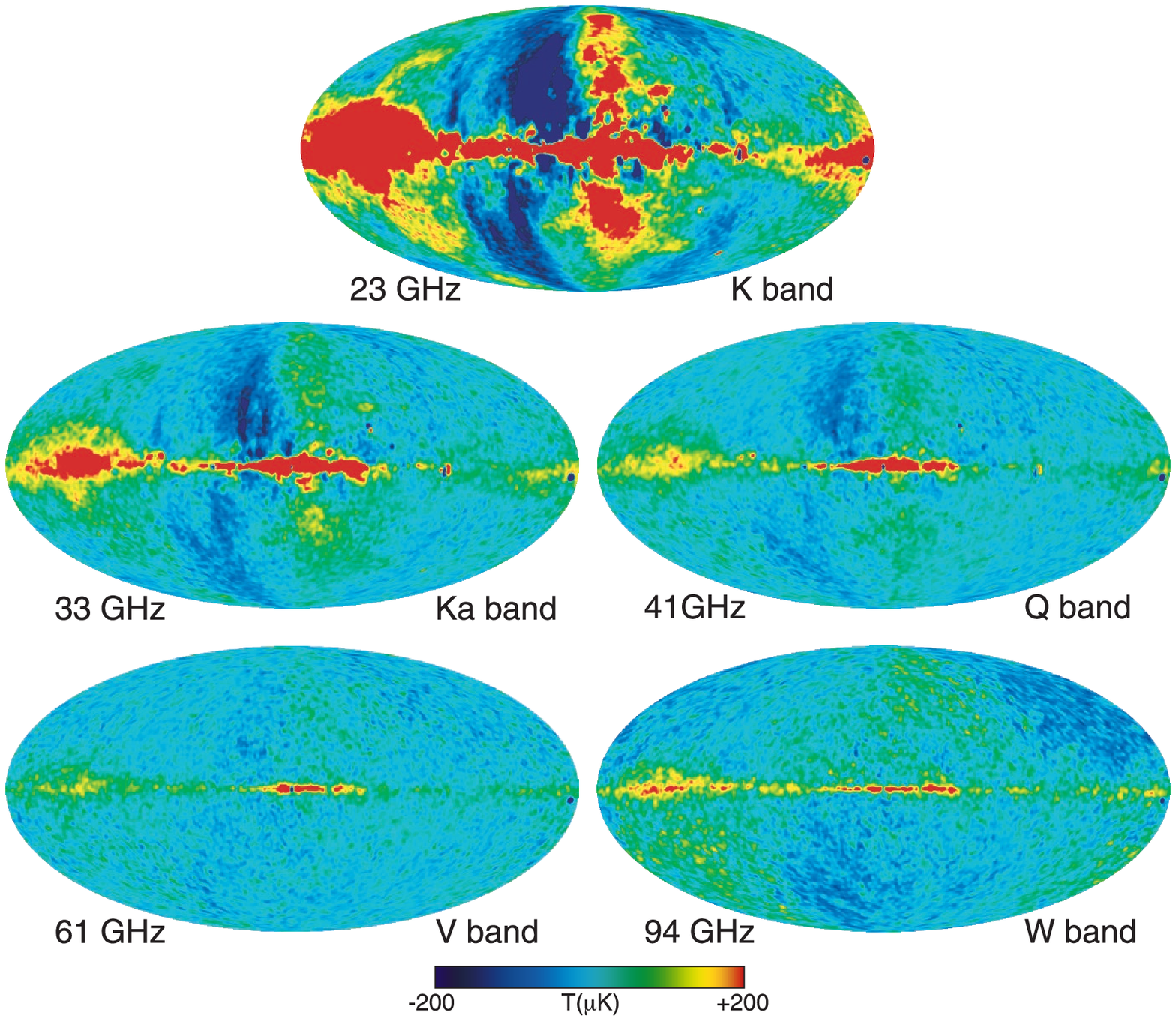}
\caption{Nine-year  Stokes Q polarization sky maps in Galactic coordinates shown in a Mollweide
projection.  Maps have been smoothed to an effective Gaussian beam of $2\ddeg0$.
The smooth large angular scale features visible in W-band, and to a lesser extent in V-band,
are the result of a pair of modes that are poorly constrained in map-making, yet properly
de-weighted in the analysis.
\vspace{1mm} \newline (A color version of this figure is available in the online journal.)} 
\label{fig:qmaps}
\end{figure}

\begin{figure}
\epsscale{0.99}
\plotone{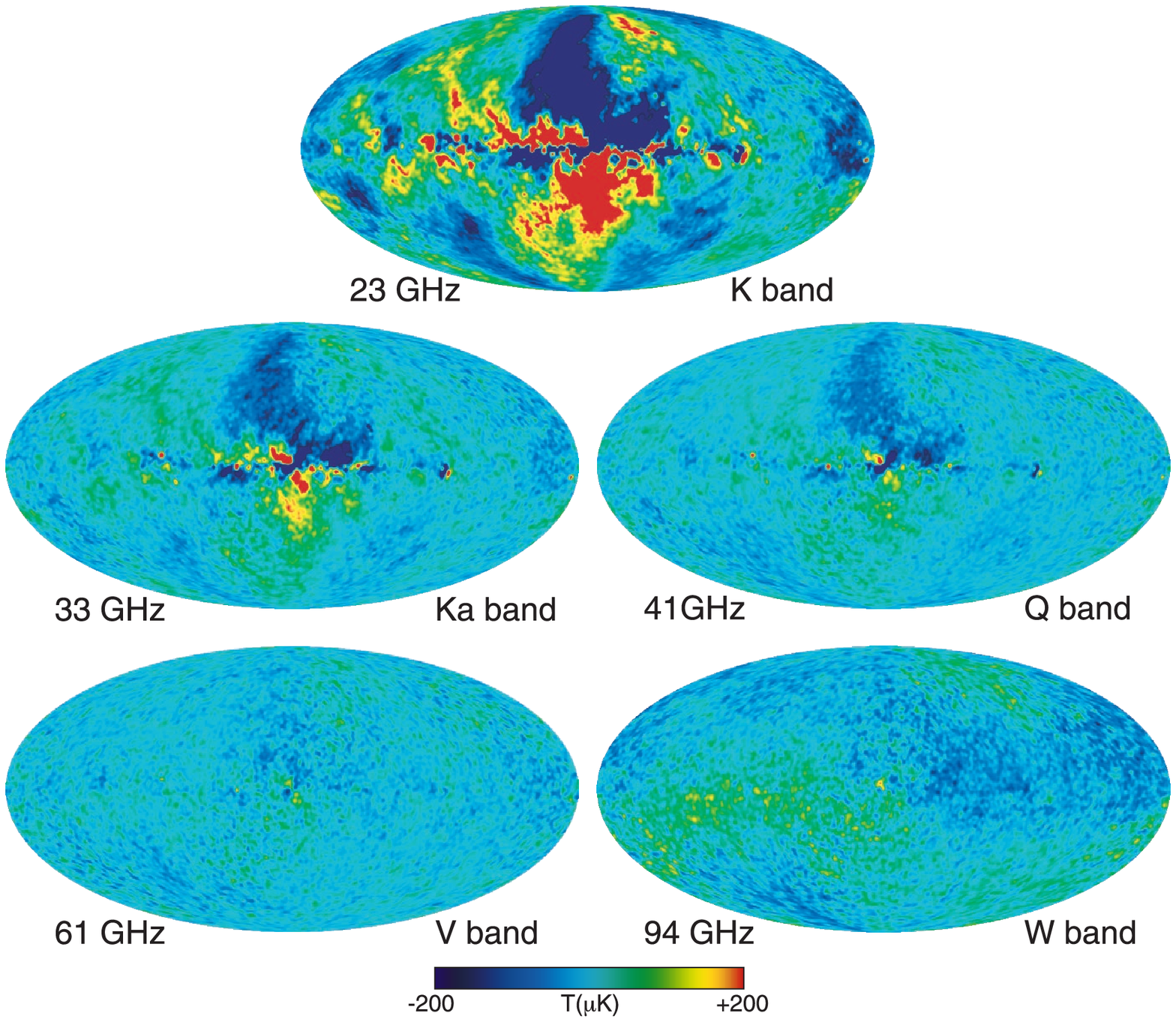}
\caption{Nine-year  Stokes U polarization sky maps in Galactic coordinates shown in a Mollweide
projection.  Maps have been smoothed to an effective Gaussian beam of $2\ddeg0$. 
The smooth large angular scale features visible in W-band, and to a lesser extent in V-band,
are the result of a pair of modes that are poorly constrained in map-making, yet properly
de-weighted in the analysis.
\vspace{1mm} \newline (A color version of this figure is available in the online journal.)}
\label{fig:umaps}
\end{figure}

\begin{figure}
\epsscale{0.99}
\plotone{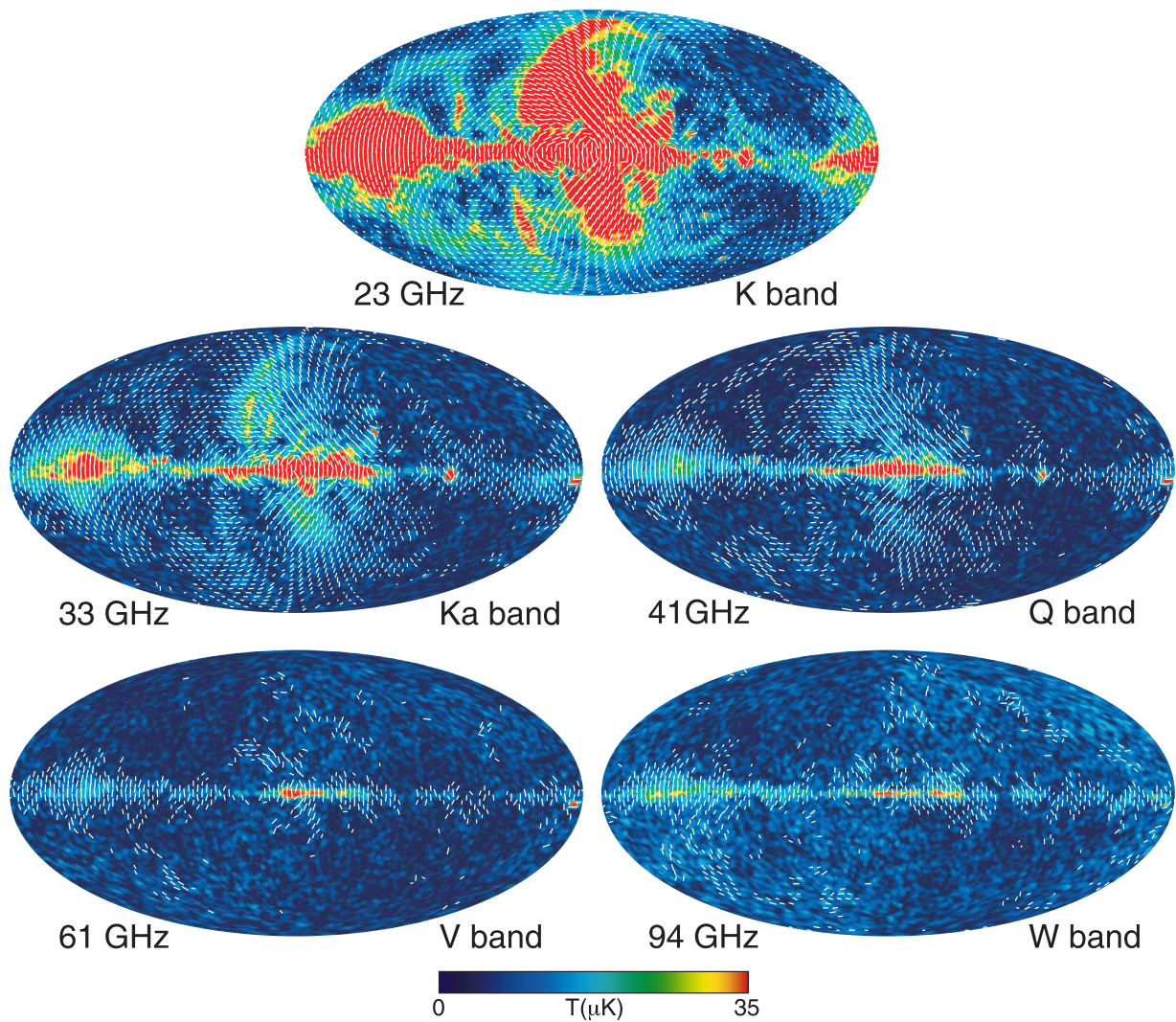}
\caption{Nine-year polarized intensity ($P$) sky maps in Galactic coordinates shown in a Mollweide
projection; $P=(Q^2 + U^2)^{0.5}$, where $Q$ and $U$ are Stokes parameters. Maps have been smoothed 
to an effective Gaussian beam of $2\ddeg0$. Plotted line segments show polarization angles for HEALPix
$\texttt{nside}=16$ pixels where the signal-to-noise exceeds 1.
\vspace{1mm} \newline (A color version of this figure is available in the online journal.)}
\label{fig:pmaps}
\end{figure}

Following the approach discussed above, we present the nine-year temperature (Stokes I) full sky maps in Figure \ref{fig:tmaps}. The corresponding Stokes Q and Stokes U full sky maps are shown in Figures \ref{fig:qmaps} and \ref{fig:umaps}, respectively. Figure \ref{fig:pmaps} shows the nine-year polarized intensity maps of $P=(Q^2+U^2)^{0.5}$ with superposed polarization angle line segments where the signal-to-noise ratio exceeds unity.

\subsubsection{Noise Characterization of the High Resolution Maps}

The noise in the r9 and r10 maps is  described assuming the radiometer noise distribution is stationary,
 has a white spectrum and is normally distributed. With these assumptions it can be shown that the noise component of a Stokes I sky map,
 ${\bf t}_n$, is given by~\citep{jarosik/etal:2011}
\begin{equation}
\tilde{\bf t}_{\rm n} = ({\bf M^TM})^{-1}\cdot{\bf M^Tn},
\end{equation}
where ${\bf M}$ is the mapping matrix as described in \S~\ref{sec:std_proc_summary} and ${\bf n}$
is a vector of normally distributed random numbers that characterizes the radiometer noise,
\begin{equation} \label{eqn:sigma_def}
  \langle {\bf n} \rangle = 0,\ \langle {\bf n n^T} \rangle = \sigma_0^2 {\bf I},
\end{equation}
where the brackets indicate an ensemble average and $\sigma_0$ describes the noise amplitude. The pixel-pixel noise correlation matrix  is then
\begin{equation} \label{eqn:sigma_sigma0}
{\bf \Sigma} = \frac{\langle {\bf t}_{\rm n} {\bf t}_{\rm n}^{T} \rangle}{\sigma_0^2} = ({\bf M^T M})^{-1}.
\end{equation}
Ideally the value of $\sigma_0$ is obtained by evaluating
\begin{equation} \label{eqn:sigma_inv_sigma0}
\sigma_0^2 N_{pix}= \langle {\bf t}_n^T {\bf \Sigma}^{-1} {\bf t}_n \rangle,
\end{equation}
where $N_{pix}$ is the number of map pixels, but such a calculation is intractable with high resolution maps. In practice only the diagonal elements of 
Equation~\refeqn{eqn:sigma_inv_sigma0} are evaluated. Since
\begin{equation}
 {\bf \Sigma}^{-1} = {\bf M^T M}
\end{equation}
the diagonal elements of  ${\bf \Sigma}^{-1}$ are simply the number of observations\footnote{The small correction terms arising from transmission 
imbalance in the radiometers, $1 \pm x_{im}$, are omitted from this equation 
for simplicity, but appear in the next, modified equation.} of each pixel, $N_{\rm obs}$.
Each data sample from a \wmap\ differential radiometer is a measure of the temperature difference between the
sky locations at the A and B side telescope boresights. Reconstructing a map from differential data involves two
different pixels for each observation, a pixel that is being updated and a reference pixel. The noise in each pixel therefore has
contributions both from the noise in the radiometric data for each sample and noise in the value of the reference pixel. If $\sigma_0$
 represents the radiometer noise for an individual sample, the noise contribution from the reference pixel is approximately
$\sigma_0/\sqrt{N_{obs}(p)}$, where $N_{obs}(p)$ is the number of observations used to calculate the value of the reference pixel, $p$.
As the map resolution increases the mean value of $N_{obs}$ decreases, making the reference pixel noise more significant relative to the radiometer noise. The omission of the 
off-diagonal terms in the evaluation of Equation~\refeqn{eqn:sigma_inv_sigma0} ignores the contribution to the noise from the reference beam pixels in the evaluation of $\sigma_0$.
This effect is evident when the $\sigma_0$ values for r9 and r10 versions of the Stokes I sky maps are compared. The $\sigma_0$ values from the r10 maps 
have values from 0.3\% (W-band) to 1.5\% (K-band) higher than those obtained form the corresponding r9 sky maps. The low sampling rate of the K-band radiometer
results in lower $N_{obs}$ values and hence the largest effect.

A more accurate determination of $\sigma_0$ can be made by equating the diagonal elements of Equation~\refeqn{eqn:sigma_sigma0} since these quantities are directly
measurable from the sky maps. The diagonal elements of ${\bf \Sigma}$ may be calculated relatively simply given two well justified 
assumptions: 1) The sky map noise is uncorrelated between pixels; and 2) 
The reference pixels associated with each main pixel are distinct.
With these assumptions diagonal elements of ${\bf \Sigma}$ are estimated as
\begin{equation}
{\bf \Sigma}_{y,y} = \left[ \sum_{i, p_A(i) = y}w(p_B(i))\frac{(1 + x_{\rm im})^2}{1 + 1/N_{\rm obs}(p_B(i))} + \sum_{i, p_B(i) = y}w(p_A(i))\frac{(1 - x_{\rm im})^2}{1 + 1/N_{\rm obs}(p_A(i))} \right]^{-1},
\end{equation} 
where $i$ is a sample index of the TOD and the  sums are over observations for which the A-side and B-side beams observe pixel $y$. The processing mask is represented by $w$, which has value zero in masked
pixels and unity elsewhere. The $1 \pm x_{im}$ factors are corrections arising from the transmission imbalance factors and $N_{obs}$ represents the number of observations contained in the reference beam pixel
of the sky map. The $1/N_{obs}$ terms in the denominators increase the value of $\Sigma_{y,y}$ to account for the additional
noise    arising from the  reference beam pixels. In the limit where $N_{obs}$ is very large for all observations the value of ${\bf \Sigma}_{y,y}$ is simply $1/N_{obs}(y) = 1/{\bf \Sigma}^{-1}_{y,y}$.
The values of  $\sigma_0$ obtained from r9 and r10 Stokes I maps, evaluated using diagonal elements of Equation~\refeqn{eqn:sigma_sigma0}, agree to $\approx 0.05\%$ with the worst discrepancy being $\approx 0.1\%$. This is a 
significant improvement relative to the former method. 

The $N_{obs}$ fields of the nine-year r9 and r10 sky maps now contain the
reciprocals of the diagonal element of the ${\bf \Sigma}$ matrix as it
is now considered a more accurate measure of the pixel noise. This
change  allows the map noise in each pixel to still be calculated as
$N = \sigma_0/\sqrt{N_{obs}}$ providing that the values of $\sigma_0$
published with the nine-year analysis are used.  Because the $\sigma_0$
values computed from r10 maps differ by less than $0.1\%$ from those computed
from r9 maps, the r9 values are adopted for all \wmap\ nine-year
analysis.

These methods have been extended and applied to the Stokes Q and U
maps and the spurious response map S. This change improved the
agreement between the $\sigma_0$ values for the temperatures and
polarization maps to $\approx 0.5\%$ from $\approx 1.1 \%$ in earlier
data releases.  Table \ref{tab:sigma0} gives the nine-year $\sigma_0$
values by DA for temperature (Stokes I) and polarization (Stokes Q,
Stokes U), and spurious response S.

\begin{deluxetable}{cccc}
  \tablewidth{0pt}
  \tablecaption{Noise per Observation in Nine-Year Maps\label{tab:sigma0}}
  \tablehead{
    \colhead{DA} &
    \colhead{$\sigma_0(I)$} &
    \colhead{$\sigma_0(Q,U)$} &
    \colhead{$\sigma_0(Q,U)$} \\
   \colhead{} &
   \multicolumn{2}{c}{Uncleaned} &
   \colhead{Template Cleaned} \\
    \colhead{} &
    \colhead{(mK)} &
    \colhead{(mK)} &
    \colhead{(mK)}}
  \startdata
K1  & 1.429 & 1.435 &    NA   \\
Ka1 & 1.466 & 1.472 & 2.166 \\
Q1  & 2.245 & 2.254 & 2.710 \\
Q2  & 2.131 & 2.140 & 2.572 \\
V1  & 3.314 & 3.324 & 3.534 \\
V2  & 2.949 & 2.958 & 3.144 \\
W1  & 5.899 & 5.912 & 6.157 \\
W2  & 6.562 & 6.577 & 6.850 \\
W3  & 6.941 & 6.958 & 7.246 \\
W4  & 6.773 & 6.795 & 7.076
\enddata
\end{deluxetable}

\subsection{Beam Symmetrized Map Processing}  \label{sec:beamsymmaps}

The \wmap\ telescope beams display varying degrees of asymmetry about the line-of-sight direction, with the amount of asymmetry
related to the position of the feed horn relative to the center of the focal plane \citep{page/etal:2003b}.
The largest asymmetries appear in the lower frequency channels since their feed horns are furthest from the center of the focal plane.
\wmap\ observes each map pixel a large number of times at various azimuthal orientations 
(rotations about the line-of-sight direction).
The degree to which the beam asymmetry is manifest in the final sky maps depends on both the intrinsic beam asymmetry
and the distribution of azimuthal beam orientations  used to observe each pixel. 
A uniform set of finely spaced azimuth angles will result in a symmetric effective beam, while any deviations from a
 uniform distribution will couple some of the beam asymmetry into the sky map. 

The \wmap\ scan pattern causes pixels near
 the ecliptic poles to to be sampled relatively uniformly over a wide range of azimuthal angles, while pixels near
 the ecliptic plane are only sampled over a $\approx \pm 22\ddeg5$ degree range. This results in the effective beam shape varying with sky position;
 regions near the ecliptic poles have more symmetric effective beam shapes than those near the ecliptic plane.
Each pixel is observed roughly the same number of times with the A-side and B-side beams,  further symmetrizing
 the effective beam shape since the axis of asymmetry for the A and B side beams project to different sky directions. 

The \wmap\ window functions are calculated from symmetrized beam profiles generated by azimuthally averaging beam maps 
obtained from observations of Jupiter. 
All \WMAP\ data releases have window function uncertainties incorporated into
the \WMAP\ likelihood code. As described in Appendices A and B of
\citet{hinshaw/etal:2007}, these are dominated by uncertainties in the shape of
the symmetrized beam profile. 

The effects of asymmetric beams 
\citep{page/etal:2003b, hinshaw/etal:2007} were confirmed  in numerical  simulations by \cite{Wehus/etal:2009}. More recently it was found with high statistical significance that the hot and cold spots near the ecliptic plane have a preferred ellipticity, while the angle-averaged small-scale power spectrum near the ecliptic plane is equal to the angle-averaged power spectrum near the ecliptic pole \citep{groeneboom/eriksen:2009, hanson/lewis/challinor:2010}. \citet{hanson/lewis/challinor:2010} and \citet{bennett/etal:2011} suggested that this was likely due largely to the spatially varying effective beam shape  and in this paper we confirm that hypothesis.

\begin{figure}
\epsscale{0.7}
\plotone{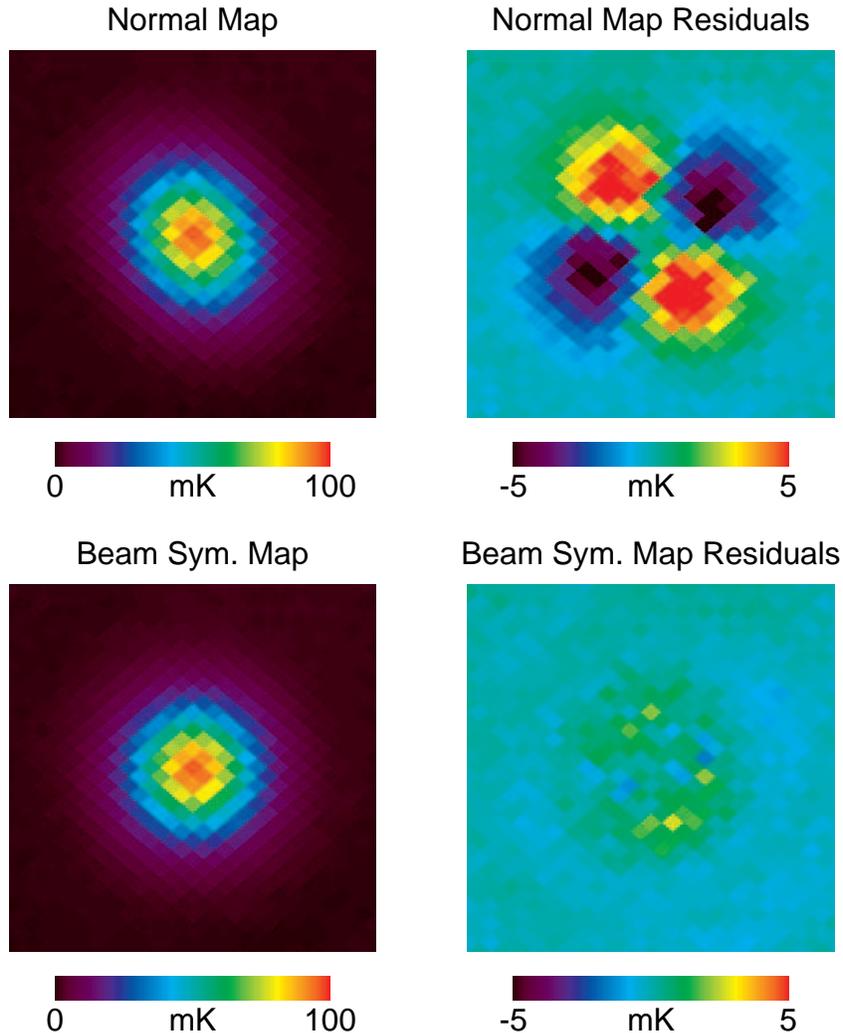}
\caption{K-band images of supernova remnant Tau A (3C 144), at [J2000.0] position ($05^{\rm h}34^{\rm m}31^{\rm s}, 22^\circ01'$) from
 the first year of \wmap\ observations. The left panels display the total  intensity and right the residuals after removal of a best fit
circularly symmetric beam profile. The maps generated with the new partial deconvolution processing (bottom) display significantly reduced
 beam asymmetry compared with those generated with the standard processing (top).
In other words, the apparent asymmetry in Tau A seen in the top left is a result of the asymmetric K-band beam and is not intrinsic to Tau A. 
The degree of a source's apparent asymmetry is dependent on its sky position
and the \wmap\ frequency at which it is observed:
the effect is most pronounced for bright K-band sources at low ecliptic
latitudes (Section~4.2).
As such, we display the K-band observations of Tau A to demonstrate the effectiveness
of the deconvolution in a worst case of beam asymmetry in the
normally processed maps.
\vspace{1mm} \newline (A color version of this figure is available in the online journal.)}
\label{fig:TauA}
\end{figure}

\reffig{fig:TauA} displays the supernova remnant Tau A as it appears in
the year-1 K-band sky map. Tau A is compact relative to the K-band beam size ($\approx 0\ddeg82$ FWHM) and relatively isolated, 
so it approximates a point source for the purpose of mapping the effective beam shape. The beam asymmetry is
clearly seen seen in both the sky map and in the residual map after removal of the best fit symmetrized beam profile.
The symmetrized beam profile was fit to the map
with 6 free parameters, 3 characterizing a baseline (x-slope, y-slope and offset), and three specifying the beam (x-position, y-position, and amplitude). 

The \wmap\ nine-year data release includes a new set of Stokes I maps that have been processed to reduce the asymmetry of the effective beam.
The processing  deconvolves only the asymmetric portion of the beam from the data resulting in a sky map containing the true sky signal
convolved with the symmetrized beam profile. 

A more accurate representation of the signal component of \wmap's TOD utilizes an observation matrix, ${\bf M}_{\rm ns}$,  parameterizing 
the total beam response, written as the sum of a component axisymmetric about the beam LOS, ${\bf M}_{\rm s}$, and a non-axisymmetric component, ${\bf M}_{\rm n}$,
\begin{eqnarray}
{\bf d} &=& {\bf M}_{\rm ns} {\bf t}, \\
{\bf M_{\rm ns} } &=& {\bf M_{\rm s} } + {\bf M_{\rm n}}. \label{eqn:ObsMat}
\end{eqnarray}
Using Equation~\ref{eqn:sym_beam_M} the observation matrix  may be expressed as
\begin{equation}
{\bf M_{\rm ns} } = ({\bf M } + {\bf M}_{\rm n} {\bf C}^{-1}){\bf C}.
\end{equation}
Given this form of the TOD it is possible to solve for the input sky map convolved by the
 axisymmetric beam response, ${\bf \tilde{t}}_{\rm c}$, by evaluating
\begin{equation} \label{eqn:master_deconv}
 {\bf \tilde{t}}_{\rm c} = {\bf Ct} = [ {\bf M}_{\rm am}^{\rm T} {\bf N}^{-1} ({\bf M } + {\bf M}_{\rm n} {\bf C}^{-1})]^{-1}~{\bf \tilde{t}_0} .
\end{equation}
The beam symmetrized  maps contain the input sky signal convolved with the symmetrized beam profile independent of 
sky position. \reffig{fig:TauA} displays a map of the Taurus A region from
a map processed in this manner. The improvement in the beam symmetry is evident in both the raw image and the residuals 
after removing the best fit symmetrized beam profile. These maps significantly improve the symmetry of the effective beam, 
but also have a larger window function uncertainty caused by the limited resolution and signal-to-noise ratio of the beam maps
and numerical approximations needed to make their computation practical. Therefore, beam symmetrized maps are generated only 
for Stokes I and are not intended for the precise fitting of cosmological parameters, but should prove useful in foreground fitting,
studying regions of compact emissions, and certain tests of non-Gaussianity. It should also be noted that deconvolving the 
asymmetric beam shape from the maps necessarily introduces additional pixel-pixel noise correlations
above those contained in the standard maps. No year-to-year correlations are introduced, so power 
spectra calculated from year-to-year cross spectra remain unbiased,
but the uncertainty of the spectra cannot be accurately calculated based on the number of observations ($N_{\rm obs}$) of each map pixel alone.  

\subsubsection{Processing Details}
The beam symmetrized maps are generated by solving Equation~\refeqn{eqn:master_deconv} iteratively using a stabilized bi-conjugate gradient method 
\citep{templates}. In this procedure the product 
\begin{equation} \label{eqn:iter}
{\bf M}_{\rm am}^{\rm T} {\bf N}^{-1} ({\bf M } + {\bf M}_{\rm n} {\bf C}^{-1})\cdot {\bf \tilde{t}}_{{\rm c},i}
\end{equation}
is evaluated repeatedly and the solution ${\bf \tilde{t}}_{{\rm c}, i}$  updated after each iteration, $i$, driving the value of this expression to
${\bf \tilde{t}}_0$. 
The product \refeqn{eqn:iter} is evaluated  by looping through the TOD; each observation corresponds to multiplying one row of 
${\bf M}+ {\bf M}_{\rm n} {\bf C}^{-1}$ by the current iteration of the solution, ${\bf \tilde{t}}_{{\rm c},i}$. The first term in each 
multiplication, ${\bf M} {\bf \tilde{t}}_{{\rm c},i}$,  is the weighted sum of the  map pixels values nearest 
the LOS directions of the two beams, corresponding to the differential sky signal smoothed by the
axisymmetric beam response.  Each row of the matrix ${\bf M}$ contains two non-zero elements with values $(1+x_{\rm im})$ and $(-1 + x_{\rm im})$,
 the weight factors for the A and B side beams. (The $x_{\rm im}$ term ($| x_{\rm im}| \ll 1$) accounts for a small imbalance in 
radiometer response to beam filling signals from the A and B sides.)  

The second term in the product of Equation~\refeqn{eqn:iter},
${\bf M}_{\rm n} {\bf C}^{-1}{\bf \tilde{t}}_{{\rm c},i}$, corresponds to the differential signal from the  non-axisymmetric beam response
 for the current LOS and azimuthal beam orientations. The nonzero elements in each row of ${\bf M}_{\rm n}$ are the pixel weights of the 
non-axisymmetric beam response of the two beams, also weighted by the $(\pm 1 + x_{\rm im})$ factors. To keep the computation time tractable 
only contributions within a radius 
$r_{\rm sl}$ (30 \mrad for K-, Ka-band, 26 \mrad for Q-, V-, and W-band) of the LOS of each beam are used. The circular regions contributing
 to the signal for the A and B beams do not overlap, so their contributions may be calculated separately then summed. 

The matrix ${\bf C}^{-1}$ performs a deconvolution by the symmetrized beam pattern. It is therefore 
rotationally symmetric and the product ${\bf M}_{\rm n} {\bf C}^{-1}$ may be evaluated once for each beam,
forming convolution kernels ${\bf K}_{\rm A}$ and ${\bf K}_{\rm B}$. The 
contribution of ${\bf M}_{\rm n} {\bf C}^{-1}{\bf \tilde{t}}_{{\rm c},i}$ for each beam is then evaluated by mapping these  kernels
 to the corresponding pixels of ${\bf \tilde{t}}_{{\rm c},i}$ for the LOS and azimuthal orientation for each observation
 and summing their products. 

\begin{figure}

\epsscale{1.00}
\plotone{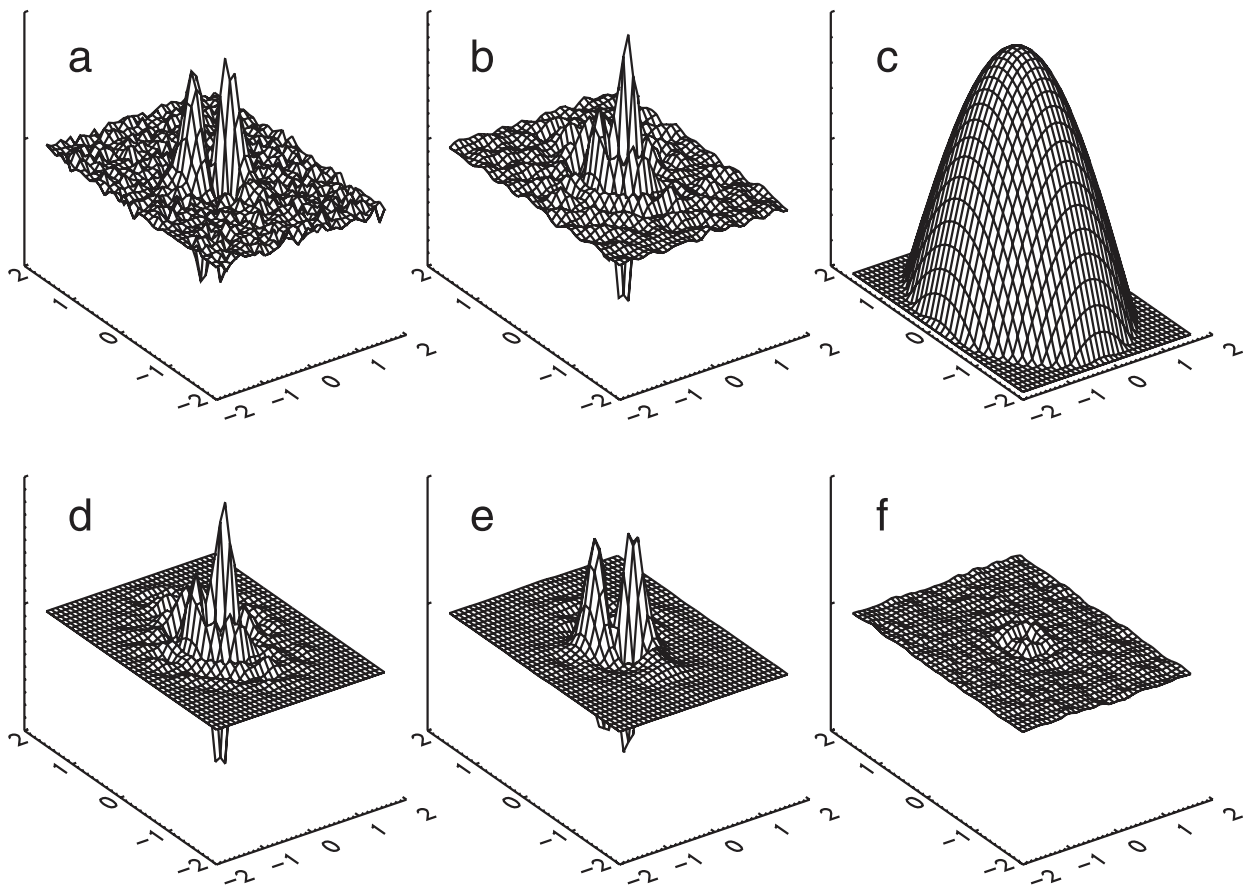}
\caption{Plots illustrating the formation of the kernel used to generate the symmetrized beam maps for the Q1 DA. The x and y axes
are in units of degrees centered on the beam LOS. The z-axis represents weight and panels (a), (e) and (f) use the same scale.
 (a) The residual (non-axisymmetric) component 
of the beam obtained by subtracting the  best fit axisymmetric beam from the total beam map. (b) The residual beam after
 Wiener deconvolution. (c) The cosine apodization function. (d) The convolution kernel used to generate the symmetrized beam maps
consisting of the cosine weighted Wiener deconvolved residual map. (e) The convolution kernel reconvolved with the axisymmetric beam.
(f) The difference between the residual beam map (a) and the map making kernel convolved with the axisymmetric beam (e).}
\label{fig:KernelSeq}
\end{figure}

\reffig{fig:KernelSeq} illustrates the steps used in forming the kernel for the Q1 A-side beam.
First (in panel {\it a}) a map of  the non-axisymmetric beam response, ${\bf M}_{\rm n}$, is formed on a Cartesian grid by subtracting the best fit symmetrized beam 
profile from the total beam profile in Equation~\refeqn{eqn:ObsMat}. Next the product ${\bf M}_{\rm n}{\bf C}^{-1}$, is evaluated by performing a
Wiener deconvolution of ${\bf M}_{\rm n}$.  A Wiener deconvolution is used to minimize the impact of noise on the deconvolved map. 
(In performing the Wiener weighting the signal component of the result was assumed to be proportional to
the input, ${\bf M}_{\rm n}$, while the noise was assumed to be white and its magnitude obtained from portions of the beam map far from the
LOS direction.) Even using the Wiener weighting, some noise remains in the deconvolved maps at relatively large radii from the LOS
 direction. A cosine apodization function is therefore introduced to smoothly taper the value of the kernel to zero at radial distance $r_{\rm sl}$
from the beam LOS. This procedure eliminates artifacts in the maps that would be caused by a sharp cutoff of the 
kernel noise at the radius $r_{\rm sl}$. The fidelity of the kernel is demonstrated in Figures~\ref{fig:KernelSeq}e and \ref{fig:KernelSeq}f that 
show the kernel re-convolved with
the symmetrized beam. After re-convolution the majority of the non-axisymmetric beam response is recovered
 without the introduction of excessive noise. 

Ideally the kernel weights representing the non-axisymmetric beam response sum to zero for each observation. This is only approximately true
in practice since the HEALPix pixelization used for the solution ${\bf \tilde{t}}_{{\rm  c},1}$ and the Cartesian
grid of the kernel are incommensurate, resulting in slightly different combinations of weights
being used for different LOS directions and azimuthal beam orientations. This results in small variations of the total weight for
observation of different points on the sky. 

The mean value of a map generated by ideal differential data is unconstrained. The non-idealities in the
radiometers parameterized by the transmission imbalance factors, $x_{im}$, weakly constrain the mean value of the maps, but occasionally
maps solutions with relative large mean values are generated. The spatially varying total weights described above can couple to these mean values
resulting in small spurious map features. This problem is remedied by subtracting  the sum of the kernel weights used for each observation from
the value in ${\bf M}$ corresponding to the weight of the LOS pixel, resulting in a uniform weight for each observation. This choice insures
 that the total weight of the A and B side observations are $(1+x_{\rm im})$ and $(-1 + x_{\rm im})$ respectively, guaranteeing that the beam 
symmetrized maps agree with the normal maps at angular scales larger than the characteristic size of the convolution kernels.
 
\begin{figure}
\epsscale{0.6}
\plotone{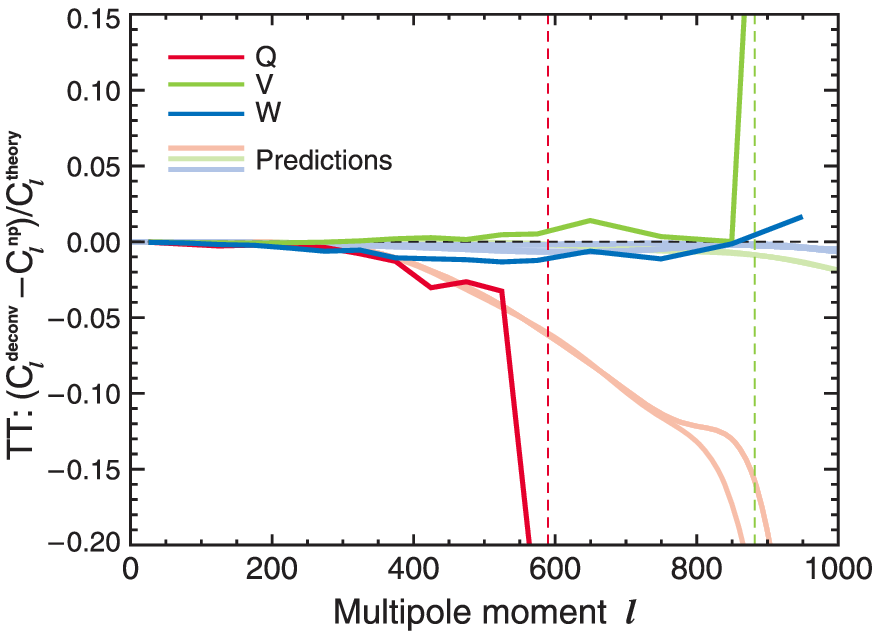}
\caption{
Verification of effects of asymmetric beams on the power spectrum.  Given
beam measurements, the formalism in Appendix B of \citet{hinshaw/etal:2007}
analytically quantifies the beam asymmetry effect on the power
spectrum.  This is plotted as a fractional deviation between an ideally
deconvolved power spectrum ($C_l^{deconv}$) and the power
spectrum of a normally processed map ($C_l^{np}$) with no correction for beam
asymmetries.  These ``predictions'' of fractional deviations are plotted per DA
in the light colored solid lines.  The Q-band effects become significant at
$\ell\sim 400$, but Q-band is not used in the \wmap\ cosmological power
spectrum.  V-band effects become significant at $\ell \sim 1000$, however,
V-band is deweighted compared to W-band at high $\ell$ because of its larger
beam size.  W-band effects from the asymmetric beams can be seen to be $\lesssim 1\%$.  
While \citet{hinshaw/etal:2007} provides an analytic prediction, we have
explicitly deconvolved the maps in pixel space, allowing for a direct
inter-comparison of the analytic with the numerical approach.  The dark red,
green and blue solid lines are the fractional deviations in power spectra for
Q-, V- and W-bands from the directly deconvolved maps.  A comparison between the
light and dark colored lines per frequency band shows close agreement up to a
multipole moment where we expect the spectra derived from the beam-symmetrized
maps to break down because the prediction does not account for 
correlations introduced by the deconvolution.  The Q-band deviations
occur after the window function has dropped below 2.5\% and the V-band
deviations below 1.5\%.  The vertical dashed lines indicate where
window functions are at 1\% of their maximum value.  The close agreement between
the predictions and explicit deconvolution verifies our understanding of
asymmetric beam effects and allows us to conclude that the spectrum from the
normally processed (i.e. not deconvolved) maps differs from the
ideally-deconvolved spectrum by $< 1\%$.  Thus the final \WMAP\ power spectrum
is based on the normally-processed V- and W- band maps.  
\vspace{1mm} \newline (A color version of this figure is available in the online journal.)
}
\label{fig:sym_map_spectra}
\end{figure}

\reffig{fig:sym_map_spectra} displays the ratio of the TT power spectra of the beam symmetrized maps to those of the normally processed maps
and ratios as predicted in \citet{hinshaw/etal:2007}. 
The spectra from the different map processings  agree exactly at low $\ell$ as expected and agree with the predictions 
within 2\% in regions of adequate signal-to-noise ratios.
 
\clearpage
\section{Foreground Fits} \label{sec:foregrounds}

\subsection{Overview}

In this section we examine the nature of the Galactic and extragalactic foreground emission. These foregrounds are important to understand so as to achieve an appropriate separation of CMB anisotropy from foreground emission, to elucidate the underlying astrophysical emission processes, and to transfer the precise \wmap\ calibration to 
astronomical emission sources that can be used by other observers for calibration purposes.

\begin{figure}
\epsscale{0.90}
\plotone{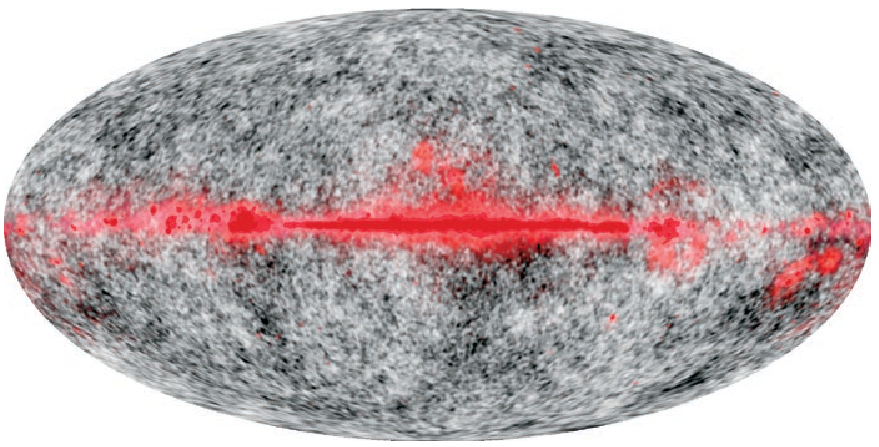}
\caption{False color image representing the spectral information from multiple \wmap\ bands. 
Q-band is red, V-band is green, and W-band is blue. In this representation, a CMB thermodynamic spectrum appears as grey.
\vspace{1mm} \newline (A color version of this figure is available in the online journal.)} 
\label{fig:qvw_3color}
\end{figure}

\begin{figure}
\epsscale{0.90}
\plotone{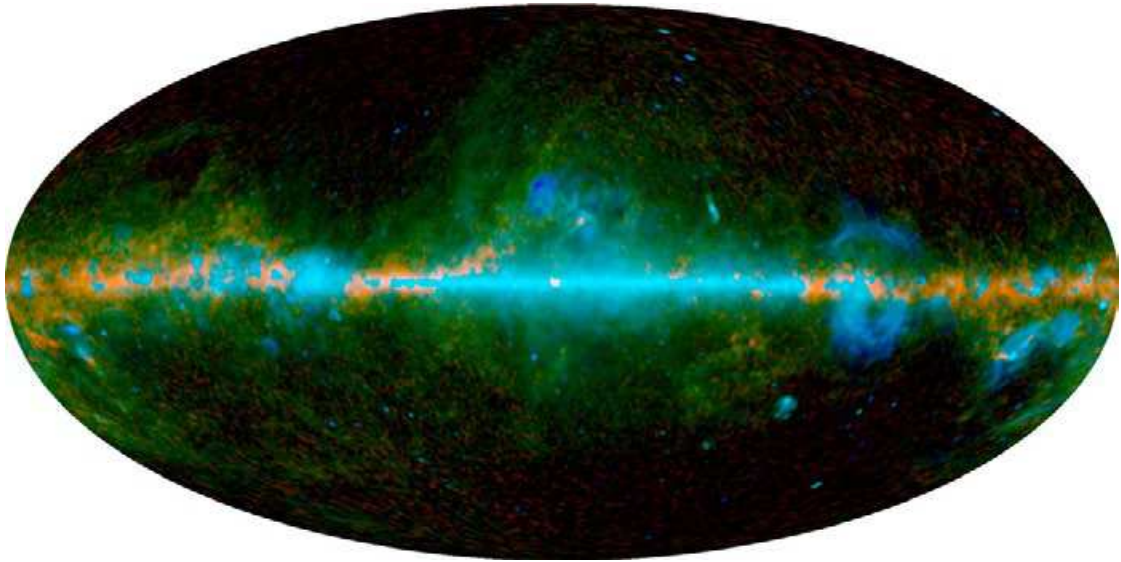}
\caption{False color image derived from a combination of \wmap\ band differences chosen to 
highlight differing spectral components.  Red (W-V) highlights regions where thermal emission
from dust is highest.  Blue (Q-W) is dominated by free-free emission.  Green ((K-Ka)-1.7(Q-W))
illustrates contributions from synchrotron and spinning dust.
\vspace{1mm} \newline (A color version of this figure is available in the online journal.)} 
\label{fig:band_diff_3color}
\end{figure}

\begin{figure}
\epsscale{0.99}
\plotone{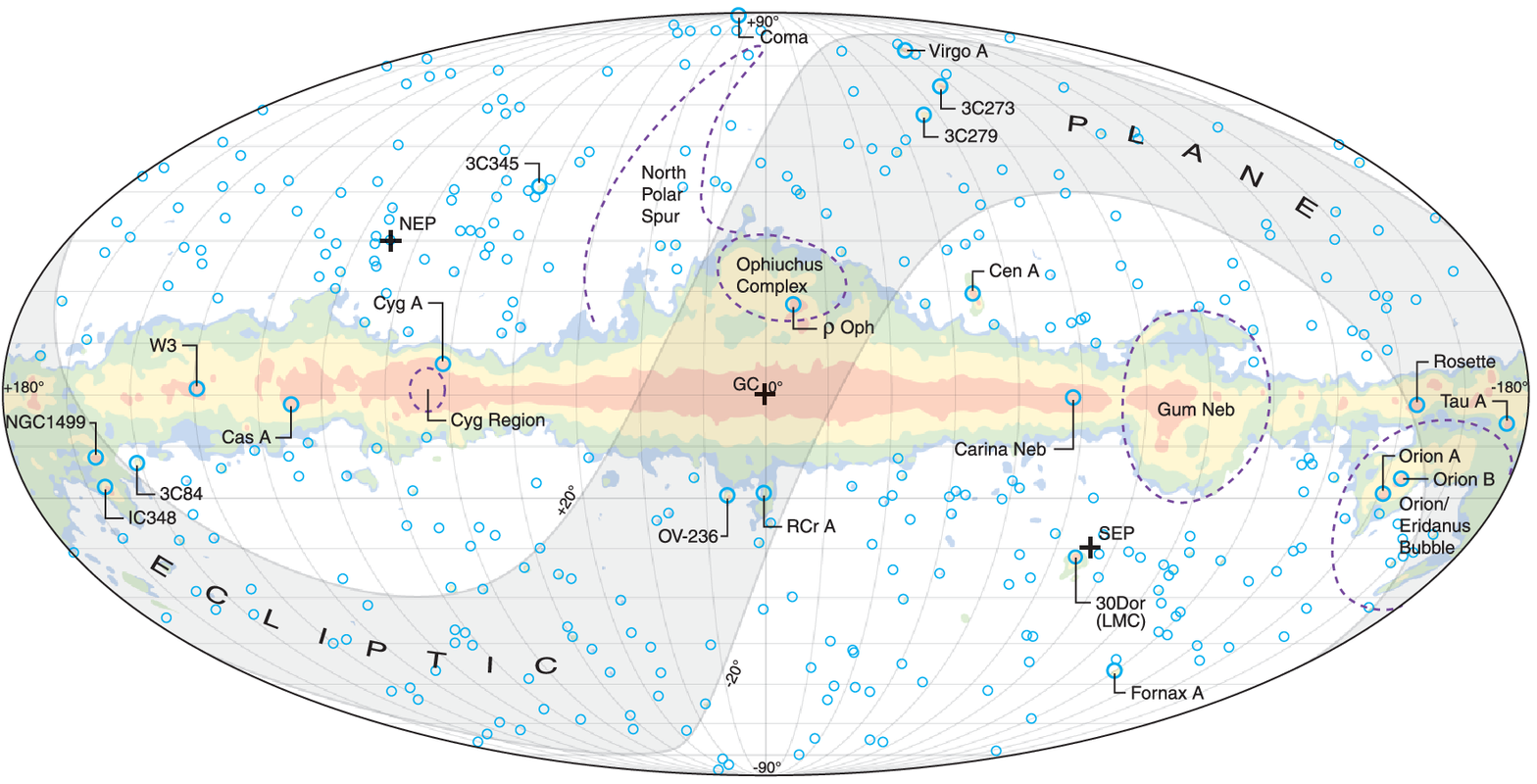}
\caption{Microwave emission near the Galactic plane is traced by a K-band minus W-band difference map, which eliminates CMB anisotropy. A log scale is used for the color region and blue circles represent the positions of the brightest point sources, as seen by \wmap.
\vspace{1mm} \newline (A color version of this figure is available in the online journal.)}
\label{fig:finder}
\end{figure}

The separation of CMB anisotropy from foregrounds depends critically upon their different spectra. This is illustrated in Figure \ref{fig:qvw_3color} where a model-free three-color display of \wmap\ data clearly differentiates the (pink) diffuse and point source foreground emission from the (gray) CMB anisotropy.  Likewise, \wmap\ maps in different frequency bands can be convolved to a common
angular resolution and subtracted to form a CMB-free, foreground emission-only map.
Three such difference maps, in turn, can be combined into a three-color display that highlights the spectral differences of the foregrounds across the sky. An example of this is shown in Figure \ref{fig:band_diff_3color}.
Figure \ref{fig:finder} provides an orientation of the microwave emission sources on the sky.

This section is divided into two major subsections: point source analyses
are presented first in \ref{ssec:point_sources}, followed by diffuse foregrounds
in \ref{ssec:diffuse_fg}. The point source subsection
begins with a discussion of \wmap\ observations of the planets Jupiter
and Saturn (Section \ref{sssec:planets}). For Saturn we separate the
emission into a disk and ring component.  In Section~\ref{sssec:point_source_cats} we
describe two techniques to identify other point sources and we provide point source
catalogs in Appendices~\ref{5bandsources} and \ref{3bandsources}.
We then go on to discuss our analysis of the diffuse foregrounds.  In
Section~\ref{sssec:tmpl_cln} we describe the approach taken to mask and clean
diffuse foregrounds for the purpose of carrying out the cosmological analysis of the CMB,
such as the angular power spectra.  In Section \ref{sec:ilc} we present the new nine-year
internal linear combination (ILC) map.  Since ILC error characterization is
dependent on a knowledge of the foregrounds, a deeper ILC discussion is deferred until
after a foreground characterization analysis.
To identify the nature of the foregrounds we describe
three different fitting techniques: the maximum entropy method (MEM) in Section~\ref{sec:MEM};
Monte Carlo Markov Chains (MCMC) in Section \ref{sec:MCMC}; and $\chi^2$ fitting in Section
\ref{sec:six_band_fits}. We conclude this section with a synthesis based
on these analysis efforts.  Section \ref{sec:cross-comparison} includes an intercomparison
of results from the three fitting techniques and a comparison of foreground component maps
averaged over the three fits with the corresponding template maps used in foreground cleaning.
Finally, Sections~\ref{sec:ilc_errors} and \ref{sec:ilc_considerations}  discuss
ILC errors. Estimates are presented of residual foreground bias in the ILC map and
ILC error due to CMB-foreground covariance.  Appendix \ref{sec:bandcenter} describes small
variations in \WMAP\ bandpasses that occurred over the nine-year mission, which are taken
into account in our foreground analyses.  They have no significant effect on the CMB or
cosmology analysis.

\subsection{Point Sources} \label{ssec:point_sources}

\subsubsection{Planets and Celestial Analysis} \label{sssec:planets}

A detailed analysis of \wmap\ seven-year observations of planets and selected celestial calibrators is
given by \citet{weiland/etal:2011}, including intercomparisons with relevant results in the literature.
Here we concentrate on updated nine-year \wmap\ results for some of these sources.    

\paragraph{Jupiter}

Mean nine-year Jupiter temperatures are derived from the $l=0$ component of the unnormalized 
beam transfer functions $B_l$.  The symmetrized beam response to Jupiter,
$T_{pk}\Omega_{beam}$, may be directly derived from $B_0$.  As described in \citet{weiland/etal:2011}, 
all Jupiter observations have 
been corrected to a fiducial solid angle $\Omega_{Jup}^{ref} = 2.481\times10^{-8} sr$. 
Mean Jupiter temperatures $T_{Jup}$ are thus computed using the relation 
$T_{Jup} = T_{pk}\Omega_{beam}/\Omega_{Jup}^{ref}$.
These temperatures are presented in Table~\ref{tab:juptemp}.  Quoted uncertainties are a quadrature
sum of estimated beam solid angle errors from Table~\ref{tab:beam_quantities} and the uncertainty
in the absolute calibration.  The mean Jupiter temperatures
derived from the five-year, seven-year and nine-year data releases are consistent with each other 
within the quoted uncertainties.

The stability of Jupiter emission over the nine-year baseline is evaluated by 
computing seasonal temperatures per DA and comparing them to their nine-year means.
We compute $\Delta T/T$
as the mean deviation of all DAs from their nine-year mean values, and include a $1\sigma$ standard
deviation as a measure of coherency.  These results are listed in Table~\ref{tab:jup_seasonal_temps}.
From the seven-year analysis, \citet{weiland/etal:2011} placed an upper limit on variability of 
$0.2 \pm 0.4 \%$.  Although consistent with this value, the Jupiter observations from the last two  
seasons introduce the statistically weak (PTE $=14\%$) possibility of a decreasing trend in temperature with time. 
Given our measurement uncertainties, a constant temperature is a very good fit to 
the data and that is what we use in our analysis.

Out of caution, we examined the hypothesis that there might be instrumental or calibration issues contributing to
slightly lower Jupiter temperatures computed for the last few seasons of data.
To determine if there might be a systematic calibration error within the last two years of the mission,
yearly flux values for celestial sources Cas A, Cyg A, and Tau A were computed and compared
against seven-year trends; no evidence for any calibration inconsistency was found.  Since
Jupiter is not a steep-spectrum source, bandpass center frequency variations are also not an important factor;
we expect an effect of less than $\pm0.05\%$ over the 9 years in the K- through V-bands.
In terms of Jupiter itself, there is no clear temperature trend with Sun-Jupiter distance or sub-\wmap\ 
latitude.  

\begin{deluxetable}{ccccc}
  \tablewidth{0pt}
  \tablecolumns{5}
  \tabletypesize{\scriptsize}
  \tablecaption{Nine-Year Mean Jupiter Temperatures \label{tab:juptemp}}
  \tablehead{
    \colhead{ } &
    \colhead{$\nu_e^\mathrm{RJ}$ \tablenotemark{a}} &
    \colhead{$\lambda$ \tablenotemark{b}} &
    \colhead{$T$\tablenotemark{c}} &
    \colhead{$\sigma(T)$\tablenotemark{d}} \\ &
    \colhead{(GHz)}  &
    \colhead{(mm)}   &
    \colhead{(K)}    &
    \colhead{(K)}}
  \startdata
  \sidehead{per DA}
  K1  & 22.82 & 13.1 & 136.1 & 0.75 \\
  Ka1 & 33.07 &  9.1 & 147.1 & 0.68 \\
  Q1  & 40.88 &  7.3 & 153.9 & 0.78 \\
  Q2  & 40.67 &  7.4 & 154.7 & 0.76 \\
  V1  & 60.37 &  5.0 & 164.9 & 0.71 \\
  V2  & 61.24 &  4.9 & 165.9 & 0.68 \\
  W1  & 93.25 &  3.2 & 172.5 & 0.84 \\
  W2  & 93.73 &  3.2 & 173.4 & 0.85 \\
  W3  & 92.72 &  3.2 & 173.1 & 0.87 \\
  W4  & 93.57 &  3.2 & 172.3 & 0.86 \\
  \sidehead{per band}
   K  &   22.82	&  13.1	& 136.1 & 0.75 \\ 
   Ka &   33.07	&   9.1	& 147.1 & 0.68 \\	  
   Q  &   40.78	&   7.3	& 154.3 & 0.59 \\	  
   V  &   60.81	&   4.9	& 165.4 & 0.54 \\	  
   W  &   93.32	&   3.2	& 172.8 & 0.52 \\  
  \enddata
  \tablenotetext{a}{nine-year values; see Appendix~\ref{sec:bandcenter}}
  \tablenotetext{b}{$\lambda = c/\nu_e^{\mathrm{RJ}}$}
  \tablenotetext{c}{Brightness temperature
    calculated for a solid angle $\Omega_\mathrm{ref}
    = 2.481 \times 10^{-8}$ sr at a fiducial distance of 5.2 AU.
    Temperature is with respect to blank sky: absolute
    brightness temperature is obtained by adding 2.2, 2.0, 1.9, 1.5 and 1.1 K in bands
    K, Ka, Q, V and W respectively \citep{page/etal:2003b}. 
    Jupiter temperatures are uncorrected for a small synchrotron emission component
    (see \citet{weiland/etal:2011}). }
  \tablenotetext{d}{Computed from errors
    in $\Omega_B$ (Table~\ref{tab:beam_quantities}) summed in quadrature with absolute
    calibration error of $0.2\%$.}
\end{deluxetable}

\begin{deluxetable}{cccrcccc}
  \tablecolumns{5}
  \tablewidth{0pt}
  \tablecaption{Jupiter Temperature Changes by Season \label{tab:jup_seasonal_temps}}
  \tablehead{
    \colhead{Season\tablenotemark{a}} & \colhead{Start} & \colhead{End} 
    & \multicolumn{2}{c}{$\Delta T/T$ (\%)}  \\
    & & & \colhead{Mean\tablenotemark{b} } & \colhead{Scatter\tablenotemark{c}}  
    }
  \startdata
  1 & 2001 Oct 08 & 2001 Nov 22 &    0.33 &  0.26 \\
  3 & 2002 Nov 10 & 2002 Dec 24 &   -0.01 &  0.33 \\
  4 & 2003 Mar 15 & 2003 Apr 29 &   -0.14 &  0.51 \\
  5 & 2003 Dec 11 & 2004 Jan 23 &    0.17 &  0.22 \\
  6 & 2004 Apr 15 & 2004 May 30 &    0.12 &  0.23 \\
  7 & 2005 Jan 09 & 2005 Feb 21 &    0.13 &  0.35 \\
  8 & 2005 May 16 & 2005 Jul 01 &    0.07 &  0.37 \\
  9 & 2006 Feb 07 & 2006 Mar 24 &    0.32 &  0.33 \\
  10 & 2006 Jun 16 & 2006 Aug 02 &   0.18 &  0.47 \\
  11 & 2007 Mar 10 & 2007 Apr 24 &   0.53 &  0.34 \\
  12 & 2007 Jul 19 & 2007 Sep 03 &  -0.04 &  0.44 \\
  13 & 2008 Apr 11 & 2008 May 27 &  -0.05 &  0.34 \\
  14 & 2008 Aug 21 & 2008 Oct 06 &  -0.11 &  0.30 \\
  15 & 2009 May 17 & 2009 Jul 03 &  -0.46 &  0.61 \\
  16 & 2009 Sep 26 & 2009 Nov 10 &  -0.39 &  0.34 \\
  17 & 2010 Jun 24 & 2010 Aug 10 &  -0.47 &  0.27 \\
  \enddata
  \tablenotetext{a}{Season 2 omitted from analysis because Jupiter is aligned with the Galactic
    plane.}
  \tablenotetext{b}{Mean of the percentage temperature change among the
    DAs for each season, relative to the nine-year mean.}
  \tablenotetext{c}{$1\sigma$ scatter in the percentage temperature change among the
    DAs for each season.}
\end{deluxetable}

\paragraph{Saturn}

As seen by the \wmap\ satellite, the spatially unresolved microwave brightness of Saturn varies with 
orbital phase as the projected area of the ring system and oblate planetary spheroid changes aspect.
\citet{weiland/etal:2011} developed an empirical, geometrically motivated model to predict
Saturn's apparent brightness at \wmap\ frequencies, based on the first seven years (14 seasons) of 
observations.
The available range of observable ring opening angles during this seven year interval falls in the range $ -28\arcdeg < B < -6\arcdeg$. 
\citet{weiland/etal:2011} found that parameter covariance and potential systematics in their model fit
permitted a determination of Saturn's disk temperature to within roughly 3-4~K, but
noted that the inclusion of lower inclination observations in the fit should decrease
the uncertainty in the derived model parameters.
\wmap\ observations from the last two mission years include four new Saturn observing seasons,
numbered 15 through 18. Since the Saturn ring system presented an ``edge-on'' 
configuration in early 2009, these four new seasons span the cross-over from viewing the rings from below
(negative $B$) to viewing them from above (positive $B$) as seen in Table~\ref{tab:saturn}. 
These new observations at low $B$ provide the opportunity to better
constrain the predictive model for \wmap\ frequencies.

We apply the analysis methods of \citet{weiland/etal:2011} to the nine-year compendium of Saturn
observations to derive mean apparent temperatures of the Saturn system per DA per observing season,
presented in Table~\ref{tab:saturn}.   The analysis can be summarized as a three-step process.
First, a time-ordered archive of Saturn observations is created, and sky signals arising from the Galaxy and CMB are removed,
either through use of sky subtraction or masking.  Second, the individual observations from this background subtracted 
archive are binned to form
mean radial Saturn response profiles for each season and DA.  Finally, the \wmap\ beam radial profile per DA
(as determined from Jupiter observations) is fit to the Saturn radial response for that DA 
and an apparent temperature is derived.  Temperature entries for the first 14 seasons listed in Table~\ref{tab:saturn} 
may be directly
compared against those in Table~9 of \citet{weiland/etal:2011}.  There are small differences of order 0.5 to 
1~$\sigma$ between some of entries in common between the seven-year analysis and the nine-year analysis presented here.  
Differences of this nature are expected and
can be traced to small variations in calibration, beam characterization and data masking between the seven-year 
and nine-year processing.
 
The temperatures in Table~\ref{tab:saturn} may be fit with an empirical model 
that predicts Saturn's unresolved microwave brightness $T$ as a function of ring 
opening angle and frequency. 
We adopt the same model formulation as in the seven-year analysis of \citet{weiland/etal:2011},
which employs a simple geometrical summation of emission from the unobscured planetary disk, emission from the ring system
and emission from those portions of the disk obscured by the rings:
\begin{equation}
\label{eq:satmodeqn}
T(\nu,B) = T_{\mathrm{disk}}(\nu) [A_{\mathrm{ud}} + \sum_{i=1}^7 e^{-\tau_{0,i}|\csc B|} A_{\mathrm{od},i}] +
T_{\mathrm{ring}}(\nu) \sum_{i=1}^7 A_{r,i}.
\end{equation}
At a given frequency $\nu$, a single temperature is assumed
for the planetary disk, $T_{\mathrm{disk}}(\nu)$.  The model allows for seven radially concentric ring divisions. 
All rings are characterized by the same temperature $T_{\mathrm{ring}}(\nu)$, but
each of the seven ring sectors has its own ring-normal optical depth
$\tau_{0,i}$, with $1 \le i \le 7$. Each $\tau_{0,i}$ is assumed to be both constant within its ring and 
frequency independent.  
$A_{\mathrm{ud}}$, $A_{\mathrm{od},i}$ and $A_{r,i}$ are the projected areas of the unobscured disk, the portion
of the disk that is obscured by ring $i$, and $i^{th}$ ring, respectively.  These areas are normalized to the total 
(obscured+unobscured) disk area.
Model fit parameters are the five Saturn uniform disk temperatures and five mean ring temperatures (one for each \wmap\ frequency).
The geometrical ring boundaries and relative ratios $\tau_{0,i}/\tau_{0,\mathrm{max}}$
are constrained as per Table~10 of \citet{weiland/etal:2011}, where  $\tau_{0,\mathrm{max}}$
is the ring-normal optical depth for the most optically thick ring (ring 3, i.e. the outer B ring).
For the nine-year fit, the value of $\tau_{0,\mathrm{max}}$ was also allowed 
to be a fit parameter, although in practice its inclusion makes very little difference in the fit results.  
  
\begin{figure}
  \begin{center}
\epsscale{0.70}
\plotone{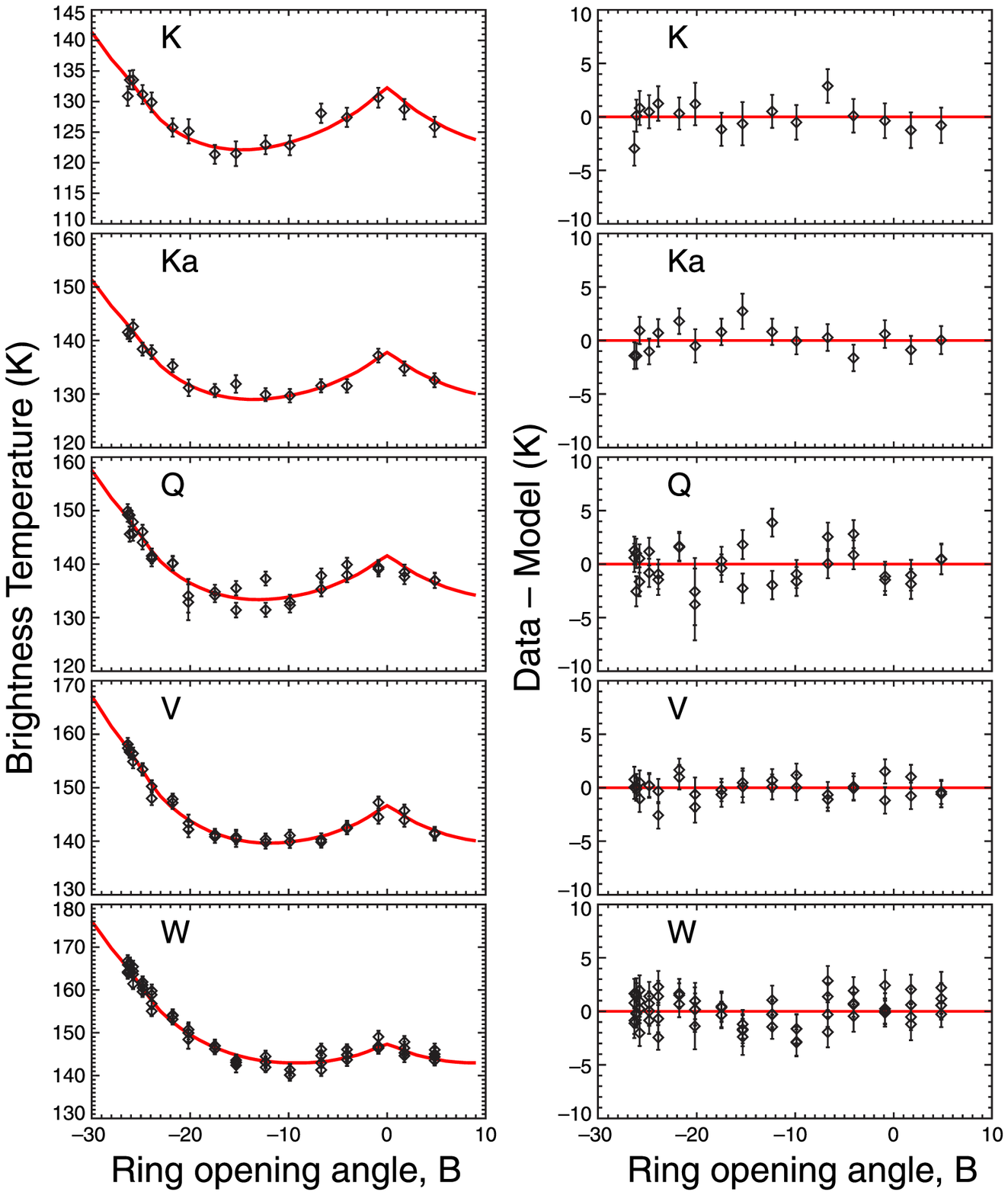}
    \caption{ Modeling results for Saturn. (Left) Brightness temperatures based on unresolved Saturn observations
    as a function of ring inclination $B$ are shown in black for each \wmap\ frequency band.  Where there are
    multiple differencing assemblies per frequency, multiple points are plotted at each inclination.
    An empirical model including both ring and disk components (see text) is plotted in red.  
    The temperature of the planetary disk predicted
    by the model occurs at $B$=0$\arcdeg$, when the rings are viewed edge-on. The model is symmetric about
    $B$=0$\arcdeg$.
    (Right) Residuals (data-model) of the model fit to the data are plotted as a function of the ring opening angle.  
    \vspace{1mm} \newline (A color version of this figure is available in the online journal.)
    \label{fig:satmod}}
  \end{center}
\end{figure}

\begin{figure}
\epsscale{0.90}
\plotone{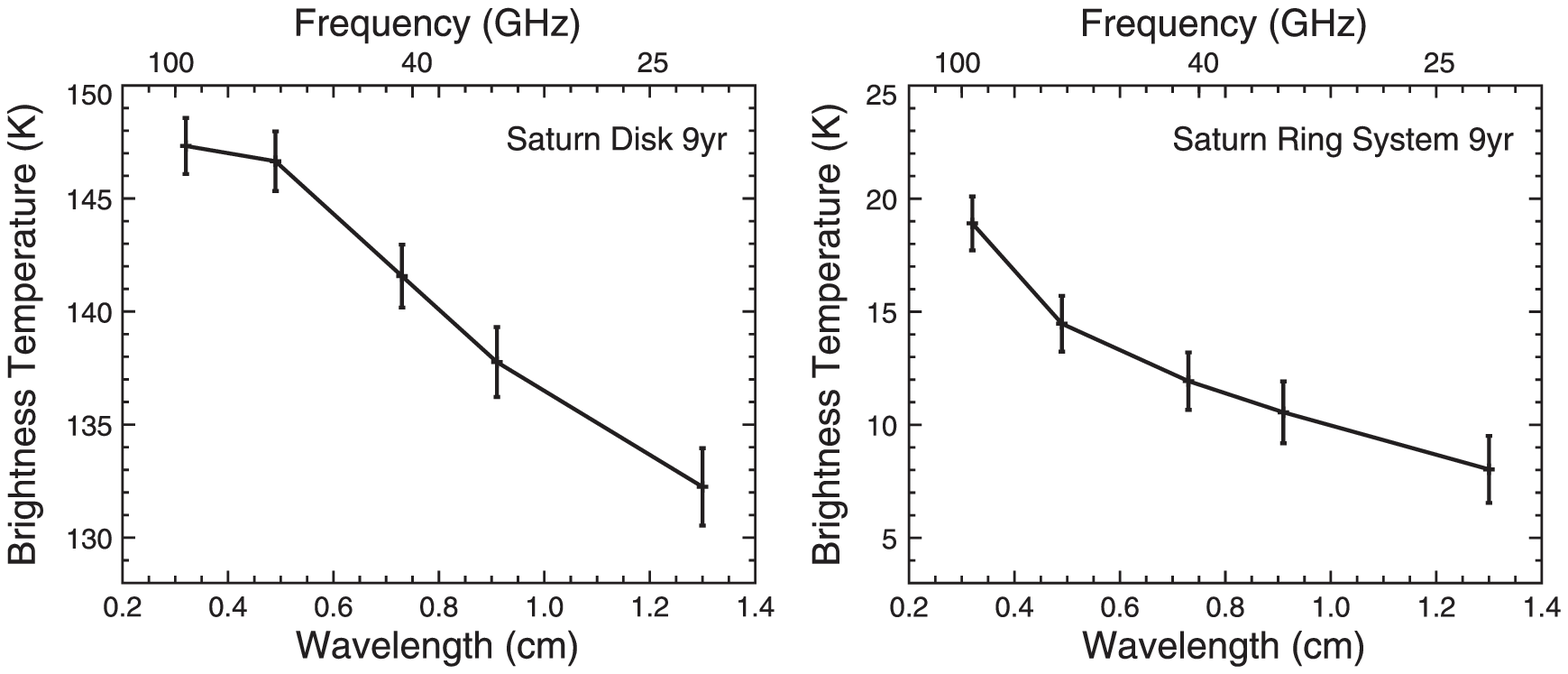}
\caption{Saturn model parameters derived from the nine-year analysis. Left: Disk temperatures for 5 \wmap\ 
frequencies. Right: Ring system temperatures. Adopted errors for the nine-year analysis have been reduced
compared to those in \citet{weiland/etal:2011}; errors for $T_{\mathrm{disk}}$ are smaller by a factor of 2.}
\label{fig:model_param_compare}
\end{figure}

The nine-year model fit returns a reduced $\chi^2$ of $\sim1.04$ for $\sim150$ degrees of freedom;
the model fit and residuals per \wmap\ frequency are shown in Figure~\ref{fig:satmod}.
On average, the rms of the residuals is $\sim1\%$ per frequency; the value for Q-band is somewhat
higher ($1.3\%$) and that for V-band is lowest ($0.7\%$).
Model parameters and their formal errors $\sigma_{\mathrm{fit}}$ are presented in Table~\ref{tab:satmodel2}.
By construction, the $T_{\mathrm{disk}}$ and $T_{\mathrm{ring}}$ model parameters are anti-correlated.  
The covariance 
between these parameters allows the possibility of systematic errors not accounted for in the fitting 
formalism.  Although the mean disk temperature is reasonably well constrained by the new \wmap\
observations from seasons 15-18, hemispheric temperature gradients or local hot spots would
negate the assumed symmetry of the empirical model, and would affect the derived mean ring temperatures.
The nine-year baseline unfortunately does not
extend far enough toward positive $B$ to assess the limits of the symmetry assumption.  Additionally,
the model's assumed ring optical depth profile may not be accurate.  As with the seven-year
analysis, we use a model variant to estimate systematic differences between models which return
similar values of $\chi^2$.  Our worst case estimate allows for differences of 0.9~K in $T_{\mathrm{disk}}$ and
0.7~K in $T_{\mathrm{ring}}$; we add these to the formal fitting errors in Table~\ref{tab:satmodel2} to 
produce the tabulated adopted error, $\sigma_{\mathrm{adopted}}$.  
The $T_{\mathrm{disk}}$ and $T_{\mathrm{ring}}$  parameters are plotted along with their
adopted errors in Figure~\ref{fig:model_param_compare}.
Within the conservative adopted errors, the nine-year
derived disk and ring temperatures are in agreement with those from the seven-year fit; 
the nine-year adopted errors for $T_{\mathrm{disk}}$ are roughly half those quoted for the seven-year fit.

\begin{deluxetable}{ccccccccccccc}
\tablecolumns{13}
\tablewidth{0pt}
\tabletypesize{\tiny}

\setlength{\tabcolsep}{0.025in}

\tablecaption{Derived Saturn Temperatures Per Observing Season Per DA
\label{tab:saturn}}

\tablehead{
   \colhead{Season\tablenotemark{a}} &
   \colhead{wRJD\tablenotemark{b}} &
   \colhead{$B$\tablenotemark{c}} &
   \multicolumn{10}{c}{$T_b$ (K)\tablenotemark{d}} \\
   \cline{4-13} \\ & & &
   \colhead{K}  &
   \colhead{Ka} &
   \colhead{Q1} &
   \colhead{Q2} &
   \colhead{V1} &
   \colhead{V2} &
   \colhead{W1} &
   \colhead{W2} &
   \colhead{W3} &
   \colhead{W4}
  }
\startdata
 1& 2172.50 &  -26& $133.5 \pm  1.5 $ & $141.0 \pm  1.2 $ & $145.6 \pm  1.4 $ & $149.2 \pm  1.4 $ & $156.9 \pm  1.2 $ & $156.7 \pm  1.1 $ & $164.2 \pm  1.1 $ & $164.4 \pm  1.4 $ & $166.2 \pm 1.4 $ & $165.9 \pm  1.3 $ \\
 2& 2302.56 &  -26& $133.6 \pm  1.6 $ & $142.6 \pm  1.3 $ & $145.7 \pm  1.4 $ & $147.9 \pm  1.3 $ & $154.9 \pm  1.2 $ & $156.4 \pm  1.1 $ & $161.4 \pm  1.2 $ & $165.5 \pm  1.4 $ & $164.3 \pm 1.4 $ & $163.8 \pm  1.3 $ \\
 3& 2551.27 &  -26& $130.9 \pm  1.6 $ & $141.6 \pm  1.2 $ & $149.2 \pm  1.3 $ & $149.9 \pm  1.3 $ & $158.1 \pm  1.2 $ & $157.4 \pm  1.1 $ & $165.9 \pm  1.2 $ & $166.9 \pm  1.4 $ & $164.0 \pm 1.4 $ & $164.3 \pm  1.3 $ \\
 5& 2928.95 &  -25& $131.2 \pm  1.5 $ & $138.4 \pm  1.2 $ & $144.1 \pm  1.3 $ & $146.1 \pm  1.3 $ & $153.4 \pm  1.2 $ & $153.4 \pm  1.1 $ & $161.2 \pm  1.2 $ & $162.0 \pm  1.4 $ & $160.5 \pm 1.4 $ & $159.8 \pm  1.3 $ \\
 7& 3305.67 &  -22& $125.8 \pm  1.5 $ & $135.3 \pm  1.2 $ & $140.2 \pm  1.3 $ & $140.1 \pm  1.3 $ & $147.2 \pm  1.1 $ & $147.9 \pm  1.1 $ & $154.0 \pm  1.1 $ & $154.2 \pm  1.4 $ & $154.2 \pm 1.4 $ & $153.2 \pm  1.2 $ \\
 8& 3437.14 &  -24& $129.9 \pm  1.6 $ & $137.8 \pm  1.3 $ & $141.0 \pm  1.5 $ & $141.7 \pm  1.4 $ & $147.9 \pm  1.3 $ & $150.2 \pm  1.1 $ & $155.0 \pm  1.2 $ & $159.3 \pm  1.5 $ & $159.8 \pm 1.5 $ & $156.9 \pm  1.3 $ \\
 9& 3685.29 &  -17& $121.4 \pm  1.5 $ & $130.6 \pm  1.2 $ & $134.8 \pm  1.3 $ & $134.1 \pm  1.3 $ & $140.9 \pm  1.2 $ & $141.3 \pm  1.1 $ & $146.2 \pm  1.1 $ & $146.9 \pm  1.4 $ & $147.1 \pm 1.4 $ & $146.3 \pm  1.3 $ \\
10& 3794.29 &  -20& $125.1 \pm  2.0 $ & $131.3 \pm  1.6 $ & $134.5 \pm  3.5 $ & $132.8 \pm  4.1 $ & $143.4 \pm  1.6 $ & $142.2 \pm  1.4 $ & $150.0 \pm  1.5 $ & $150.0 \pm  2.1 $ & $148.7 \pm 2.2 $ & $150.7 \pm  1.7 $ \\
11& 4061.48 &  -12& $122.9 \pm  1.5 $ & $129.9 \pm  1.2 $ & $131.5 \pm  1.3 $ & $137.3 \pm  1.3 $ & $139.8 \pm  1.2 $ & $140.4 \pm  1.1 $ & $141.9 \pm  1.1 $ & $144.6 \pm  1.4 $ & $143.1 \pm 1.4 $ & $143.2 \pm  1.2 $ \\
12& 4189.02 &  -15& $121.5 \pm  2.0 $ & $132.1 \pm  1.7 $ & $131.4 \pm  1.4 $ & $135.5 \pm  1.4 $ & $140.4 \pm  1.5 $ & $140.8 \pm  1.4 $ & $143.1 \pm  1.5 $ & $143.7 \pm  1.3 $ & $143.1 \pm 1.3 $ & $142.4 \pm  1.7 $ \\
13& 4436.82 &   -7& $128.1 \pm  1.6 $ & $131.5 \pm  1.2 $ & $135.3 \pm  1.4 $ & $137.8 \pm  1.3 $ & $140.3 \pm  1.2 $ & $139.9 \pm  1.1 $ & $143.0 \pm  1.2 $ & $146.2 \pm  1.4 $ & $141.3 \pm 1.4 $ & $144.8 \pm  1.3 $ \\
14& 4570.98 &  -10& $122.8 \pm  1.6 $ & $129.7 \pm  1.3 $ & $132.3 \pm  1.3 $ & $133.0 \pm  1.3 $ & $139.9 \pm  1.2 $ & $141.1 \pm  1.1 $ & $140.0 \pm  1.2 $ & $141.4 \pm  1.4 $ & $141.4 \pm 1.4 $ & $140.1 \pm  1.4 $ \\
15& 4814.77 &   -1& $130.6 \pm  1.6 $ & $137.2 \pm  1.3 $ & $139.1 \pm  1.4 $ & $139.4 \pm  1.4 $ & $144.5 \pm  1.2 $ & $147.2 \pm  1.1 $ & $146.6 \pm  1.2 $ & $149.4 \pm  1.5 $ & $146.8 \pm 1.5 $ & $146.5 \pm  1.3 $ \\
16& 4949.58 &   -4& $127.4 \pm  1.6 $ & $131.5 \pm  1.2 $ & $138.0 \pm  1.3 $ & $139.9 \pm  1.3 $ & $142.6 \pm  1.2 $ & $142.4 \pm  1.1 $ & $144.8 \pm  1.2 $ & $143.7 \pm  1.5 $ & $144.9 \pm 1.5 $ & $146.1 \pm  1.3 $ \\
17& 5191.93 &    5& $125.9 \pm  1.7 $ & $132.6 \pm  1.3 $ & $136.9 \pm  1.4 $ & $136.9 \pm  1.4 $ & $141.4 \pm  1.2 $ & $141.6 \pm  1.1 $ & $143.5 \pm  1.2 $ & $145.0 \pm  1.5 $ & $146.0 \pm 1.5 $ & $144.2 \pm  1.3 $ \\
18& 5326.82 &    2& $128.8 \pm  1.7 $ & $134.7 \pm  1.3 $ & $138.5 \pm  1.4 $ & $137.6 \pm  1.4 $ & $143.9 \pm  1.2 $ & $145.7 \pm  1.1 $ & $145.2 \pm  1.2 $ & $146.5 \pm  1.5 $ & $144.6 \pm 1.5 $ & $148.0 \pm  1.4 $ \\
\enddata
  \tablenotetext{a}{Seasons 4 and 6 omitted from analysis because Saturn is aligned with the Galactic plane.}

  \tablenotetext{b}{Approximate mean time of observations in
                    each season: wRJD = Julian Day $-2450000$.}
  \tablenotetext{c}{Approximate mean ring opening angle for
                    each season, degrees.}
  \tablenotetext{d}{Brightness temperature calculated for a solid angle $\Omega_{\mathrm{ref}}= 5.096 \times 10^{-9}$ sr at a
               fiducial distance of 9.5 AU.  A correction for planetary disk oblateness has not been applied,
              as that is accounted for in modeling.
              Temperature is with respect to blank sky: absolute
              brightness temperature is obtained by adding 2.2, 2.0, 1.9, 1.5 and 1.1 K in bands
              K, Ka, Q, V and W respectively \citep{page/etal:2003b}.}

\end{deluxetable}

\begin{deluxetable}{ccccccc}
\tablecolumns{7}
\tablewidth{0pt}
\tablecaption{Nine-Year Saturn Model Fit Parameters\label{tab:satmodel2}\tablenotemark{a}}
\tablehead{
    \colhead{Freq} &
    \multicolumn{3}{c}{Disk} &
    \multicolumn{3}{c}{Rings} \\ 
    \colhead{Band}&
    \colhead{$T_{\mathrm{disk}}$} &
    \colhead{$\sigma_{\mathrm{fit}}$} &
    \colhead{$\sigma_{\mathrm{adopted}}$} &
    \colhead{$T_{\mathrm{ring}}$} &
    \colhead{$\sigma_{\mathrm{fit}}$} &
    \colhead{$\sigma_{\mathrm{adopted}}$} \\ &
    \colhead{[K]} &
    \colhead{[K]} &
    \colhead{[K]} &
    \colhead{[K]} &
    \colhead{[K]} &
    \colhead{[K]} }
\startdata
\hline
 K   &  132.2 & 0.8 & 1.7 &   8.0   &    0.8  &  1.5 \\
 Ka  &  137.8 & 0.6 & 1.5 &  10.6   &    0.7  &  1.4 \\
 Q   &  141.6 & 0.5 & 1.4 &  11.9   &    0.6  &  1.3 \\
 V   &  146.6 & 0.4 & 1.3 &  14.5   &    0.5  &  1.2 \\
 W   &  147.3 & 0.3 & 1.2 &  18.9   &    0.3  &  1.0 \\
\hline
\enddata
\tablenotetext{a}{ A frequency independent maximum ring-normal optical depth, $\tau_{0,\mathrm{max}}$ is also a 
                 fit parameter.  Its fit value is 2.1, with a statistical error $\sigma_{\mathrm{fit}} = 0.3$;
		 the seven-year model used a fixed value of $2.0$.}
\end{deluxetable}

\subsubsection{Point Source Catalogs} \label{sssec:point_source_cats}

As for the seven-year analysis, two separate methods have been used for the identification of
point sources from \wmap\ maps and two separate point source tables have been produced.
Both methods are largely unchanged from the seven-year analysis \citep{gold/etal:2011}.
Since the use of beam-symmetrized maps would result in only minor
changes to the recovered source fluxes and since there is benefit to
continuity with previous \wmap\ point source analyses, we have
generated the source catalogs from maps that are not deconvolved.
The first method searches for point sources in each of the five \WMAP\ wavelength bands.
The nine-year signal-to-noise ratio map in each band is filtered in harmonic space by
$b_l/(b^2_lC^{\rm cmb}_l + C^{\rm noise}_l)$, where $b_l$ is the transfer function of the \map\ beam response,
$C^{\rm cmb}_l$ is the CMB angular power spectrum, and $C^{\rm noise}_l$ is the noise power 
\citep{tegmark/deoliveira-costa:1998,
refregier/spergel/herbig:2000}.
The filtering suppresses CMB and Galactic foreground fluctuations relative to point
sources.  For each peak in the filtered maps that is $\gt 5\sigma$ in any band, the unfiltered
temperature map in each band is fit with the sum of a planar base level and a beam template
formed by convolving an azimuthally symmetrized beam profile with a skymap pixel.
(This method was previously used by \citet{weiland/etal:2011} for selected celestial calibration
sources and is more accurate than the Gaussian fitting that was used for the seven-year and
earlier point source analyses.)  The peak temperature from each beam template fit is converted
to a source flux density using the conversion factor $\Gamma$ given in Table~\ref{tab:beam_quantities}.
The flux density uncertainty is calculated from the $1 \sigma$ uncertainty in the peak temperature,
and does not include any additional uncertainty due to Eddington bias.
Uncertainty due to beam asymmetry effects has been found to be negligible, about
0.1\% or less, by comparing results from beam template fits to the normally
processed K-band map with those to the beam-symmetrized K-band map for Tau A,
Cas A, and Cyg A.
Flux density values are entered into the catalog for bands where they exceed $2\sigma$ and
where the source width from an initial Gaussian fit is within a factor of two of the beam width.
A point source catalog mask is used to exclude sources in the Galactic plane and Magellanic cloud
regions. This mask has changed from the seven-year analysis in accordance with changes
in the KQ85 temperature analysis mask.  A map pixel is outside of the nine-year point source catalog mask
if it is either outside of the diffuse component of the nine-year KQ85 temperature analysis mask
or outside of the seven-year point source catalog mask.  The new catalog mask admits 83\% of the sky.

The second method of point source identification is the CMB-free method originally applied
to one-year and three-year V- and W-band maps by \citet{chen/wright:2008} and to five-year
V- and W-band maps by \citet{wright/etal:2009}.  The method used here is that applied to
five-year Q-, V-, and W-band maps by \citet{chen/wright:2009} and to seven-year Q-, V-, and W-band maps
by \citet{gold/etal:2011}.  The V- and W-band maps are smoothed to Q-band resolution.  An
internal linear combination (ILC) map (see Section \ref{sec:ilc} ) is then formed from the three maps using weights such
that CMB fluctuations are removed, flat-spectrum point sources are retained with fluxes
normalized to Q-band, and the variance of the ILC map is minimized.  The ILC map is filtered
to reduce noise and suppress large angular scale structure.  Peaks in the filtered map
that are $\gt 5\sigma$ and outside of the nine-year point source catalog mask are identified as
point sources, and source positions are obtained by fitting the beam profile plus a baseline to
the filtered map for each source.  For the nine-year analysis, the position of the brightest
pixel is adopted instead of the fit position in rare instances where they differ
by $>0.1\degrees$.  Source fluxes are estimated by integrating the Q, V, and W
temperature maps within $1.25\degrees$ of each source position, with a weighting
function to enhance the contrast of the point source relative to background fluctuations,
and applying a correction for Eddington bias due to noise (sometimes called ``deboosting'').

We identify possible 5 GHz counterparts to the \WMAP\ sources found by both methods by
cross-correlating with the GB6
\citep{gregory/etal:1996}, PMN \citep{griffith/etal:1994, griffith/etal:1995, wright.a/etal:1994,
wright.a/etal:1996}, \citet{kuehr/etal:1981}, and \citet{healey/etal:2009} catalogs.  
A 5 GHz source is identified as a counterpart if it lies within $11\arcmin$ of
the \map\ source position (the mean \map\ source position uncertainty is $4\arcmin$).
When two or more 5 GHz sources are within $11\arcmin$, the brightest is assumed to be
the counterpart and a multiple identification flag is entered in the catalog.

The nine-year five-band point source catalog is presented in 
Appendix \ref{5bandsources}
and the nine-year QVW point source catalog is presented in 
Appendix \ref{3bandsources}.
The five-band
catalog contains 501 sources, the QVW catalog contains 502 sources, and the two catalogs have
387 sources in common.  As noted by \citet{gold/etal:2011}, differences in the source populations
detected by the two search methods are largely caused by Eddington bias in the five-band source
detections due to CMB fluctuations and noise.  At low
flux levels, the five-band method tends to detect point sources located on positive CMB fluctuations
and to overestimate their fluxes, and it tends to miss sources located in negative CMB fluctuations.
Other point source detection methods have been applied to {\WMAP} data and have identified sources
not found by our methods (e.g., \citet{scodeller/etal:2012, lanz/etal:2012, ramos/etal:2011}, and
references therein).

\subsection{Diffuse Foregrounds} \label{ssec:diffuse_fg}

\subsubsection{Introduction to diffuse foreground analysis}\label{sssec:diffuse_fg_intro}

In this section we evaluate the diffuse foreground emission both for the purpose of separation from the CMB anisotropy and for characterizing the nature of the foreground components.  As a prelude to our cosmological analyses we fit and remove external foreground template map data from the \wmap\ maps and we mask remaining regions estimated to be significantly contaminated.  We discuss this temperature and polarization cleaning, and the masks, below.  To elucidate the characteristics and nature of the diffuse foregrounds we implement four techniques: internal linear combination (ILC) technique; Maximum Entropy Method (MEM); Markov Chain Monte Carlo (MCMC) fits; and $\chi^2$ fits.

Our analysis of the diffuse foregrounds generally uses the five bands of \wmap\ data in conjunction with other data sets.  \wmap\ was designed to observe in the spectral region where the ratio of the CMB to the foregrounds is at its maximum. This minimizes the amplitude of contamination and needed corrections or masking, which is good for cosmology.  To achieve an improved signal-to-noise ratio of the foregrounds themselves, it is sometimes useful to use external data where the foreground emission is weak.  

Foreground analyses are done using $1\arcdeg$ smoothed
beam-symmetrized nine-year temperature maps in the five \WMAP\ bands.
As in our previous foreground studies, the zero level of each
map is set such that a fit to the ILC-subtracted map of the form
$T(\vert b \vert) = T_p \, \csc \vert b \vert + c$, over the range
$-90^{\circ} < b < -15^{\circ}$, yields $c=0$.  This assumes a
plane-parallel slab model for the Galactic emission.  Formal
$1\sigma$ uncertainties in the map zero levels (calculated as
the quadrature sum of (1) the uncertainty in the fit intercept $c$
and (2) the difference in intercepts from southern and northern
Galactic hemisphere fits) are 7.2, 5.9, 3.6, 1.8, and 0.76 $\mu$K
in thermodynamic units for K-, Ka-, Q-, V-, and W-bands respectively.
The South Galactic pole brightness $T_p$ from the fitting is
$77.9 \pm 1.5$, $30.1 \pm 0.6$, $17.7 \pm 0.4$, $8.6 \pm 0.2$, and $9.4 \pm 0.3\, \mu$K
in thermodynamic units for K-, Ka-, Q-, V-, and W-bands respectively. 
The Stokes Q and U maps have well-defined zero levels and
no monopole corrections are applied to them.

Previous \wmap\ team analyses have used the \citet{finkbeiner:2003} H$\alpha$
map corrected for extinction as a template for free-free emission
\citep{bennett/etal:2003c}.  The Finkbeiner map is a composite of the 
Virginia Tech Spectral line Survey \citep{dennison/simonetti/topasna:1998},
the Southern H-alpha Sky Survey Atlas \citep{gaustad/etal:2001},
and the Wisconsin H alpha Mapper survey \citep{haffner/etal:2003}.  The
extinction correction assumes that H$\alpha$ emission and extinction
are uniformly mixed along each line of sight,
\begin{equation} \label{eq:extcorr}
I(\rm H\alpha)_{\rm extinction-corrected} = \it I(\rm H\alpha) \: \tau/(1-e^{-\tau}).
\end{equation}
Here $\tau$ is the dust optical depth at the wavelength of H$\alpha$ and
was calculated from the $E(B-V)$ map of \citet{schlegel/finkbeiner/davis:1998} as
\begin{equation}
\tau = 2.2 \thinspace E(B-V),
\end{equation}
which assumes an extinction law for $R_V = 3.1$, characteristic of
the diffuse interstellar medium.

\begin{figure}
\epsscale{0.70}
\plotone{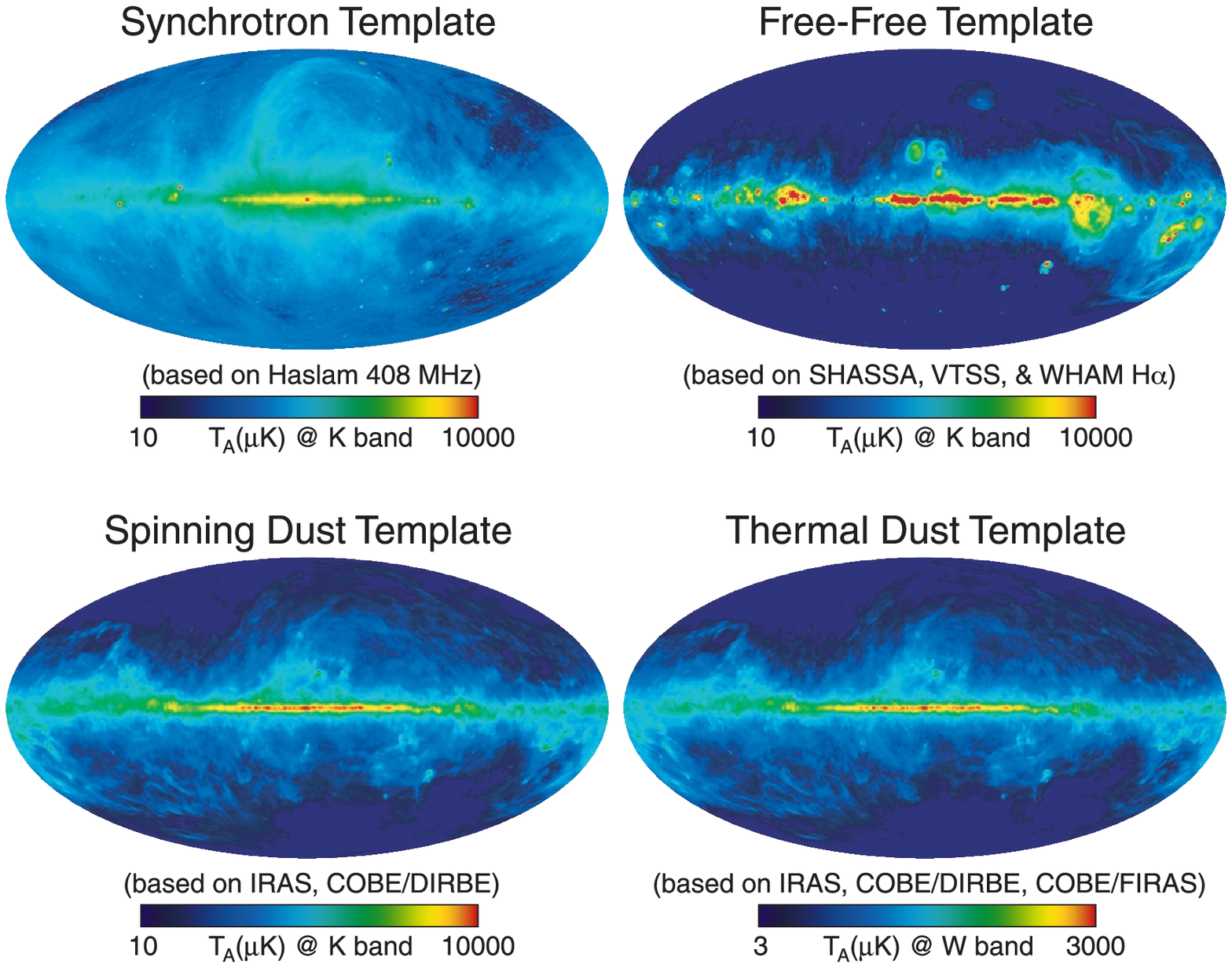}
\caption{Foreground evaluation is generally based on a combination of the data from the five \wmap\ bands and external observations where the CMB contamination is negligible.  The external observations used for foreground fitting and template removal are shown. These provide approximate probes of the synchrotron, free-free, spinning dust, and thermal dust emission.
\vspace{1mm} \newline (A color version of this figure is available in the online journal.)
}
\label{fig:fgpriors}
\end{figure}

Recent studies of selected dust clouds at $20\arcdeg < \vert b \vert < 40\arcdeg$
have shown that scattered
H$\alpha$ can make a significant contribution to the observed H$\alpha$
brightness for some lines of sight \citep{mattila/etal:2007,
lehtinen/etal:2010, witt/etal:2010}.  Here we apply an approximate
correction to our previous H$\alpha$ template for the contribution of
scattered H$\alpha$, based on correlations between H$\alpha$ and 100 $\mu$m
emission found by \citet{witt/etal:2010} for four selected clouds and by
\citet{brandt/draine:2012} for Sloan Digital Sky Survey blank sky regions at
intermediate to high Galactic latitudes.
Brandt and Draine noted that $I(100 \mu \rm m)$ varies in proportion to the product of
the dust column density and the radiation field that heats the dust.  If the
spatial variation of the illuminating H$\alpha$ radiation field in the Galaxy is
similar to that of the radiation responsible for dust heating, $I(100 \mu \rm m)$ may
be a good tracer of scattered H$\alpha$.  The scattering correction we adopt is
\begin{equation} \label{eq:scattcorr}
I(\rm H\alpha)_{\rm scattering-corrected} = \it I(\rm H\alpha)_{\rm extinction-corrected} - 0.11 \thinspace \it I(\rm 100 \mu m),
\end{equation}
where $I(\rm H\alpha)$ is in Rayleighs, $I(100 \mu \rm m)$ is the \citet{schlegel/finkbeiner/davis:1998}
100 $\mu$m map in MJy sr$^{-1}$, and the $I(100 \mu \rm m)$ coefficient is a mean of the values of
$0.129 \pm 0.015$ R/(MJy sr$^{-1}$) found by \citet{witt/etal:2010} and $0.090 \pm 0.017$ R/(MJy sr$^{-1}$)
found by \citet{brandt/draine:2012}.  These
correlation slopes were measured for regions with $\tau < 1$, but we apply
Equation~\refeqn{eq:scattcorr}
over the entire sky.  This assumes that the Equation~\refeqn{eq:extcorr} extinction correction
is valid for the scattered component (\it i.e.\rm, the scattered H$\alpha$ emissivity and the dust
extinction are uniformly mixed along each line of sight) and it neglects
effects of multiple scattering that may be important for lines of sight
with high optical depth.  The H$\alpha$ template is made by applying the corrections for 
extinction and scattering to version 1.1 of the Finkbeiner H$\alpha$ map, smoothing 
from 6$\arcmin$ FWHM to 1$\arcdeg$ FWHM, and setting a small number of negative pixels to zero.
The resulting H$\alpha$-based microwave template is shown in Figure \ref{fig:fgpriors} as the 
``Free-Free Template''.

Uncertainties in both the extinction correction and
the scattering correction are large for high $\tau$, but we find that results of
our analyses using the template are not sensitive to these uncertainties.
For the foreground cleaning of the temperature maps, the mask used in
template fitting is chosen to minimize the combined effects of template
error and foreground-CMB covariance (Section \ref{sssec:tmpl_cln}).  
For the MEM foreground fitting, the free-free prior is
formed from the H$\alpha$ template, but for high $\tau$ lines of sight the 
observed brightness in the \wmap\ bands is great enough that the MEM results are not
strongly affected by the free-free prior.

Prior to the nine-year analysis, the Haslam map used in the MCMC fitting and as
a prior in the MEM fitting was the Fourier-filtered version available from LAMBDA.
This version mitigates scan striping in the Haslam map, but also removes
many strong point sources.  Removal of the point sources affected the
quality of some foreground fits for pixels in the Galactic plane.
For this reason, the unfiltered Haslam map (also available on LAMBDA) is
now used for these applications and its projection to K-band is shown in Figure \ref{fig:fgpriors}. 

\clearpage

\subsubsection{Template Cleaning and Masks\label{sssec:tmpl_cln}}

All-sky templates of Galactic foregrounds or combinations of foregrounds which are 
``CMB-free'' are fit in a least-squares sense to the \wmap\ sky maps
to construct a foreground model at each frequency.  The foreground model is subtracted from the
\wmap\ sky maps to produce reduced foreground, or ``cleaned'' maps, which are used in turn for 
power spectrum analysis.  The cleaning is applied to sky maps from the standard map-making
procedure, not to beam-symmetrized sky maps.  Cleaning of temperature and polarization maps is
treated independently. 

\paragraph{Temperature cleaning}
\label{sec:temperature_cleaning}

A limited set of all-sky foreground templates is available for use in
modeling potential contributions from synchrotron, free-free and dust 
emission.  After testing a number of different template combinations, 
we continue to adopt a foreground model map, $M(\nu,p)$, of the form
\begin{equation}\label{eqn:kkahadust}
M(\nu,p) = c_\mathrm{1}(\nu)[T_\mathrm{K}(p) - T_\mathrm{Ka}(p)] + c_\mathrm{2}(\nu)I_\mathrm{H\alpha}(p) +
              c_\mathrm{3}M_\mathrm{dust}(p),
\end{equation}
where $p$ indicates the pixel, the frequency dependence is entirely contained in the coefficients $c_i$,
and the spatial templates are the \wmap\ K-Ka temperature difference map in thermodynamic mK ($T_K - T_{Ka}$),
an H$\alpha$ map ($I_{H\alpha}$) in units of Rayleighs, and dust model 8 from \citet{finkbeiner/davis/schlegel:1999}
evaluated at 94 GHz in units of mK antenna temperature ($M_{dust}$).  The K-Ka template is formed using
standard (not beam symmetrized) maps. The values of the coefficients $c$
are such that the model map $M(\nu,p)$ is in thermodynamic mK.

However, although the form of the model is the same as that used in previous \wmap\ analyses, there
are modifications in the details of its application.  
As described in Section~\ref{sssec:diffuse_fg_intro}, the nine-year extinction corrected H$\alpha$ template
incorporates a scattering correction, a refinement not present in the seven-year analysis.
Also, in recognition of the possible contribution of spinning dust to the
Galactic emission and the uncertain synchrotron behavior with frequency, 
spectral and coefficient positivity constraints are no longer imposed in the template fitting.  
This allows maximum freedom in the fit, but makes physical
interpretation of model coefficients more difficult.  

There has also been a change in the portion of sky used in computing the foreground model fit.
Derived model coefficients are dependent on the fraction of the sky 
which is fit:  a full sky fit minimizes the covariance of the 
templates with the CMB signature in the \wmap\ data, but maximizes potential template cleaning residuals (bias) 
by including sky regions where the templates are more uncertain (generally close to the Galactic plane).
For example, the extinction correction applied to the H$\alpha$ map is only approximate and 
this template is an imperfect tracer of free-free emission in optically thick regions.
In general, as more sky is excluded from the fit, CMB-template covariance increases, while
template cleaning bias decreases.
The ``optimal'' sky cut for template fitting may be determined by examining 
these two competing errors as a function of sky cut, and choosing the mask for
which the sum of the two errors is a minimum.
For this purpose, several simulated five-band Galaxy models 
of differing complexity were constructed.  Each model is added to a CMB realization, and
then cleaned using the algorithm in Equation~\refeqn{eqn:kkahadust} and a chosen sky cut.  This is performed for 
100 CMB realizations per sky cut;
the mean bias is the template cleaning error and the variance is the CMB covariance.
We have used the ``KpX'' series of Galactic masks, described by
\citet{bennett/etal:2003} as a graduated set of sky cuts.  The masking in the ``KpX'' series is based on 
K-band intensity: higher values of X indicate a smaller portion of bright sky is cut.
For each simulation, the sum of both errors were plotted as a function of sky cut and a rough minimum chosen.
Prior to the nine-year analysis, we had used the Kp2 mask for template fitting.  However, the
simulations indicated a more conservative choice would employ a smaller sky cut.  The Kp8 mask was adopted
for the nine-year cleaning.

Template cleaning coefficients derived using the updated procedure are shown in
Table~\ref{tab:kkahadust_coeff} for the Q,V and W DAs.  As noted previously, the ability of
the fit to trade freely among the three templates makes physical interpretation difficult.
Monte Carlo simulations have shown that the negative coefficients $c_1$ derived for W-band result from 
template covariance with the CMB. 
The change of template cleaning method from the seven-year to the nine-year
analysis has little effect on power spectrum analysis.  There is a slight 
change in the evaluated low-$\ell$ power spectrum.
For $2\leq \ell \leq 16$, using the MASTER method with the KQ85y9 mask,
the absolute value of the change in $\ell(\ell+1)/(2\pi)C_\ell$ due to
the change in template cleaning is typically $4\%$ of cosmic variance per $\ell$.

\begin{deluxetable}{cccc}
\tablewidth{0pt}
\tablecaption{Template Cleaning Temperature Coefficients\label{tab:kkahadust_coeff}}
\tablehead{
\colhead{DA\tablenotemark{a}} &
\colhead{$c_1$\tablenotemark{b}}  &
\colhead{$c_2$ ($\mu$K/R$^{-1}$)\tablenotemark{b}} & 
\colhead{$c_3$\tablenotemark{b}}
}
\startdata
  Q1   &     0.284    &    0.890    &     0.231 \\	     
  Q2   &     0.284    &    0.898    &     0.226 \\	     
  V1   &     0.0630   &    0.554    &     0.686 \\	     
  V2   &     0.0567   &    0.541    &     0.716 \\	     
  W1   &    -0.0179   &    0.351    &     1.609 \\	     
  W2   &    -0.0182   &    0.349    &     1.617 \\	     
  W3   &    -0.0146   &    0.342    &     1.587 \\	     
  W4   &    -0.0153   &    0.345    &     1.594 \\	     
\enddata
\tablenotetext{a}{\wmap\ has two differencing assemblies (DAs) for the Q- and V-bands, and four for
              the W-band; the high signal-to-noise total intensity allows each DA to be
	      fit independently.}
\tablenotetext{b}{The $c_i$ coefficients produce model maps in thermodynamic mK.}
\end{deluxetable}

\paragraph{Polarization cleaning}

\begin{figure}
\epsscale{0.70}
\plotone{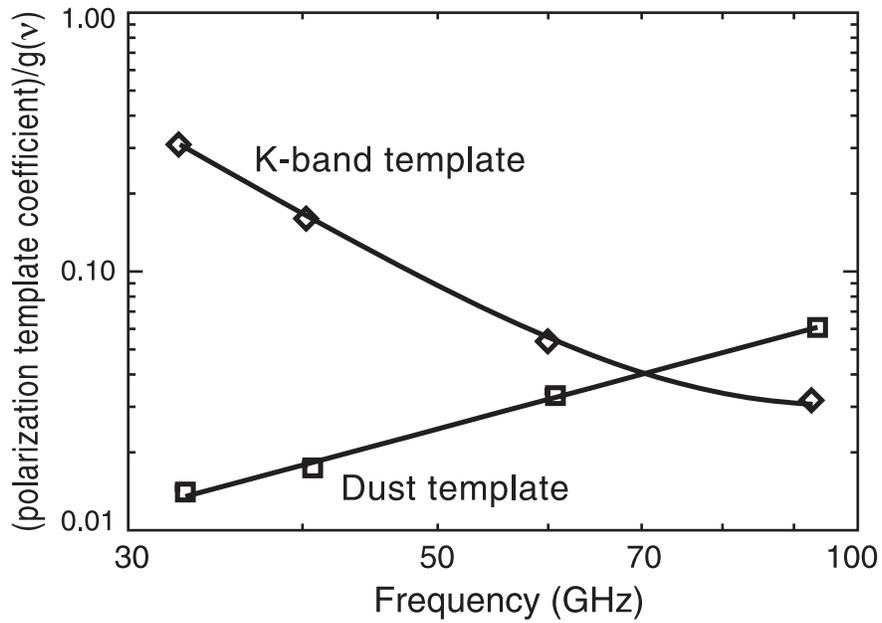}
\caption{Polarization template coefficients, scaled to produce model maps in antenna temperature,
as a function of frequency.  The curves show the predictions of a simple model with synchrotron and 
thermal dust polarization in which about
2/3 of the dust polarization is traced by the dust template and about 1/3 is traced by the
K-band template.}
\label{fig:kdpol_coefs}
\end{figure}

The polarization cleaning method is unchanged from the seven-year analysis.
The nine-year Stokes Q and U maps are degraded to low resolution ($N_\mathrm{side}$=16)
and the data for pixels outside of the Q-band processing mask are fit to a linear combination
of low resolution templates.  The fit has the form
\beq
  [Q(\nu), U(\nu)]  = a_1(\nu) \thinspace [Q, U]_\mathrm{K} + a_2(\nu) \thinspace [Q, U]_\mathrm{dust}\label{eqn:kdpol1}.
\eeq
The template used for synchrotron is the nine-year \WMAP\ K-band polarization, $[Q,U]_\mathrm{K}$.
The template for dust, $[Q,U]_\mathrm{dust}$, is constructed from the \cite{finkbeiner/davis/schlegel:1999}
dust model 8 evaluated at 94 GHz together with a polarization direction map derived from starlight
measurements and a geometric suppression map to account for the magnetic field geometry, as described in
\cite{page/etal:2007}.  The coefficients of the fit to the nine-year data are listed in
Table \ref{tab:poltemplate} and plotted against frequency in Figure \ref{fig:kdpol_coefs}.

\begin{deluxetable}{ccccc}
\tablewidth{0pt}
\tablecaption{Template cleaning polarization coefficients\label{tab:poltemplate}}
\tablehead{
\colhead{Band} & \colhead{$a_1$\tablenotemark{a}}  &
\colhead{$\beta_s(\nu_\mathrm{K},\nu)$\tablenotemark{b}} & \colhead{$a_2$\tablenotemark{a}} & 
\colhead{$\beta_d(\nu,\nu_\mathrm{W})$\tablenotemark{b}}
}
\startdata
Ka &   0.3204  &   -3.13  &  0.0145 &    1.41 \\
Q  &  0.1682  &  -3.13 &  0.0182    &   1.50 \\
V  &  0.0594  & -2.97   &  0.0364  &   1.41 \\
W  &  0.0398   &  -2.43  &  0.0758 &   \nodata \\
\enddata
\tablenotetext{a}{The $a_i$ coefficients are dimensionless and produce model maps in thermodynamic mK.}
\tablenotetext{b}{The spectral indices refer to antenna temperature.}
\end{deluxetable}

Full-resolution ($N_\mathrm{side} = 512$) foreground-reduced Stokes Q and U maps were produced using the
coefficients in Table \ref{tab:poltemplate} with full-resolution versions of the K-band and dust
polarization templates smoothed to $1\arcdeg$ FWHM.  In making the full resolution dust template,
the starlight polarization map used to determine polarization direction was upgraded to full resolution
using nearest neighbor sampling.  Smoothing of the templates to $1\arcdeg$ FWHM potentially leaves
artifacts in the foreground-reduced maps due to small-scale power or beam asymmetries, but previous analyses
have found no sign of these effects \citep{gold/etal:2011}.  Data sets for all templates are available
on the LAMBDA website.

The spectrum of K-band polarization template coefficients flattens significantly with increasing
frequency, which is unexpected for synchrotron emission.  This flattening can be understood if,
due to shortcomings of the dust polarization template, some fraction of the dust polarization
is traced by the K-band template.  We illustrate this using a simple model.
The polarization maps are modeled as a sum of synchrotron and thermal dust components,
\begin{equation}
  [Q(\nu), U(\nu)]  = [Q(\nu), U(\nu)]_\mathrm{synch} + [Q(\nu), U(\nu)]_\mathrm{dust}\label{eqn:kdpol2}.
\end{equation}
Assuming the synchrotron polarization has a power law spectrum and is
traced exactly in all bands by the K-band polarization template,
the synchrotron component is
\begin{equation}
  [Q(\nu), U(\nu)]_\mathrm{synch} = \frac {g(\nu)} {g(\nu_K)} \left (\frac {\nu} {\nu_K} \right)^{\beta_{synch}} [Q, U]_\mathrm{K}\label{eqn:kdpol3},
\end{equation}
where the antenna temperature to thermodynamic temperature conversion factors $g$ are needed
because the polarization maps and K-band template are in thermodynamic units.
Assuming the dust polarization has a power law spectrum and is traced by a combination
of the dust polarization template and the K-band polarization template, with the relative
contributions of the two templates independent of frequency, the dust component is
\begin{equation}
 [Q(\nu), U(\nu)]_\mathrm{dust} = \frac {g(\nu)} {g(\nu_W)} \left (\frac {\nu} {\nu_W} \right)^{\beta_{dust}} (f_1 \thinspace [Q,U]_\mathrm{dust} + f_2 \thinspace [Q, U]_\mathrm{K})\label{eqn:kdpol4},
\end{equation}
where $f_1$ and $f_2$ are constants.
Inserting Equations \refeqn{eqn:kdpol3} and \refeqn{eqn:kdpol4} in Equation~\refeqn{eqn:kdpol2} and
comparing with Equation~\refeqn{eqn:kdpol1} gives expressions for the template fit coefficients,
\begin{equation}
a_1(\nu) = \frac {g(\nu)} {g(\nu_K)} \left (\frac {\nu} {\nu_K} \right)^{\beta_{synch}} + f_2 \thinspace \frac {g(\nu)} {g(\nu_W)} \left (\frac {\nu} {\nu_W} \right)^{\beta_{dust}}
\end{equation}
and
\begin{equation}
 a_2(\nu) = f_1 \thinspace \frac {g(\nu)} {g(\nu_W)} \left (\frac {\nu} {\nu_W} \right)^{\beta_{dust}} .
\end{equation}
Fitting these expressions to the $a_1(\nu)$ and $a_2(\nu)$ values in Table \ref{tab:poltemplate}
gives $\beta_{synch}=-3.13$, $\beta_{dust}=1.44$, $f_1=0.076$, and $f_2=0.024$.  The fits are
shown by the curves in Figure \ref{fig:kdpol_coefs}.  They match the template coefficients very well
with no need for an additional emission mechanism such as spinning dust or magnetic dust polarization.
In this simple model, the K-band template component contributes about 1/3 of the rms dust polarization
and the dust template component contributes about 2/3.

This suggests that there is room for improvement in the dust polarization template.  Some alternate dust
templates were tested in fitting the polarization maps, but none of them gave significant improvement in
$\chi^2$.  These include a template based on K-band polarization directions instead of directions from
starlight measurements, a template based on a geometric
suppression map calculated from the ratio of observed K-band polarized intensity to K-band synchrotron
total intensity from the seven-year MCMC shifted spinning dust model \citep{gold/etal:2011}, and two
templates from \cite{Odea/etal:2012} based on different Galactic magnetic field models.

\paragraph{Masks}

Sky masks for CMB temperature analysis are generated as described
by \citet{gold/etal:2011}.  The process begins with K- and Q-band
maps smoothed to $1\deg$ resolution, from which an estimate of the CMB
is subtracted to leave maps that effectively consist of foreground emission
alone.  The CMB is estimated using the internal linear combination (ILC) method
\citep{hinshaw/etal:2007}.  Both the K and the Q maps are masked at a flux
contour that leaves either $75\%$ or $85\%$ of the sky unmasked.  The K
and Q-band sky masks for each cut level are combined so that any pixel excluded by either
cut is excluded by the combination.  The resulting combinations, dominated by
diffuse Galactic emission, are called KQ75 and KQ85, labeled by the admitted
sky fraction ($f_\mathrm{sky}$) of the input masks.

These masks are intended primarily to be applied to the foreground-cleaned
versions of the sky maps for power spectrum and non-Gaussian analysis.  They
are made more effective for this purpose by extending them to include regions
where the cleaning algorithm is subject to possible systematic error.  A
$\chi^2$ analysis is done using differences $\textrm{Q}-\textrm{V}$ and
$\textrm{V}-\textrm{W}$ between cleaned band maps at a reduced HEALPix
resolution of $N_\mathrm{side}=32$ \citep{gorski/etal:2005}, or ``res 5'' in
\wmap\ terminology.  Regions of four or more contiguous pixels with $\chi^2$
higher than 4 times that of the polar caps are used to define the mask
extensions, after $3\deg$ smoothing and cleanup steps to remove small
``islands'' caused by noise.

A point source mask is added to each of the diffuse sky masks.  The point source
mask from the seven-year analysis is updated to include newly detected \wmap\ point
sources and the 100 GHz sources in the Planck early release compact source catalog.
An exclusion radius of $1.2\arcdeg$ is used for sources stronger than 5 Jy in any
band and an exclusion radius of $0.6\arcdeg$ is used for weaker sources. 

\begin{figure}
\epsscale{0.99}
\plotone{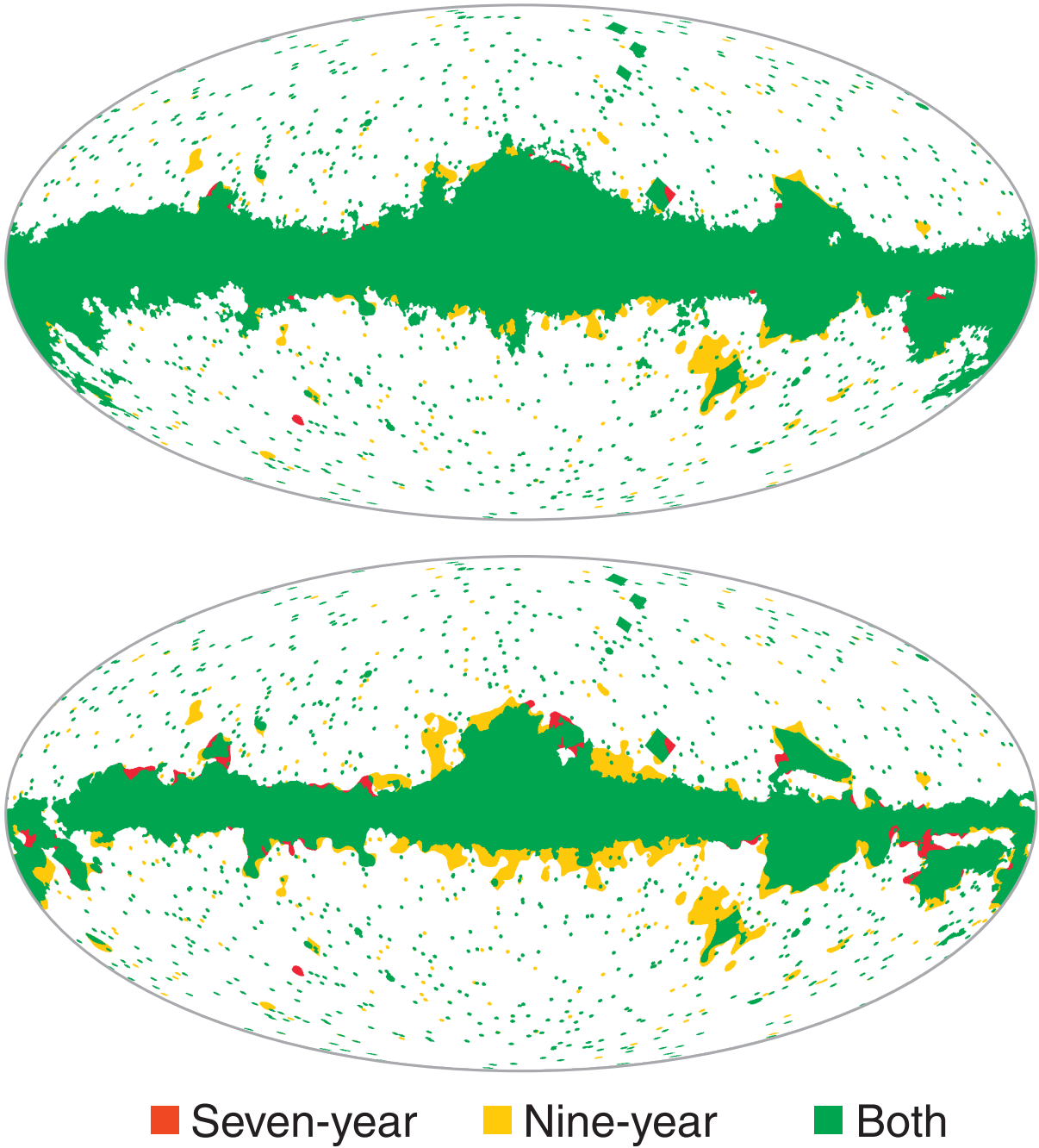}
\caption{Comparison of nine-year masks to seven-year masks.  At the
top KQ75y7 and KQ75y9 are compared, and at the bottom KQ85y7 and
KQ85y9.  Green regions are masked in both the nine-year and seven-year
masks, yellow regions are newly masked in the nine-year masks, and
red regions are masked in the seven-year masks but no longer in the
nine-year masks.
\vspace{1mm} \newline (A color version of this figure is available in the online journal.)
\label{fig:masks}}
\end{figure}

The nine-year versions of the final KQ85 and KQ75 sky masks, called KQ85y9 and
KQ75y9, respectively, are compared to the seven-year versions in Figure
\ref{fig:masks}.  Changes in the foreground cleaning residuals have altered the
morphology of the mask in the vicinity of the Gum Nebula, the Large Magellanic
Cloud, and the Galactic center, with the largest relative changes occurring in
the KQ85 mask.  For KQ75, $f_\mathrm{sky}$ is decreased from $70.6\%$ to
$68.8\%$, a difference of $1.8\%$ of the sky, and for KQ85, $f_\mathrm{sky}$
is decreased from $78.2\%$ to $74.8\%$, a difference of $3.7\%$ of the sky.

The sky mask for CMB polarization analysis is generated using cuts in K-band
polarized intensity and a model of polarized dust emission, together with
masking of point sources, as described by \citet{page/etal:2007} and
\citet{gold/etal:2009}.  The nine-year polarization mask is the same as the
seven-year version except that three additional point sources were added using
a 1 degree exclusion radius - Hydra A, HB89 1055+018, and BL Lac.

\clearpage

\subsubsection{Internal Linear Combination (ILC)} \label{sec:ilc}

The internal linear combination (ILC) method seeks to produce a map of the CMB anisotropy
from a linear combination of the five \wmap\ frequency bands.  A first application of the
method is described by \citet{bennett/etal:2003c}. The algorithm was revised slightly 
by \citet{hinshaw/etal:2007}; we refer to this version of the algorithm as the ``classic ILC", 
it has remained unchanged throughout subsequent \wmap\ data releases. 
As described in \citet{hinshaw/etal:2007}, the algorithm 
divides the sky into 12 regions -- a larger high latitude region and 11 smaller regions spread
across the galactic plane.  Use of the smaller regions along the plane allows for
spatially varying foreground complexity.  For each of these smaller regions, five band-weights are
computed by minimizing the temperature variance in the region, under the constraint that 
common-mode CMB signal is unaffected.  Weights for the larger high latitude region are computed 
in a similar manner, but using pixels from locations near the outer-Galactic plane.  Exact
definitions of these regions are provided on LAMBDA.

We compute the nine-year classic ILC using the coadded deconvolved band maps which have been
smoothed to a FWHM of $1\arcdeg$.  The weights applied to the 5 frequency maps for each of the 
12 sky regions are shown in Table~\ref{tab:ilc_coeffs}.  Values for the weights change slightly
compared to previous \wmap\ releases as pixel noise, calibration and beam profiles have been
refined. 

To the eye, the ILC presents a reasonably foreground-free image of the CMB anisotropy.
The beauty of the algorithm is that it is relatively independent of assumptions 
about foregrounds.  However, assessing the underlying uncertainty in the resultant anisotropy
map is a difficult problem which heavily relies on knowledge of the Galactic foregrounds.
In subsequent sections, we will discuss efforts to improve the classic ILC, as well as characterize
the level to which foreground residuals remain.

\begin{deluxetable}{cccccc}
\tablecolumns{6}
\tablewidth{0pt}
\tablecaption{ILC Coefficients per Region\label{tab:ilc_coeffs}\tablenotemark{a}}
\tablehead{ 
\colhead{Region} &
\colhead{K-band} &
\colhead{Ka-band} & 
\colhead{Q-band} &
\colhead{V-band} &
\colhead{W-band} 
}
\startdata
    0	&     0.1555	&  -0.7572    &  -0.2689    &  2.2845	&   -0.4138  \\
    1	&     0.0375	&  -0.5137    &   0.0223    &  2.0378	&   -0.5839  \\
    2	&     0.0325	&  -0.3585    &  -0.3103    &  1.8521	&   -0.2157  \\
    3	&    -0.0910	&   0.1741    &  -0.6267    &  1.5870	&   -0.0433  \\
    4	&    -0.0762	&   0.0907    &  -0.4273    &  0.9707	&    0.4421  \\
    5	&     0.1998	&  -0.7758    &  -0.4295    &  2.4684	&   -0.4629  \\
    6	&    -0.0880	&   0.1712    &  -0.5306    &  1.0097	&    0.4378  \\
    7	&     0.1578	&  -0.8074    &  -0.0923    &  2.1966	&   -0.4547  \\
    8	&     0.1992	&  -0.1736    &  -1.8081    &  3.7271	&   -0.9446  \\
    9	&    -0.0813	&  -0.1579    &  -0.0551    &  1.2108	&    0.0836  \\
   10	&     0.1717	&  -0.8713    &  -0.1700    &  2.8314	&   -0.9618  \\
   11	&     0.2353	&  -0.8325    &  -0.6333    &  2.8603	&   -0.6298  \\ 
\enddata
\tablenotetext{a}{ The ILC temperature (in thermodynamic units) at pixel $p$ of region $n$ is
$T_n(p) = \Sigma_{i=1}^{5}\zeta_{n,i}T^{i}(p)$, where $\zeta$ are the coefficients above
and the sum is over \wmap's frequency bands.  In addition (and as has been done before), 
the region smoothing from \citet{hinshaw/etal:2007} has been applied and an ILC bias has been
subtracted.}
\end{deluxetable}

\subsubsection{Maximum Entropy Method (MEM)}
\label{sec:MEM}

A MEM-based approach originally developed by \cite{bennett/etal:2003c} and \cite{hinshaw/etal:2007} is used
to model the Galactic foreground emission spectrum in the \WMAP\ bands on a pixel-by-pixel basis.  Spatial
templates of different emission components from external data are used as priors, and the model is designed
to revert to the priors in regions of low signal-to-noise ratio.  Thus the analysis is of most interest
for separating and characterizing the different emission components in high signal-to-noise regions.  The
model foreground maps that are produced have complicated noise properties so they are not useful for
foreground removal in cosmological analyses.

The nine-year MEM analysis differs from previous analyses \citep{bennett/etal:2003c,hinshaw/etal:2007,
gold/etal:2009,gold/etal:2011} in that spinning dust emission is treated as a separate emission component.
Previously, synchrotron emission and spinning dust emission were treated together as a single component
and an iterative method was used to solve for the spectrum of this component for each pixel.

The analysis is done using $1\arcdeg$ smoothed beam-symmetrized nine-year sky maps in the five \WMAP\ bands,
with the ILC map subtracted from each map and conversion to antenna temperature applied.  The zero level
of each map is set such that a $\csc |b|$ fit, for HEALPix $N_\mathrm{side} = 512$ pixels at $b < -15\degrees$
and outside of the KQ85 mask, yields a value of zero for the intercept.
The maps are degraded to HEALPix $N_\mathrm{side} = 128$ pixelization, and a model is fit for each pixel $p$
by minimizing the function
\begin{equation}
   H = \chi^2 + \lambda(p) \sum_c T_c(p) \ln\left[\frac{T_c(p)}{eP_c(p)}\right].
\end{equation}
Here $T_c$ and $P_c$ are the model brightness and template prior brightness for foreground component $c$
($e$ is the base of natural logarithms).  The form of the second term ensures positivity of the solution 
$T_c$ for each component, which alleviates degeneracy between the components.  The parameter $\lambda$
controls the relative weight of the data and the priors in the fit.  As in previous analyses, we base 
$\lambda(p)$ on the foreground signal strength: $\lambda(p) =  0.6\;[T_{\rm K}(p)]^{-1.5}$, where 
$T_{\rm K}(p)$ is the K-band ILC-subtracted map in mK antenna temperature.  

\begin{figure}
\epsscale{0.50}
\plotone{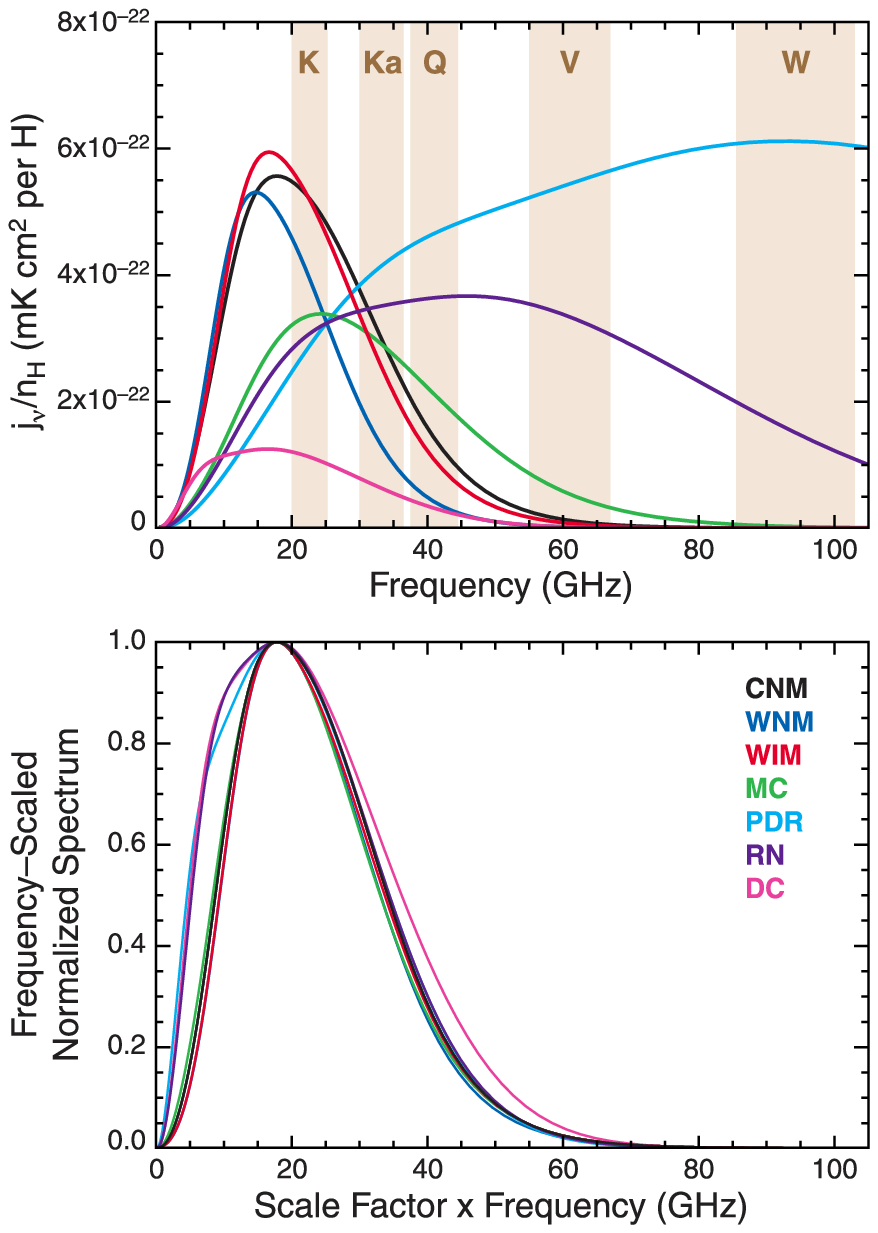}
\caption{The top panel shows spinning dust emissivity spectra predicted
by the model of \cite{ali-hamoud/etal:2009} and \cite{silsbee/etal:2011}
for the nominal physical conditions that they adopted for different ISM
environments - cold neutral medium (CNM), warm neutral medium (WNM),
warm ionized medium (WIM), molecular cloud (MC), photodissociation region
(PDR), reflection nebula (RN), and dark cloud (DC). The spectra were
calculated using version 2.01 of the code SpDust provided by the authors,
for the case where dust grains are allowed to rotate around non-principal axes.
The spectra are in units of brightness temperature per H column density.
The bottom panel shows the same spectra normalized to a peak of unity and
scaled to a common peak frequency (that of the CNM spectrum, 17.8 GHz).
The predicted spectral shapes for the different environments are similar.
We adopted the CNM case for the shape of the spinning dust spectrum in 
our foreground fitting.  We used this as an externally provided spectral
template in our fits, usually with our own arbitrary amplitude and frequency
scaling.  The fit results in no way imply
that the underlying physical mechanisms or the line-of-site conditions have
been established.
\vspace{1mm} \newline (A color version of this figure is available in the online journal.)}
\label{fig:spindust_spectra}
\end{figure}

The MEM foreground model is a sum of synchrotron, free-free, spinning dust, and thermal dust components.
The adopted spectra for synchrotron, free-free, and thermal dust emission are fixed power laws with
$\beta = -3.0, -2.15$, and +1.8, respectively.  The adopted synchrotron spectral index is consistent with
measurements of K- to Ka-band spectral index from \WMAP\ polarization data, for which free-free and spinning dust
contributions are expected to be negligible.  For spinning dust emission, we adopt a spectral shape predicted
by the model of \cite{ali-hamoud/etal:2009} and \cite{silsbee/etal:2011}.  The top panel of Figure 
\ref{fig:spindust_spectra} compares predictions of this model for different interstellar environments.
We adopt the spectral shape for their nominal cold neutral medium conditions.  The bottom panel shows
that the predicted shape does not vary much for different conditions if a multiplicative frequency shift
is allowed for.  The MEM model includes a frequency scale factor for the spinning dust spectrum
for pixels where the spinning dust prior is brighter than 0.1 mK.  This is constrained such that the peak
frequency is in the range from 10 to 30 GHz.  For other pixels, the peak frequency is fixed at 14.4 GHz,
a typical value found for the Galactic plane region.

The adopted priors are shown in Figure \ref{fig:fgpriors}.  The synchrotron prior is based on the 408 MHz
map of \cite{haslam/etal:1982}.  We use the original version of this map; our previous MEM analyses used a filtered
version in which striping and point sources are suppressed.  We add a zero level offset of 3.9 K, as suggested
by \cite{tartari/etal:2008} based on absolute measurements of sky brightness at 600 and 820 MHz.  We subtract
the 2.725 K CMB monopole and an extragalactic contribution of 12.96 K, from the analysis of ARCADE 2 and other
data by \cite{fixsen/etal:2011}.  The 408 MHz map is then scaled to form the prior in K-band using a spectral
index of -2.9.  (The ARCADE 2 extragalactic background is used instead of a source count based value such as
2.6 K from \cite{gervasi/etal:2008} because it gives a prior that is more consistent with the csc $b$ normalized
K-band map at high latitudes.)  The free-free prior is the scattering-corrected,
extinction-corrected H$\alpha$ template described in \ref{sssec:diffuse_fg_intro}, scaled to free-free brightness
temperature in K-band using 11.4 $\mu$K R$^{-1}$ \citep{bennett/etal:2003c}.
The spinning dust prior is the temperature-corrected dust map of \cite{schlegel/finkbeiner/davis:1998}, scaled to
spinning dust brightness temperature in K-band using 9.5 $\mu$K MJy$^{-1}$ sr.  This is the slope of the correlation
between the dust map and a map of spinning dust brightness from fits to \cite{haslam/etal:1982} 408 MHz,
\cite{duncan/etal:1995} 2.4 GHz, and ILC-subtracted \WMAP\ data in the Galactic plane.  The thermal dust prior is
the prediction of model 8 of \cite{finkbeiner/davis/schlegel:1999} at 94 GHz.  All of the prior maps have been
smoothed to $1\arcdeg$ FWHM.

\begin{figure}
\epsscale{1.00}
\plotone{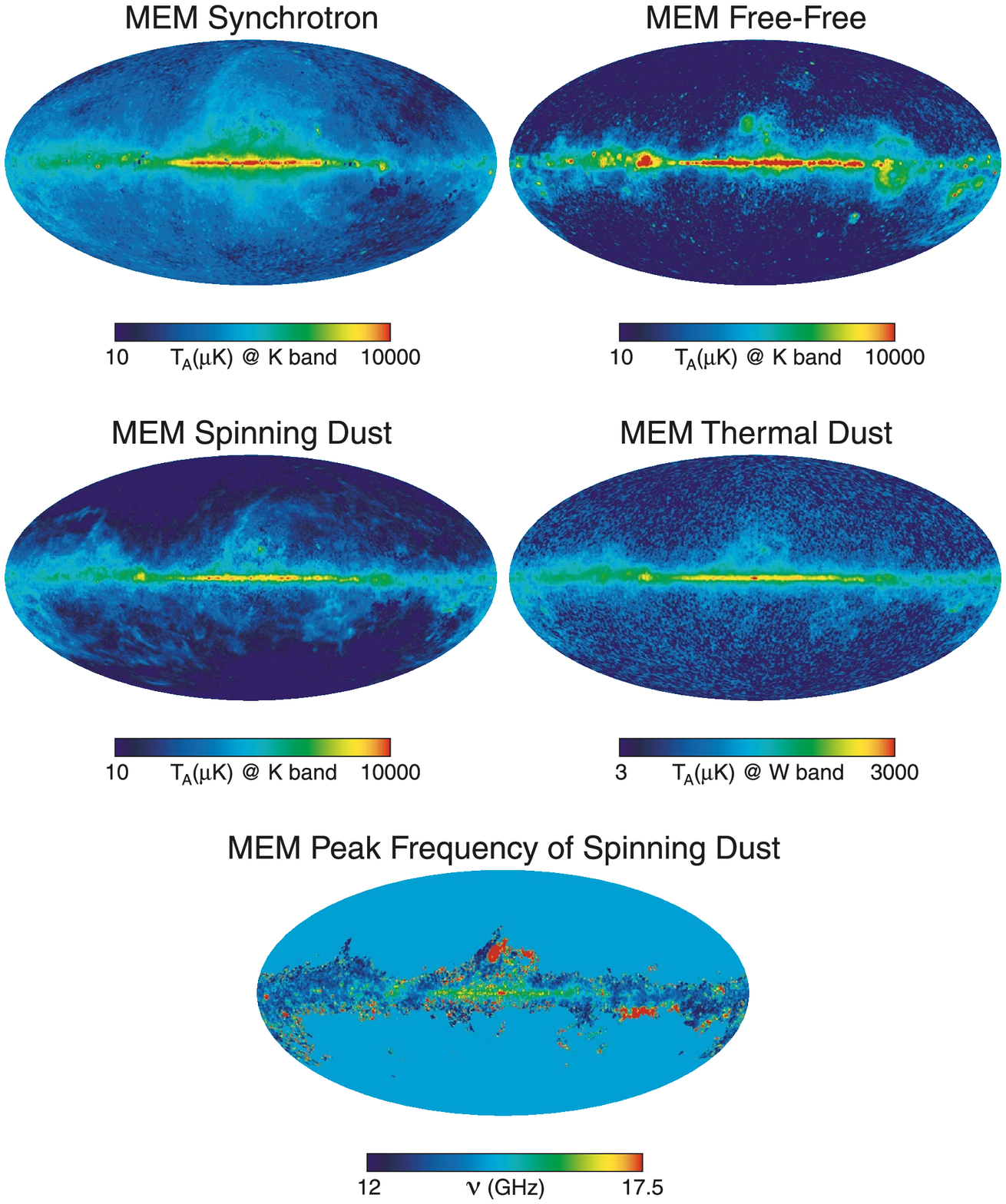}
\caption{Parameter maps from the MEM model fit. The top four maps are shown
on logarithmic scales and the others are on linear scales.
\vspace{1mm} \newline (A color version of this figure is available in the online journal.)}
\label{fig:mem_results}
\end{figure}

The adopted model provides good fits to the data without iterative adjustment of the synchrotron component
spectrum as used in previous analyses.  For pixels at $|b| < 5\arcdeg$, absolute residuals are typically less
than 0.01, 0.34, 1.2, 2.1, and 0.7 \% in K-, Ka-, Q-, V-, and W-bands, respectively.
Maps of the foreground components and peak frequency of spinning dust from the MEM analysis are shown in
Figure~\ref{fig:mem_results}.

\subsubsection{Markov Chain Monte Carlo Fitting\label{sec:MCMC}}

We again perform a pixel-based Markov chain Monte Carlo (MCMC) fitting
technique to the five bands of \WMAP\ data.  Our method is similar to that of
\cite{eriksen/etal:2007}, but we focus more on Galactic foregrounds rather than
CMB. The fit results of the prior releases have been reproduced, with the
``base'' model, which uses three power-law foregrounds, producing virtually the
same reduced $\chi^2$ per pixel.  We have again also included the 408 MHz map
of \cite{haslam/etal:1981} with a zero-point determined using the same $\csc |b|$
method as for the \WMAP\ data.  

There are two main changes from the previous release.
The first is that the MCMC fit now uses the spinning dust spectrum for grains
in a ``cold neutral medium'' as computed by \cite{silsbee/etal:2011}, with
an optional frequency shift parameter described below.  This change was made so
that the MCMC fit uses the same spinning dust spectrum as the rest of the
nine-year analysis.  The second significant change is that the spinning-dust
model is now run with the synchrotron spectral index as a free parameter.  This
was done to improve the quality of the fit, also discussed below.

The MCMC fit is performed on one-degree smoothed maps downgraded to HEALPix
$N_\mathrm{side}=64$. A MCMC chain is run for each pixel, where the basic model
is
\begin{equation} \label{eq:basemodel}
	T(\nu) =
	T_{s} \left( \frac{\nu}{\nu_\mathrm{K}} \right)^{\beta_s(\nu)}
	+ T_{f} \left( \frac{\nu}{\nu_\mathrm{K}} \right)^{\beta_f}
	+ a(\nu) T_\mathrm{cmb} + T_{d} \left( \frac{\nu}{\nu_\mathrm{W}}\right)^{\beta_d}
\end{equation}
for the antenna temperature.  The subscripts $s,f,d$ stand for synchrotron,
free-free, and dust emission, $\nu_\mathrm{K}$ and $\nu_\mathrm{W}$ are the
effective frequencies for K- and W-bands ($22.5$ and $93.5$ GHz), and $a(\nu)$
accounts for the slight frequency dependence of a $2.725$ K blackbody using the
thermodynamic to antenna temperature conversion factors found in
\cite{bennett/etal:2003c}.
The fit always includes polarization data as well, where the model is
\begin{equation}
	Q(\nu) = Q_{s} \left( \frac{\nu}{\nu_\mathrm{K}} \right)^{\beta_s(\nu)}
	+ Q_{d} \left( \frac{\nu}{\nu_\mathrm{W}}\right)^{\beta_d}
	+ a(\nu) Q_\mathrm{cmb}
\end{equation}
\begin{equation}
	U(\nu) = U_{s} \left( \frac{\nu}{\nu_\mathrm{K}} \right)^{\beta_s(\nu)}
	+ U_{d} \left( \frac{\nu}{\nu_\mathrm{W}}\right)^{\beta_d}
	+ a(\nu) U_\mathrm{cmb}
\end{equation}
for Stokes Q and U parameters.  Thus there are a total of 15 pieces of data for
each pixel ($T$, $Q$, and $U$ for five bands).

As for the previous two releases, the noise for each pixel at
$N_\mathrm{side}=64$ is computed from maps of $N_\mathrm{obs}$ at
$N_\mathrm{side}=512$.  To account for the smoothing process, the noise is then
rescaled by a factor calculated from simulated noise maps for each frequency
band.  The MCMC fit treats pixels as independent, and does not use pixel-pixel
covariance, which leads to small correlations in $\chi^2$ between neighboring
pixels.  This has negligible effect on results as long as goodness-of-fit is
averaged over large enough regions.

K-band is used as a template for the polarization angle of synchrotron and dust
emission, so $U_s$ and $U_d$ are not independent parameters, identical to the
previous analyses.  All models also fix the free-free spectral index to
$\beta_f = -2.16$, the same as in the seven-year release.

Results from the models discussed below are listed in \reftbl{tab:mcmc_chi2}.
The ``base'' model uses three power-law foregrounds, where the synchrotron
spectral index $\beta_s(\nu)$ is taken to be independent of frequency but may
vary spatially, and the dust spectral index $\beta_d$ is allowed to vary
spatially.  We assume the same spectral indices for polarized synchrotron and
dust emission as for total intensity emission.  This model has a total of 10
free parameters per pixel: $T_s$, $T_f$, $T_d$, $T_\mathrm{cmb}$, $\beta_s$,
$\beta_d$, $Q_s$, $Q_d$, $Q_\mathrm{cmb}$, and $U_\mathrm{cmb}$.

For models with a spinning dust component, another term is added to
Equation~\ref{eq:basemodel}
\begin{equation}
	\label{eq:sdfunc}
	 T_{sd}(\nu) = T_{sd}(\nu_\mathrm{K}) S_{sd}(\nu) ,
\end{equation}
Where $S_{sd}(\nu)$ parameterizes the shape of the spinning dust spectrum, and
is interpolated from values for the ``cold neutral medium'' spectrum given by
\cite{silsbee/etal:2011}.  An optional shift parameter can be used to
rescale the frequency dependence before interpolation.  This shift parameter
applies to the full sky and does not vary per pixel.  After shifting and
interpolation, the spectrum $S_{sd}(\nu)$ is normalized to unity at K-band,
leaving $T_{sd}(\nu_\mathrm{K})$ as the only spinning dust parameter for each
pixel.  Independent fits were performed to determine the best-fit shift
parameter, which for the averaged sky was found to be $0.84$. Inside the Kp12
mask (within a few degrees of the galactic plane) the preferred shift
parameter may be somewhat lower ($0.77$), but the evidence is not strong.

The spinning dust component is assumed to have negligible polarization, as
theoretical expectations for the polarization fraction are low compared to
synchrotron radiation \citep{lazarian/draine:2000}, and polarization data
thus far show no evidence that such a component is necessary (Section 
\ref{sssec:tmpl_cln}, \cite{lopez-caraballo/etal:2011a}, \cite{dickinson/etal:2011},
\cite{rubino-martin/etal:2012}).   
This model then has 11 free parameters per pixel: the 10
parameters of the base model, plus the spinning dust amplitude.

MCMC fits for the nine-year release were performed with the addition of the 408
MHz data compiled by \cite{haslam/etal:1981}.  The error on the zero point for
this data was estimated in that work to be $\pm 3$ K, with an overall
calibration error of $10\%$. As the MCMC method treats all input maps equally, for
consistency we estimate and subtract a nominal zero point offset of 7.4 K,
as determined by the same $\csc|b|$ method we use for the \WMAP\ sky maps.
For comparison, \cite{lawson/etal:1987} used a comparison with 404
MHz data to find a uniform (presumably extragalactic) component with a
brightness of $5.9$ K.  

\begin{deluxetable}{llccc}
      \tablewidth{0pt}
      \tabletypesize{\footnotesize} 
      \tablehead{
      \colhead{Dataset} &
      \colhead{Model} &
      \colhead{Galactic plane} &
      \colhead{outside Galactic plane} &
      \colhead{full-sky average}
      }
      \tablecaption{Reduced $\chi^2$ per pixel of MCMC fits\label{tab:mcmc_chi2}}
      \startdata
      \WMAP\ five-band  & (a) base                 & 2.38 & 1.17 & 1.29 \\
                        & (b) sd096                & 1.00 & 1.06 & 1.05 \\
      \hline
      \WMAP\ \& 408 MHz & (c) base                 & 2.46 & 1.13 & 1.25 \\
                        & (d) sd096                & 6.27 & 1.42 & 1.88 \\
                        & (e) sd070                & 1.76 & 1.33 & 1.37 \\
                        & (f) bsfree sd084         & 1.24 & 1.03 & 1.05 \\
                        & (g) bsfree Strong sd084  & 1.05 & 1.01 & 1.01 \\
      \enddata
\end{deluxetable}

We find that to best fit the 408 MHz data, the spinning dust spectrum needs to
have its peak frequency adjusted downward by approximately 15\%, similar to the
case in the previous release.  We also find that a much better fit is achieved
in the plane by varying the synchrotron spectral index, which for that region
allows a $\chi^2_\nu = 1.24$ versus $\chi^2_\nu = 1.76$ with
fixed index, for 8.5 effective degrees of freedom.
The mean spinning dust fraction inside the KQ85 mask is somewhat lower than in
the seven-year fit, at $10\%$ of 22 GHz flux compared to $18\%$ in the
seven-year fit.  

\begin{figure}
\epsscale{0.7}
\plotone{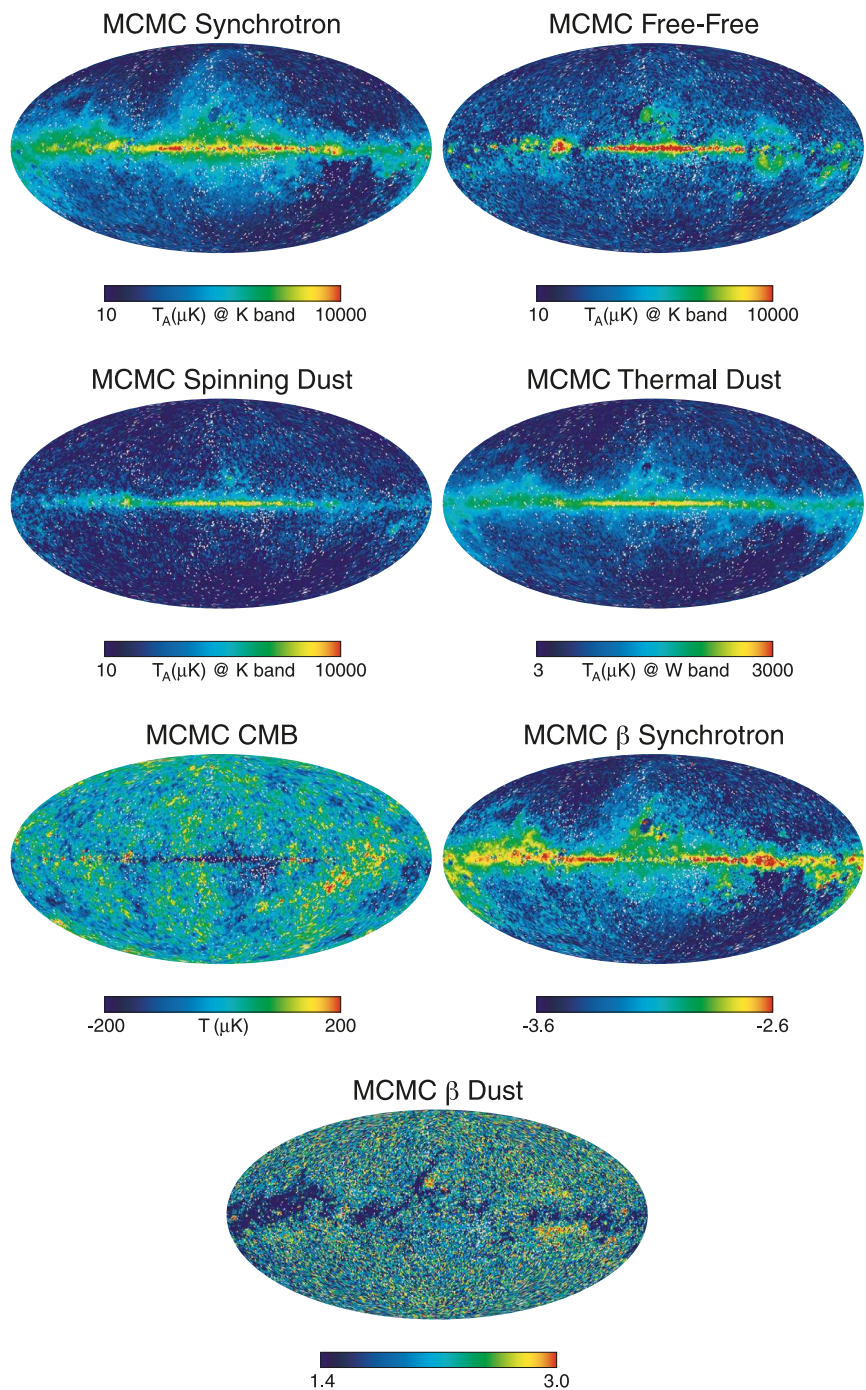}
\caption{Parameter maps from the MCMCg model fit. The top four maps are
shown on logarithmic scales and the others are on linear scales.
Accurate determination of the CMB close to the Galactic plane is inhibited by CMB-foreground covariance.
The map for $\beta$ synchrotron is evaluated at 40 GHz.
\vspace{1mm} \newline (A color version of this figure is available in the online journal.)}
\label{fig:mcmc}
\end{figure}

We also find that the fit is improved by taking into account some mild
steepening of the synchrotron spectrum from 408 MHz to \WMAP's frequency
range. \cite{strong/orlando/jaffe:2011} have compared mid-latitude synchrotron
measurements and estimates from 22 MHz to 94 GHz with predictions of cosmic ray propagation models
based on cosmic ray and gamma ray data.  We adopted their best fit pure diffusion
model (``galdef\_ID\_54\_z04LMPD\_g0\_1.3\_withsecS'') to compute an
effective synchrotron spectral index between 408 MHz and 23 GHz (\WMAP\ K-band),
as well as the index from 23 GHz to 94 GHz over which range it remains nearly
constant.  We calculate the difference in these two indices to be $-0.12$.
Our model g (hereafter MCMCg, and listed on the last line of \reftbl{tab:mcmc_chi2})
then uses this difference, so that while the model
parameter $\beta_s$ is used as the synchrotron index for the \WMAP\ bands, the
value $\beta_s+0.12$ is used to extrapolate the synchrotron component down to
408 MHz for comparison to the map of Haslam et al.  The parameters from this
fit are shown in Figure~\ref{fig:mcmc}.

\subsubsection{Six Band Minimal Prior Chi-Squared Fitting}
\label{sec:six_band_fits}

In this section we attempt to find a best fit foreground model that is
consistent with both the \WMAP\ data and Haslam.  This is intended to be a
faster fit than was done with the MCMC method in Section~\ref{sec:MCMC}, and so
it allows us to experiment with models more rapidly.  Because this method simply
finds the maximum likelihood point of the foreground model, it does not provide
errors bars as the MCMC method does.  Also, we avoid priors in the form of
foreground component sky maps, which were used in the MEM fitting in
Section~\ref{sec:MEM}.  The priors we use in this section are mostly in the form
of the foreground spectral shapes (relative antenna temperature as a function of
frequency) instead of in the form of sky maps.  This is a complementary form of
analysis to the MEM fitting.

\paragraph{Data and noise}
\label{sec:fg_data}

Our data consists of maps smoothed to a common resolution of $1^\circ$ FWHM,
which we pixelize at r6.  We use 6 maps: 408 MHz and
the five \WMAP\ bands.  
We use the original Haslam map (408 MHz) as in Section~\ref{sec:MEM} with the
same offsets, except in this case we do not use the ARCADE extragalactic background.  
Instead of subtracting 12.96~K, we subtract 2.6~K \citep{tartari/etal:2008}, so the
Haslam map used in this section is 10.36~K brighter in antenna temperature in
all pixels.
The rms noise in each pixel of the 408 MHz map
is taken to be 10\% of the antenna temperature, added in quadrature with a 0.6~K
uncertainty in zero point \citep{haslam/etal:1982, tartari/etal:2008}. 

We consider three noise components for the \WMAP\ bands in this foreground
fitting: the 0.2\% overall gain uncertainty, the $\sigma_0/\sqrt{N_{\rm obs}}$
instrument noise, and the uncertainty in the diffuse foreground monopole
corrected with the $\csc |b|$ offsets, discussed previously.
Because our fitting is done on a
per-pixel basis, we approximate these errors as uncorrelated between pixels,
and we add them in quadrature.

The instrument noise can be treated carefully to account for the smoothing to 
$1^\circ$ FWHM.  Typically it is inaccuracies in the foreground model that cause
$\chi^2$ to be large and not the details of the noise.  However, a detailed
treatment of the noise smoothed to $1^\circ$ in r6 pixels is given in
Appendix~\ref{sec:smooth_noise}.  Again, because we fit on a per-pixel
basis, we ignore the correlations in noise between nearby pixels.

\paragraph{Foreground Models}

We start with a simple foreground model consisting of several simple power laws,
and progressively add complexity to the model to improve the fit.
The foreground model we use involves temperature only; we did not try to fit
polarization.  The sequence of foreground models we use is listed in
Table~\ref{tab:fg_models}, and details are discussed below.

\begin{deluxetable}{cccccccccc}
    \tablewidth{0pt}
    \tablecolumns{10}
    \tablecaption{$\chi^2$ Minimal Prior Fits of Foreground Models}
    \tabletypesize{\footnotesize} 
    \tablehead{
        \colhead{Model} &
        \colhead{synchrotron\tablenotemark{a}} &
        \colhead{$\Delta \beta_{\rm sync}$\tablenotemark{b}} & 
        \colhead{ff\tablenotemark{c}} &
        \colhead{$\beta_{\rm dust}$} &
        \colhead{SD\tablenotemark{d}} &
        \colhead{SD peak\tablenotemark{e}}  &
        \colhead{$\nu$\tablenotemark{f}} &
        \colhead{$\chi^2$/pixel\tablenotemark{g}} &
        \colhead{$f_{\rm bad}$\tablenotemark{h}}
    }
\startdata
1  &   power    & 0    & yes & 1.8                    &  no  &  -                & 4 & 6.1  & 37\% \\
2  &   power    & vary & yes & 1.8                    &  no  &  -                & 5 & 2.5  & 11\% \\
3  &   power    & vary & yes & 1.6--2.0               &  no  &  -                & 6 & 2.3  &  9.5\% \\
4  &   power    & vary & yes & 1.8                    &  yes &  15.1             & 6 & 1.5  &  4.4\% \\
5  &   power    & vary & yes & 1.8                    &  yes &  12.5--17.8       & 7 & 0.64 &  0.59\% \\
6  &   Strong   & 0    & yes & 1.8                    &  yes &  15.1             & 5 & 5.4  & 30\% \\
7  &   Strong   & 0    & yes & 1.8                    &  yes &  12.5--17.8       & 6 & 4.1  & 20\% \\
8  &   Strong   & vary & yes & 1.8                    &  yes &  15.1             & 6 & 1.2  &  2.1\% \\
9  &   Strong   & vary & yes & 1.8                    &  yes &  12.5--17.8       & 7 & 0.60 &  0.48\% \\
\enddata
\tablenotetext{a}{Whether the synchrotron is treated as a pure power law or
modeled according to a model from \citet{strong/orlando/jaffe:2011}.}
\tablenotetext{b}{For both power law and Strong et al. synchrotron models, we
either set the spatial variation in spectral index $\Delta \beta_{\rm sync}$ 
to zero or allow it to vary: $-0.5 \le \Delta \beta_{\rm sync} \le 0.5$.
In the case of a power law, $\Delta \beta_{\rm sync}$ is a perturbation added to 
$\beta_{\rm sync} = -3.0$.
}
\tablenotetext{c}{The free-free spectrum is given by \citet{oster:1961}; we use
an electron temperature of 8000 Kelvin.}
\tablenotetext{d}{Whether a spinning dust spectrum in the shape of the cold neutral medium
is used.}
\tablenotetext{e}{Range of available peak frequencies for the spinning dust
spectrum, in GHz.  This is either fixed at 85\% of the peak frequency 17.8 GHz
for the cold neutral medium (which is 15.1 GHz), or allowed to be a range from
70\% to 100\% of the CNM peak frequency (which is 12.5 GHz to 17.8 GHz).  }
\tablenotetext{f}{Degrees of freedom in the model.  Most degrees of freedom are
constrained: foreground amplitudes must all be positive, for example.
The highly constrained CMB amplitude is included as a degree of freedom. }
\tablenotetext{g}{The mean $\chi^2$ per pixel, averaged over the whole sky 
(for temperature only, not polarization),
where $\chi^2$ values greater than 10 are set to exactly 10 so that a few
extremely bad pixels don't throw off the whole fit.  This $\chi^2$
value includes deviations of the model from Haslam and \WMAP\ bands, but not
deviations from the ILC prior.  Since there are 6 measurements in each pixel 
(and an ILC prior) and $4\le\nu\le 7$ degrees of freedom in the model, we would expect 
$\chi^2$/pixel $\approx 6-\nu$ for a good fit if we had unconstrained variables.}
\tablenotetext{h}{The fraction of the pixels where $\chi^2 > 10$.  This is an
estimate of the sky fraction where the fit is bad.  Again, the $\chi^2$
used here includes the difference of the model from the six bands, but does not include
deviations from the ILC prior.}
\label{tab:fg_models}
\end{deluxetable}

The synchrotron spectrum is either taken to be a pure power law in antenna
temperature, $T_A \propto \nu^{\beta_{\rm sync}}$, 
or derived from assuming the spectral index curve from
\citet{strong/orlando/jaffe:2011}, Figure 6, upper right corner. 
This is the curve for a low-energy electron injection index of 1.3 and is the
same spectrum as used in the MCMC fitting.
To this spectral index curve we optionally add an offset in $\beta_{\rm sync}$,
$-0.5 \le \Delta \beta_{\rm sync}\le 0.5$
independent of frequency.
We numerically integrate this spectral index curve to obtain synchrotron antenna
temperature.

The free-free spectrum is the slightly curved model given by \citet{oster:1961} and
rearranged for antenna temperature by \citet{bennett/etal:2003c}.
This is
\be
T_A^{WMAP}(\nu) \propto
\frac{1 + 0.2218 \ln(T_e / 8000 {\rm K}) - 0.1479 \ln(\nu / 41 {\rm GHz})}
{(\nu / 41 {\rm GHz})^2 (T_e / 8000 {\rm K})^{1/2}}.
\ee
For simplicity we use an electron temperature of 8000 K.  We expect 
variations in electron temperature, but these do not strongly affect the
shape of the spectrum.

The dust spectrum is given by a pure power law, typically with a fixed spectral index of 
$\beta_{\rm dust} = 1.8$.

Finally, we add a spinning dust component.  This is an antenna temperature
spectrum from the \cite{silsbee/etal:2011} model prediction for cold neutral
medium (CNM) conditions, 
with an optional frequency scale factor.  If the spectrum is plotted
as antenna temperature as a
function of log frequency, the frequency scale factor simply shifts the spectrum
left or right.  However, instead of quoting the frequency scale factor, we
instead quote the peak frequency, when the spectrum is measured in antenna
temperature as a function of frequency.  The peak frequency of the CNM spectrum
is 17.8 GHz.

All of these foregrounds are assumed to have a positive scale factor associated 
with them.  Synchrotron, free-free, and spinning dust are normalized to K-band
antenna temperature, and dust is normalized to W-band antenna temperature.

The CMB is modeled as a blackbody with constant thermodynamic temperature.
To make the CMB fit look statistically isotropic, we add a prior that
the CMB must be within 5 $\mu$K rms of the nine-year ILC.  Without this prior,
the data do not constrain the CMB very tightly in the galactic plane,
and we find the CMB preferring values lower than -250 $\mu$K.

To approximate the finite width of the \WMAP\ bandpasses, we calculate these
spectra at three frequencies per band and determine the \WMAP\ response from a
weighted average, as described in Appendix~\ref{sec:bandpass_integration}. 

\paragraph{Fitting Code}
Fitting the foregrounds is a least squares problem.  However, we modify
the simple linear least squares problem in two ways: we constrain the
coefficients, and we allow nonlinear foreground spectra.  Constraining the
coefficients is essential, because we know the foregrounds are always positive.
Unconstrained least squares fitting will frequently give a very negative and
therefore unphysical foreground.  Secondly, we allow nonlinear foregrounds,
in the sense that the total foreground is not simply a linear combination of
fixed foreground spectra.  We allow the spectra to vary, for example by allowing
the synchrotron spectral index to be a fit parameter, or by allowing the peak
frequency of spinning dust to be a fit parameter.

There are several codes which can be used to solve this problem.  We have not
made a thorough search of all available software, 
and we only considered code in IDL since that is the language in which much of
our other software is written.
We have found two codes to be
useful: a bound variable least squares routine and a Levenberg-Marquardt solver.

We found a Bound Variable Least Squares (BVLS) routine\footnote{bvls.pro, available
from \url{http://www-astro.physics.ox.ac.uk/~mxc/idl/}} to be very fast, but it is
restricted to linear foreground models and so it cannot solve for varying
spectral indices or spinning dust frequency scale parameters.  Because of this
constraint we do not use it to report results in this paper.
However, this code does have the
advantage that parameters can be constrained to be positive, so it can provide
physically reasonable fits.   

For the results reported in this section (in Table~\ref{tab:fg_models}) 
we use the mpfitfun.pro routine\footnote{available from
\url{http://cow.physics.wisc.edu/~craigm/idl/idl.html}}, 
which uses the Levenberg-Marquardt algorithm
and was written by Craig Marquardt, for the constrained nonlinear least squares
fitting.  This is somewhat slower than the BVLS code because it cannot use the assumption that the
$\chi^2$ function is precisely quadratic in all of the fit coefficients.
The ability to calculate foreground spectra quickly is an important factor in
the speed of these calculations.  We discuss a useful rapid method of calculating the
integral over the \WMAP\ bandpasses in Appendix~\ref{sec:bandpass_integration}.

\paragraph{Results}

\begin{figure}
\epsscale{0.9}
\plotone{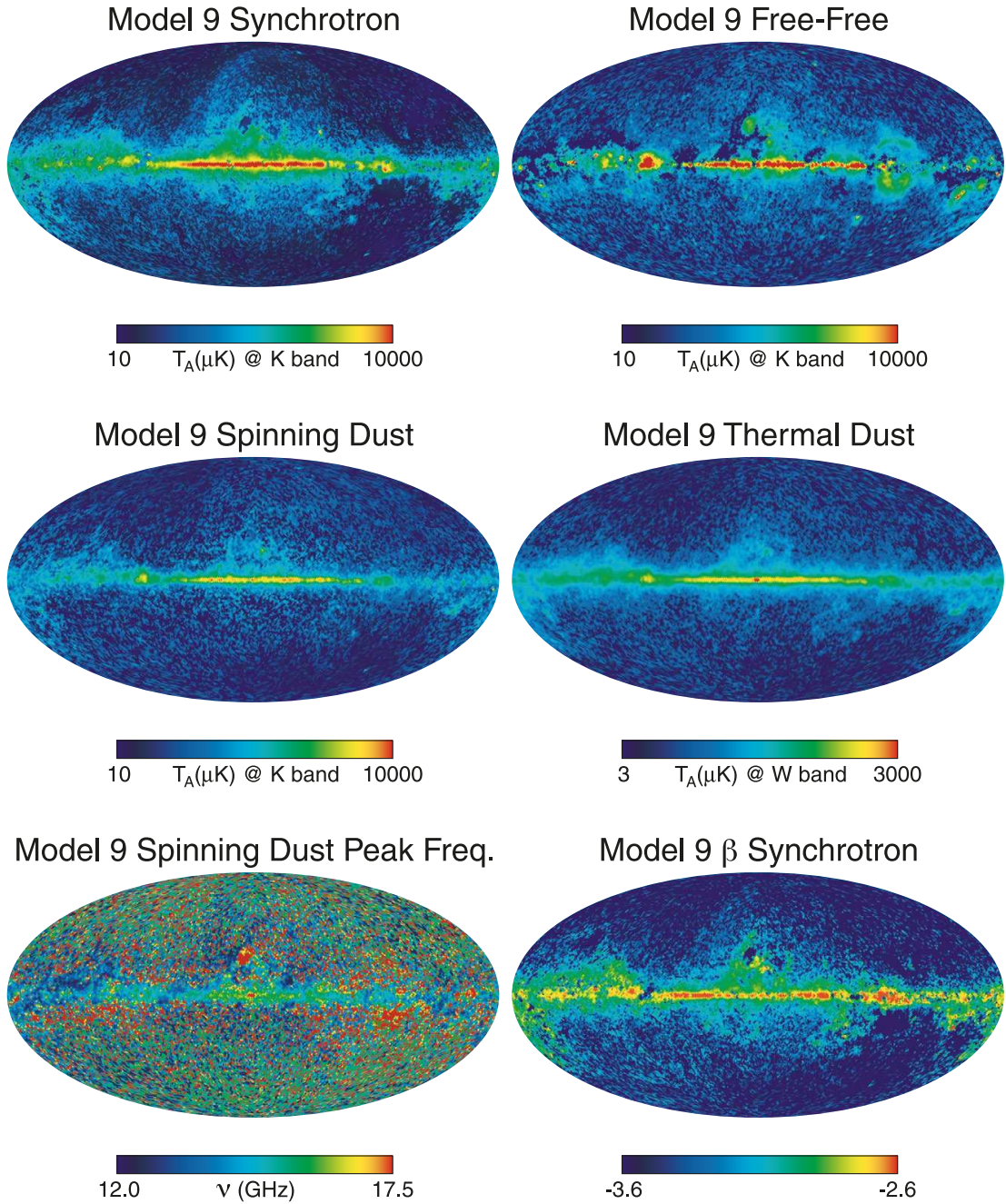}
\caption{Parameter maps from the Model 9 fit. The top four maps are shown
on logarithmic scales and the others are on linear scales.
The map for $\beta$ synchrotron is evaluated at 40 GHz.
\vspace{1mm} (A color version of this figure is available in the online journal.)}
\label{fig:model9}
\end{figure}

The results of this simple foreground fitting are shown in the last columns of
Table~\ref{tab:fg_models}.  
Additionally, maps from the Model 9 fit are shown in Figure~\ref{fig:model9}.
A set of three fixed power laws in Model~1 does not fit the
data well.  Allowing spatial variation of the synchrotron power-law spectral index
in Model~2 substantially
improves this, but 11\% of the sky is still fit very poorly.  Allowing spatial variation
of the dust spectral index in Model~3 does not substantially improve the number of
well fit pixels,
so we fix the dust spectral index to $\beta=1.8$.  Adding a spinning dust
component with peak frequency of 15.1 GHz (which is 0.85 times the CNM peak
frequency of 17.8 GHz) does improve the fit, and allowing that peak frequency to
vary between 12.5 GHz and 17.8 GHz helps even more.  See Models 4 and 5.

Because it is probable that the synchrotron is not a pure power law and because
we use the Haslam data at 408 MHz, which is much lower in frequency than the
\WMAP\ data, we test a curved synchrotron model from \citet{strong/orlando/jaffe:2011}.
If we do not allow the spectral index to vary, we again get bad fits in Models 6
and 7.  However, a varying spectral index combined with a spinning dust
component produces results that are fractionally better than a pure power law
with the same spinning dust components, as can be seen by comparing models 5 and 9, and comparing
models 4 and 8.

None of these fits is perfect.  Even in Model 9, there remain a few pixels
that are not fit well.  These are primarily in Ophiuchus, the galactic plane,
and the Gum nebula.  

\subsubsection{Diffuse Foreground Results}

\paragraph{Cross-Comparison of Foreground Fits} \label{sec:cross-comparison}

Maps of parameters from the MEM, MCMCg, and six-band $\chi^2$ Model 9 fits
are shown in Figures \ref{fig:mem_results}, \ref{fig:mcmc}, and 
\ref{fig:model9}.  A summary of the parameter treatment for each of these three models
is provided in Table~\ref{tab:fg_models_summary}.

\begin{deluxetable}{cccc}
    \tablewidth{0pt}
    \tablecolumns{4}
    \tablecaption{Summary of Foreground Decomposition Model Assumptions}
    \tabletypesize{\footnotesize} 
    \tablehead{
        \colhead{Parameter} &
        \colhead{MEM} &
        \colhead{MCMCg} & 
        \colhead{$\chi^2$ Model 9}
    }
\startdata
$\beta_\mathrm{sync}\tablenotemark{a}$            &   $-3.0$, fixed          & Strong, $|\Delta\beta|<0.5$ &  Strong, $|\Delta\beta|<0.5$ \\
$\beta_\mathrm{dust}$                             &   $+1.8$, fixed          &   free                      &  $+1.8$, fixed  \\
$\beta_\mathrm{ff}$                               &   $-2.15$, fixed         &  $-2.16$, fixed             &  $-2.09$ -- $-2.17$, fixed\tablenotemark{b} \\
$\nu_\mathrm{peak}^\mathrm{sd}\tablenotemark{c}$  &   $10$--$30$, constrained  &   $14.95$, fixed          &  $12.5$--$17.8$, constrained \\
CMB                                               &    ILC subtracted      &   free                      &  ILC prior     \\
polarization data fit                             &    no                  &   yes                       &  no   \\
external foreground spatial priors    &   Haslam, SFD, FDS, H$\alpha$\tablenotemark{d}       &   no                        &  no  \\
\enddata
\tablenotetext{a}{Synchrotron is assumed to be a power law with a fixed spectral index, $\beta_\mathrm{sync}$, or a variable power law based on a \citet{strong/orlando/jaffe:2011} model, 
with a best fit value of $\Delta\beta$ added to the spectral index.}
\tablenotetext{b}{The free-free spectrum for the $\chi^2$ Model 9 fit is given by \citet{oster:1961} with a fixed electron temperature $T_e = 8000$~K.  The spectral index, $\beta_{\rm ff} = -2.14$ at K-band and $-2.17$ at W-band.  It increases to $-2.09$ at 408 MHz.}
\tablenotetext{c}{A spinning dust cold neutral medium spectral shape is used with an allowed range of a peak frequency shift, specified in GHz.
}
\tablenotetext{d}{``Haslam'': the 408 MHz survey of \cite{haslam/etal:1982}; 
 ``SFD'': the temperature-corrected dust map of \cite{schlegel/finkbeiner/davis:1998};
 ``FDS'': thermal dust model 8 from \citet{finkbeiner/davis/schlegel:1999};
 ``H$\alpha$'': H$\alpha$ all-sky mosaic from \citet{finkbeiner:2003}.
 }
\label{tab:fg_models_summary}
\end{deluxetable}

Results from these three models are a sampling of the possible
parameter space which can be used to produce a total foreground model in
each \wmap\ band.  Each of these models possesses strengths and weaknesses,
which can be used to offset one another.  Included in these considerations are
the treatment of the CMB component, the use of spatial priors,
and the use of spectral constraints.

{\em Treatment of the CMB. } Both the MEM and Model 9 make use of the ILC as the CMB estimator:
the MEM subtracts the ILC from the \wmap\ data before fitting, and
Model 9 uses the ILC as a strong prior.  As discussed in Section~\ref{sec:ilc_errors},
the ILC is an imperfect estimate of the true CMB, containing a
residual foreground bias signal.  MCMCg, on the other hand, treats
the CMB as a free parameter in its fit solution.  While this is
a strength for MCMCg at high latitudes, CMB-foreground 
covariance is strongest in the Galactic plane, and MCMCg does not
separate the CMB and foregrounds well there.  Use of the ILC
provides a better constraint in that case.

{\em Use of spatial priors.}  The MEM uses spatial templates to constrain its fitting solution at high
latitudes where signal-to-noise is lower than in the Galactic plane.
This produces a less noisy parameter solution at high latitudes when
compared to the MCMCg and Model 9 $\chi^2$ fit.
This is valuable to the extent that one trusts those priors, both
in terms of zero levels and spatial structure.    

{\em Use of spectral constraints.}  The synchrotron spectral index $\beta_s$
is a pivotal parameter in model fitting, since
its behavior influences the model allocation between synchrotron, free-free and 
spinning dust.  The MEM assumes a constant value of $\beta_s = -3$ at \wmap\
frequencies.  Model 9 and MCMCg allow each
pixel to fit for this parameter independently, within the constraints
of a \citet{strong/orlando/jaffe:2011} spectral dependence.  Positional gradients,
including a latitudinal gradient, are probably closer to physical
reality than a constant value \citep{kogut/etal:2007}.  However,
with this degree of freedom comes the possibility for degeneracies
with the free-free and spinning dust parameters.
In Figure~\ref{fig:model9_degen} we show results from a foreground degeneracy analysis 
for a representative pixel in the six-band Model 9 fit.  There are
significant degeneracies between parameter pairs that include either synchrotron
amplitude or $\beta_s$.  (A similar result was presented by \citet{gold/etal:2009}
for the five-year MCMC analysis, although that lacked a spinning dust component).
We believe degeneracies are a factor in the appearance of the
MCMCg and Model 9 $\beta_s$ maps, which show a strong latitudinal
gradient and a dust-like morphology in some regions, e.g., extending
south of the plane over $150\arcdeg < l < 190\arcdeg$ and in the North Celestial Pole 
HI loop that extends north of the plane over $120\arcdeg < l < 150\arcdeg$ 
\citep{meyerdierks/etal:1991}.
All three models share a common spectral
shape for the spinning dust spectrum.  This shape is allowed to shift
peak frequencies for MEM and Model 9, while MCMCg adopts a fixed peak
frequency.  Although the use of a common shape seems well motivated
(see Figure~\ref{fig:spindust_spectra}), there is no
guarantee that it is correct for all pixels. This is an additional
source of uncertainty in the fits, as observational deviations from this shape 
will be distributed primarily among free-free and synchrotron components.
We note an apparent power deficit in the Model 9 free-free map, present to a lesser
extent in the MCMCg result, which is dust-like in signature. 
Finally, we note that all models assume a fixed $\beta_{ff}$, and 
only MCMCg allows for a free $\beta_{dust}$.  These are less uncertain values, but
errors in fixed values can ripple into other components.

\begin{figure}
\epsscale{1.0}
\plotone{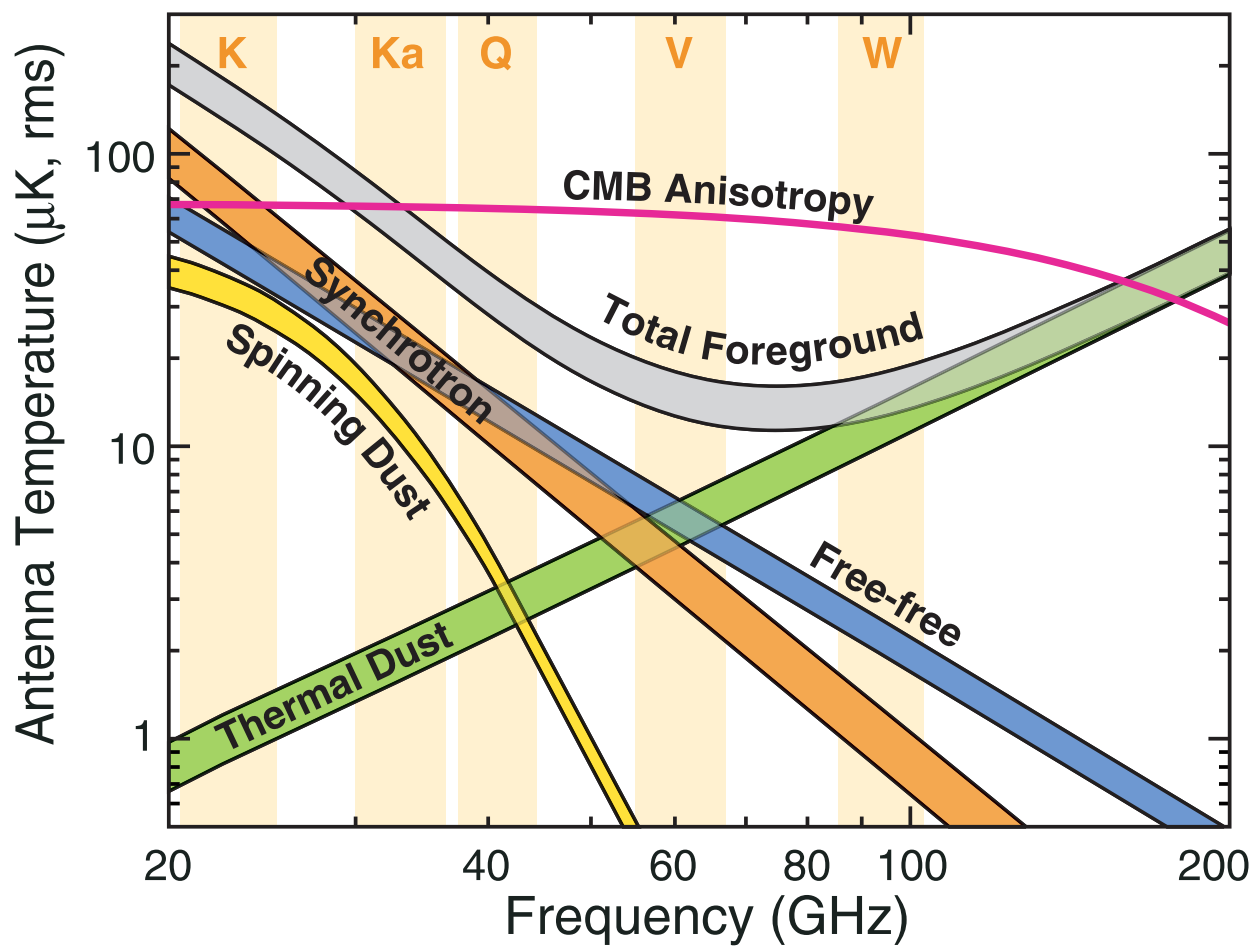}
\caption{Spectra of CMB and foreground anisotropy.  The foreground anisotropy
results are averages over the three foreground models (MCMCg, MEM, and Model 9).
The upper curve for each foreground component shows results for pixels outside of
the KQ85 mask, and the lower curve shows results outside of the KQ75 mask.  The
different foreground models are in good agreement for the total foreground
anisotropy.  Results for the individual foreground components depend on model
assumptions discussed in the text, and typically differ among the three models by
5\% to 25\%.
\vspace{1mm} \newline (A color version of this figure is available in the online journal.)}
\label{fig:spectral_overview}
\end{figure}

\begin{figure}
\epsscale{1.0}
\plotone{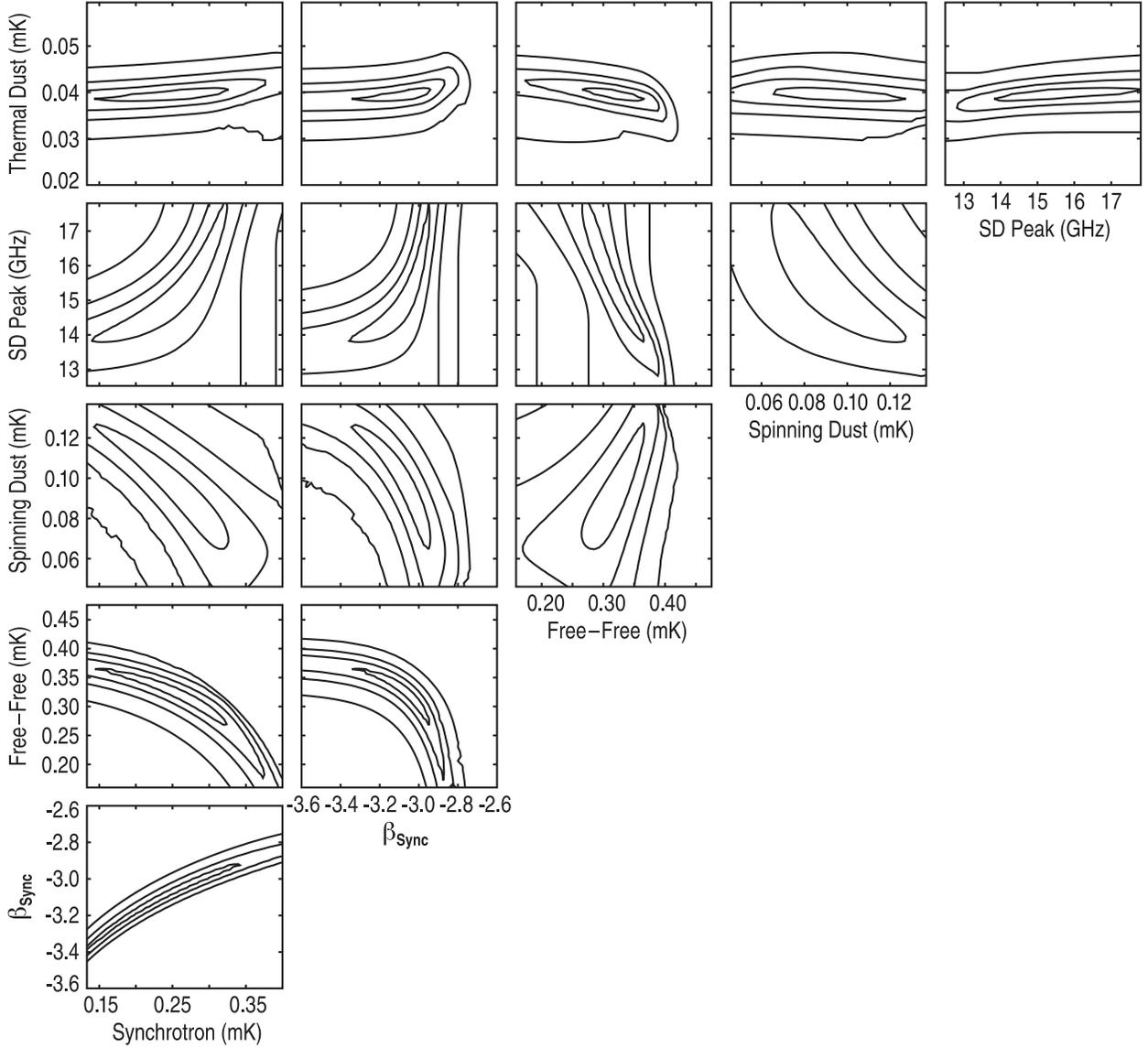}
\caption{Results from foreground degeneracy analysis for six-band
Model 9 fitting.  The contour plots illustrate the
degeneracy between model parameters for a
representative single pixel foreground spectrum.
Each panel shows the change in $\chi^2$ as the selected
pair of parameters are varied from their best-fit values
while marginalizing over the other parameters.  Contours
are shown for $\Delta \chi^2$ values of 0.2, 1, 3, and 10,
except values of 0.5, 3, and 10 are used for $\beta_{\rm sync}$
vs. synchrotron amplitude. 
There are significant degeneracies between parameter pairs
that include either synchrotron amplitude or synchrotron
spectral index, except for those that include thermal dust
amplitude.}
\label{fig:model9_degen}
\end{figure}

It is nevertheless possible to find relative agreement among these models,
especially at higher latitudes. The high latitude foreground spectral components in the \wmap\ bands are shown in Figure \ref{fig:spectral_overview} and all of the fitting techniques support this spectral decomposition of the foregrounds.

The actual foregrounds, especially at low Galactic latitudes, are clearly more complex
than our parameterizations allow, since variations in physical conditions
exist along any line of sight.  There are some sky locations that
were not well fit even with all of the degrees of freedom allowed by the $\chi^2$
fitting, such as in Ophiuchus.  Given the complexity of the foreground
emission mechanisms sampled by the \wmap\ bands, separating the CMB from the 
total observed foreground
is a more straightforward and reliable process than the decomposition of that total
foreground into physical components.  Although we have found imperfections
in the dust and free-free templates we use for foreground cleaning, those 
imperfections are primarily confined to regions which are masked from use 
in the cosmological analysis, and the use of foreground cleaned maps
in the power spectrum analysis is robust. 

A remaining item of interest is the microwave ``haze''.  
The first claim of a haze \citep{finkbeiner:2004} suggested an excess of
free-free emission compared to the expectation from H$\alpha$, and was dubbed a
``free-free haze''.
No longer believed to be free-free emission, its
exact shape and attribution has evolved in the literature.  In general the haze
is described 
as an excess extended diffuse emission near the Galactic center. 
This excess appeared as a residual from the decomposition 
of \wmap\ K, Ka and Q maps using external templates 
\citep{finkbeiner:2004,dobler/finkbeiner:2008a}.  
The templates most often used for this purpose are the
Haslam 408 MHz map, a de-extincted form of the \citet{finkbeiner:2003} H$\alpha$ all-sky mosaic and the
\citet{finkbeiner/davis/schlegel:1999} thermal dust models.  

While the excess compared to
external templates is clear, the attribution to a physical mechanism 
associated with Galactic emission is not. 
One interesting possibility characterizes the haze as a separate hard spectrum synchrotron component 
associated with the Gamma-ray bubbles \citep{planckintermediate:IX, dobler/etal:2010}.
\citet{planckintermediate:IX} uses a Gibbs sampler to fit a foreground model
outside a Galactic mask that assumes separate hard and soft power-law spectra.
The cut-sky maps with these spectra are further decomposed, using external
templates, into emission components with a distinct residual identified as a
$\beta_s \sim -2.55$ synchrotron haze.
It is also possible to find reasonable models which
adequately describe the data without the invocation of a haze component, as in e.g. \citet{dickinson/etal:2009}.
In these cases, the haze excess is absorbed and distributed amongst other low frequency
Galactic components.
For example, a typical K-band haze intensity at roughly $\pm20\arcdeg$ latitude near the Galactic center is 
$\sim100\, \mu$K \citep{planckintermediate:IX},
whereas K-band residuals in those locations for the MEM, MCMCg, and Model 9 models are roughly 
zero with a $1\sigma$ deviation of a few $\mu$K.
Existence of the haze as a 
separate spatial component is model dependent.  It depends on 
foreground spectral assumptions, which affect the emission allocation between the  
CMB and the decomposition of the Galactic foregrounds into physical components.
Because the haze is easily absorbed into other model components if not explicitly
accounted for, and a number of remaining uncertainties exist in the morphology and behavior of
low-frequency emissions in general (e.g. spinning dust), we feel this is a topic which remains 
open.  Additional observations would be beneficial, especially at frequencies below K-band.

\begin{figure}
\epsscale{0.90}
\plotone{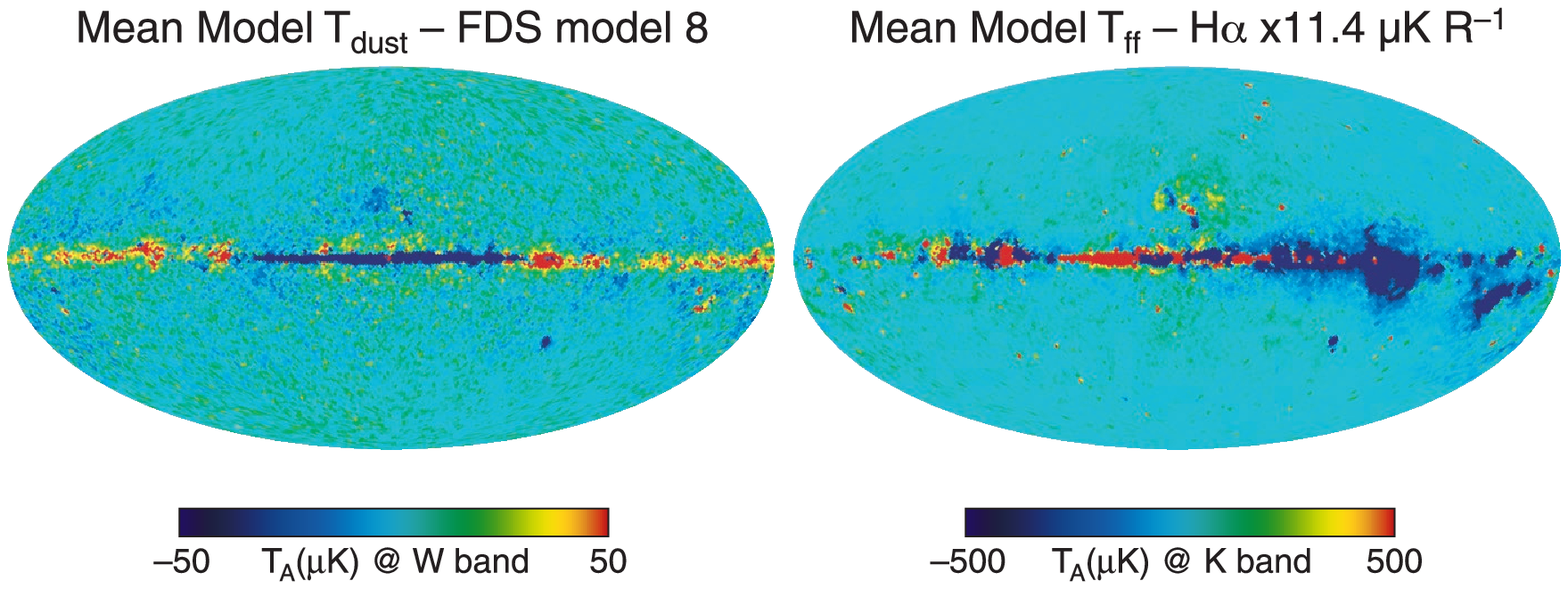}
\caption{ 
({\it Left}): Thermal dust amplitude at W-band averaged over the MCMCg, MEM and Model~9 fits minus
the thermal dust model 8 from \citet{finkbeiner/davis/schlegel:1999}.
({\it Right}): Free-free amplitude at K-band averaged over the same three models, minus the free-free
template estimated from H$\alpha$ observations.
\vspace{1mm} \newline (A color version of this figure is available in the online journal.)
}
\label{fig:models_vs_templates}
\end{figure}

\begin{figure}
\epsscale{0.90}
\plotone{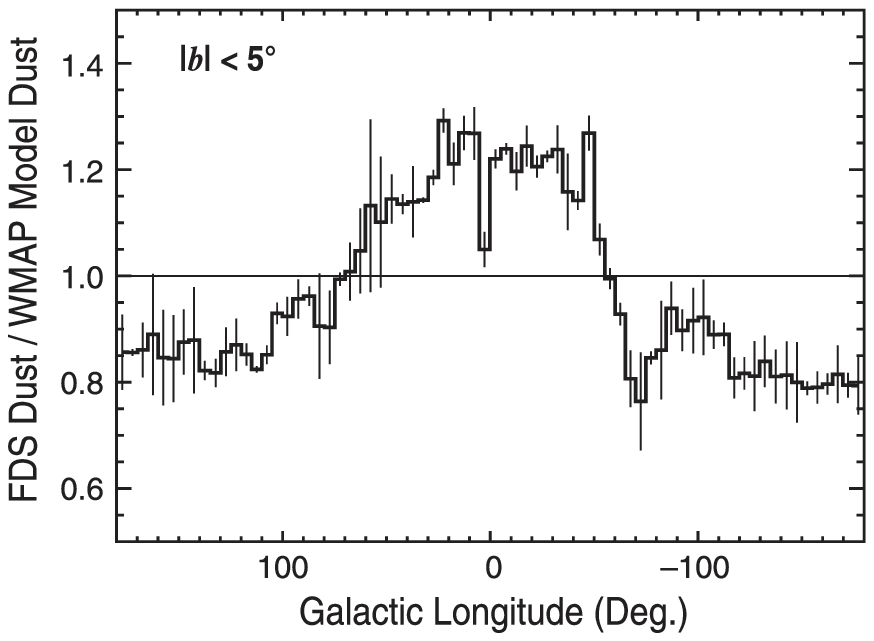}
\caption{ 
The ratio of W-band predicted thermal dust emission (\cite{finkbeiner/davis/schlegel:1999} model 8) 
to the mean over three models (MCMCg, MEM, Model~9) as a function of longitude for $|b|< 5\arcdeg$.
Error bars are derived from the rms scatter of the three models about the mean.  A line is
a plotted at 1.0 to guide the eye.  Modeled emission shows systematic variations from the
FDS prediction by up to 20\%.   
}
\label{fig:dust_tt_plot}
\end{figure}

Although the thermal dust and free-free parameter amplitudes differ between the models
presented here in details, there are clear common-mode similarities when they are
compared against their externally derived equivalents (which we have used in Section~\ref{sssec:tmpl_cln} for 
template cleaning).
Figure~\ref{fig:models_vs_templates} illustrates these common-mode features by taking the mean parameter
amplitudes from three models presented in this paper (MCMCg, MEM and chi-square fitting Model~9),
and differencing them against their template counterparts.  On the left in Figure \ref{fig:models_vs_templates} is
the mean thermal dust amplitude at W-band minus the 94 GHz estimate derived 
from IRAS and COBE data by \citet{finkbeiner/davis/schlegel:1999}.  We have chosen to difference against
their model~8, but a similar result is obtained for their other two-component dust model, model~7.   
In the Galactic plane, all of the three \wmap\ models show more emission in the outer plane and less in the inner plane
than that predicted from the FDS models.  A more quantitative representation of the planar 
differences is shown in Figure~\ref{fig:dust_tt_plot}.  Correlations between MEM, MCMCg and Model~9
have roughly unity slopes, whereas correlations against FDS model~8 indicate FDS is brighter by
up to $\sim20\%$ in high intensity regions in the inner Galaxy.

The right-hand image in Figure~\ref{fig:models_vs_templates} shows the difference between
a mean K-band free-free emission estimate from the same three models in this paper and that from
scattering-corrected de-extincted H$\alpha$ using a conversion factor of $11.4~\mu{\rm K}$ R$^{-1}$.  
Scatter between models in the plane generally disallows a definitive free-free mapping there.  However,
differences between the free-free
emission predicted from H$\alpha$ and the free-free model estimates in this paper consistently indicate
that the H$\alpha$ prediction is higher by roughly 20-30\% in the Gum and Orion regions.
Free-free differences for the Gum away from the plane, where the optical depth is $<1$, can be explained
by a low electron temperature for this region \citep{dickinson/davies/davis:2003, woermann/etal:2000}.  
Differences for other regions
are most likely due to errors in the extinction correction, since the assumption of uniformly mixed
dust and gas may not be valid.  Although W-band Galactic emission is primarily either from thermal dust
or free-free, linear combinations of the FDS dust model and H$\alpha$ predicted free-free have
consistently been unable to describe the \wmap\  data in the plane; these apparent errors in both templates
are consistent with those fitting errors.

\paragraph{ILC Errors}
\label{sec:ilc_errors}

Here we consider two types of error in the ILC: error due to CMB-foreground
covariance, and error due to an incorrect estimate of the bias.  See for example
\citet{hinshaw/etal:2007}.  These are errors which leave residual foreground
signatures in the ILC estimate of the CMB.
\footnote{The ILC also has the three types of errors in the band maps
mentioned in Section~\ref{sec:fg_data}: gain calibration error, instrument
noise, and $\csc |b|$ foreground monopole errors.  These can be propagated
through to the ILC using the ILC regions and the weights given in
Table~\ref{tab:ilc_coeffs}.}

The bias correction is directly related to the foreground model.  To determine
the ILC bias, we take maps of our foreground-only estimate (without CMB) in each
of the five \WMAP\ bands and construct an ILC directly.  The specific attribution
of the foregrounds to individual components (synchrotron, free-free, etc.) is
not needed in this step; we only require maps of the total foreground in each
band.  If the foregrounds are sufficiently complex (if they are not a linear
combination of 4 or fewer spectra in each region), then there will be residuals
in this foreground-only ILC, and this is the ILC bias.  The ILC bias consists of
foregrounds that cannot be removed by any set of ILC weights.  With enough
diversity in foreground spectral components, we can find a linear combination of foreground spectra that mimics
the CMB, and we cannot remove the CMB signature from the ILC by construction,
because the ILC weights must sum to 1.  To deal with the ILC bias, we construct
a foreground model, compute the ILC bias, and subtract it directly from the ILC.
Inaccuracies in the foreground model will translate to an incorrect subtraction
of the ILC bias. 

\begin{figure}
\epsscale{0.90}
\plotone{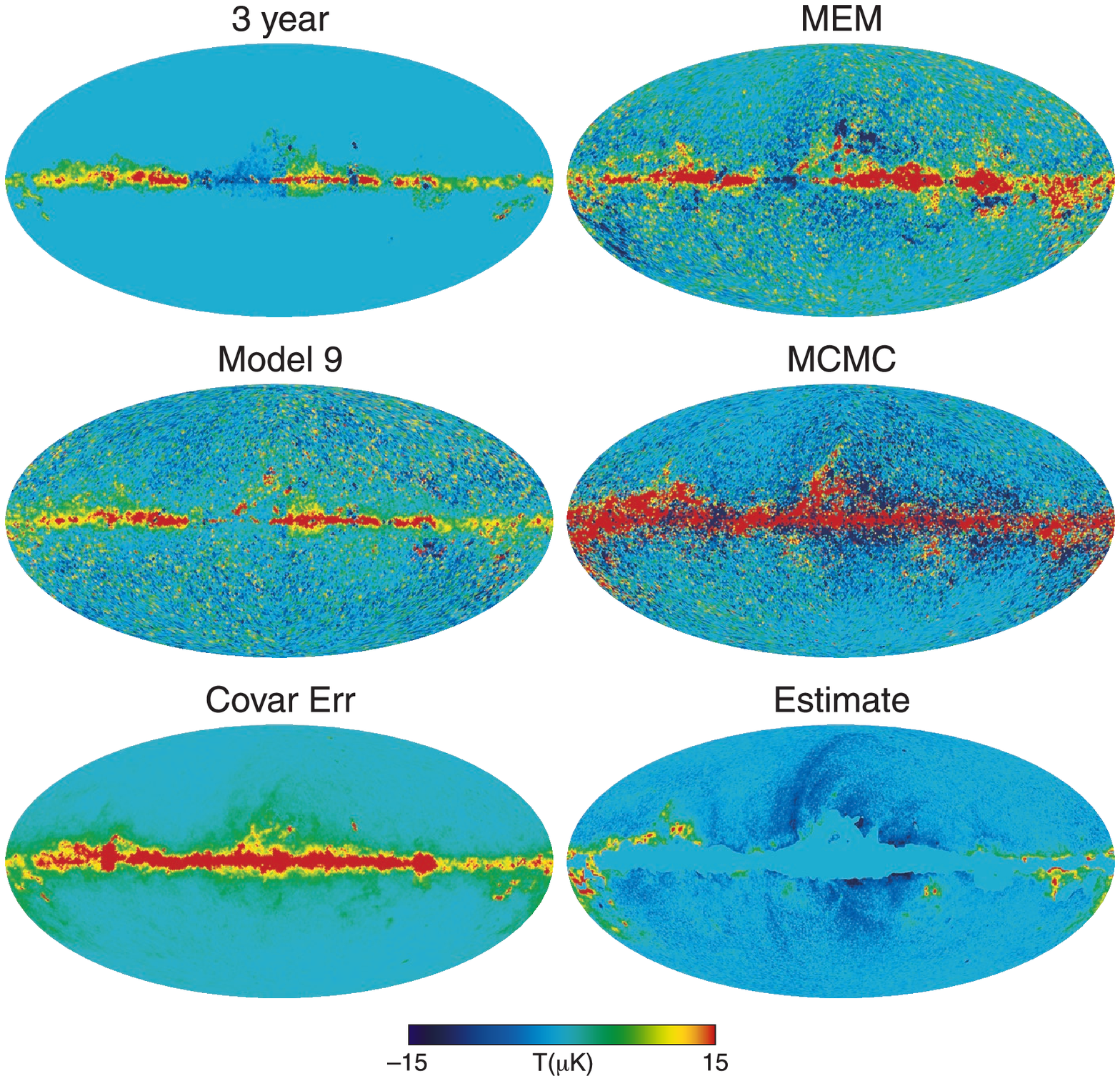}
\caption{Estimates of foreground bias error remaining in the ILC map, on a scale of
$\pm15~\mu{\rm K}$.
{\it Top left}: Bias map from the three-year analysis of \citet{hinshaw/etal:2007}. The map is
zeroed outside the Kp2 cut.
{\it Top right and middle}: Bias estimates resulting from the application of the nine-year ILC
coefficients to the Galaxy models from MEM, Model 9 and MCMCg analysis.  The bias map from
the MCMCg analysis is overestimated in the plane (see text).
{\it Bottom left}: ILC error from foreground-CMB covariance.  Within the Kp2 cut, this error and
the foreground bias are of comparable magnitude.
{\it Bottom right}: An estimate of the potential magnitude of ILC foreground bias outside the
Kp2 cut, based on the various model results, with heavy weight given to the MCMCg model. 
Bias errors of $10\mu{\rm K}$ or less are indicated.
\vspace{1mm} \newline (A color version of this figure is available in the online journal.)
}
\label{fig:ilc_bias}
\end{figure}

An estimate of the ILC bias was computed by \citet{hinshaw/etal:2007} from
simulations and three-year data.    We revisit the bias computation using
the Galactic emission estimates in the five \wmap\ bands from Model 9, MEM and MCMCg.
If these models perfectly describe the total Galactic emission at \wmap\ frequencies,
then a bias map can easily be constructed by applying the flight ILC weights (given in
Table~\ref{tab:ilc_coeffs}) to these
foreground maps.  Such an application is shown in Figure~\ref{fig:ilc_bias}.  For comparison,
Figure~\ref{fig:ilc_bias} also shows  
the bias correction from the three-year analysis, which is non-zero within the Kp2 mask
and zero everywhere outside the mask. 

Close to the Galactic plane, the bias computed from the MCMCg model is larger than that for
the other two models.  Removal of this bias from the uncorrected \wmap\ data ILC  
shows a clear negative residual in the plane for $|l| < 120\arcdeg$, indicating over-correction. 
In addition, ILC regional weights computed for the MCMCg model are sufficiently different from
flight data values to render the model ``goodness'' suspect near the plane within the Kp2 cut.
This is in part due to poorly constrained apportionment between CMB and Galactic signals in the
plane. In particular there is an inverse correlation between CMB and dust spectral index,
resulting in higher fractional residuals in portions of the plane for the MCMCg fit to V-band.
V-band typically has the highest
ILC weight, so these residuals lead to a higher bias for this model. 
Within the Kp2 cut, both Model 9 and the MEM bias 
maps show similar behavior
to the three-year bias map, although details vary.  Both models also return foreground ILC
regional weights similar to data values, with the MEM showing the closest correspondence.
Bias levels within the Kp2 cut are estimated from these two models as near $20$ $\mu{\rm K}$ or less.
These levels are either of similar magnitude or smaller compared to those computed for
the CMB-foreground covariance in the same location (see below). 

Estimating the foreground bias at higher latitudes is more difficult than for the Galactic plane regions.  
Since classic ILC weights are primarily determined using sky pixels within the Kp2 cut
(even for the high latitude region 0), correspondence between derived model and data weights
is only a useful diagnostic for pixels within the Kp2 mask.  In addition,
both the MEM and Model 9 
results are ILC dependent: MEM subtracts the ILC from the data as a prelude to foreground fitting, 
and the six-band $\chi^2$ Model 9 fit relies on the ILC as a strong prior.
Since the classic ILC algorithm applies no bias correction outside the Kp2 cut, it is
possible for any existing high-latitude ILC foreground bias to either remove or add power
to the high latitude sky which is being fit to a Galaxy model.  Since Galactic signals are
generally weaker here than in the plane, the fractional error is potentially higher.
Here the MCMC method provides
the most objective model for estimating high latitude bias, since the CMB contribution is 
determined independently as part of the fitting process. We have used an amalgam of the
three model bias maps to construct a very crude estimate of ILC bias outside of the Kp2 cut,
giving the most weight to the MCMCg result.    All three bias maps show a common
characteristic dust-like excess in the outer Galaxy near the edges of the Kp2 cut.  Two of
the three bias maps show a low-level inner Galaxy deficit with a synchrotron-like signature.
Noise in the bias maps makes a clear determination of the morphology difficult; we have used
templates to represent the spatial structure, but the fine structural detail of the templates
should not be taken as truth. 
Our rough estimate of the high latitude ILC bias is
shown at the bottom right of Figure~\ref{fig:ilc_bias}. High-latitude ILC bias is estimated
at $10$ $\mu{\rm K}$ or less.

The CMB-foreground covariance was discussed in \citet{hinshaw/etal:2007}.
Because the ILC weights are constructed by minimizing the variance in a region,
the weights adjust to allow foreground fluctuations to cancel CMB fluctuations
as much as possible.  This is more of a problem for small regions.
Because the total foreground level is well measured in the plane (even
if we allow complete uncertainty in the CMB for an error term of
$\sigma\approx70\,\mu{\rm K}$, the foregrounds are bright enough to make this
term small), we can estimate how much the foregrounds could correlate with a
random CMB sky with a given power spectrum.  This estimate will not change
substantially with different foreground models (different estimates of how much
of the \WMAP\ data is CMB and how much is foreground) because it only requires
knowledge of the total foreground level, which is well constrained by the data.
We can experimentally determine the CMB-foreground covariance by generating many
CMB simulations, adding a foreground model to each CMB simulation, 
making a bias-subtracted ILC, and
forming an error map by subtracting the true CMB from the ILC in each simulation.
This gives us an ensemble of error maps, which span a 48 dimensional space.
Since the CMB simulation is perfectly subtracted by any set of weights that add
to 1, our error maps contain no CMB from the simulation.  They only contain errors
from residual foregrounds.  Since there are 60 weights (going into the 12 regions
of the ILC) and 12 constraints where sets of weights must add to 1, there are 48
degrees of freedom in the ILC error.  As with the ILC bias, the results do depend
on foreground model, but not nearly as strongly, as mentioned above.

We construct the 48 maps showing the ILC foreground-CMB covariance modes at res
6 as follows.  We take the foreground Model 9 from
Section~\ref{sec:six_band_fits} and prograde it directly to r9 (with no extra
smoothing), where the ILC regions are defined.  Then we form ILCs by the usual
method, except that we do not smooth between regions as described in Equation~(18)
of \citet{hinshaw/etal:2007} because we next degrade back to r6, which has a
similar effect.  We do this for 1000 CMB realizations, and form a $49152\times
1000$ matrix of the maps, of which we take a singular value decomposition to
determine the most common modes, taking care to normalize properly.  There are
only 48 singular values that are not effectively zero; we use the 1000
simulations to better sample these 48 modes and better determine their
eigenvalues.

\begin{figure}
\epsscale{0.70}
\plotone{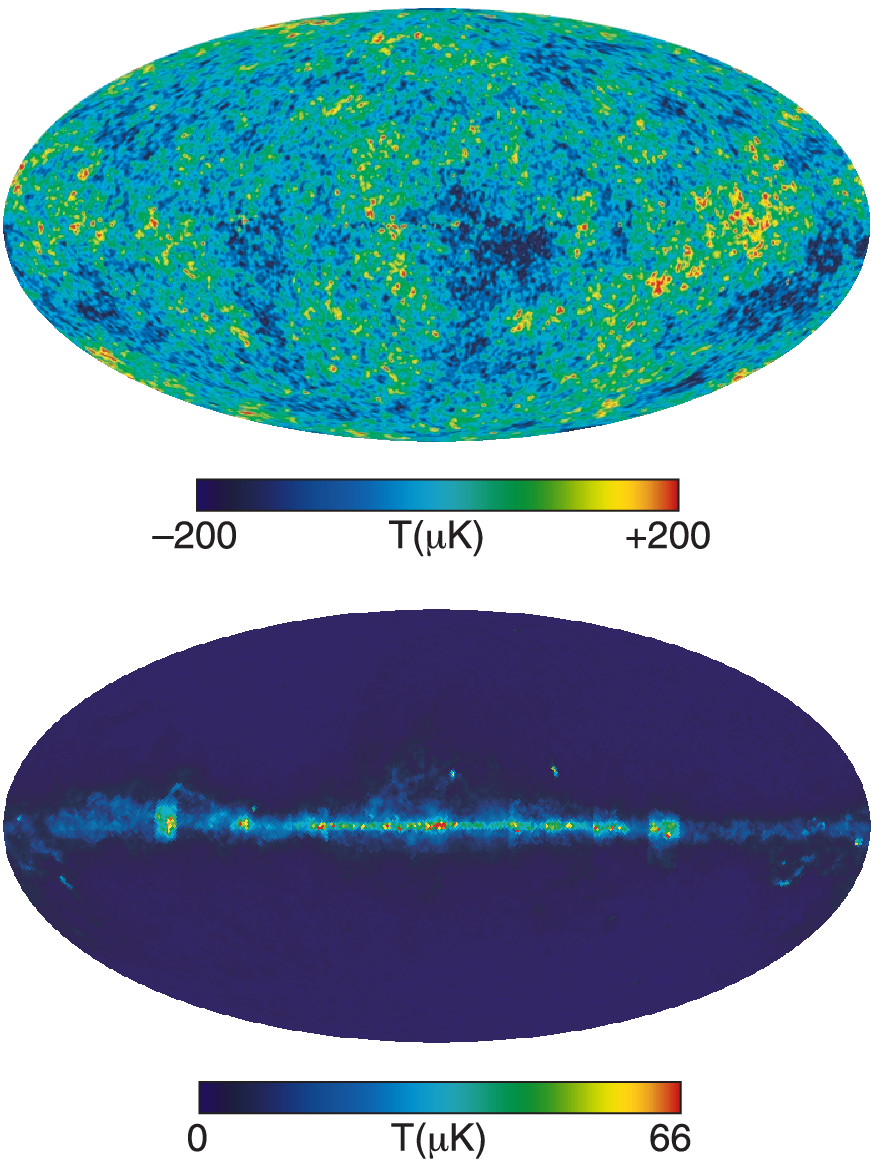}
\caption{
The top map is the nine-year ILC.
The bottom sky map displays the part of the ILC error in each pixel due to foreground-CMB covariance, using the Model 9
foreground estimate from Section~\ref{sec:six_band_fits}.
This shows the square root of the diagonal of the covariance matrix, on a linear 
color scale.  Therefore it shows the standard deviation of expected
error fluctuations, marginalizing over correlations between pixels.
The color scale range was chosen because the r6 ILC map has a
CMB standard deviation of 66 $\mu$K.  Thus, full scale on this map has equal variance 
with the CMB, and at the halfway point on this color scale the
foreground-CMB error variance is down to a quarter of the CMB variance.  
\vspace{1mm} \newline (A color version of this figure is available in the online journal.)
}
\label{fig:ilc_error}
\end{figure}

These modes provide the eigenvalues with nonzero eigenvectors of the
foreground-CMB covariance error matrix.  We compute the square root of the
diagonal elements of this matrix to provide a visual estimate (that ignores
correlations) of this error.  The nine-year ILC map and this error map
are shown in Figure~\ref{fig:ilc_error}.

We demonstrate the use of this error description by propagating the
foreground-CMB error to the quadrupole-octupole alignment, which we describe
in Section~\ref{sec:quad_oct}.

\paragraph{ILC Considerations}
\label{sec:ilc_considerations}

The primary difficulty with any method of extracting the CMB from the data is 
determining how much of the temperature in each pixel is foreground and how much
is CMB.  The data only constrain the sum of these two, and we must make other
assumptions in order to separate them.

The ILC specifically assumes that the CMB has a blackbody spectrum while the
foregrounds do not.  In addition, the ILC assumes that while the foregrounds may
change amplitude across a region, an individual foreground does not change its
spectral shape (proportional to antenna temperature as a function of frequency),
so that a set of ILC weights can null a given foreground everywhere in a region.
Along with this, the ILC assumes that there are four or fewer foreground spectral
shapes, since if there were more, we would not be able to remove them all with
only the five bands of \wmap\ data.  If there were five foreground
spectra, some linear combination of them would be able to mimic a blackbody
spectrum, which the ILC has been designed to keep.

\begin{figure}
\epsscale{0.90}
\plotone{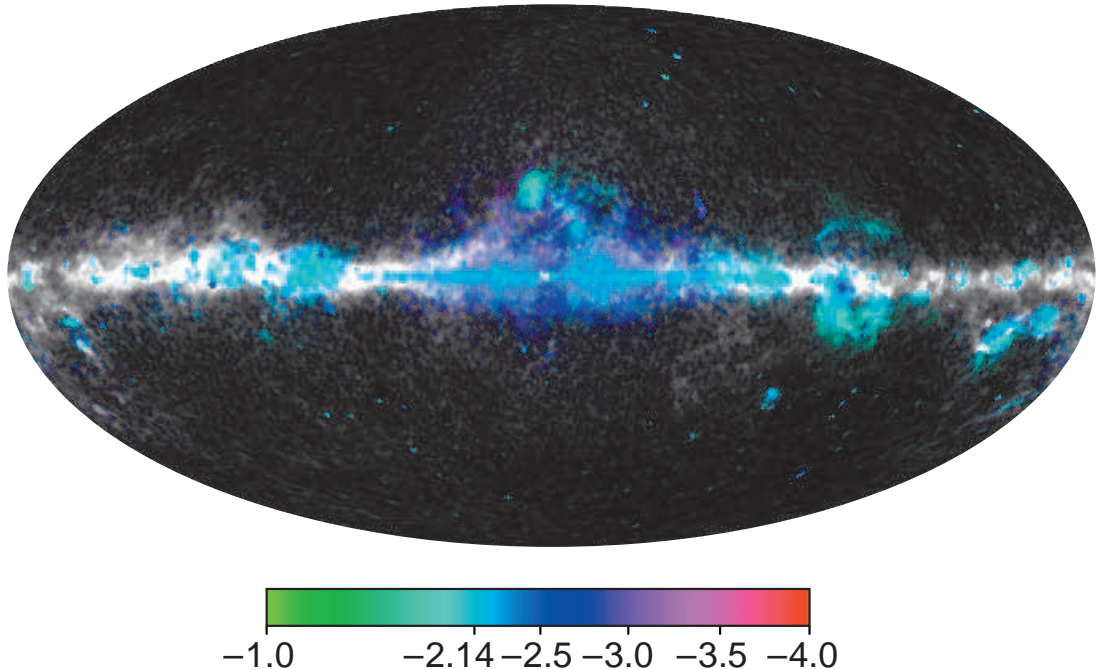}
\caption{The dominant power law in a pixel, combined with information about whether the data in
that pixel look like a pure power law, over the \WMAP\ bands. 
This image was generated by individually specifying the hue, saturation, and
value (HSV) for each pixel.
The hue, shown in the color scale, describes which power law best fits the data.  
It is labeled with values of $\beta$, where the power law in antenna temperature
is $T_A(\nu) \propto \nu^\beta$.
The saturation describes how well the data fit a power law,
so that desaturated (white, gray, black) pixels are not well fit by any power law.
Specifically, let $n_A$ be a 5-vector of the \WMAP\ thermodynamic temperatures, rescaled to be a
unit vector,  and let $n_p$ be a
5-vector of the best fit power law in antenna temperature, converted to
thermodynamic and then also rescaled to be a unit vector.
Then the saturation is $n_A \cdot n_p$, which is just the cosine of the angle
between these two vectors.  The scale is from 0.995 (unsaturated) to 1.0
(completely saturated), so if the two 5-vectors are more than 5.73 degrees apart,
the pixel is unsaturated.  
The value in the HSV color space is the magnitude of the data 5-vector, so it is
the square root of the sum of the squares of the \WMAP\ thermodynamic
temperatures, on a scale of 0 to 2 mK.
Therefore blacker pixels have less emission in all bands; lighter pixels have more
emission. 
The nine-year ILC was subtracted from the \WMAP\ data, before computing the
above image. 
\vspace{1mm} \newline (A color version of this figure is available in the online journal.)
}
\label{fig:dominant_power_law}
\end{figure}

Figure~\ref{fig:dominant_power_law} is one way to visualize the foreground
complexity of the \wmap\ data.  It shows in color the regions that are
approximate power laws, and it shows in grayscale regions that are not well fit
by a single power law.  The ILC methodology can handle more than a single power
law foreground (it can remove up to four of them), so this is not directly a map
of where the ILC will work well.  However, this figure does show the varying
nature of foreground spectra across the sky.

Choosing the ILC region size is a trade-off between foreground complexity and
foreground-CMB covariance.  By choosing small regions, we give the foregrounds
less chance to vary their shape over a region (such as by changing a synchrotron
spectral index).  But small regions are more susceptible to foreground-CMB
covariance, as discussed in \cite{hinshaw/etal:2007}, which suppresses the
variance of the ILC to the extent that the foregrounds and CMB correlate.

We could, for example, take minimum variance to be our figure of merit for an
ILC map and allow arbitrary gerrymandering of the regions on a pixel-by-pixel
basis.  This could be done with a simulated annealing algorithm adjusting some
small number of regions (e.g., 4) within a galactic mask.  However, this would
result in an ILC with variance inside the mask well below the expected CMB
variance, because the regions optimize the foreground-CMB covariance to
artificially suppress the ILC fluctuations.  More knowledge than just the ILC
variance is needed for intelligent region selection.

The foreground-CMB covariance can be estimated moderately well, since it only
depends on an approximate foreground model and knowledge of the CMB power
spectrum.  We estimate this error in Section~\ref{sec:ilc_errors} and propagate
it to the quadrupole-octupole alignment in Section~\ref{sec:quad_oct}.  Other
errors, such as those due to foregrounds changing spectral shape over a region
or more than 4 foreground spectra in a region (these cause the ILC bias), are
harder to estimate because they require an accurate separation of CMB from
foregrounds in the first place.  The demands on this foreground model accuracy
depend on the amplitude of the foregrounds.  For a pixel dominated by CMB, a
slight foreground correction need not be extraordinarily accurate in a
fractional sense. Yet for an extremely bright foreground location on the plane
(say, a bright H II region), the foreground model must have supreme fractional
accuracy to distinguish meaningfully a tiny CMB contribution from the dominating
foregrounds.  

A more accurate ILC would require either a better bias subtraction or better
region selection designed to minimize the needed bias correction; both of these 
require a highly accurate foreground model.  A foreground model that separates
out different components (such as synchrotron, free-free, etc.) is not needed,
only a model that gives the total foreground in each band.  The ILC bias can be
directly calculated by making an ILC of this foreground-only data set, and 
regions could be selected to minimize the bias correction needed in each region.
However, if we already have an accurate separation of the CMB from foregrounds,
then the ILC method is no longer necessary, since we already have a map of the CMB.

\clearpage
\section{Nine-Year Angular Power Spectra}  \label{sec:powerspectra}

In this section we present the nine-year \wmap\ intensity and polarization angular power spectra.  We describe changes in methodology from earlier analyses, and discuss the new results.

The nine-year temperature-temperature (TT) power spectrum computation
uses the full set of V-band and W-band cross-power-spectra. For $2 \leq
l \leq 32$ the TT power spectrum relies on the Gibbs sampled pixel
likelihood, as was the case with the five-year and seven-year data releases.  
New for this nine-year analysis, the $32 < l \leq 1200$ TT
power spectrum is calculated using unbiased and optimal $C^{-1}$
estimation. Earlier releases provided power spectra computed using the
Monte Carlo Apodised Spherical Transform EstimatoR (MASTER) method, an
unbiased but non-optimal quadratic estimator \citep{hivon/etal:2002}. As
was the case for the seven-year \wmap\ analysis, the polarization power
spectra continue to be computed using MASTER.

For the $2 \leq \ell \leq 32$ Gibbs sampling, we use a slightly different ILC
map than we have in the past.  We use a bias-corrected one-region ILC map.  The
same weights are used for the whole sky; these weights are chosen to minimize
the variance of the ILC outside of the combination of the first-year Kp8 mask
and the seven-year point source mask.  The data used for this low-resolution
analysis are the deconvolved one-degree-smoothed nine-year maps for K- through
W-bands.  The coaddition over nine years was done using a slightly older version
of $N_{\rm obs}$ that was available at the time we did the calculation; this has
a small effect on the final nine-year temperature maps.

The bias correction for this ILC requires a foreground model.  We determine the
foreground model by fitting four one-degree smoothed templates and a monopole
term to the one-degree smoothed W-band data.  We do the fit outside the
combination of a Kp22 mask and seven-year source mask, to avoid requiring that
the templates be highly accurate in the brightest portion of the galactic plane.
The four templates are as follows.  We use the FDS model 8, evaluated at 94 GHz,
as described in Section~\ref{sec:temperature_cleaning}; a de-extincted H$\alpha$
map with scattering correction applied, described in detail in
Section~\ref{sssec:diffuse_fg_intro}; a dust model emission ``delta correction''
map, computed as FDS model 8 multiplied by 
$(T_{\rm dust} - \langle T_{\rm dust} \rangle) / \langle T_{\rm dust} \rangle$,
where $T_{\rm dust}$ is the dust temperature map from SFD and the average dust
value $\langle T_{\rm dust} \rangle$ was calculated outside the Kp2 mask; and a
map of discrete HII region emission (primarily along the plane), evaluated at
2.7 GHz and 1 degree beam width using data from the \citet{paladini/etal:2003}
catalog of 1442 Galactic HII regions.  This last map was scaled to 93 GHz
assuming an optically thin free-free spectrum for each source. After removal of
these foregrounds from the W-band map, we consider the remainder to be a pure
CMB map.  To obtain our foreground model of the galaxy, we subtract this CMB
estimate from each band of the flight data.  Our foreground model therefore has
information about how much temperature comes from the CMB and how much from
foregrounds, but it does not break the foreground temperature into physical
components, since this is not necessary to estimate ILC bias.  

The ILC bias can then be calculated as the error in an ILC map, averaged over
many CMB realizations but using the same foreground model.  It can be directly
computed by making an ILC of the foreground-only data, without adding in a CMB
simulation.  We subtract this ILC bias from the one-region ILC described above.

We do use CMB simulations to determine the foreground-CMB covariance error
modes.  Using a power spectrum from a set of seven-year simulations, we generate
100 CMB realizations, add our foreground model, and generate a one-region ILC as
above.  There are four error modes, since we generate the ILC from five weights
with the single constraint that they must sum to 1.  We determine these modes
from the covariance matrix of errors.  We find that one mode is negligible
outside of the KQ85y9 mask that is used for Gibbs sampling, so we only
marginalize over the three most important CMB-foreground covariance modes in the
Gibbs sampler.

We smooth the ILC map to $5^\circ$ FWHM (Full Width at Half Maximum) before any
masking; this is the map over which we Gibbs sample.  
Since the ILC is already smoothed to $1^\circ$ FWHM, this
requires an additional smoothing by $\sqrt{24} \approx 4\ddeg9$.  We
then degrade the map to r5, and add $2$ $\mu$K rms noise per pixel to
the r5 ILC, as was done in the five-year and seven-year data releases.
The Gibbs sampler uses a mask based on degrading the KQ85y9 mask to r5,
and leaving unmasked only those r5 pixels for which $>50\%$ of the r9
pixels are unmasked.  The KQ85y9 mask allows through 2353196 out of
3145728 pixels, or 74.8\% of the sky.  After degrading to r5 by the
above method, the mask lets through 9496 out of 12288 pixels, or 77.3\%
of the sky.  According to our newly estimated ILC errors, the pixels
near the edge of this mask may fluctuate randomly up to about $\sim 11$
$\mu$K, so residual foregrounds are a small fraction of the CMB variance
when the masked ILC is used.

\subsection{High $\ell$ TT summary}

The optimal (i.e.~minimum variance) power spectrum estimator has been known for
many years~\citep{Tegmark:1996qt,Bond:1998zw} but has appeared to be
computationally intractable for a large ($\gsim 10^6$ pixel) experiment such as
\wmap.  As a result, standard practice is to use estimators
that do not achieve optimal statistical errors, in exchange for reduced
computational cost.  For the nine-year \wmap\ data, we replace the MASTER power
spectrum estimator by the optimal unbiased quadratic estimator.
This optimal estimator has now been implemented in a
computationally affordable way.  We report the first \wmap\ power spectrum with
optimal error bars on the TT spectrum across the entire observed range of scales 
$2 \lsim \ell \lsim 1200$.

The basic building block is a fast algorithm~\citep{Smith:2007rg} for multiplying
a temperature map (thought of as a length-$\Npix$ vector $x$) by the
$\Npix$-by-$\Npix$ inverse covariance matrix $C^{-1}$.  Here, the covariance
matrix $C = S+N$ consists of signal and instrumental noise contributions, and
incorporates the Galactic mask, the instrument beam size, and marginalization 
over the monopole and dipole.  The multigrid algorithm from~\cite{Smith:2007rg} allows a
single multiplication operation of the form $x \rightarrow C^{-1}x$ to be
performed for \wmap\ in $\approx 10$ core-minutes, although it is impossible to
compute (or even store) the matrix $C^{-1}$ in dense form.  This means that all
computations involving $C^{-1}$ must be formulated so that they are based on a
(reasonably small) number of multiplications of the form $x \rightarrow C^{-1}x$.

In practice, we need to modify the optimal estimator
$\hC_\ell$ by removing auto-correlations, which are highly sensitive to the
instrumental noise model.  For an all-sky experiment such as \wmap\, the noise
must be known to $\lsim 0.1$\% to avoid a statistically significant
additive bias to $\hC_\ell$.  This level is impractical to achieve, but 
sensitivity to the noise model can be mitigated by constructing a
modified estimator, $\hC_\ell^\times$, that only includes terms calculated from cross-spectra.

The unnormalized estimator written out for a single map $d$ is 
\be
\hE_\ell[d] = \frac{1}{2} d^T C^{-1} A \Pi_\ell A^T C^{-1} d  \label{eq:hE_def}
\ee
where $A$ is the $a_{\ell m}$-to-map operator that includes beam convolution,
and $\Pi_\ell$ projects out all modes not at a given multipole $\ell$.
The optimal power spectrum estimator $\hC_\ell$ is constructed from
\be
\hC_\ell[d] = F_{\ell\ell'}^{-1}\left(\hE_\ell[d] - \N_\ell \right),
\ee
where $\N_\ell$ is the noise bias and the Fisher matrix $F_{\ell\ell'}$ is given by
\be
F_{\ell\ell'} = \frac{1}{2} \Tr\Big( A^T C^{-1} A \, \Pi_\ell \, A^T C^{-1} A \, \Pi_{\ell'} \Big).
\ee
We also construct a cross-correlation-only power spectrum estimator $\hC_\ell^\times$ with zero noise bias,
by only keeping cross-correlations between maps with independent noise.
More specifically, we divide the data into maps $d_\alpha$, where $\alpha=(c,y)$
indexes a combination of a differencing assembly
$c=\mbox{V1},\mbox{V2},\mbox{W1},\mbox{W2},\mbox{W3},\mbox{W4}$ 
and a specific single year of \wmap\ data, $y$.
The unnormalized estimator $\hE_\ell$ defined in~(\ref{eq:hE_def}) can then be written as a double sum
over pairs $(\alpha,\beta)$; we simply keep the terms with $\alpha\ne\beta$ to
define an unnormalized cross-correlation estimator $\hE^\times_\ell$.
(In implementation, it is more computationally efficient to subtract the terms
with $\alpha=\beta$.)
We then define the cross-correlation estimator $\hC_\ell^\times$  by
$\hC_\ell^\times = (F_{\ell\ell'}^\times)^{-1} \hE^\times_{\ell'}$,
where $F_{\ell\ell'}^\times$ is an appropriately modified Fisher matrix.

The \wmap\ $C^{-1}$ TT pipeline provides a power spectrum estimate
and an estimate for the covariance matrix $\Cov(C_\ell,C_{\ell'})$.
To account for the slight non-Gaussianity of the
likelihood at $\ell > 32$, our likelihood remains the combination of a
Gaussian and offset log-normal distribution in $\mathscr{C}_\ell^{\rm th}$, as
discussed in \citet{verde/etal:2003}.  Discussion of the log-normal distribution
for cosmological likelihoods is also in \citet{bond/jaffe/knox:2000} and
\citet{sievers/etal:2003}.
We use a noise estimate to provide the offset in our offset log-normal
distribution, $\mathscr{N}_\ell$.  
This is the error in the power spectrum due to instrument noise, in the form of
$\ell(\ell+1)C_\ell/(2\pi)$.
Additional variables to describe the likelihood include
\be
\widehat{\mathscr{C}}_\ell \equiv \frac{\ell(\ell+1)\widehat{C}_\ell}{2\pi}
\qquad
\mathscr{C}^{\rm th}_\ell \equiv \frac{\ell(\ell+1)C^{\rm th}_\ell}{2\pi}
\ee
\be
\widehat{z}_\ell \equiv \ln(\widehat{\mathscr{C}}_\ell + \mathscr{N}_\ell)
\qquad
z^{\rm th}_\ell \equiv \ln(\mathscr{C}^{\rm th}_\ell + \mathscr{N}_\ell)
\ee
\be
\mathscr{Q}_{\ell\ell'} \equiv (\mathscr{C}_\ell^{\rm th} + \mathscr{N}_\ell) \mathcal{Q}_{\ell\ell'}
(\mathscr{C}_{\ell'}^{\rm th} + \mathscr{N}_{\ell'}),
\ee
where $\mathcal{Q}_{\ell\ell'}$ is the inverse covariance matrix of the power
spectrum estimate $\widehat{\mathscr{C}}_\ell$ provided by the optimal
estimator.  Finally, we write the \wmap\ likelihood as a combination of a
Gaussian and offset log-normal distribution.
\ba
\ln \mathscr{L}_{\rm Gauss} & = & -\frac12 \sum_{\ell\ell'} 
(\mathscr{C}_\ell^{\rm th} - \widehat{\mathscr{C}}_\ell) \mathcal{Q}_{\ell\ell'} 
(\mathscr{C}_{\ell'}^{\rm th} - \widehat{\mathscr{C}}_{\ell'}) 
+ {\rm const}. \\
\ln \mathscr{L}_{\rm LN} & = & -\frac12 \sum_{\ell\ell'} 
(z_\ell^{\rm th} - \widehat{z}_\ell) \mathscr{Q}_{\ell\ell'} 
(z_{\ell'}^{\rm th} - \widehat{z}_{\ell'})  \\
\ln \mathscr{L}_{\rm WMAP} & = & \frac13 \ln \mathscr{L}_{\rm Gauss} + 
\frac23 \ln \mathscr{L}_{\rm LN}
\ea

\subsection{The $C^{-1}$ Pipeline}
\label{sec:wmap}

We first applied the new $C^{-1}$ pipeline to the seven-year \WMAP\ data after its publication.
We performed end-to-end tests to arrive at the first \wmap\ power spectrum that is optimal 
for all values of $\ell$. We then compared the new power spectrum with the pseudo-$C_\ell$ 
MASTER spectrum from the \wmap\ seven-year release.
We did not propagate the optimal power spectrum to cosmological parameter
constraints for the seven-year data.  Based on the seven-year power spectrum comparisons, we decided to 
implement the $C^{-1}$ power spectrum for what are now the nine-year \wmap\ results.

The \wmap\ seven-year data $C^{-1}$ evaluation used foreground-cleaned maps from
the six V- and W-band differencing assemblies, further subdivided by individual
year data $y=1,2,\ldots 7$, for a total of 42 cross-correlations.  We masked
regions of high Galactic foreground emission and bright point sources by using
the KQ85 mask~\citep{Gold:2010fm}.  We report a power spectrum to
$\ellmax=1200$, but we ran the pipeline to $\ellmax=1500$ to avoid edge
artifacts near the maximum multipole of the reported power spectrum.

Unless otherwise specified, all results are based on the power spectrum estimator $\hC_\ell^\times$, which 
only contains cross-correlations.
After estimating the power spectrum, we subtract an estimate of the bias due to unresolved point sources, assuming
a single population of radio sources with frequency dependence $g_{\rm ant}(\nu) \propto \nu^{-2.09}$ in antenna
temperature, or equivalently
\be
g(\nu) \propto
    \left( \frac{h\nu}{kT_{\rm CMB}} \right)^{-2}
    \frac{ \big( \exp(h\nu/kT_{\rm CMB}) - 1 \big)^2 }{ \exp(h\nu/kT_{\rm CMB}) } \nu^{-2.09}
\ee
in thermodynamic temperature units, where $h$ is Planck's constant, $k$
is the Boltzmann constant, and $T_{\rm CMB}$ is the CMB monopole temperature.

\subsubsection{$C^{-1}$ Pipeline Tests}
\label{ssec:tests}

\begin{figure}
\plotone{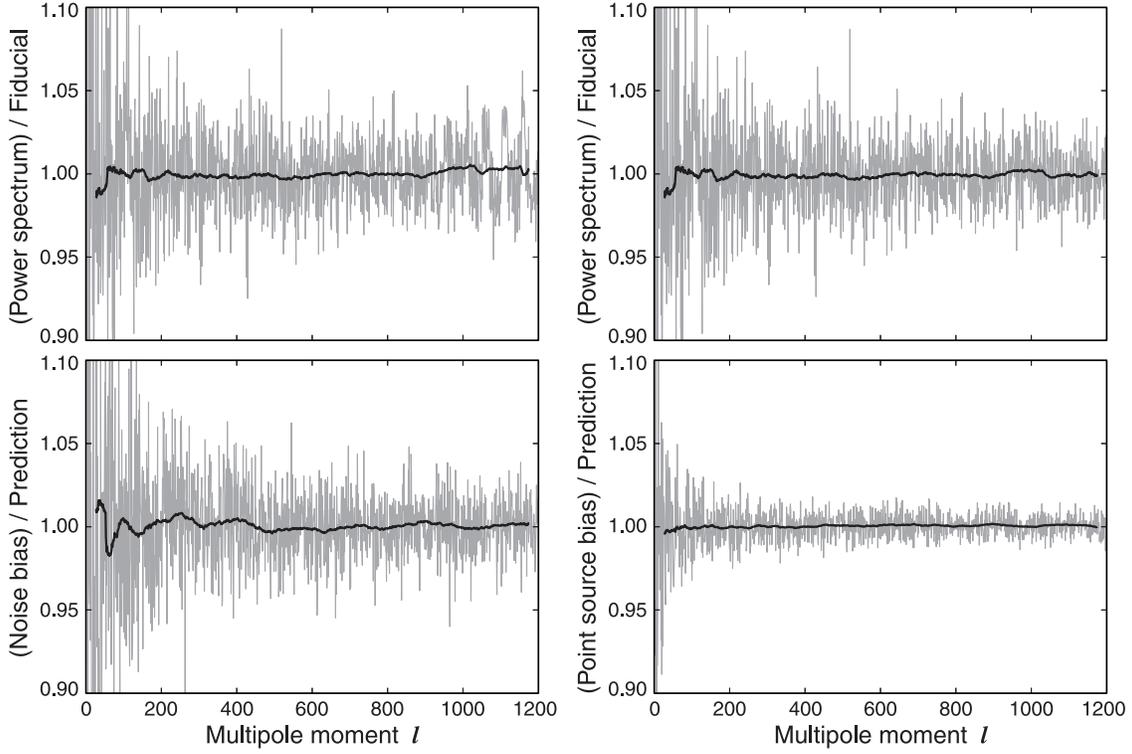}
\caption{End-to-end Monte Carlo pipeline tests.  The gray lines are individual $\ell$'s
and the black lines are boxcar smoothed with $\Delta\ell=50$.  In all four cases, the
ratio of the Monte Carlo estimated power spectrum and the predicted value is consistent with
unity.
 {\em Top left.} Ratio $\langle \hC_\ell^\times \rangle_{\rm sig} / C_\ell^{\rm fid}$
between mean estimated power spectrum of CMB-only simulations and the fiducial
input spectrum.
 {\em Top right.} Same as top left panel, but using the auto-correlation estimator $\hC_\ell$ instead
of the no-auto estimator $\hC_\ell^\times$.
 {\em Bottom left.} Ratio
between mean estimated power spectrum of noise-only simulations and the predicted noise bias,
using the auto-estimator $\hC_\ell$.
  {\em Bottom right.} Ratio between mean estimated power spectrum of point source simulations
and predicted bias.}
\label{fig:end_to_end}
\end{figure}

In our power spectrum pipeline, we precompute three quantities:
a transfer matrix $F_{\ell\ell'}$ that represents the mean response of the unnormalized estimator
at multipole $\ell$ to CMB power at multipole $\ell$'; the bias of the power spectrum estimator
due to unresolved point sources; and the noise bias,
for the auto-correlation estimator $\hC_\ell$ (but not
for the cross-correlation estimator $\hC_\ell^\times$).
In Figure \ref{fig:end_to_end}, we present end-to-end Monte Carlo tests of these precomputations
using three simulated ensembles:
CMB-only simulations, point source simulations, and noise-only simulations.
In all cases the ratio of the recovered power spectrum (averaged over many Monte Carlo
realizations) to the expected power spectrum is consistent with unity.

Our pipeline uses interpolation in $\ell$ to estimate transfer matrices, noise bias,
and point source bias.  We did an end-to-end test of the interpolation accuracy as follows.
We reran the pipeline with half the interpolation step size, treated the difference
between the two estimates as a power spectrum bias, and then we did a Fisher matrix forecast to determine whether
the resulting bias was statistically significant.  In all three cases, we found that the resulting
bias is $\lsim 0.02\sigma$, i.e.~much too small to be important.

We estimate the power spectrum covariance matrix $\Cov(\hC^\times_\ell, \hC^\times_{\ell'})$
using Monte Carlo simulations.
A direct Monte Carlo estimation of a 1200-by-1200 covariance matrix
would require a prohibitive number of simulations, but this can be sped up using
computational tricks: 
(1) the covariance $\Cov(\hC_\ell,\hC_{\ell'})$ of the auto-estimator is
equal to the inverse Fisher matrix $F_{\ell\ell'}^{-1}$, so we only need Monte Carlos
for the estimator difference $(\hC_\ell^\times-\hC_\ell)$;
(2) we only estimate variances and assume that off-diagonal covariances are given by
appropriately rescaling Fisher matrix elements;
and (3) we smooth the variance estimates in $\ell$.
These tricks allow the covariance matrix to be accurately estimated from a small number of simulations.
As an end-to-end convergence test, we compared covariance matrices
$C_{256}$, $C_{512}$ constructed using 256 and 512 Monte Carlo simulations respectively.
We found that all matrix entries were nearly identical in that all Karhunen-Lo\`eve eigenvalues of the
matrix pair $(C_{256},C_{512})$ are between 0.999 and 1.001.

\subsubsection{$C^{-1}$ Versus MASTER Comparison}

\begin{figure}
\plotone{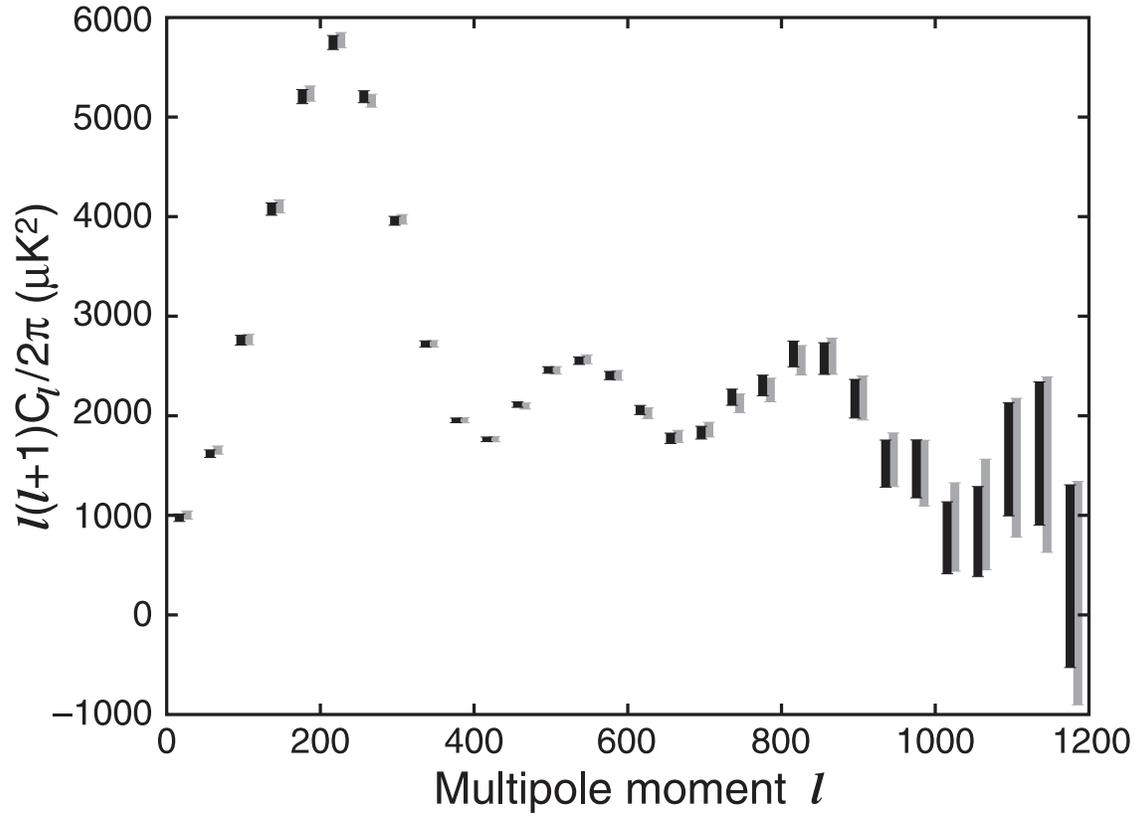}
\label{fig:errorbars}
\caption{Binned \WMAP7 power spectrum estimates using the optimal pipeline from this paper (left/black error bars),
with the estimates from the \WMAP7 release~\citep{Larson:2010gs} shown for comparison (right/grey error bars).}
\end{figure}

\begin{figure}
\epsscale{0.75}
\plotone{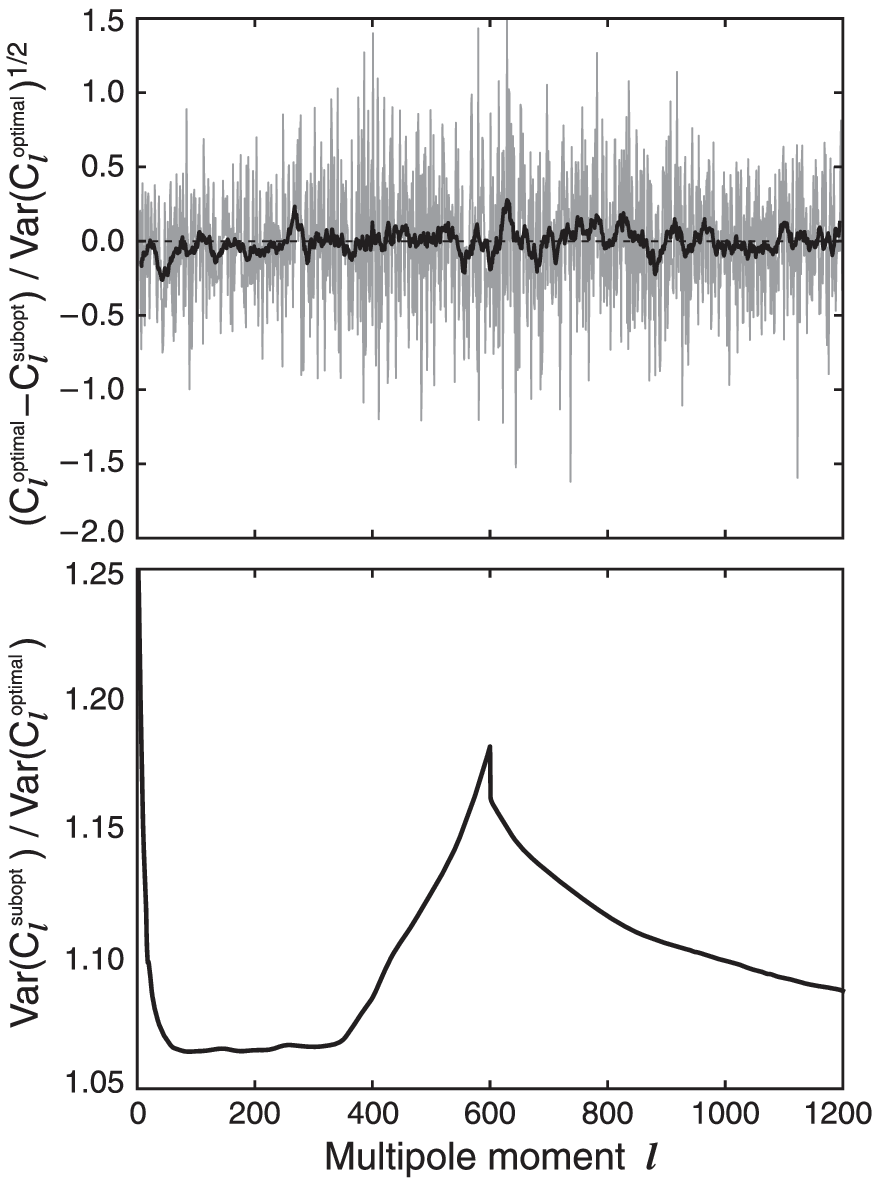}
\caption{Detailed comparison between \WMAP7 optimal power spectrum estimator and suboptimal estimator from~\citet{Larson:2010gs}. 
  {\em Top:} Difference $(\hC_\ell^{\rm optimal} - \hC_\ell^{\rm subopt}) / \Var(\hC_\ell^{\rm optimal})^{1/2}$ 
between the two estimators in ``sigmas'', for every $\ell$, and boxcar-smoothed with $\Delta\ell=10$.
  {\em Bottom:} Variance ratio between suboptimal and optimal estimators.
}
\label{fig:ratios}
\end{figure}

In Figure \ref{fig:errorbars}, we show the binned power spectrum estimates for the seven-year \wmap\ data obtained using the optimal
pipeline, described above, with the sub-optimal MASTER results used in the seven-year \wmap\  release~\citep{Larson:2010gs} shown for comparison.
The agreement is excellent; the two estimators agree to better than 1$\sigma$
in every $\ell$-bin, as expected when comparing an optimal and near-optimal analysis of the same
data.

To compare the two estimators more closely, in the left panel of Figure \ref{fig:ratios} we show the difference between
the optimal and sub-optimal estimators, before and after smoothing in $\ell$.
No systematic trends are seen, as expected if the difference is pure statistical scatter.
There is a small region near $\ell=50$ where the optimal estimator fluctuates to a lower value
of $C_\ell$ than the sub-optimal estimator.
This fluctuation slightly shifts the best-fit value of the spectral index $n_s$, as discussed by \citet{hinshaw/etal:prep}. This appears to be the most important difference between the two estimators for purposes of cosmological parameter estimation, aside from the effective sensitivity improvement discussed below.

The right panel of Figure \ref{fig:ratios} shows the ratio between the power spectrum variance $\Var(C_\ell)$
obtained using the optimal and sub-optimal estimators.  The optimal estimator improves the variance
by 7--17\% depending on the value of $\ell$.  This level of improvement is roughly comparable to the improvement
in going from seven-year to nine-year data (which varies from no improvement at low $\ell$ to a factor of $9/7=1.28$ in $C_\ell$ 
at high $\ell$).

\subsection{\wmap\ Power Spectra}

\begin{figure}
\epsscale{1.00}
\plotone{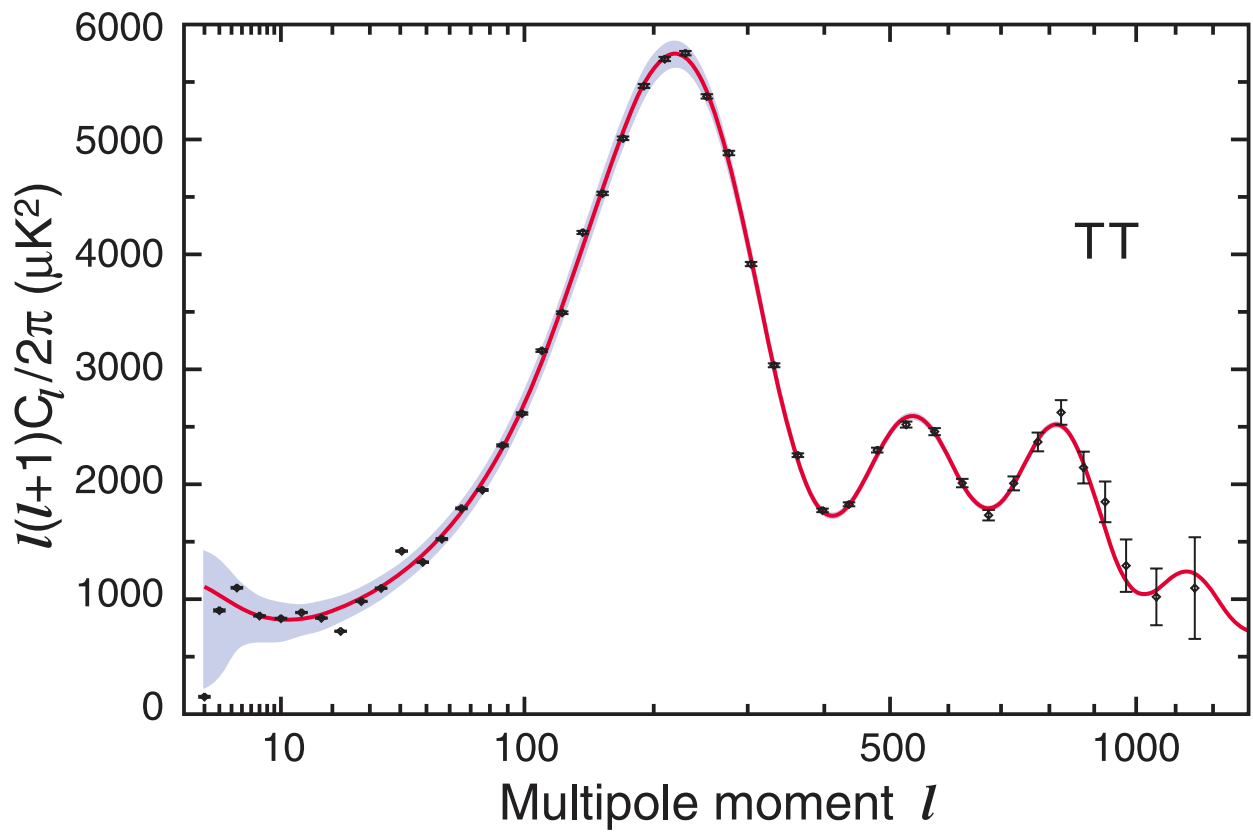}
\caption{
The nine-year \wmap\ TT angular power spectrum.  The \wmap\ data are in black,
with error bars, the best fit model is the red curve, and the smoothed binned
cosmic variance curve is the shaded region. The first three acoustic peaks are well-determined.
\vspace{1mm} \newline (A color version of this figure is available in the online journal.)
}
\label{fig:tt_spectrum}
\end{figure}

The nine-year TT angular power spectrum is shown in Figure \ref{fig:tt_spectrum}. The cosmic variance curve on the power spectrum has been adjusted to more
accurately reflect cosmic variance.  In the past, the value of $f_{\rm sky}$
that we used to expand the error bars was generated by the MASTER code, and it was
roughly the geometric area of the observed sky, which was not optimal.  With the
$C^{-1}$ method of estimating the power spectrum, such as was used in the Gibbs
sampler, one can reconstruct the low \ell\ multipoles on the full sky more
accurately than one might naively expect.  Doing so makes $f_{{\rm sky},\ell}$ 
close to unity at very low \ell.  
In Figure \ref{fig:tt_spectrum}, we use the value of $f_{{\rm sky},\ell}$ generated
by the high-\ell\ $C^{-1}$ code, which is applicable at all lower $\ell$.

The shaded region represents the $1 \sigma$ error bar from cosmic variance, which is
the region where 68\% of binned power spectra that are randomly sampled from the
theory curve would appear.  We form
the error bars around the 68\% with highest probability density per unit
$C_\ell$.  These are determined by sampling $10^6$ power spectra from the theory
spectrum and binning them.  At each multipole \ell, the value of the power
spectrum is sampled from a $\chi^2_\nu$ distribution (which has a mean of $\nu$) 
with $\nu = (2\ell + 1) f_{{\rm sky},\ell}^2$ 
degrees of freedom.  The spectrum is then scaled
by $\ell(\ell+1)C_\ell/(2\pi \nu)$ to give it the correct mean.  Sampling
from the $\chi^2_\nu$ distribution rapidly is done by choosing random numbers in
the interval $[0,1]$ and then using an interpolated cumulative density function
to determine the value of $\chi^2_\nu$.  After binning the power spectra, we
determine the location of the error bars for each bin by finding the pair of
samples that enclose 68\% of the other samples in the bin and are closest
together.

After determining the bin error bars, we consider how to plot the cosmic
variance error bar for a binned angular power spectrum.  Due to the abrupt 
change in binning, from a bin size
of 1 at $\ell=2,3$ to a bin size of 2 for the bin containing $\ell=3$ and
$\ell=4$, the cosmic variance error bar drops significantly.

Despite using a binning scheme, we opt to plot the theory power spectrum as a curve at
each \ell, instead of a binned quantity.  Recall that for the random
distribution of $\ell(\ell+1)C_\ell/(2\pi)$ values, the mean of the theory
spectrum values in a bin is the mean of the binned cosmic variance samples.
Binning the mean of the distribution at each \ell\ gives the mean of bin.  (This
is not true for the median or the mode.)  Likewise, we want to put an unbinned
error bar on the curve with the height of the upper error bar as the height of 
the upper error bar on the binned value.  In this way, the
average height of the cosmic variance curve over the bin is the correct upper
error bar for that bin.  We then use a spline interpolation of the upper
and lower error bars between each bin center.  This makes the above
statement fractionally less true, but prevents abrupt changes in the height of
the cosmic variance curve at the bin edges.  The measurements are cosmic variance limited for $l < 457$ and have a signal-to-noise ratio above unity for $l < 946$.  

The change of the template cleaning method from the seven-year to the nine-year
analysis results in a slight change in the low-$\ell$ power spectrum.
For $2\leq \ell \leq 16$, using the MASTER method with the KQ85y9 mask,
the absolute value of the change in $\ell(\ell+1)/(2\pi)C_\ell$ due to
the template cleaning is typically $4\%$ of cosmic variance per $\ell$.

\begin{figure}
\epsscale{0.90}
\plotone{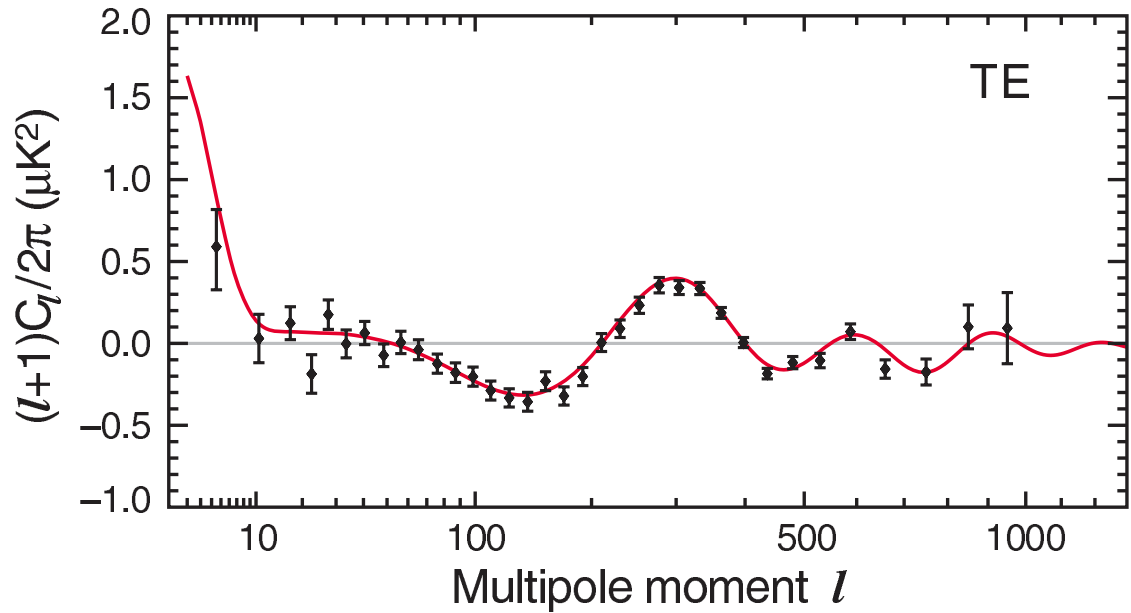}
\caption{
The TE spectrum.  The \wmap\ data points and error bars are in black.  The red theory curve is fit to the full \wmap\ data, including the TT angular power spectrum data.  
Note that the vertical axis on these spectra is $(\ell+1)C_\ell/(2\pi)$ instead
of $\ell(\ell+1)C_\ell/(2\pi)$; this vertical scale differs from that
of the TT spectrum plot by a factor of \ell.
The lowest \ell\ TE bin where $2 \le \ell \le 7$ has been adjusted using
a pixel likelihood code.    
\vspace{1mm} \newline (A color version of this figure is available in the online journal.)
}
\label{fig:te_spectrum}
\end{figure}

Figure \ref{fig:te_spectrum} shows the temperature cross-power spectrum with the E-mode polarization (TE) spectrum.  
This angular cross-power spectrum is computed using the MASTER likelihood code, 
with the lowest $2 \le \ell \le 7$ bin determined using
the more accurate pixel likelihood code.  This was conditioned on
the maximum likelihood power spectrum, and varied the value
$(\ell+1)C_\ell^{TE}/(2\pi) = B_{2\text{--}7}$.  The value $B_{2\text{--}7}$
is independent of \ell.  To maintain the requirement that
$C_\ell^{TE} \le \sqrt{C_\ell^{EE} C_\ell^{TT}}$ 
for a given bin value $B_{2\text{--}7}$,
we adjust the $C_\ell^{EE}$ spectrum upward from the best fit theory
only as much as needed, on an \ell\ by \ell\ basis.
As we vary $B_{2\text{--}7}$, the error bar is based on the minimum
$\chi^2$ value, and where $\Delta\chi^2 = 1$ in either direction.
This gives an asymmetric error bar.  Note that this 
would be a $1\sigma$ error bar for a Gaussian distribution, but it does
not necessarily contain 68\% of the likelihood due both to conditioning
on the higher \ell\ TT, TE and EE power spectra, as well as to 
the non-Gaussian shape of the power spectrum meaning that 
$\Delta\chi^2 = 1$ does not correspond exactly to a 68\% confidence interval.

\begin{figure}
\epsscale{0.90}
\plotone{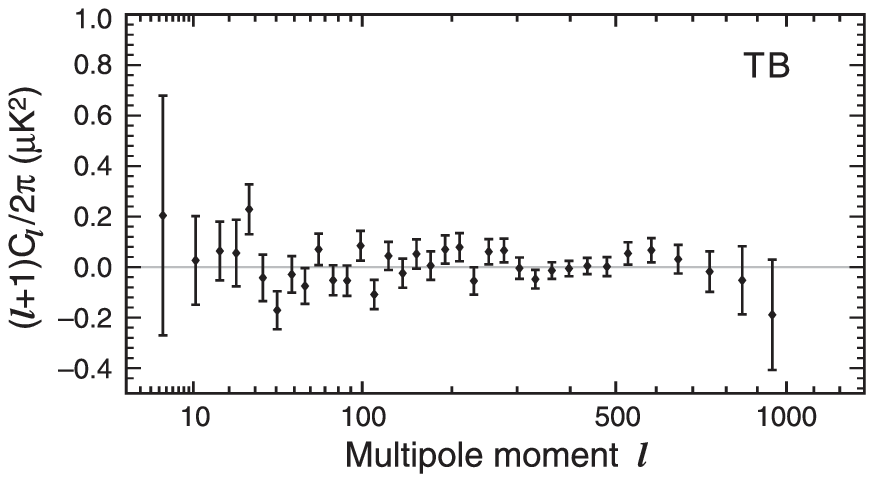}
\caption{
The TB spectrum.
The TB spectrum uses the MASTER likelihood code.
Note that the vertical axis on these spectra is $(\ell+1)C_\ell/(2\pi)$ instead
of $\ell(\ell+1)C_\ell/(2\pi)$; this vertical scale differs from that of the TT spectrum plot by a
factor of \ell.
}
\label{fig:tb_spectrum}
\end{figure}

\begin{figure}
\epsscale{0.6}
\plotone{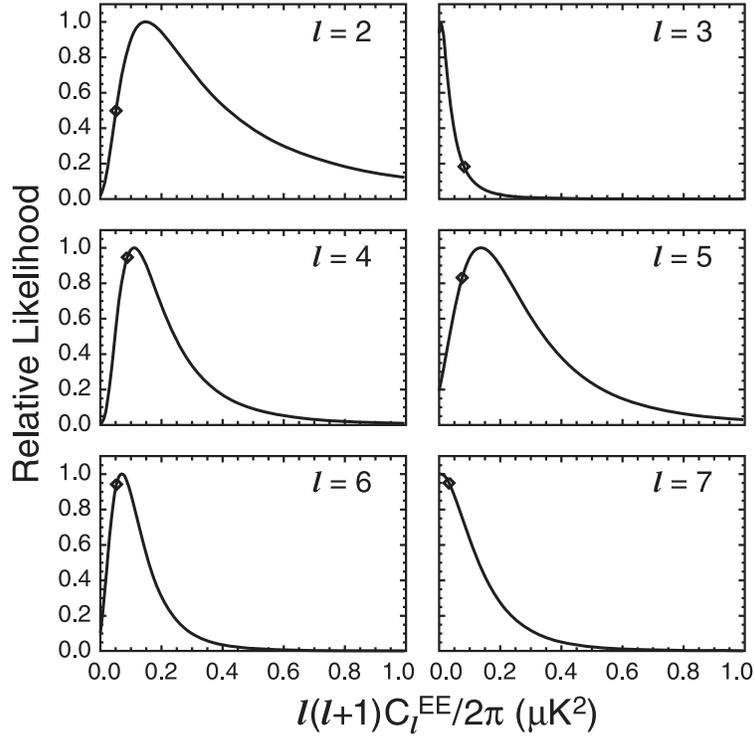}
\caption{
Individual likelihood functions of the low $\ell$
EE polarized power are shown for $\ell=2$ through $7$.  When fitting
at a particular $\ell$, we set $C_\ell$ at all other values of $\ell$
to the value in the best fit \wmap\ power spectrum. In addition, at the
$\ell$ in question we set $C_\ell^{TE}=0$ to maintain that
$C_\ell^{TE} \le \sqrt{C_\ell^{TT} C_\ell^{EE}}$. The black diamonds
denote the best fit \wmap\ EE power spectrum. These likelihood functions include sample variance.  
}
\label{fig:low_ell_EE} 
\end{figure}

\begin{figure}
\epsscale{0.60}
\plotone{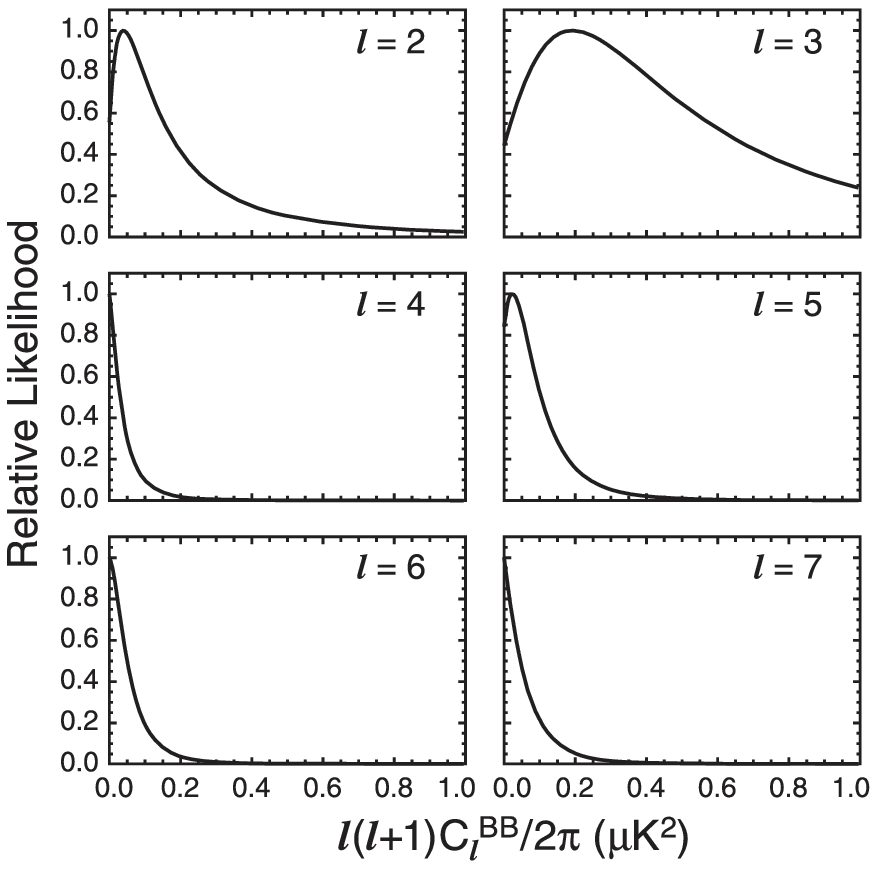}
\caption{
Low $\ell$ BB spectra. Other $C_\ell$  values are fixed to the best fit \wmap\ power spectrum.  
}
\label{fig:low_ell_BB}
\end{figure}

Figure \ref{fig:tb_spectrum} shows the temperature cross-power spectrum with the B-mode polarization (TB) spectrum.  
This angular cross-power spectrum is computed using the MASTER likelihood code. The TB angular power spectrum is expected to be zero and the data are consistent with this expectation.
The $2\le l\le 7$ EE power spectrum is shown in Figure \ref{fig:low_ell_EE}.  
The $2\le l\le 7$ BB power spectrum is shown in Figure \ref{fig:low_ell_BB}. 

For running chains, we update the Sunyaev Zel'dovich spectrum template to the
spectrum given by \cite{battaglia/etal:2011}. 
Their thermal SZ spectrum is multiplied by 3.61 to 
scale from 150 GHz to V-band (61 GHz).  To convert from 150 to 148 GHz for ACT,
we multiply by 1.05.  The kinetic SZ spectrum does not need to be rescaled.
The sum of kinetic and thermal spectra is used as the SZ template, for the
frequency corresponding to each experiment; it is this sum that is multiplied 
by the SZ amplitude which is varied in the Markov chains.

\clearpage
\section{Power Spectrum Goodness of Fit and Map Anomalies} \label{sec:goodnessoffit}

\subsection{Goodness of Fit}

The likelihood code we release comes with a test code that runs on the \wmap\
nine-year best fit $\Lambda$CDM power spectrum (with no extra priors).  This
splits up the likelihood into several parts.  We first look at each part and
then combine the results for an overall estimate of goodness of fit.  The
high-$\ell$ TT spectrum in the $\ell$ range 33--1200 has 1168 degrees of
freedom, and a $\chi^2$ value of 1200.  This gives a reduced $\chi^2$ value of
1.027, and the probability to exceed this is 25.1\%, which indicates a good fit
to the data.  The high-$\ell$ TE spectrum in the $\ell$ range 24--800 has 777
degrees of freedom and a $\chi^2$ value of 815.4 for the same model.  The probability to exceed
this $\chi^2$ value is 16.5\%, which again indicates a good fit.  The low-$\ell$
polarized pixel-based likelihood contains 585 unmasked res 3 pixels each with a
Q and U Stokes parameter, for 1170 degrees of freedom.  The $\chi^2$ value for
this part of the likelihood is 1321.  The probability to exceed this $\chi^2$
value is 0.13\%, which is unusually low.

We have not yet mentioned the low $\ell$ TT and TE spectra.  Recall that the low
$\ell$ polarized pixel likelihood decorrelates the temperature and polarization
maps of the sky using the ILC and TT and TE spectra, as described in Appendix D
of \citet{page/etal:2007}.  After doing this, one obtains a $\chi^2$ for the
pixelized QU likelihood that incorporates information about TE, which is why we
do not have a separate TE $\chi^2$ value for $\ell \le 23$.  The $\ell \le 32$
TT likelihood is computed by a Blackwell-Rao estimator, based on Gibbs samples.
This code does not naturally generate a value comparable to a $\chi^2$ quantity.
However, it does provide a likelihood function which can be applied to any low
$\ell$ TT spectrum, and in the process of doing the sampling one obtains many
spectra (not smooth, typically) which have been sampled from this likelihood
function.  One can look at the distribution of likelihoods resulting from these
spectra and determine whether our best fit spectrum creates an unusually low 
likelihood.  We do this and find that our best fit power spectrum generates
an acceptable likelihood value.

Adding the three $\chi^2$ values mentioned above gives 3115 degrees of freedom
with a total $\chi^2$ value of 3336.4.  The probability to exceed this $\chi^2$
value is 0.3\%, which is still unusually low.  This is driven completely by the
low $\ell$ polarized likelihood.

We investigated the origin of the excess $\chi^2$ in the low-$l$ polarization data. To see if there is any evidence for systematic effects in difference maps, we computed $\chi^2$ from six combinations of difference maps involving Ka-, Q-, and V-bands: Ka$-$Q, Ka$-$V, Ka$-$QV, Q$-$V, Q$-$KaV, and V$-$KaQ, where QV, KaV and KaQ are the corresponding weighted-averages of maps in two different frequency bands. We find that none of these combinations show an anomalous $\chi^2$. The average and standard deviation of $\chi^2$ is $1180\pm 47$ for 1170 degrees of freedom. The largest value of $\chi^2$ is 1236 from Ka$-$QV, and the probability to exceed (PTE) is 8.8\%. We then computed the optical depth, $\tau$, from Ka$-$QV, finding that it is consistent with zero (the maximum likelihood value lies in $\tau<0.002$, well below the 68\%~CL statistical uncertainty of $\delta\tau=0.014$). Therefore, we conclude that the low-$l$ polarization data pass the null test, and any residual systematic error we do not detect in difference maps has a negligible impact on our estimation of $\tau$. This null test also shows that the residual polarized synchrotron emission in Ka, if any, has a negligible impact on $\tau$.

To get an idea of how much additional noise we would need to include in the noise covariance matrix of the co-added KaQV map to explain the $\chi^2$, we add an uncorrelated noise variance to each r3 pixel ($N_{side}=8$), $N_{ij}\to N_{ij}+\sigma^2_{r3}\delta_{ij}$. We find $\sigma_{r3}=0.27~\mu{\rm K}$ brings the reduced $\chi^2$ to unity. The instrumental noise per r3 pixel of the co-added KaQV map ranges from 0.43 to 1.57~$\mu$K, with the average and standard deviation of $0.86\pm 0.17~\mu{\rm K}$. Therefore, an additional noise variance, $\sigma_{r3}^2$, required to explain the excess $\chi^2$ is an order of magnitude smaller than a typical instrumental noise variance per r3 pixel of the co-added KaQV map.

Next, we computed the tensor-to-scalar ratio, $r$, from the low-$l$ B-mode polarization data only. We found that $r$ was consistent with zero, with the 95\%~CL upper bound of $r<2.0$. The maximum likelihood value occurs at $r=0.40$, which is already ruled out by the limit from the CMB temperature power spectrum, $r<0.17$~(95\%~CL); thus, it cannot be due to inflationary B-modes. For $r=0.4$, the low-$l$ B-mode power spectrum amplitude is less than the scalar E-mode amplitude by a factor of six, and thus it is a small signal (and is consistent with zero).

We next examined residual foregrounds. By enlarging the edges of the polarization P06 mask by 1, 2, and 3 pixels, we found that the PTE increased from 0.1\% to 0.9\%, 5\%, and 12\%, respectively. While this may suggest the presence of residual foregrounds in the polarization data, this may also be partly due to the reduction of degrees of freedom (the degrees of freedom decrease from 1170 to 850, 582, and 344, respectively), as fewer degrees of freedom are more forgiving for larger values of the reduced $\chi^2$. Indeed, changes in the values of the reduced $\chi^2$ are modest: it drops from 1.13 to 1.12, 1.10, and 1.09, respectively. 

Therefore, we conclude that the excess $\chi^2$ likely to be at least partially due to residual foregrounds, which we do not include in the noise covariance matrix.  These foregrounds may not mostly be from the regions near the mask edges. However, the effect on our estimation of $\tau$ is negligible compared with the statistical uncertainty.

\subsection{Power Spectra Goodness of Fit with Even-Odd multipoles}

The analysis of the even excess effect seen in the seven-year TT power spectrum 
\citep{bennett/etal:2011} has been repeated using the nine-year data.  The even excess statistic
compares the mean $C_\ell$ at even values of $\ell$ with the mean $C_\ell$ at
odd values of $\ell$ within a defined $\ell$ domain.  More formally, we define
\begin{displaymath} 
\mathcal{E}_\ell = 
\frac{\langle \mathcal{C}^\mathrm{obs}_\ell 
- \mathcal{C}^\mathrm{th}_\ell\rangle_\mathrm{even}
- \langle \mathcal{C}^\mathrm{obs}_\ell 
- \mathcal{C}^\mathrm{th}_\ell\rangle_\mathrm{odd}}
{\langle \mathcal{C}^\mathrm{th}_\ell\rangle},
\end{displaymath}
where $\mathcal{C}_\ell = \ell(\ell+1)C_\ell/2\pi$, the superscript
``obs'' refers to the observed power spectrum, and the superscript
``th'' refers to a fiducial theoretical power spectrum used for
normalization.  In this paper, as before, we bin
$\mathcal{E}_\ell$ by $\Delta\ell = 50$.

The seven-year analysis used a set of more than 11000 Monte Carlo CMB simulations
to probe the significance of the even excess.  This large set was
computationally inexpensive because the TT power spectra were estimated using
the Monte Carlo Apodised Spherical Transform EstimatoR
\citep[MASTER;][]{hivon/etal:2002}.
However, in the nine-year analysis, the TT power spectra are computed
using a new estimator weighted using the $C^{-1}$ matrix, and the Monte Carlo
realizations are much slower.  Consequently, we now use a smaller 
set of 512 simulations of the full nine-year $C^{-1}$-weighted
power spectrum 

\begin{figure}
\epsscale{0.99}
\plotone{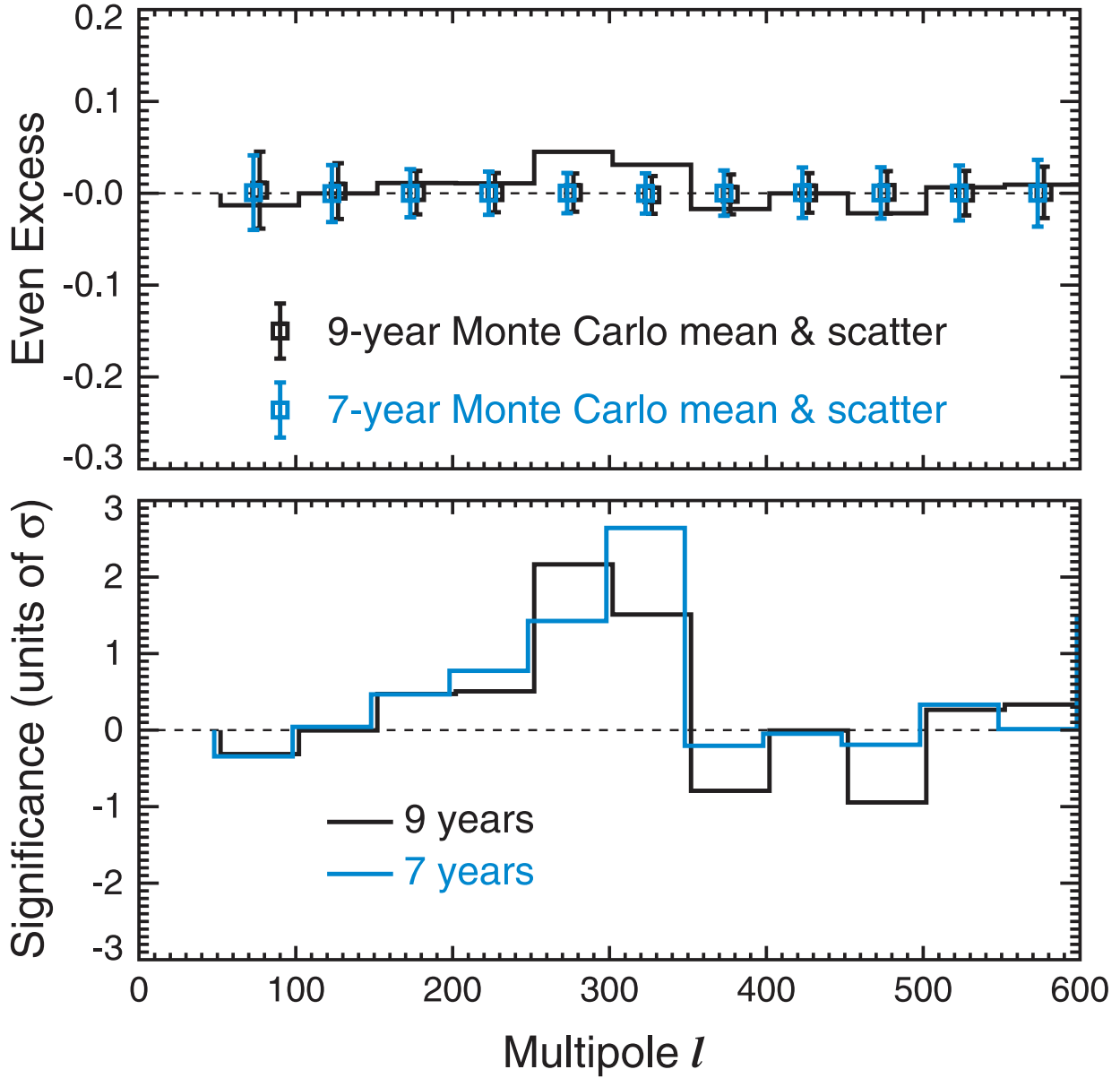}
\caption{Top: Even excess $\mathcal{E}_\ell$ in the observed power
spectrum, in bins of $\Delta\ell=50$, compared to the mean and
scatter from 512 Monte Carlo realizations.    Bottom:
$\mathcal{E}_\ell$ as in the top plot, converted to significance
units by normalizing to the Monte Carlo scatter in each bin.  Only
the $\ell=250-299$ and $\ell=300-349$ bins show a significance
greater than $1\sigma$.  Black: nine-year results; blue: seven-year results
from \cite{bennett/etal:2011}.
\vspace{1mm} \newline (A color version of this figure is available in the online journal.)
\label{fig:evenbin}}
\end{figure}

Figure \ref{fig:evenbin} shows $\mathcal{E}_\ell$ as a function of $\ell$
within bins of $\Delta\ell=50$.  Results from the nine-year analysis are shown in
black, and those from the seven-year analysis are shown in blue
\citep[see][Figure 9]{bennett/etal:2011}.  The overall trend of the results with $\ell$ is 
similar in
the nine-year analysis to what it was in the seven-year analysis, except that the rise in
$\mathcal{E}_\ell$ over the domain $50 \leq \ell < 350$ is no longer monotonic.
Also, in the nine-year analysis, two of the three negative values of
$\mathcal{E}_\ell$, which denote excess power at \emph{odd} values of $\ell$,
have higher absolute value than in the seven-year analysis.

\cite{bennett/etal:2011} examined a combined $\ell$ bin for $250 \leq \ell < 350$
as an example of a posteriori analysis. The value of $\mathcal{E}_\ell$
in this bin was $0.0446$, as compared to a Monte Carlo scatter of $\sigma=0.0155$,
for a $2.9\sigma$ level of significance.  The equivalent values for the
nine-year analysis using the $C^{-1}$ power spectrum estimator are $\mathcal{E}_\ell=0.0381$,
with a Monte Carlo scatter of $\sigma=0.0144$, for a reduction in the level
of significance to $2.6\sigma$.

The de-biased $\mathcal{E}_\ell$ test described by \cite{bennett/etal:2011} has also been
repeated for the nine-year analysis. This test chooses the maximum value of the bin-by-bin
statistical significance $\mathcal{E}_\ell/\sigma(\mathcal{E}_\ell)$ from the $\ell$ bins
being considered, rather than focusing on only one bin, so that the a posteriori character
of the test is weakened \citep[see][Figure 11]{bennett/etal:2011}.  We use bins of width
$\Delta\ell=50$ for $50\leq\ell<600$.  The nine-year test gives similar results to the seven-year
test, but at a reduced significance.  In the seven-year test, the de-biased $\mathcal{E}_\ell$
test gave a probability to exceed (PTE) of $5.11\%$ for the observed spectrum as compared to the
Monte Carlo distribution, whereas in the nine-year test, the PTE is $14.3\%$, equivalent to a
$1.1\sigma$ result.  Similarly, bins with a high value of the odd excess
($-\mathcal{E}_\ell$) were less frequent than expected in the seven-year power spectrum, with a
PTE of $98.9\%$ in the de-biased test. This effect is also weaker in the nine-year power
spectrum, which gives a PTE of $90.2\%$, equivalent to a $1.3\sigma$ result.

The even-odd effect in the observed power spectrum does not appear to be an
artifact of the power spectrum estimator, since it is seen both with the MASTER
method (seven years) and with the $C^{-1}$ method (nine years).  However, in the
nine-year analysis, the superficial test for $250\leq\ell<350$ yields a result
with reduced significance as compared to nine years, and the de-biasing strategy
further reduces the significance of both the even power excess and the odd
power deficit to $\sim 1\sigma$.  The conclusion of \citet{bennett/etal:2011} that
the even-odd effect is probably a statistical fluke stands, and indeed is
strengthened, after the nine-year tests.

\subsection{Quadrupole Amplitude}

\begin{figure}  
\epsscale{1.00}
\plotone{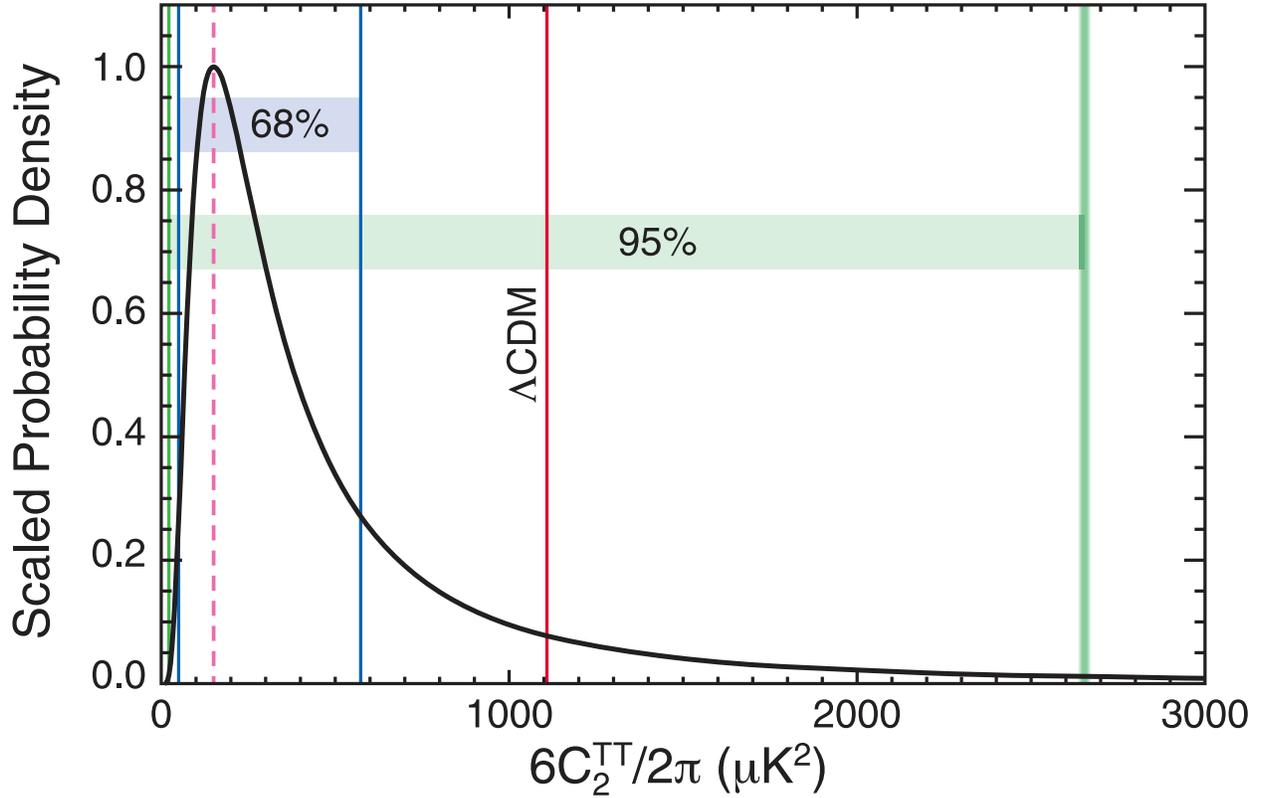}
\caption{
The likelihood of the true value of 
$\ell(\ell+1)C^{TT}_\ell/(2\pi) = 6 C^{TT}_2/(2\pi)$ for $\ell=2$, 
based on our measured sky.  This is computed using the
Blackwell-Rao estimator run on Gibbs samples, and it marginalizes
over all other values of $C^{TT}_\ell$.
The maximum likelihood point is shown as the pink line; one and two sigma
regions are shown as blue and green lines.  The best fit $\Lambda$CDM
theory spectrum computed on \wmap\ nine-year data only is shown in red.
\vspace{1mm} \newline (A color version of this figure is available in the online journal.)
}
\label{fig:quad1}
\end{figure}

Since the first-year \wmap\ data release there has been speculation about the low 
value of the $l=2$ quadrupole moment.  As concluded in the \cite{bennett/etal:2011} seven-year results paper, while the quadrupole amplitude is below the mean expected amplitude for the model, it is not surprisingly or disturbingly low. Figure \ref{fig:quad1} illustrates the likelihood of the true value of 
$\ell(\ell+1)C^{TT}_\ell/(2\pi) = 6 C^{TT}_2/(2\pi)$ for $\ell=2$, 
based on our measured sky.  A Blackwell-Rao estimator run on Gibbs samples and marginalized over all other values of $C^{TT}_\ell$ results in the maximum 
likelihood quadrupole amplitude shown by the pink line.  The $1\sigma$ and $2\sigma$ 
regions are shown as blue and green horizontal bands.  The best fit $\Lambda$CDM
theory spectrum computed on \wmap\ nine-year data only is shown in red.  We conclude from this that the theoretically expected quadrupole amplitude (based on a $\Lambda$CDM fit to the full angular power spectrum is well between $1\sigma$ and $2\sigma$, hardly an unlikely event. 

\begin{figure} 
\epsscale{1.00}
\plotone{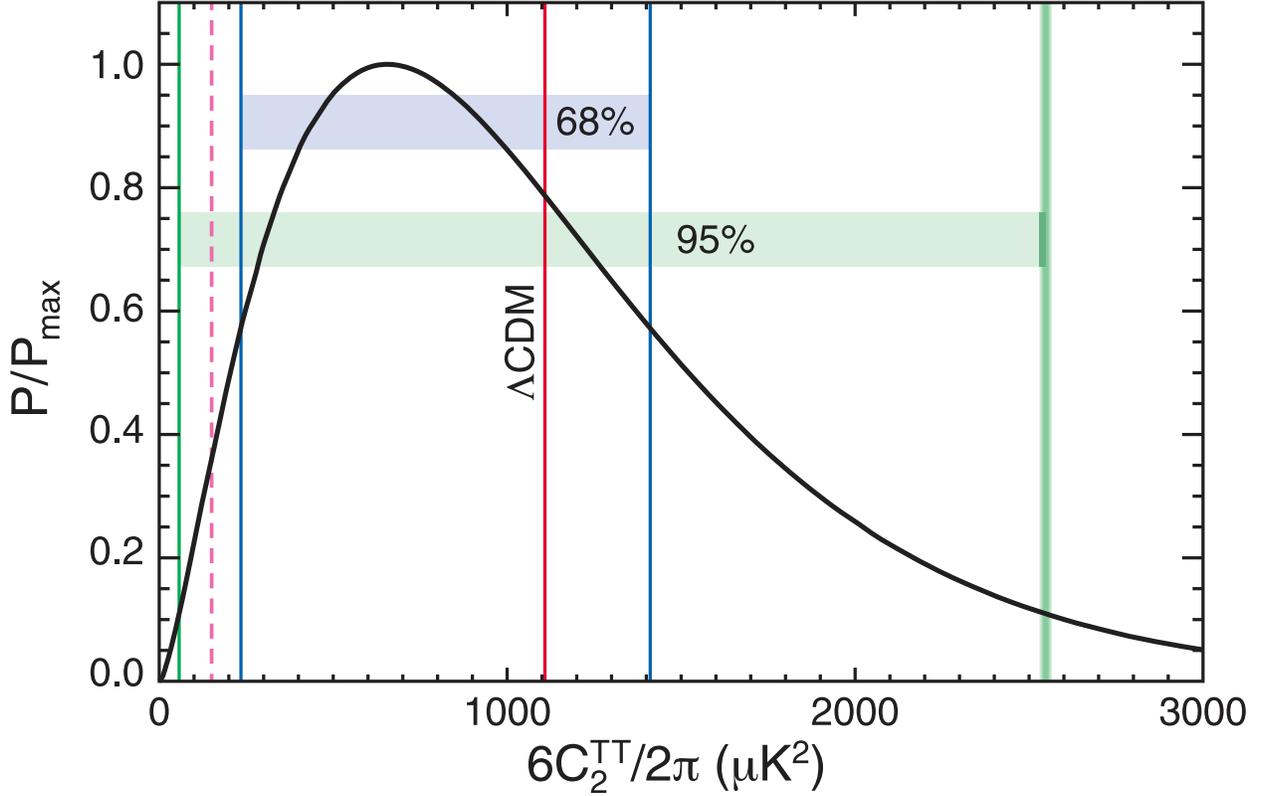}
\caption{
The cosmic variance probability distribution for the quadrupole, given the 
theory power spectrum.  
This assumes we know 
$\ell(\ell+1)C^{TT}_\ell/(2\pi) = 6 C^{TT}_2/(2\pi) = 1109\; \mu\text{K}^2$
(red line) and plots the distribution of quadrupole power values we could measure
for random Hubble volumes.  
Note that $6 C^{TT}_2/(2\pi)$ is the mean of the distribution; due to the skewness of the
$\chi^2$ distribution, the peak of the distribution is substantially lower.
One and two-sigma regions are shown.
The quadrupole cosmic variance distribution has $\nu = 2\ell+1=5$ degrees of 
freedom.  Assuming $f_{\rm sky} \approx 0.99$, we plot a $\chi^2$ distribution
based on $\nu = (2\ell+1)f_{\rm sky}^2 \approx 4.9$ degrees of freedom.
The peak of the distribution is then lower than the mean by
a factor of $(\nu-2)/\nu$, putting it at 656 $\mu\text{K}^2$.
\vspace{1mm} \newline (A color version of this figure is available in the online journal.)
}
\label{fig:quad2}
\end{figure}

Looked at the other way, we can ask the relative probability of observing the particular
quadrupole value given the mean expected value based again on a $\Lambda$CDM fit to the full angular power spectrum.  This is shown in Figure \ref{fig:quad2}.  Again, one can see that the distribution is far from Gaussian and that the peak of the likelihood function is well displaced from its mean, such that the single most likely value for the expected quadrupole is close to half of the mean value.  The observed quadrupole value is a relative probability of 40\%, more than $1\sigma$ but less than $2\sigma$ away from expectations.  The quadrupole value thus cannot be said to be anomalously low; it is well within the expected statistical variance.

\subsection{Alignment of the Quadrupole and Octupole}
\label{sec:quad_oct}

The quadrupole and octupole, expected to have independent and random orientations, 
were aligned to $<0.5^\circ$ in the seven-year ILC
map \citep{bennett/etal:2011}.  In the nine-year ILC map, we find that the orientations of the 
quadrupole and octupole differ by $\sim 3^\circ$.  Most of this change
is due to the fact that the nine-year ILC map has been improved by the use of
the asymmetric beam deconvolution described in Section~\ref{sec:beamsymmaps}.
Other minor changes are due to small improvements of the gain model and window
functions from two years of additional data, as well as the updated foreground
mask (which slightly changes the $\csc \beta$ fits and hence the monopole offset
in each ILC region).  A nine-year ILC made without the beam deconvolution has a
quadrupole-octupole misalignment of $\sim 1^\circ$, confirming that the
improvement of the use of deconvolution is the dominant source of the change
from seven to nine years of data.

We now address the significance of $\sim 3^\circ$ octupole-quadrupole alignment
in the nine-year map by examining its sensitivity to the separation of the CMB
from the foregrounds.  To do this, we use the error description of the
CMB-foreground covariance, discussed in Section~\ref{sec:ilc_errors}.
The CMB-foreground covariance in the ILC is described in terms of 48 error modes
(computed at r6), which provide the eigenvectors with nonzero eigenvalues of the
$49152\times49152$ pixel space covariance matrix.  We first change bases from
pixel space into the 12-dimensional space spanned by the quadrupole and octupole
modes (5 for the quadrupole, 7 for the octupole).  This results in a $12\times12$ 
covariance matrix for the error in the quadrupole and octupole $a_{\ell m}$
coefficients.  For convenience, we use real-valued harmonics and so we have a
real-valued covariance matrix.  Then, we generate many Gaussian random
realizations of perturbations to the quadrupole and octupole 
(i.e. realizations of CMB quadrupole and octupole errors) based on this
covariance matrix.  We add these to the quadrupole and octupole from the
nine-year ILC, and check the alignment for each, using the same method as
described in \citet{bennett/etal:2011}.

Among these realizations, we find the median quadrupole-octupole misalignment to
be $6^\circ$.  The probability of a $\le 6^\circ$ alignment is 0.55\%.  This means that the
significance of the octupole-quadrupole alignment is $< 3\sigma$, i.e.\ it is
not significant.  Occasional perturbations to the ILC realign the quadrupole and
octupole perfectly, and about 5\% of the perturbations misalign them by more
than $20^\circ$.  Note also that this encompasses only one of the types of error
in the ILC.  Including an estimate of the ILC bias error will further degrade
the significance of any observed alignment.

We conclude that our ability to remove foregrounds is the limiting factor in the
measurement of the cosmological quadrupole-octupole alignment.  The already low
statistical significance ($< 3\sigma$) of the estimated alignment must be
further degraded by the posterior selection made to examine this particular
quantity.  Given that there is no evidence of experimental systematic effects,
and that the foreground-CMB separation contributes substantially to the
alignment uncertainty, the estimated alignment appears to be a low-significance
chance occurrence. 

\clearpage
\section{Cosmological Results and Implications}
\label{sec:cosmology}

We have seen that the \wmap\ power spectrum is well fit by only six parameters.  The quadrupole amplitude is not anomalously low, 
and the quadrupole-octupole alignment cannot be considered anomalous as it is within the range allowed by cosmic variance and foreground subtraction uncertainties.

The bipolar power spectrum of the final nine-year maps shows a large signal
similar to the one we reported in the seven-year results. This signal exhibits a
strong ecliptic latitude dependence, in both the seven and nine-year data.  The
bipolar power spectrum of the new beam-symmetrized (deconvolved) maps shows that this signal has
largely gone away, but there now appears a high-$\ell$ signal with the opposite
sign.  This is expected since the deconvolution process correlates pixel noise
in a way that we do not correct for in the estimation process.  Our primary
motivation was to check that the latitude-dependent signal at low-$\ell$ was due
to beam asymmetry, and we believe that is now well established.  There is little
motivation to correct the side-effects at high-$\ell$, since doing so would be
non-trivial, and there was no hint of an anomaly there to begin with.  In
summary, our new analysis demonstrates that
the latitude dependent signal in the bipolar power spectrum seen in both the
seven and nine-year non-deconvolved maps was real and caused by \WMAP's beam
asymmetry. Further, since beam asymmetry has negligible effect on the angular
power spectrum, $C_\ell$, we adopt the simpler non-deconvolved maps for power
spectrum estimation and cosmological parameter studies.

The power spectrum contains all of the cosmological information in the map if, and only if, the fluctuations are Gaussian with random phases across the non-masked portion of the map.  In this section we show that this is indeed the case within the estimated measurement and analysis uncertainties. We then summarize the cosmological parameter discussion of \cite{hinshaw/etal:prep} with cosmological parameters derived using only \wmap\ data and derived when combined using external data as well.

\subsection{Non-Gaussianity}

The simplest model of inflation, namely single-field slow-roll inflation 
with canonical kinetic term and a nearly flat potential $V(\phi)$, predicts
that the initial adiabatic curvature $\zeta({\bf k})$ has only tiny deviations
from Gaussianity~\citep{Acquaviva:2002ud,Maldacena:2002vr}.
However, alternate models of the early universe predict several possible types
of deviations from Gaussian statistics, making the search for non-Gaussianity 
in the CMB a powerful, multifaceted probe of the early universe.

\subsubsection{$\fnlloc$, $\fnleq$, and $\fnlorth$}

We will limit our search for non-Gaussianity to the 3-point function or
bispectrum, and parameterize it by:
\be
\langle \zeta_{{\bf k}_1} \zeta_{{\bf k}_2} \zeta_{{\bf k}_3} \rangle = 
  \left( \fnlloc B_{\rm loc}(k_1,k_2,k_3) + \fnleq B_{\rm eq}(k_1,k_2,k_3) + \fnlorth B_{\rm orth}(k_1,k_2,k_3) \right)
  (2\pi)^3 \delta^3\left( \sum {\bf k}_i \right)
\ee
where $\fnlloc$, $\fnleq$, $\fnlorth$ are free parameters to be estimated, and
the local, equilateral, and orthogonal template bispectra are defined by:
\ba
B_{\rm loc}(k_1,k_2,k_3) &=& \frac{6}{5} \left( P_\zeta(k_1) P_\zeta(k_2) + \mbox{2 perm.} \right) \\
B_{\rm eq}(k_1,k_2,k_3) &=& \frac{3}{5} \Big( 6 P_\zeta(k_1) P_\zeta(k_2)^{2/3} P_\zeta(k_3)^{1/3} 
  - 3 P_\zeta(k_1) P_\zeta(k_2) \nn \\
  && \hspace{1cm} - 2 P_\zeta(k_1)^{2/3} P_\zeta(k_2)^{2/3} P_\zeta(k_3)^{2/3} + \mbox{5 perm.} \Big) \\
B_{\rm orth}(k_1,k_2,k_3) &=& \frac{3}{5} \Big( 18 P_\zeta(k_1) P_\zeta(k_2)^{2/3} P_\zeta(k_3)^{1/3} 
  - 9 P_\zeta(k_1) P_\zeta(k_2) \nn \\
  && \hspace{1cm} - 8 P_\zeta(k_1)^{2/3} P_\zeta(k_2)^{2/3} P_\zeta(k_3)^{2/3} + \mbox{5 perm.} \Big)
\ea
The $\{ \fnlloc, \fnleq, \fnlorth \}$ basis for the three-point function is large
enough to encompass a range of interesting models.
Local-type non-Gaussianity is generic to some multi-field inflation models,
for example curvaton models~\citep{Linde:1996gt,Lyth:2002my} and
variable reheating models~\citep{Dvali:2003em,Zaldarriaga:2003my},
and also to some alternatives to inflation, such as
``new'' ekpyrosis~\citep{Creminelli:2007aq,Buchbinder:2007tw} 
and cyclic \citep{lehners/steinhardt:2008b,lehners/steinhardt:2008} models.
Also, there is a theorem \citep{Creminelli:2004yq} that implies
that no single-field model of inflation can generate detectable 
$\fnlloc$.
Equilateral-type and orthogonal-type non-Gaussianity can be generated in single-field
models, and generically appear when there are non-negligible interaction terms in the 
inflationary Lagrangian.

We constrain the $f_{NL}$ parameters using the optimal (i.e.~minimum variance unbiased)
bispectrum estimator implemented in~\citet{Smith:2009jr}, which builds on previous work 
\citep{Komatsu:2003iq,Creminelli:2005hu,Smith:2006ud}.
The estimator optimally combines channels with different noise maps
and beams by filtering the data with the inverse signal+noise covariance $C^{-1} = (S+N)^{-1}$,
and includes a one-point term (in addition to a three-point term) which reduces
the variance.
Unless otherwise specified, we use the V-band and W-band differencing assemblies from \wmap\ (six maps total),
remove regions of high Galactic foreground and point source emission using the nine-year KQ75 mask,
and marginalize three foreground templates corresponding to synchrotron, free-free, and dust emission.
With foreground marginalization enabled, the same $f_{NL}$ estimates are obtained on raw and template-cleaned
maps.

Our ``bottom line'' constraints on non-Gaussianity are as follows:
\ba
\fnlloc = 37.2 \pm 19.9 & \hspace{1cm} & (-3 < \fnlloc < 77 \mbox{ at 95\% CL}) \nn \\
\fnleq = 51 \pm 136 & \hspace{1cm} & (-221 < \fnleq < 323 \mbox{ at 95\% CL}) \nn \\
\fnlorth = -245 \pm 100 & \hspace{1cm} & (-445 < \fnlorth < -45 \mbox{ at 95\% CL}) \label{eq:fnl}
\ea
The $\fnlloc$ constraint includes a correction for the ISW-lensing contribution to the bispectrum,
which arises from the large-scale correlation between the CMB temperature and the CMB lensing potential.
We find that the ISW-lensing bispectrum biases the $\fnlloc$ estimator by $\Delta\fnlloc = 2.6$;
this bias has been subtracted from the estimate in Equation~\refeqn{eq:fnl}.
The ISW-lensing bias was computed using the Fisher matrix approximation, but this has been 
shown to be an excellent approximation to the exact
result~\citep{Hanson:2009kg,Lewis:2011fk}.

The constraint on each $f_{NL}$ parameter in Equation~\refeqn{eq:fnl} assumes that the other two $f_{NL}$ parameters
are zero.  For a joint analysis of all three parameters, we need the bispectrum Fisher matrix:
\be
F = 
\left( \begin{array}{ccc}
  25.25 & 1.06 & -2.39 \\
  1.06 & 0.54 & 0.20 \\
  -2.39 & 0.20 & 1.00
\end{array} \right)
\times 10^{-4} \label{eq:fnl_fisher}
\ee
where the ordering of the rows and columns is $\fnlloc, \fnleq, \fnlorth$.
The statistical error on each $f_{NL}$ parameter in Equation~\refeqn{eq:fnl},
with the other two $f_{NL}$ parameters fixed to zero, is $(F_{ii})^{-1/2}$,
and the correlation between two estimators in Equation~\refeqn{eq:fnl} is equal to the
rescaled off-diagonal matrix element $F_{ij} / (F_{ii}F_{jj})^{1/2}$.\footnote{This
estimator covariance is appropriate for our convention that each $f_{NL}$ estimator
is defined to be the optimal estimator assuming that the other two $f_{NL}$ parameters
are zero.  There is an alternate definition in which each $f_{NL}$ estimator is defined
with the other two $f_{NL}$ parameters marginalized; in this case the estimator covariance
matrix would be the inverse Fisher matrix $(F^{-1})_{ij}$.
The two definitions are linear combinations of each other, and therefore give identical
results in a joint analysis, provided that the off-diagonal correlations are properly incorporated.}
An example of a two-parameter joint analysis is shown in Figure \ref{fig:single_field} below.

\subsubsection{$\fnlorth$ Diagnostic Tests and Interpretation}

The most striking result in Equation~\refeqn{eq:fnl} is the estimate for $\fnlorth$, which is non-zero at 2.45$\sigma$.   
The (two-sided) probability of obtaining a value with this statistical significance in a Gaussian fiducial cosmology
is 1.4\%. This is not significant enough by itself to consider it a detection, but even further caution is required.
When interpreting this probability, it must be kept in mind that we look for multiple deviations from the vanilla
$\Lambda$CDM model\footnote{A partial list includes the three $f_{NL}$ parameters, the spatial curvature $\Omega_K$,
tensor-to-scalar ratio $r$, running of the spectral index $(dn_s/d\log k)$, dark energy equation of state $w$,
isocurvature amplitudes $\alpha_0,\alpha_{-1}$, and neutrino mass $m_\nu$.}, so it is statistically unsurprising that
{\em one} such deviation is at this significance level.
The rest of this section will be devoted to consistency checks and interpretation of the $\fnlorth$ result.

One possible source of systematic error is contamination by residual foregrounds.
Since we marginalize over synchrotron, free-free and dust templates in our bispectrum estimator,
any foreground contribution that is a linear combination of these spatial templates does not
contribute to $\fnlorth$.
However, since the templates are not perfect, there will be residual contributions at some level.
A simple procedure that gives the rough order of magnitude
is to disable template marginalization in the estimator,
and compute the foreground contribution to $\fnlorth$ in an ensemble of simulated {\em raw}
maps without any foreground cleaning.
We simulate raw maps using random CMB and noise realizations,
and a fixed dust realization given by model 8 of~\citet{finkbeiner/davis/schlegel:1999}.
We do not include synchrotron and free-free foregrounds since dust dominates in W-band and is a significant fraction of the V-band foreground. 
In each simulation, we compute the difference $(\Delta\fnlorth)$ between the $\fnlorth$ estimate obtained from
the raw map, and the $\fnlorth$ estimate that would be obtained from the CMB+noise contribution
alone.  We find that the mean value of $(\Delta\fnlorth)$ is 1.1 and the RMS scatter is 5.2.
This presumably overestimates the dust contribution since we are not attempting to remove
foregrounds at all.
Since the shift $(\Delta\fnlorth)$ seen in these simple simulations
is much smaller than the statistical error $\sigma(\fnlorth)$, we conclude that residual foregrounds are unlikely to be a significant contaminant.

\begin{figure}
\plotone{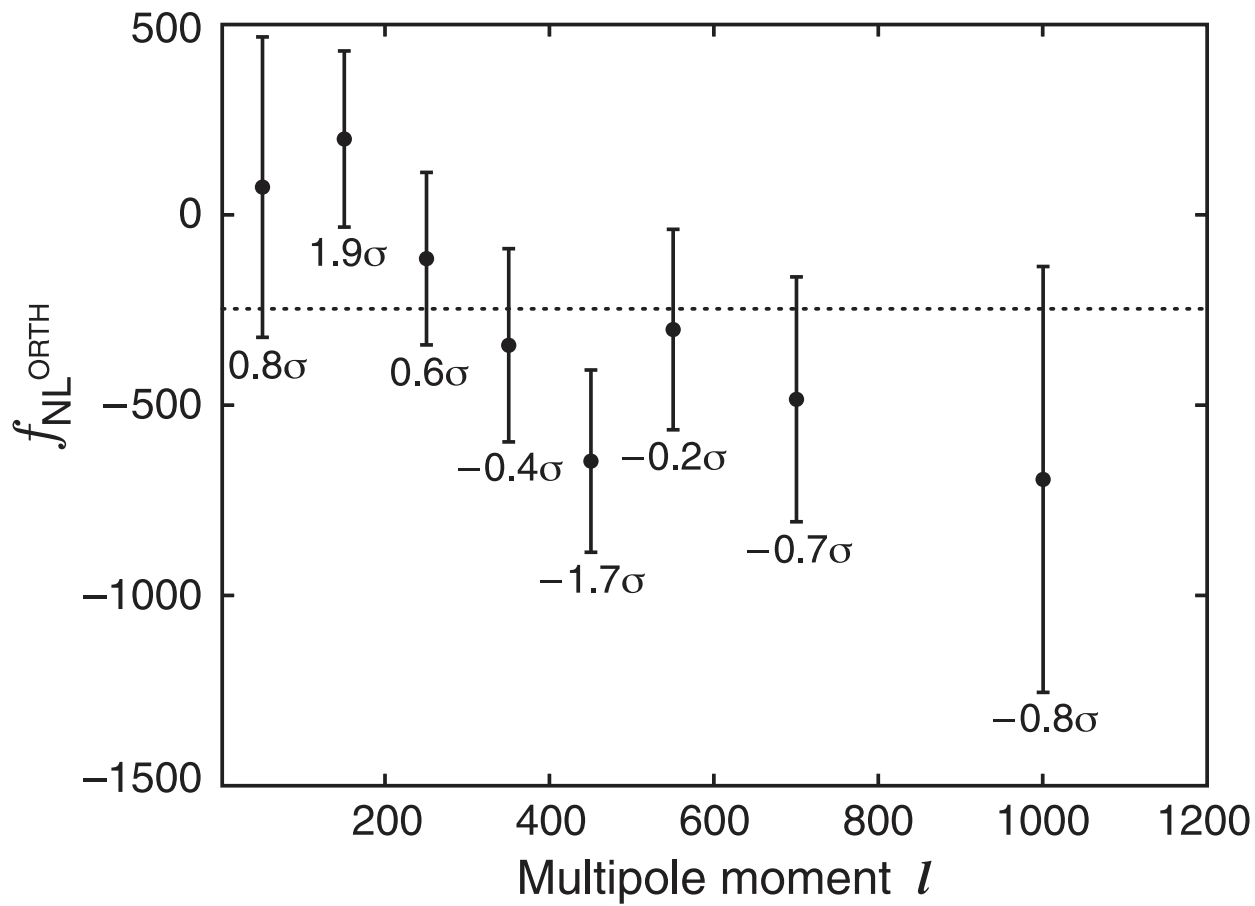}
\caption{A test for scale-dependent systematics: $\fnlorth$ estimates in $\ell$-bands, with cumulative best-fit value
$\fnlorth=-245$ shown by the dotted horizontal line.  Each error bar is labeled with the statistical significance
of the deviation from the cumulative best-fit value (not the deviation from zero).  No
evidence for scale-dependent systematics is seen.}
\label{fig:fnl_orthog_in_lbands}
\end{figure}

As a first test for instrumental systematic effects, we check for consistency between different angular scales by splitting
the $\fnlorth$ estimator in $\ell$-bands.
Our procedure is as follows:
we write the $\fnlorth$ estimator as a sum over triangles, restrict the sum to triangles whose
maximum multipole $\max(\ell_1,\ell_2,\ell_3)$ is in a given bin $(\ell_{\rm min}, \ell_{\rm max})$, and then
appropriately normalize so that the band-restricted sum is an unbiased estimator of $\fnlorth$.
This prescription for binning the $\fnlorth$ estimator has the property that if we combine $\fnlorth$ estimates
in all bins up to some multipole $\ell_{\rm max}$, the result agrees with simply rerunning the $\fnlorth$ estimator
with maximum multipole $\ell_{\rm max}$.
It also has the property that $\fnlorth$ estimates in different $\ell$-bands are nearly uncorrelated.

In Figure \ref{fig:fnl_orthog_in_lbands}, we show the $\fnlorth$ estimate in $\ell$-bands,
with the cumulative best-fit value $\fnlorth=-245$ shown for comparison.  Each bin is consistent with 
the cumulative best-fit value at 2$\sigma$, and the overall $\chi^2$ of the fit to a constant $\fnlorth$ value is
good ($\chi^2=8.8$ with seven degrees of freedom).
We therefore conclude that there is no evidence for scale-dependent systematic contamination.

\begin{figure}
\plotone{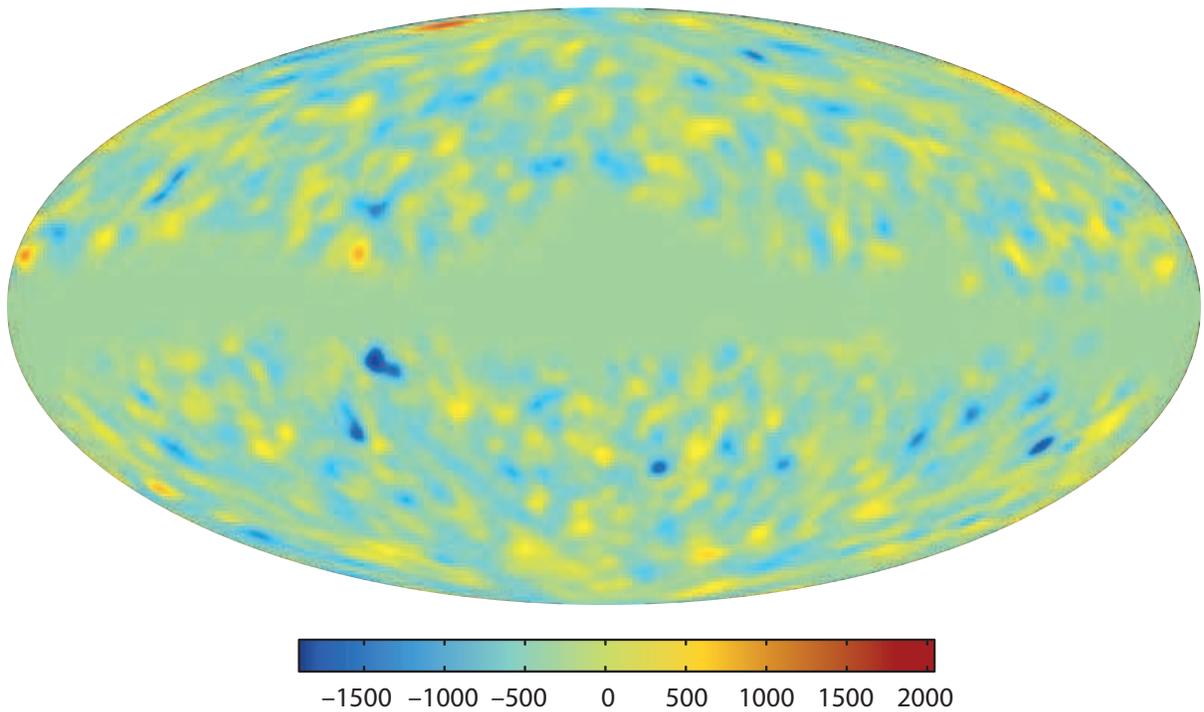}
\caption{A visual test for sky location dependent systematics:
skymap showing the contribution of different parts of the sky to the $\fnlorth$
estimator, in units of ``$\fnlorth$ per steradian''. 
We do not detect any significant localized features in this map.
\label{fig:fnl_orthog_skymap}
\vspace{1mm} \newline (A color version of this figure is available in the online journal.)}
\end{figure}

As a second test for systematics, we can ask whether estimates of $\fnlorth$
in different parts of the sky are consistent.
The bispectrum estimator is naturally written as an integral over position on the sky, so a
convenient way to visualize the position dependence is to simply plot the integrand as
a skymap (Figure \ref{fig:fnl_orthog_skymap}).
This skymap is in units of ``$\fnlorth$ per steradian'' and has the property that its integral
over the whole sky is precisely equal to the estimated $\fnlorth=-245$.  If we restrict
the integral to a subregion $\Omega$ of the sky, the value of the integral
will roughly equal the value that  would be obtained if we re-ran the estimator 
using masking to isolate the subregion $\Omega$ (appropriately rescaled by the
area of $\Omega$).
Visual inspection of the skymap is a convenient way to look for an unexpected feature
(e.g., a large contribution near the Galactic plane would
suggest foreground contamination), although it might be difficult to assess the statistical
significance of an {\em a posteriori} feature if found.
Our interpretation of Figure \ref{fig:fnl_orthog_skymap} is that no visually striking features
are seen; the skymap looks qualitatively similar to skymaps obtained from Gaussian simulations.

As a more quantitative test for consistency between different parts of the sky, we 
estimated $\fnlorth$ in the portions of the following regions that lie outside
the KQ75 mask:
the northern Galactic hemisphere, 
the southern Galactic hemisphere, 
within $30^\circ$ of the ecliptic plane,
and the ecliptic poles ($> 30^\circ$ from the ecliptic plane).  
We find that for any pair of these regions, the estimated $\fnlorth$ values are
consistent at 2$\sigma$, relative to an ensemble of
Monte Carlo simulations.
The $\fnlorth$ estimates in these four subregions are
$-139 \pm 139$,
$-361 \pm 142$,
$-132 \pm 144$,
and $-336 \pm 138$, respectively.

As a final test for systematics, we can compare $\fnlorth$ estimates from different channels,
or combinations of channels.
In the first two columns of Table~\ref{tab:fnl_channels}, we show the result of applying the
$\fnlorth$ estimator for several combinations of channels.
To assess whether the $\fnlorth$ estimates from a given pair of rows are statistically consistent,
we subtract the two estimates, and compare the result to the same quantity (the difference of two
$\fnlorth$ estimates) evaluated in an ensemble of Monte Carlo simulations.
This way of assessing consistency fairly incorporates the correlation between $\fnlorth$ estimates
that arises because the CMB realization (and the noise realizations, if the two rows have channels
in common) is shared.
The matrix in the rightmost columns of Table~\ref{tab:fnl_channels} shows the result of doing
this consistency test for all pairs of rows in the table.

\begin{table}
\begin{center}
\begin{tabular}{|c|c||ccccccccc|}
\hline
channels & $\fnlorth$ & \multicolumn{9}{|c|}{Discrepancy in ``sigmas''} \\
  & & VW & V & W & V1 & V2 & W1 & W2 & W3 & W4 \\
\hline
All VW channels & $-245.5 \pm 99.6$
 & --  & 2.2$\sigma$  & 1.5$\sigma$  & 1.5$\sigma$  & 2.1$\sigma$  & 0.7$\sigma$  & 1.4$\sigma$  & 1.1$\sigma$  & 2.2$\sigma$ \\
All V-band channels & $-125.9 \pm 112.7$
 & 2.2$\sigma$  & --  & 2.3$\sigma$  & 0.1$\sigma$  & 0.7$\sigma$  & 0.4$\sigma$  & 0.3$\sigma$  & 0.1$\sigma$  & 1.1$\sigma$ \\
All W-band channels & $-320.2 \pm 112.1$
 & 1.5$\sigma$  & 2.3$\sigma$  & --  & 2.1$\sigma$  & 2.5$\sigma$  & 1.7$\sigma$  & 2.2$\sigma$  & 2.0$\sigma$  & 3.2$\sigma$ \\
V1 only & $-119.3 \pm 129.1$
 & 1.5$\sigma$  & 0.1$\sigma$  & 2.1$\sigma$  & --  & 0.3$\sigma$  & 0.5$\sigma$  & 0.3$\sigma$  & 0.1$\sigma$  & 1.0$\sigma$ \\
V2 only & $-91.3 \pm 124.2$
 & 2.1$\sigma$  & 0.7$\sigma$  & 2.5$\sigma$  & 0.3$\sigma$  & --  & 0.8$\sigma$  & 0.0$\sigma$  & 0.2$\sigma$  & 0.8$\sigma$ \\
W1 only & $-172.1 \pm 140.1$
 & 0.7$\sigma$  & 0.4$\sigma$  & 1.7$\sigma$  & 0.5$\sigma$  & 0.8$\sigma$  & --  & 0.7$\sigma$  & 0.5$\sigma$  & 1.4$\sigma$ \\
W2 only & $-88.1 \pm 152.2$
 & 1.4$\sigma$  & 0.3$\sigma$  & 2.2$\sigma$  & 0.3$\sigma$  & 0.0$\sigma$  & 0.7$\sigma$  & --  & 0.2$\sigma$  & 0.7$\sigma$ \\
W3 only & $-111.0 \pm 154.2$
 & 1.1$\sigma$  & 0.1$\sigma$  & 2.0$\sigma$  & 0.1$\sigma$  & 0.2$\sigma$  & 0.5$\sigma$  & 0.2$\sigma$  & --  & 0.9$\sigma$ \\
W4 only & $-5.7 \pm 147.7$
 & 2.2$\sigma$  & 1.1$\sigma$  & 3.2$\sigma$  & 1.0$\sigma$  & 0.8$\sigma$  & 1.4$\sigma$  & 0.7$\sigma$  & 0.9$\sigma$  & -- \\
\hline
\end{tabular}
\end{center}
\caption{A test for consistency between channels.  The first two columns show $\fnlorth$ estimates obtained from
different subsets of \wmap\ channels.  The matrix on the right shows the level of discrepancy between each pair of
channel subsets, in ``sigmas'' after comparing to an ensemble of Monte Carlo simulations.}
\label{tab:fnl_channels}
\end{table}

This ``two-way'' null test can be generalized to an $N$-way null test that tests mutual
consistency between $\fnlorth$ estimates obtained in all $N$ rows of the table.
We represent the $\fnlorth$ estimates as a length-$N$ vector $f_i$, and compute the $N$-by-$N$
covariance matrix $C_{ij}$ using Monte Carlo simulations with shared CMB and noise realizations.
We then compute an overall best-fit $\fnlorth$ value $F$ which minimizes 
$\chi^2 = (f_i-F) C^{-1}_{ij} (f_j-F)$.
If the $N$ estimates are mutually consistent, then the value of $\chi^2$ at the minimum
will be distributed as a $\chi^2$ random variable with $(N-1)$ degrees of freedom.

We find that the channel-channel null tests are marginal.
The $N$-way null test gives $\chi^2 = 16.3$ with 8 degrees of freedom, corresponding to one-sided
probability $p=0.038$.
The most discrepant pair of rows in Table~\ref{tab:fnl_channels} is (W,W4), which differ by 3.2$\sigma$
relative to Monte Carlo simulations.
This statistical significance should not be taken at face value since there are 36 matrix entries in
Table~\ref{tab:fnl_channels}, and we have chosen the most anomalous one.
However, if we construct the same matrix for each member of an ensemble of simulations, we find that
the probability that at least one pair of rows is discrepant by $> 3.2\sigma$ is 2.6\%.
Finally, we observe that the discrepancy between V-band and W-band channels, which is in some sense
the most natural split, is 2.3$\sigma$, corresponding to probability $p=0.021$.

We conclude that there is some tension in the channel-channel null tests, with $p$-value around a few
percent depending on which test is chosen.
Since we have also considered null tests that pass cleanly (i.e.~the tests based on scale 
dependence and sky location), our interpretation is that one failure at the few-percent level does not indicate
systematic contamination, although the discrepancy between V-band and W-band is of some concern.
We therefore cautiously proceed to discuss the physical implications of the non-Gaussianity
constraints.

We opt to work in the context of single-field inflation, and use the effective field theory
developed in~\citet{Cheung:2007st,Cheung:2007sv}.
The EFT provides a master Lagrangian which is general enough to describe almost all single-field 
models of inflation.  See also \citet{gruzinov:2005,chen/etal:2007}. 
The action consists of a standard kinetic term, plus small interaction
terms whose coefficients parameterize allowed non-Gaussianity:
\be
S = \int d^4x\, \sqrt{-g} \left[ -\frac{\Mpl^2 \dot H}{c_s^2} \left( \dot\pi^2 - c_s^2 \frac{(\partial_i\pi)^2}{a^2} \right)
   + (\Mpl^2 \dot H) \frac{1-c_s^2}{c_s^2} \left( \frac{\dot\pi(\partial_i\pi)^2}{a^2} + \frac{A}{c_s^2} \dot\pi^3  \right) + \cdots \right]  \label{eq:eft}
\ee
Non-Gaussianity is parameterized by a dimensionless sound speed $c_s$, and a dimensionless parameter $A$
that represents the ratio between the coefficients the operators of $\dot\pi^3$ and $\dot\pi(\partial_i\pi)^2$.
We treat $c_s$ and $A$ as free parameters, but specific models will make predictions.
For example, in DBI inflation~\citep{Alishahiha:2004eh}, $c_s$ is a free parameter (but related to the 
tensor-to-scalar ratio) and $A=-1$.

The coefficients in the action~(\ref{eq:eft}) can be related to the parameters $\fnleq$, $\fnlorth$
by calculating the bispectra generated by the cubic operators $\dot\pi^3$ and $\dot\pi (\partial_i\pi)^2$,
and projecting them onto the basis of template bispectra~\citep{Senatore:2009gt}.  The result is:
\ba
\fnleq &=& \frac{1-c_s^2}{c_s^2} (-0.276 + 0.0785 A) \nn \\
\fnlorth &=& \frac{1-c_s^2}{c_s^2} (0.0157 - 0.0163 A)  \label{eq:cs_A}
\ea
where the numerical coefficients are specific to the nine-year \wmap\ results and have been computed using the exact Fisher matrix, including
CMB transfer functions and \wmap\ noise properties. 
For generic values of $A$, $\fnleq$ is larger than $\fnlorth$ (by an order of magnitude) 
and equilateral non-Gaussianity is generated.
However, there is an order-unity window of values (roughly $3.1 \lsim A \lsim 4.2$) where
$\fnlorth$ is larger than $\fnleq$, and orthogonal non-Gaussianity is generated.

\begin{figure}[!ht]
\epsscale{0.6}
\plotone{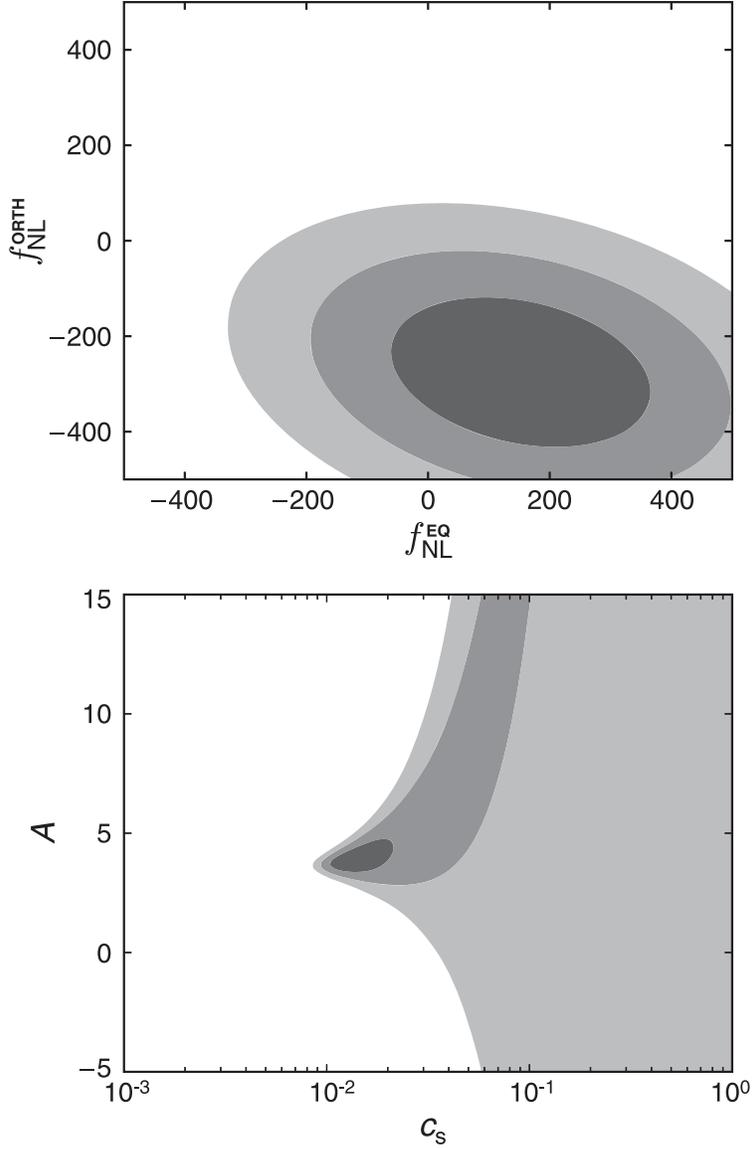}
\caption{\wmap\ nine-year constraints on non-Gaussianity in single-field inflation.  {\em Upper panel.}  68\%, 95\%, and 99.7\% confidence regions in
the $\fnleq,\fnlorth$ plane, defined by threshold $\chi^2$ values 2.28, 5.99, 11.62, as appropriate for a $\chi^2$
random variable with two degrees of freedom.  
$(f_{NL}^{\rm eq}, f_{NL}^{\rm orth})=(0,0)$ is consistent with the data to within 99\%~CL. 
{\em Lower panel.} Confidence regions on the dimensionless sound speed $c_s$ and
interaction coefficient $A$ (defined in Equation~\refeqn{eq:eft}), obtained from the top panel via the change of variables
in Equation~\refeqn{eq:cs_A}.
The upper bound on $f_{NL}^{\rm eq}$ gives a lower bound on $c_s$, which is consistent with $c_s = 1$.
}
\label{fig:single_field}
\end{figure}

Since single-field models that produce $\fnlorth$ are also expected to produce $\fnleq$ at
some level, it is natural to analyze joint constraints in the two-parameter space $\{ \fnleq, \fnlorth \}$.
To set up a joint analysis, we define notation as follows.
Let $f_i = (\fnleq,\fnlorth)$ be a two-component vector containing model parameters, let
${\hat f}_i = (51,-245)$ be the values of the associated estimators (i.e.~the last two rows
of Equation~\refeqn{eq:fnl}), and let $F_{ij}$ be the associated $2\times 2$ Fisher matrix (i.e.~the lower
right corner of Equation~\refeqn{eq:fnl_fisher}.
Then for given model parameters $f_i$, we define a $\chi^2$ statistic,
\be
\chi^2 = \sum_{ij} f_i F_{ij} f_j - 2 \sum_i F_{ii} f_i \hat f_i + \sum_{ij} {\hat f}_i F_{ii} F_{ij}^{-1} F_{jj} {\hat f}_j.
\ee
We threshold this $\chi^2$ to obtain confidence regions in the $(\fnleq,\fnlorth)$ plane.
These confidence regions are shown in the left panel of Figure \ref{fig:single_field}.
We note that the point $(\fnleq,\fnlorth)=0$ is just outside the 2$\sigma$ contour, which means that it 
is just barely a $>2\sigma$ event when $\fnleq$ is included in the parameter 
space.
More precisely, the relevant $\Delta\chi^2$ is 7.16 with two degrees of freedom; the probability of
getting a $\Delta\chi^2$ this large in a Gaussian cosmology is 2.8\%.

In the right panel of Figure \ref{fig:single_field}, we change variables to show confidence regions
in the parameter space $(c_s,A)$.
These confidence regions were obtained under the assumption that the
single-field bispectra are well-approximated by the equilateral and orthogonal
template shapes.   However, we have checked that nearly identical confidence
regions are obtained if the exact tree-level bispectra for the operators
${\dot\pi}^3$ and $\dot\pi (\partial_i \pi)^2$ are used throughout the analysis.

\clearpage

\subsection{Cosmological parameters}

\begin{figure}  \label{f:lambda_matter}
\epsscale{1.0}
\plotone{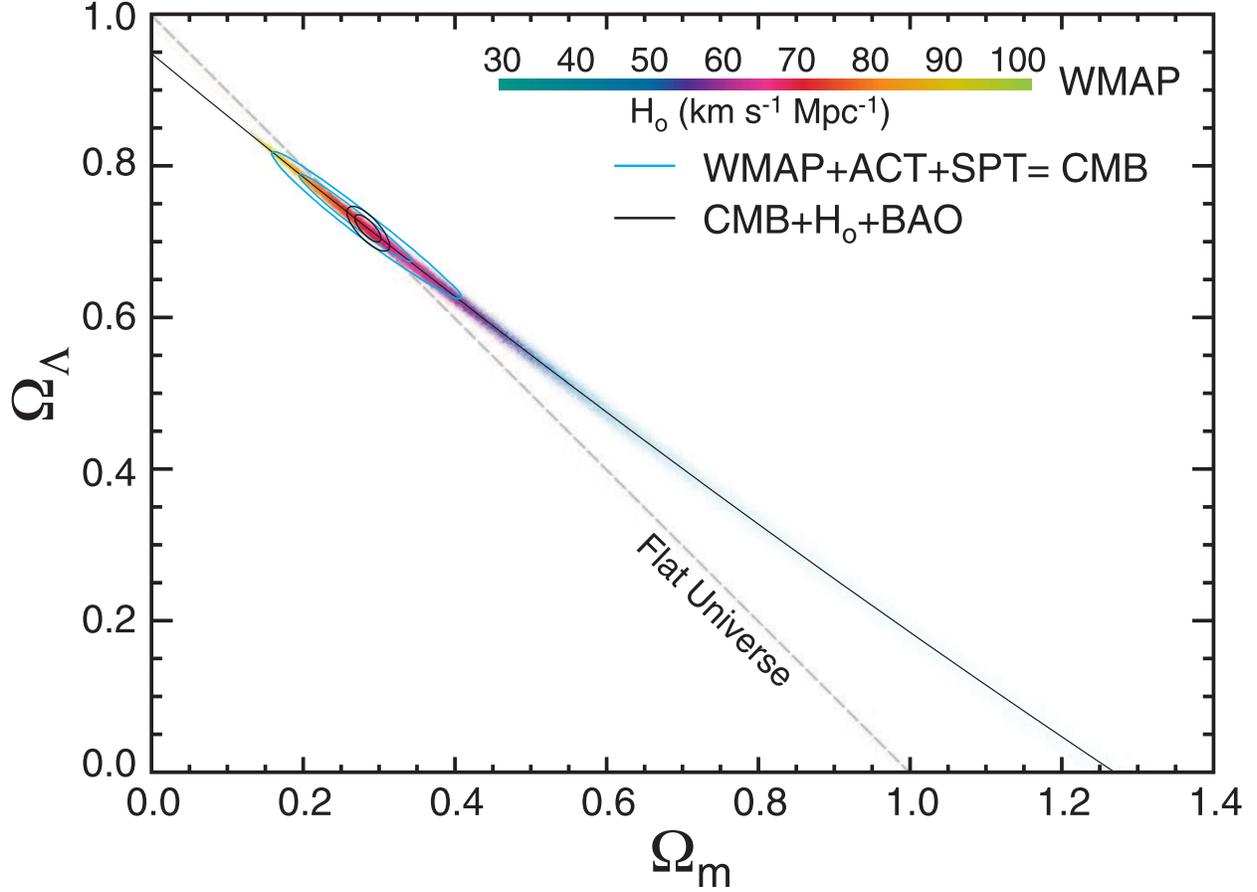}
\caption{
Constraints on curvature.
Flat universes fall on the $\ensuremath{\Omega_m} + \ensuremath{\Omega_\Lambda}=1$ line. 
Allowed regions are shown for \WMAP, CMB,
and CMB combined with BAO and \ensuremath{H_0} data, all 
with a hard prior of $\ensuremath{H_0} < 100\, {\rm km}\,{\rm s}^{-1}\,{\rm Mpc}^{-1}$.
\wmap\ data is represented by 290,000 Markov chain points, colored by their
value of \ensuremath{H_0}.
The \wmap\ data follow a geometric degeneracy ridge represented by the slightly
curved line, a parabola with equation 
$\ensuremath{\Omega_\Lambda} = 0.0620\, \ensuremath{\Omega_m}^2 - 0.825\,
\ensuremath{\Omega_m} + 0.947$.
The most likely point in the \WMAP-only chain has 
$\ensuremath{\Omega_\Lambda} = \ensuremath{0.721}$ and
$\ensuremath{\Omega_m} = \ensuremath{0.279}$, 
which is flat to three significant figures, even though this constraint was not enforced.
The \WMAP\ data alone require $\ensuremath{\Omega_\Lambda} > 0.58$ at 68\% CL and 
$\ensuremath{\Omega_\Lambda} > 0.22$ at 95\% CL.  The
contours show constraints when adding high-\ell\ CMB data (blue) and BAO and
\ensuremath{H_0} data (black).  These constraints are consistent with those from 
\WMAP\ alone, with the tightest constraint being 
\ensuremath{\Omega_{\rm tot} = 1.0027^{+ 0.0038}_{- 0.0039}}
\citep{hinshaw/etal:prep}.
\vspace{1mm} \newline (A color version of this figure is available in the online journal.)
\label{fig:lambda_matter}}
\end{figure}

\cite{hinshaw/etal:prep} examine various versions of cosmological models fit to select combinations of cosmological data.  These combinations are all rooted in \wmap\ data, which strongly limits possible cosmological models. There is, however, a narrow ridge of geometric degeneracy that applies to CMB measurements.  This is seen in Figure \ref{f:lambda_matter}.  Assuming a flat geometry breaks the degeneracy and forces a precise value for the Hubble constant. Alternatively, non-CMB cosmological measurements generally also break the CMB degeneracy and also result in a precise value for the Hubble constant. The fact that these Hubble constant values are consistent within their uncertainties is equivalent to concluding that the universe is flat within the measurement errors.

Table \ref{tab:lcdm_wmap_ext} gives the cosmological values for a six parameter flat $\Lambda$CDM model and a list of derived parameters that follow from it. Also tabulated are results from an additional seventh parameter added to the model. For example, if the number of relativistic degrees of freedom is allowed to vary beyond the standard three neutrinos, if tensor modes are allowed, or if the universe is allowed to deviate from a flat geometry. In addition, we summarize select constraints on non-$\Lambda$CDM models, such as deviating from a cosmological constant by allowing for a dark energy equation of state parameter $w\ne 1$.

In the last column of Table \ref{tab:lcdm_wmap_ext} we provide values for the same parameters described above but now arrived at by combining \wmap\ data with data from finer scale CMB measurements from ACT and SPT (extended CMB, or ``eCMB''), baryon acoustic oscillation (BAO) data, and data from the direct measurements of the Hubble constant ($H_0$). If we assume that all of these data sets are well-described by their published uncertainties, then these parameters provide a precise and accurate description of our universe. 

In an effort to provide a quantitative estimate of the overall impact of nine
years of \wmap\ data on cosmological parameters, we compare the final \wmap\
nine-year likelihood with pre-\wmap\ CMB data.
A paper entitled ``Last Stand Before \WMAP'' \citep{wang/etal:2002}  
provides a likelihood using only CMB data, just prior to \WMAP's initial 2003 results.
We find that the six parameter cosmological volume determined by \wmap\ data alone is a
factor of 68,000 times smaller than the allowed volume 
before \wmap.
To compute this factor, we take the cosmological volume to be proportional
to the square root of the determinant of the covariance matrix of the
parameters. 
Since the optical depth to last scattering was ill-constrained before \wmap, we
assign to it a constraint of $\tau < 0.3$.
We ensure that the parameter distributions are well-sampled by the
\wmap\ nine-year and pre-\wmap\ parameter chains by running over a half million
points in all of the relevant chains and verifying convergence, so the chains
sample the likelihoods well.  
We use six parameters in our volume-determining covariance matrix and those same
six parameters are sampled in Markov chains.
With flat priors on each, the six parameters are:
\ensuremath{\Omega_bh^2}, \ensuremath{\Omega_ch^2}, \ensuremath{\Omega_\Lambda},
\ensuremath{10^9 \Delta_{\cal R}^2}, \ensuremath{n_s}, and \ensuremath{\tau}. 
(Technically, we also include \ensuremath{A_{\rm SZ}} 
in both the pre-\wmap\ and \wmap\ chains and the
covariance matrix. 
\ensuremath{A_{\rm SZ}} is largely unconstrained by both data sets and is instead constrained by the
hard prior of $0 \le \ensuremath{A_{\rm SZ}} \le 2$,  so it has
negligible effect on the parameter volume and is only included so we can
marginalize over it.)  
Overall, we conclude that 99.9985\% of the allowed pre-\wmap\ six-parameter $\Lambda$CDM
models have been ruled out by WMAP data alone.  Only 0.0015\% remain. 
In addition to the large improvement in CMB measurement precision, the accuracy improvement arising from the reduction in systematic error afforded by \wmap\ is considerable.

Departing from the
simplest $\Lambda$CDM model, we consider a $\Lambda$CDM model with tensors,  
by adding the tensor-to-scalar ratio, \ensuremath{r}.
For this seven-parameter model, the reduction of
the cosmological volume is a factor of 117,000.  

Of course, when \wmap\ data are combined with a rich array of other significant cosmological data the stress-test for $\Lambda$CDM has been extraordinary. It is notable that only six parameters are required to achieve a sufficient fit to all cosmological data and that the underlying $\Lambda$CDM has not broken. Quite the contrary, a set of precise and accurate parameters now form a standard model of cosmology within the framework of the big bang theory (an expanding and cooling universe) and inflation (an underlying tilted power spectrum of primordial Gaussian-random adiabatic fluctuations).

\begin{deluxetable}{llcc}
\label{tab:lcdm_wmap_ext}
\tablecaption{Cosmological Parameter Summary}
\tablewidth{0pt}
\tabletypesize{\scriptsize}
\tablehead{\colhead{Parameter} & \colhead{Symbol} & \colhead{\wmap\tablenotemark{a}} &
\colhead{\wmap+eCMB+BAO+$H_0$\tablenotemark{a}\tablenotemark{b}}}
\startdata
\multicolumn{2}{c}{\bf 6-parameter \lcdm\ fit parameters\tablenotemark{c}} \\[1mm]
Physical baryon density  
& \quad \ensuremath{\Omega_bh^2}
& \ensuremath{0.02264\pm 0.00050}
& \ensuremath{0.02223\pm 0.00033} \\ 
Physical cold dark matter density
& \quad \ensuremath{\Omega_ch^2} 
& \ensuremath{0.1138\pm 0.0045}
& \ensuremath{0.1153\pm 0.0019} \\
Dark energy density ($w=-1$)
& \quad \ensuremath{\Omega_\Lambda} 
& \ensuremath{0.721\pm 0.025}
& \ensuremath{0.7135^{+ 0.0095}_{- 0.0096}} \\
Curvature perturbations ($k_0=0.002$ Mpc$^{-1}$)\tablenotemark{d}
& \quad \ensuremath{10^9 \Delta_{\cal R}^2} 
& \ensuremath{2.41\pm 0.10}
& \ensuremath{2.464\pm 0.072} \\
Scalar spectral index
& \quad \ensuremath{n_s} 
& \ensuremath{0.972\pm 0.013}
& \ensuremath{0.9608\pm 0.0080} \\
Reionization optical depth
& \quad \ensuremath{\tau} 
& \ensuremath{0.089\pm 0.014}
& \ensuremath{0.081\pm 0.012} \\
Amplitude of SZ power spectrum template
& \quad \ensuremath{A_{\rm SZ}} 
& \ensuremath{< 2.0\ \mbox{(95\% CL)}}
& \ensuremath{< 1.0\ \mbox{(95\% CL)}} \\
\multicolumn{2}{c}{\bf 6-parameter \lcdm\ fit: derived parameters\tablenotemark{e}} \\[1mm]
Age of the universe (Gyr)
& \quad \ensuremath{t_0}
& \ensuremath{13.74\pm 0.11}
& \ensuremath{13.772\pm 0.059} \\
Hubble parameter, $H_0 = 100h$ km/s/Mpc
& \quad \ensuremath{H_0}
& \ensuremath{70.0\pm 2.2}
& \ensuremath{69.32\pm 0.80} \\
Density fluctuations @ 8$h^{-1}$ Mpc
& \quad \ensuremath{\sigma_8} 
& \ensuremath{0.821\pm 0.023}
& \ensuremath{0.820^{+ 0.013}_{- 0.014}} \\
Velocity fluctuations @ 8$h^{-1}$ Mpc
& \quad \ensuremath{\sigma_8 \Omega_m^{0.5}} 
& \ensuremath{0.434\pm 0.029}
& \ensuremath{0.439\pm 0.012} \\
Velocity fluctuations @ 8$h^{-1}$ Mpc
& \quad \ensuremath{\sigma_8 \Omega_m^{0.6}} 
& \ensuremath{0.382\pm 0.029}
& \ensuremath{0.387\pm 0.012} \\ [3mm]
Baryon density/critical density
& \quad \ensuremath{\Omega_b} 
& \ensuremath{0.0463\pm 0.0024}
& \ensuremath{0.04628\pm 0.00093} \\
Cold dark matter density/critical density
& \quad \ensuremath{\Omega_c} 
& \ensuremath{0.233\pm 0.023}
& \ensuremath{0.2402^{+ 0.0088}_{- 0.0087}} \\
Matter density/critical density ($\Omega_c+\Omega_b$)
& \quad \ensuremath{\Omega_m} 
& \ensuremath{0.279\pm 0.025}
& \ensuremath{0.2865^{+ 0.0096}_{- 0.0095}} \\
Physical matter density
& \quad \ensuremath{\Omega_mh^2} 
& \ensuremath{0.1364\pm 0.0044}
& \ensuremath{0.1376\pm 0.0020} \\
Current baryon density (cm$^{-3}$)\tablenotemark{f}
& \quad \ensuremath{n_b} 
& \ensuremath{(2.542\pm 0.056)\times 10^{-7}}
& \ensuremath{(2.497\pm 0.037)\times 10^{-7}} \\
Current photon density (cm$^{-3}$)\tablenotemark{g}
& \quad $n_{\gamma}$
& $410.72 \pm 0.26$
& $410.72 \pm 0.26$ \\
Baryon/photon ratio
& \quad \ensuremath{\eta}
& \ensuremath{(6.19\pm 0.14)\times 10^{-10}}
& \ensuremath{(6.079\pm 0.090)\times 10^{-10}} \\ [3mm]
Redshift of matter-radiation equality
& \quad \ensuremath{z_{\rm eq}} 
& \ensuremath{3265^{+ 106}_{- 105}}
& \ensuremath{3293\pm 47} \\
Angular diameter distance to $z_{\rm eq}$ (Mpc)
& \quad \ensuremath{d_A(z_{\rm eq})} 
& \ensuremath{14194\pm 117}
& \ensuremath{14173^{+ 66}_{- 65}} \\
Horizon scale at $z_{\rm eq}$ ($h$/Mpc)
& \quad \ensuremath{k_{\rm eq}} 
& \ensuremath{0.00996\pm 0.00032}
& \ensuremath{0.01004\pm 0.00014} \\
Angular horizon scale at $z_{\rm eq}$
& \quad \ensuremath{\ell_{\rm eq}} 
& \ensuremath{139.7\pm 3.5}
& \ensuremath{140.7\pm 1.4} \\ [3mm]
Epoch of photon decoupling
& \quad \ensuremath{z_{*}} 
& \ensuremath{1090.97^{+ 0.85}_{- 0.86}}
& \ensuremath{1091.64\pm 0.47} \\
Age at photon decoupling (yr)
& \quad \ensuremath{t_{*}} 
& \ensuremath{376371^{+ 4115}_{- 4111}}
& \ensuremath{374935^{+ 1731}_{- 1729}} \\
Angular diameter distance to $z_{\ast}$ (Mpc)\tablenotemark{h}
& \quad \ensuremath{d_A(z_{*})}
& \ensuremath{14029\pm 119}
& \ensuremath{14007^{+ 67}_{- 66}} \\
Epoch of baryon decoupling
& \quad \ensuremath{z_{d}} 
& \ensuremath{1020.7\pm 1.1}
& \ensuremath{1019.92\pm 0.80} \\
Co-moving sound horizon, photons (Mpc)
& \quad \ensuremath{r_s(z_*)} 
& \ensuremath{145.8\pm 1.2}
& \ensuremath{145.65\pm 0.58} \\
Co-moving sound horizon, baryons (Mpc)
& \quad \ensuremath{r_s(z_d)} 
& \ensuremath{152.3\pm 1.3}
& \ensuremath{152.28\pm 0.69} \\
Acoustic scale, $\theta_* = r_s(z_*)/d_A(z_*)$ (degrees)
& \quad \ensuremath{\theta_{*}} 
& \ensuremath{0.5953\pm 0.0013}
& \ensuremath{0.59578\pm 0.00076} \\
Acoustic scale, $l_* = \pi/\theta_*$
& \quad \ensuremath{\ell_{*}} 
& \ensuremath{302.35\pm 0.65}
& \ensuremath{302.13^{+ 0.39}_{- 0.38}} \\
Shift parameter
& \quad \ensuremath{R} 
& \ensuremath{1.728\pm 0.016}
& \ensuremath{1.7329\pm 0.0058} \\ 
Conformal time to recombination
& \quad \ensuremath{\tau_{\rm rec}} 
& \ensuremath{283.9\pm 2.4}
& \ensuremath{283.2\pm 1.0} \\ [3mm]
Redshift of reionization
& \quad \ensuremath{z_{\rm reion}} 
& \ensuremath{10.6\pm 1.1}
& \ensuremath{10.1\pm 1.0} \\ 
Time of reionization (Myr)
& \quad \ensuremath{t_{\rm reion}} 
& \ensuremath{453^{+ 63}_{- 64}}
& \ensuremath{482^{+ 66}_{- 67}} \\ 
\multicolumn{2}{c}{\bf 7-parameter \lcdm\ fit parameters\tablenotemark{i}} \\[1mm]
Relativistic degrees of freedom\tablenotemark{j}
& \quad \ensuremath{N_{\rm eff}} 
& \ensuremath{> 1.7\ \mbox{(95\% CL)}}
& \ensuremath{3.84\pm 0.40} \\
Running scalar spectral index\tablenotemark{k}
& \quad \ensuremath{dn_s/d\ln{k}} 
& \ensuremath{-0.019\pm 0.025}
& \ensuremath{-0.023\pm 0.011} \\
Tensor to scalar ratio ($k_0=0.002\,{\rm Mpc}^{-1}$)\tablenotemark{l}
& \quad \ensuremath{r} 
& \ensuremath{< 0.38\ \mbox{(95\% CL)}}
& \ensuremath{< 0.13\ \mbox{(95\% CL)}} \\
Tensor spectral index\tablenotemark{l}
& \quad \ensuremath{n_t} 
& \ensuremath{> -0.048\ \mbox{(95\% CL)}}
& \ensuremath{> -0.016\ \mbox{(95\% CL)}} \\
Curvature ($1 - \ensuremath{\Omega_{\rm tot}}$)\tablenotemark{m} 
& \quad \ensuremath{\Omega_k} 
& \ensuremath{-0.037^{+ 0.044}_{- 0.042}}
& \ensuremath{-0.0027^{+ 0.0039}_{- 0.0038}} \\
Fractional Helium abundance, by mass
& \quad \ensuremath{Y_{\rm He}} 
& \ensuremath{< 0.42\ \mbox{(95\% CL)}}
& \ensuremath{0.299\pm 0.027} \\
Massive neutrino density\tablenotemark{n}
& \quad \ensuremath{\Omega_\nu h^2} 
& \ensuremath{< 0.014\ \mbox{(95\% CL)}}
& \ensuremath{< 0.0047\ \mbox{(95\% CL)}} \\
Neutrino mass limit (eV)\tablenotemark{n}
& \quad \ensuremath{\sum m_\nu} 
& \ensuremath{< 1.3\ \mbox{(95\% CL)}}
& \ensuremath{< 0.44\ \mbox{(95\% CL)}} \\[1mm]
\multicolumn{2}{c}{\bf Limits on parameters beyond \lcdm} \\[1mm]
Dark energy (const.) equation of state\tablenotemark{o}
& \quad \ensuremath{w}
& \ensuremath{-1.71<w<-0.34\ \mbox{(95\% CL)}}
& \ensuremath{-1.073^{+ 0.090}_{- 0.089}} \\
Uncorrelated isocurvature modes
& \quad \ensuremath{\alpha_{0}}
& \ensuremath{< 0.15\ \mbox{(95\% CL)}}
& \ensuremath{< 0.047\ \mbox{(95\% CL)}} \\
Anticorrelated isocurvature modes
& \quad \ensuremath{\alpha_{-1}}
& \ensuremath{< 0.012\ \mbox{(95\% CL)}}
& \ensuremath{< 0.0039\ \mbox{(95\% CL)}}
\enddata

\tablenotetext{a}{Unless otherwise stated, the values given are the mean of the
parameter in the Markov chain, and the 1-$\sigma$ region determined by removing
the lowest and the highest 15.87\% probability tails of the Markov chain to leave the
central 68\% region.}

\tablenotetext{b}{The \wmap+eCMB+BAO+\ensuremath{H_0} data set
\citep{hinshaw/etal:prep} includes the following.  The
\ensuremath{H_0} data consists of a Gaussian prior on the present-day value of
the Hubble constant, \ensuremath{H_0} = 73.8 $\pm$ 2.4 km $\rm s^{-1}$ $\rm
Mpc^{-1}$\citep{riess/etal:2011}.  The BAO priors...}

\tablenotetext{c}{The 6 parameters in this section are the parameters varied in
the chain.  A seventh parameter, \ensuremath{A_{\rm SZ}}, is also varied but is
constrained to be between 0 and 2.  The \wmap\ data do not strongly constrain
\ensuremath{A_{\rm SZ}}, which is why the 95\% CL interval simply returns the prior.
The eCMB data set does constrain the SZ effect, and prefers lower amplitudes of
the SZ template.  We call this a 6-parameter fit because only 6 parameters are
needed to fit the data well; the \ensuremath{A_{\rm SZ}} parameter is used only to
marginalize over the SZ effect and therefore include it in the error
bars.  All parameters varied in the Markov chains have flat priors, and in
this chain only the \ensuremath{A_{\rm SZ}} parameter requires hard constraints
limiting how much it can fluctuate.}

\tablenotetext{d}{$k=0.002$ Mpc$^{-1}$  $\longleftrightarrow$  $l_{\rm eff} \approx 30$.}

\tablenotetext{e}{These additional parameters are determined by the parameters 
being varied in the Markov chain.  Because these are not the parameters directly
being sampled, we are not necessarily assuming flat priors on these parameters.}

\tablenotetext{f}{Baryon density is given in units of proton masses
per cubic centimeter.}

\tablenotetext{g}{$T_{\rm CMB} = 2.72548\pm0.00057$ K, from \citet{fixsen:2009}.  This
parameter $n_\gamma$ is not varied in the Markov chains; the error bar is determined
directly from the error in CMB temperature.}

\tablenotetext{h}{Comoving angular diameter distance.}

\tablenotetext{i}{The parameters reported in this section place limits on
deviations from the simple 6-parameter $\Lambda$CDM model.  A complete listing of all parameter
values and uncertainties for each of the extended models studied is available on
LAMBDA.}

\tablenotetext{j}{Allows $N_{\rm eff}$ number of relativistic species, with the prior $0< \ensuremath{N_{\rm eff}} <10$.}

\tablenotetext{k}{Allows running in scalar spectral index but no tensor modes.}

\tablenotetext{l}{Allows tensor modes but no running in scalar spectral index.
We constrain the tensor to scalar ratio at $k=0.002$~Mpc$^{-1}$ to be $\ensuremath{r}>0$, 
and the tensor spectral index is related to the
tensor to scalar ratio by $\ensuremath{n_t} = -\ensuremath{r}/8$.}

\tablenotetext{m}{Allows non-zero curvature, $\Omega_k \ne 0$.}

\tablenotetext{n}{Allows a massive neutrino component, $\Omega_{\nu} > 0$.}

\tablenotetext{o}{Allows $w \ne -1$, but constrains it to be $-2.5 \le w \le 0$
and assumes $w$ is constant with redshift and $\Omega_k = 0$.}
\end{deluxetable}

\clearpage
\section{Conclusion} \label{sec:conclusions}

1) We have updated the raw data archive to include the full nine years of \wmap\ data. We have updated the pointing, calibration, and transmission imbalance factor solutions.

2) We have updated our beam maps and window functions based on the full nine years of \wmap\ data. We have made full sky maps of the five-band data in temperature and polarization, and we characterize the noise.

3) In addition to the standard map-making, we have implemented a new beam-symmetrized set of maps designed to reduce the effects of the asymmetric beams. These maps reduce the latitude dependence of the power spectrum and thus we confirm that the power asymmetry was largely due to the asymmetric beams, as expected.  This has no effect on the overall power spectrum and cosmological parameters, but is important to the notion of statistical isotropy, which is now more rigorously supported. The beam-symmetrized maps are not used for most cosmological analyses due to the complexity of the resulting noise, but they are used in foreground analysis.

4) We solve for new calibrations of Jupiter and Saturn, and we improve our model that separates the Saturn spheroid and ring components.  The final two years of \wmap\ observations include Saturn data with the rings nearly edge-on.

5) We provide new point source catalogs, using previous methods. One is based on filtering all five \wmap\ bands, and the other is based on removing the CMB from the Q-, V-, and W-band maps and then searching for peaks.

6) a) Our analysis of the diffuse foregrounds generally uses the five bands of \wmap\ data in conjunction with other data sets.  \wmap\ was designed to observe in the spectral region where the ratio of the CMB to foreground anisotropy is at its maximum while not allowing strong spectral lines to fall within any \wmap\ bandpass.  It is clear that the choice of \wmap\ frequencies succeeded in reaching these goals.  The five widely spaced \wmap\ bands and especially the low-frequency K-band radiometer have been invaluable in characterizing foregrounds. 

b) For most cosmological analyses we apply a Galactic cut and make a small correction for remaining emission using templates, but the ILC method is helpful and effective in separating the full sky CMB from foregrounds.  This separation can be done more accurately than the separation of foreground emission components from each other, for which there are degeneracies.  We present a new ILC map.  For the first time we now also provide an error estimate for this map that includes bias and foreground-CMB covariance.

c) To elucidate the characteristics and nature of the diffuse foreground components, we implement the Maximum Entropy Method (MEM), Markov Chain Monte Carlo (MCMC) fits, and $\chi^2$ fits. These are implemented with differing assumptions and priors.  Each of these methods has strengths and weaknesses, but the combination provides insight. Methods with less reliance on external templates make for noisier fits with greater degeneracy between emission components. Methods with greater reliance on external templates help to reduce noise and break degeneracies, but introduce errors, because the templates are not of the same quality as the \wmap\ data.

d) We decompose the foreground emission into synchrotron, free-free, spinning dust, and thermal dust components. The peak of the spinning dust spectrum lies below the K-band frequency (the lowest frequency \wmap\ radiometer) and is generally a sub-dominant emission component. The theoretically predicted Cold Neutral Medium (CNM) peak is at 17.8 GHz, but we solve for a peak frequency scale factor of $\approx 0.85$ that places the fitted peak frequency near 15 GHz.   The physical parameters that define the CNM are certainly only approximate, and their variation across the Galaxy is almost certainly responsible for complex spectral shape variations beyond just an amplitude and frequency shift. (Throughout this paper we use the term ``spinning dust'' without regard to the accuracy of the implied underlying physical model, but simply as the origin of a spectral template form to fit, where we allow both frequency and intensity adjustments. The actual physical emission mechanism(s) of this component may not yet be fully understood.)

e) Free-free emission is generally strong in the \wmap\ bands and the dominant foreground at high latitude in Q- and V-bands, but free-free emission is not as well traced by H$\alpha$ emission maps as one might have hoped or expected. This is true even when the H$\alpha$ emission is corrected for reflection and optical depth effects.  

f) We find a systematic Galactic plane discrepancy at the 20\% level between the thermal dust template map based on a model fit to {\it IRAS} and  {\it COBE} data and extrapolated to the \wmap\ bands, compared with our \wmap\ thermal dust fits with an inner plane/outer plane error morphology.  At high Galactic latitude the thermal dust template appears to be reasonable. The dust spectral index appears to be $\approx 1.8$ (for antenna temperature).

g) We find strong evidence that the synchrotron emission spectral index varies across the sky and is generally flatter in the plane and steepens with Galactic latitude.  In addition, the synchrotron spectral index appears to steepen with frequency. Within the \wmap\ bands the spectra of free-free, synchrotron, and spinning dust (which generally peaks at about 15 GHz and steepens at K- and Ka-bands) are far from orthogonal.  Yet, there is no spinning dust emission in the Haslam 408 MHz map, so that radio map is helpful for removing degeneracies. The foreground contributions at K-band are roughly 50\% synchrotron, 35\% free-free, and 15\% for a spinning dust like component. Free-free emission dominates in Q- and V-bands, and thermal dust emission dominates in W-band.

h) The original claim of discovery of a ``haze'' of free-free emitting gas with diminished H$\alpha$ \citep{finkbeiner:2004} has been ruled out.  Evidence of a distinct synchrotron haze feature depends on model choices in fitting, and no \WMAP\ model requires a haze component to provide a good fit to the data.  \WMAP\ MCMCg and Model 9 foreground fits show a general hardening (flattening) of the synchrotron spectral index from the Galactic poles to the plane, without a distinct haze feature.  K-band fit residuals in the haze region are $\lesssim10\%$ of the brightness identified by the \citet{planckintermediate:IX} as a $\beta_s \sim -2.55$ synchrotron haze.  However, a real haze could have been inappropriately absorbed into other components of the \WMAP\ decomposition, which has degeneracies. Likewise, the Planck haze could result from modeling assumptions, which are different from the assumptions of each of the three \WMAP\ models. Based on currently available data, we conclude that the existence of a distinct localized haze depends on the fitting and analysis methods used.  Additional data, particularly at frequencies below K-band, would help constrain model degeneracies.

i) We define a Galactic cut for fitting and removing template-traced emission for the high latitude sky and then a small additional cut for safety. The remaining high latitude sky is used for power spectrum calculation and parameter determination. This portion of the template-corrected sky is strongly dominated by CMB anisotropy. 

7) We implemented a new unbiased and optimal estimation of the TT power spectrum that uses $C^{-1}$ weighting, as opposed to the unbiased MASTER quadratic estimator. We also present the TE, EE, TB, and BB power spectra. A six parameter flat $\Lambda$CDM model is fit to these power spectra. 

8) We examined the goodness-of-fit of the $\Lambda$CDM model to the power spectrum data. The $\chi^2$ of the high-$l$ TT power spectrum is dominated by an even-$l$ versus odd-$l$ effect, as seen in the seven year analysis.  This is notable since the seven-year power spectrum was determined by MASTER and the nine-year by $C^{-1}$.  Therefore the even-odd effect cannot be an artifact of the computation method. We continue to believe that the effect is not significant as we have made posterior choices to select and examine the effect (such as a particular range of multipole moments) and there exists no known theory to produce it, especially since even sharp features in $k$-space do not remain sharp in $l$-space.

9) The quadrupole amplitude is below of the median expectation of the best fit power spectrum by $< 2\sigma$, so it is not anomalously low. No new theory could be significantly preferred (i.e., by more than 2$\sigma$) based on the quadrupole value alone. The quadrupole-octupole alignment remains approximately the same in the nine-year as seven-year data, but a new estimate of the uncertainties based on the underlying ILC map indicates that we cannot reliably remove foregrounds to the level needed to demonstrate a significant alignment. Having addressed the quadrupole value, the quadrupole-octupole alignment, and the general goodness-of-fit, we find no convincing evidence of CMB anomalies beyond the normal statistical ranges that should be anticipated to occur in a rich dataset.

10) An analysis of the CMB maps find no compelling evidence for deviations from Gaussianity. 
We find 
$f^{\rm loc}_{\rm NL} = 37.2 \pm 19.9$, with 
$-3 < f^{\rm loc}_{\rm NL} < 77$ at 95\% CL.
We also find 
$f^{\rm eq}_{\rm NL} = 51 \pm 136$, with 
$-221 < f^{\rm eq}_{\rm NL} < 323$ at 95\% CL, and 
$f^{\rm orth}_{\rm NL} = -245 \pm 100$, with
$-445 < f^{\rm orth}_{\rm NL} < -45$ at 95\% CL.
We do not find any of these quantities differ significantly from zero.  It should be noted that three quantities are computed, increasing the chance of an otherwise less likely outcome. 

11) Cosmological models are fit to the power spectrum \citep{hinshaw/etal:prep}.
A six parameter flat $\Lambda$CDM model continues to fit all of the \wmap\ data
well.  These parameters also appear to be consistent with a wide range of other
cosmological data as well. The six parameter cosmological volume determined by
\wmap\ data alone is a factor of 68,000 times smaller that the CMB constraints
before \wmap\, as assessed by the ``Last Stand Before \WMAP'' paper of
\cite{wang/etal:2002}. (Since the optical depth to scattering was not
constrained at all in that assessment, we assigned to it a constraint of $\tau < 0.3$ 
in carrying out the volume calculation.) Adding a seventh parameter
suggests a reduction of the cosmological volume by even more, a factor of
117,000.   
   
12) When \wmap\ data are combined with a rich array of other significant cosmological data the stress-test for $\Lambda$CDM is extraordinary. It is notable that only six parameters are required to achieve a sufficient fit to all cosmological data and that the underlying $\Lambda$CDM has not broken. Quite the contrary, a set of precise and accurate parameters now form a standard model of cosmology within the framework of the big bang theory (an expanding and cooling universe) and inflation (an underlying tilted power spectrum of primordial Gaussian-random adiabatic fluctuations).  General relativity combined with the Friedmann-Lema\^itre-Robertson-Walker metric leads to the Friedmann equation, which provides the background cosmology. Inflation can provide the initial conditions, including the generation of primordial perturbations via fluctuations of the inflaton and gravitational fields.  
Inflation predicts that the universe is nearly flat. We find 
\ensuremath{\Omega_k = -0.0031^{+ 0.0038}_{- 0.0039}} and 
$\vert \Omega_k \vert < 0.0094$ at 95\% confidence, within 0.95\% of flat/Euclidean.
If restricted to $\Omega_k>0$ (a negative curvature open universe) as suggested by the creation of our universe from the landscape, then $\Omega_k < 0.0062$ at 95\% CL.  A small deviation from flatness is expected and is worthy of future searches.  Inflation is also strongly supported by the observed features that the fluctuations are adiabatic, with Gaussian random phases.  The detection of a deviation of the scalar spectral index from unity reported earlier by \wmap\ now has high statistical significance (\ensuremath{n_s = 0.9608\pm 0.0080}).  The CMB has been central to posing the horizon, flatness, and structure problems for which inflation and general relativity provide solutions. 

13) Within the horizon, acoustic waves modify the primordial perturbations in a manner that depends on the values of the cosmological parameters.  
The sub-horizon CMB measurements drive the determination of the cosmological parameters and the degeneracies are broken with the addition of other cosmological observations, such as measurements of the Hubble constant and the baryon acoustic oscillations as a function of redshift determined from large galaxy surveys. Using this fact, we find that Big Bang nucleosynthesis is well supported and there is no compelling evidence for a non-standard number of neutrino species (\ensuremath{N_{\rm eff} = 3.84\pm 0.40}).

14) The requirement for both cold dark matter, which gravitates but does not interact with photons, and a substantial mass-energy component consistent with a cosmological constant, which causes an accelerated expansion of the universe as characterized by Type Ia supernovae measurements, is unavoidable because of the precision of the available data and the multiple methods of measurement.  The CMB fluctuations require dark matter and dark energy.  The inability to predict a value for vacuum energy was a pre-existing physics problem, but particle physics has no problem positing massive particles that do not interact with photons as candidates for the CDM. If the massive particles do not decay or annihilate, their identity makes little difference to cosmology. It may well turn out that the dominant mass-energy component of our universe is a cosmological constant arising from vacuum energy, and that the vacuum energy is fundamentally not a specifically predictable quantity.  It will be exciting to see how current theories develop, and especially fascinating how well these theories can be tested with data.  The CMB is a unique remnant of the early universe which has been our primary cosmological observable. It continues to be imperative to learn all that we can from it. 

\acknowledgments
The \wmap\ mission was made possible by the support of NASA.  We are grateful to Marian Pospieszalski of the National Radio Astronomy Observatory (NRAO) for his design of the microwave amplifiers that enabled the mission, and to NRAO for the development of the flight amplifiers. We also thank the project managers, Rich Day and Liz Citrin, and system engineers, Mike Bay, and Cliff Jackson, who were both expert and effective in leading the mission to launch, on-schedule and on-budget. It was a special pleasure for the science team to work closely with Cliff Jackson from the earliest times of the proposal development through to the post-launch activities. NASA has never had a finer engineer and we wish him well in his retirement. We also recognize the extraordinary efforts of the engineers, technicians, machinists, data analysts, budget analysts, managers, administrative staff, and reviewers who were all key parts of the team that created the \wmap\ spacecraft. 

C.L.B. was supported, in part, by the Johns Hopkins University.  K.M.S. was supported at the Perimeter Institute  by the Government of Canada through Industry Canada and by the Province of Ontario through the Ministry of Research \& Innovation. E.K. was supported in part by NASA grants NNX08AL43G and NNX11AD25G and NSF grants AST-0807649 and PHY-0758153.  We acknowledge use of the HEALPix \citep{gorski/etal:2005}, CAMB \citep{lewis/challinor/lasenby:2000}, and CMBFAST \citep{seljak/zaldarriaga:1996} packages. 
Some computations were performed on the GPC supercomputer at the SciNet HPC Consortium. We thank SciNet, which is funded by the Canada Foundation for Innovation under
the auspices of Compute Canada, the Government of Ontario, Ontario Research Fund -- Research Excellence, and the University of Toronto.
We acknowledge the use of the Legacy Archive for Microwave Background Data Analysis (LAMBDA). Support for LAMBDA is provided by NASA Headquarters.

\appendix
\section{Band Center Frequencies \label{sec:bandcenter}}
\label{sec:band_center_frequencies}

\begin{figure}
\epsscale{0.70}
\plotone{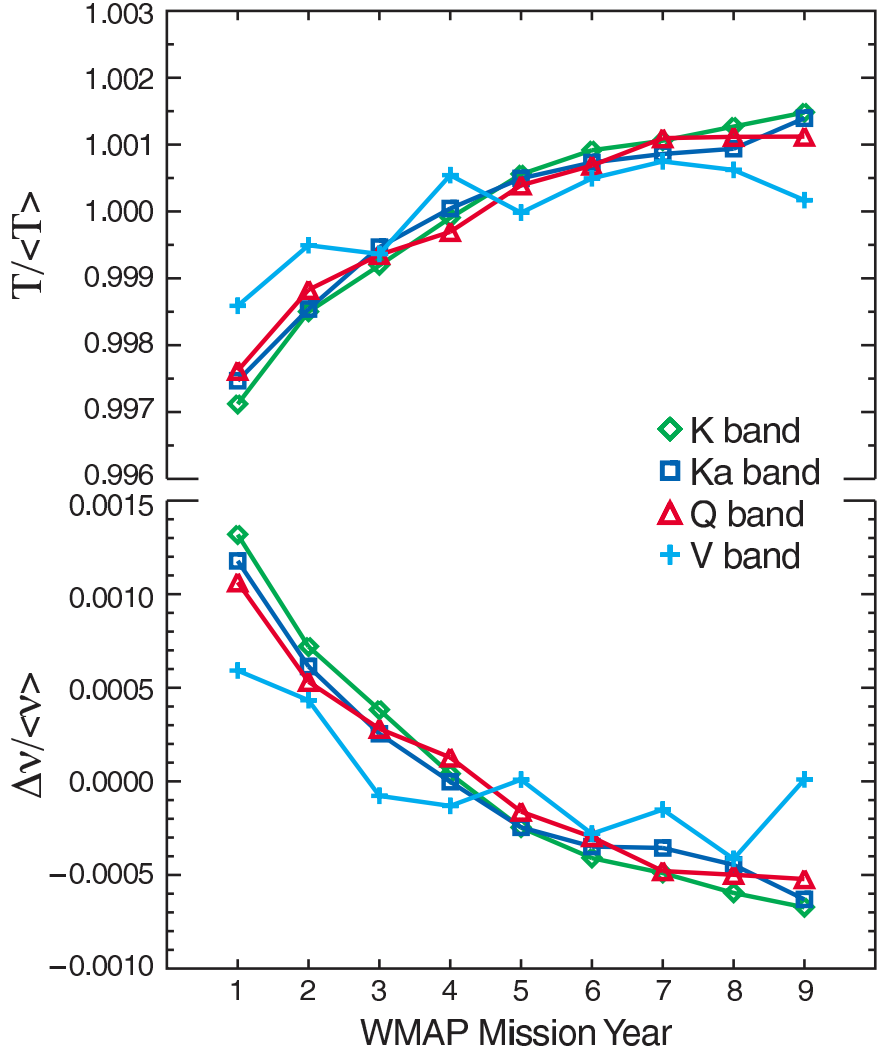}
\caption{Top - Measurements of the year-to-year fractional brightness variation of
the Galactic plane in \WMAP\ skymaps, obtained by correlating Galactic plane
signal in each single year map with Galactic plane signal in the nine-year
map.  There is a small dependence of these variations on spectral index,
which shows that they are caused by variations in effective \WMAP\ band center
frequencies over the mission.
Bottom - The year-to-year fractional variation of \WMAP\ band center frequency
derived from Galactic plane brightness variations measured for selected
spectral index bins.
\vspace{1mm} \newline (A color version of this figure is available in the online journal.)}
\label{fig:banddrift}
\end{figure}

Figure \ref{fig:banddrift} shows small year-to-year variations of
Galactic plane brightness measured from yearly maps in K-, Ka-, Q-, and V-bands.
Each yearly map was correlated against the nine-year map for pixels at
$\vert b \vert < 10\arcdeg$.  A linear slope and offset was fit to each correlation,
and the slope values are shown in Figure \ref{fig:banddrift}.  Results for W-band
are not shown because the scatter in the yearly slopes is large and no significant
variation was detected.  Analysis of DA maps has shown that the measured variation
is consistent in Q1 and Q2, and in V1 and V2.

The K$-$Q band brightness variations were previously presented for the seven-year
data in \citet{jarosik/etal:2011}, where they were described as variations in
the \WMAP\ calibration.  Further analysis has shown that the CMB signal in
yearly maps does not show such variation.  Yearly variations of the CMB
dipole amplitude in year 1-7 maps are less than $\pm 0.025 \%$
for many DAs.  We have also found that the Galactic plane brightness
variations depend on spectral index, with greater variation for
regions of steeper spectral index, so we conclude that they
are caused by variations in the effective center frequencies of the \WMAP\
bandpasses over the mission.  As the observatory's thermal control
surfaces age, a gradual warming of the \WMAP\ instrument's physical
temperature occurs \citep{greason/etal:prep}.  Given the instrument amplifier
fixed voltage bias scheme, an increase in temperature (or device aging)
can induce corresponding changes in the drain current and gain, and an
associated perturbation in the effective bandpass.

We determine the fractional variation in center frequency for each band as
follows.  Assuming the sky signal in a given pixel $p$ can be characterized
by a power law spectrum with thermodynamic temperature spectral index $\beta_p$,
the measured sky brightness for a given year $i$ is
\beq
T_i(p) = T_0(p)\left(\frac{\nu_i}{\nu_0}\right)^{\beta_p},
\eeq
where $T_0(p)$ is the sky brightness at a fiducial frequency $\nu_0$
and $\nu_i$ is the effective frequency for year $i$.
We assume $T_0(p)$ is constant in time.  For small frequency drifts,
$\Delta\nu_i /\nu_0 \equiv (\nu_i-\nu_0)/\nu_0 \ll 1$, it is useful to work
with the linearized form,
\beq
T_i(p) = T_0(p)\left[1 + \beta_p(\Delta\nu_i / \nu_0)\right].
\eeq
If we choose $\nu_0 \equiv \langle \nu_i \rangle$, where the mean is over years $i$,
then $T_0(p) = \langle T_i(p) \rangle$ and the fractional variation in frequency
is
\beq
\frac{\Delta\nu_i}{\langle \nu_i \rangle} = \left(\frac{T_i(p)}{\langle T_i(p) \rangle} - 1 \right)/ \beta_p
\eeq
For each band and each year, we calculate the pixel averaged $T_i/\langle T_i \rangle$
for Galactic plane pixels in selected spectral index ranges as the $T_i(p)$ vs $\langle T_i(p) \rangle$
correlation slope.  Spectral index was calculated using the neighboring \WMAP\ band or bands,
e.g., $\beta$(K-Ka) was used for K-band and the mean of $\beta$(K-Ka) and $\beta$(Ka-Q)
was used for Ka-band.  Each spectral index bin for a given band gives a result for the
variation of $\Delta\nu_i / \langle \nu_i \rangle$ over the mission.  These results were
found to be consistent with each other, and an average (excluding bins with high scatter)
was adopted for the variations shown for each band in Figure \ref{fig:banddrift}.  

No correction for bandpass drift is applied in our map-making.  Since the \WMAP\ observations
are made simultaneously in the different bands, the map-making always forms band maps that
have a common epoch, and each band map can be treated as having a single effective band center
frequency valid for that epoch.
Our previously published band center frequencies (see Table 4 of \citet{jarosik/etal:2011}
for point sources and Table 11 of \citet{jarosik/etal:2003} for diffuse emission) are based on
pre-flight measurements, so presumably are valid for year 1 of the flight data.
For nine-year data, a correction based on Figure \ref{fig:banddrift} should be applied.  The
correction is a reduction of the pre-flight center frequency by 0.13, 0.12, 0.11, and 0.06$\%$ for
K-, Ka-, Q-, and V-band, respectively.  This correction is included in the center frequencies
for point sources listed in Table~\ref{tab:beam_quantities}.

\section{\WMAP\ Nine-Year Five-band Point Source Catalog}
\label{5bandsources}

\begin{deluxetable}{ccccccccccc}
\tablecaption{\label{tab:fivebandptsrc} \WMAP\ Nine-Year Five-band Point Source Catalog}
\tabletypesize{\scriptsize}
\tablewidth{0pt}
\tablehead{
 \colhead{RA [hms]}
&\colhead{Dec [dm]}
&\colhead{ID}
&\colhead{K [Jy]}
&\colhead{Ka [Jy]}
&\colhead{Q [Jy]}
&\colhead{V [Jy]}
&\colhead{W [Jy]}
&\colhead{$\alpha$}
&\colhead{5 GHz ID}
&
}
\startdata
 00 04 08
&$-$47 43
&   
&$0.6\pm0.02$
&$0.7\pm0.04$
&$0.5\pm0.05$
&$0.5\pm0.08$
&\nodata
&$-0.1\pm0.3$
&PMN J0004-4736
&
\\
 00 06 06
&$-$06 23
&060
&$2.2\pm0.04$
&$1.6\pm0.05$
&$1.8\pm0.07$
&$1.9\pm0.1$
&$1.3\pm0.2$
&$-0.3\pm0.1$
&PMN J0006-0623
&
\\
 00 10 33
&\phs11 01
&   
&$0.8\pm0.03$
&$1.0\pm0.05$
&$1.2\pm0.06$
&$1.6\pm0.1$
&$1.1\pm0.2$
&$0.5\pm0.2$
&GB6 J0010+1058
&
\\
 00 12 46
&$-$39 53
&202
&$1.3\pm0.03$
&$1.1\pm0.04$
&$1.1\pm0.05$
&$0.9\pm0.09$
&\nodata
&$-0.3\pm0.2$
&PMN J0013-3954
&
\\
 00 25 24
&$-$26 03
&   
&$0.9\pm0.03$
&$0.8\pm0.05$
&$0.6\pm0.06$
&\nodata
&\nodata
&$-0.4\pm0.3$
&PMN J0025-2602\tablenotemark{a}
\\
 00 26 06
&$-$35 10
&   
&$0.7\pm0.03$
&$0.9\pm0.05$
&$0.9\pm0.05$
&$0.8\pm0.09$
&$0.9\pm0.2$
&$0.2\pm0.2$
&PMN J0026-3512
&
\\
 00 29 33
&\phs05 54
&   
&$1.1\pm0.03$
&$1.2\pm0.05$
&$1.2\pm0.06$
&$0.9\pm0.1$
&$1.4\pm0.2$
&$0.1\pm0.2$
&GB6 J0029+0554B\tablenotemark{a}
\\
 00 38 15
&$-$25 01
&   
&$0.8\pm0.03$
&$0.8\pm0.05$
&$0.7\pm0.06$
&$0.8\pm0.1$
&\nodata
&$-0.1\pm0.2$
&PMN J0038-2459
&
\\
\enddata
\tablenotetext{a}{Indicates the source has multiple possible identifications.}
\tablenotetext{b}{Source J0322-3711 (Fornax A) is extended, and the fluxes listed were obtained by aperture photometry.}
\tablenotetext{c}{Source J1356+7644 is outside of the declination range of the GB6 and PMN catalogs.  It was identified as QSO NVSSJ135755+764320 by Trushkin (2006, private communication).}
\tablenotetext{d}{Source J1632+8227 is outside of the declination range of the
GB6 and PMN catalogs. It was identified as NGC 6251 by \citet{trushkin:2003}.}
\end{deluxetable}

\section{\WMAP\ Nine-Year CMB-free QVW Point Source Catalog}
\label{3bandsources}

\begin{deluxetable}{cccccccc}
\tablecaption{\label{tab:qvwptsrc} \WMAP\ Nine-Year CMB-free QVW Point Source Catalog}
\tabletypesize{\footnotesize}
\tablewidth{0pt}
\tablehead{
 \colhead{RA [hms]}
&\colhead{Dec [dm]}
&\colhead{ID}
&\colhead{Q [Jy]}
&\colhead{V [Jy]}
&\colhead{W [Jy]}
&\colhead{5 GHz ID}
&
}
\startdata
 00 04 29
&$-$47 35
&   
&$0.4\pm0.1$
&$0.5\pm0.2$
&$-0.3\pm0.3$
&PMN J0004-4736
&
\\
 00 06 14
&$-$06 25
&060
&$2.0\pm0.2$
&$1.7\pm0.2$
&$0.7\pm0.4$
&PMN J0006-0623
&
\\
 00 10 29
&\phs10 59
&   
&$1.0\pm0.2$
&$1.2\pm0.2$
&$0.8\pm0.3$
&GB6 J0010+1058
&
\\
 00 13 23
&\phs40 55
&   
&$0.6\pm0.2$
&$0.5\pm0.2$
&$0.9\pm0.3$
&GB6 J0013+4051
&
\\
 00 19 41
&\phs25 58
&   
&$0.6\pm0.2$
&$0.3\pm0.2$
&$0.4\pm0.3$
&GB6 J0019+2602
&
\\
 00 26 07
&$-$35 12
&   
&$1.3\pm0.2$
&$0.6\pm0.2$
&$0.6\pm0.3$
&PMN J0026-3512
&
\\
 00 29 44
&\phs05 54
&   
&$0.8\pm0.2$
&$0.4\pm0.2$
&$0.9\pm0.3$
&GB6 J0029+0554B
&\tablenotemark{a}
\\
 00 38 13
&$-$02 05
&   
&$0.6\pm0.2$
&$0.4\pm0.2$
&$0.7\pm0.3$
&PMN J0038-0207
&
\\
\enddata
\tablenotetext{a}{Indicates the source has multiple possible identifications.}
\end{deluxetable}

\section{Smoothed Noise}
\label{sec:smooth_noise}

\newcommand{\npix}{\ensuremath{N_{\rm pix}}}
\newcommand{\Nobsi}{\ensuremath{N_{{\rm obs},i}}}
\newcommand{\Nobs}{\ensuremath{N_{\rm obs}}}
\newcommand{\nobsi}{\ensuremath{n_{{\rm obs},i}}}

We use maps that have been smoothed to a common resolution for several
\WMAP\ analyses. This appendix discusses how much the smoothing reduces
the random instrument noise.  This smoothing also correlates the noise
between pixels.  Here, we only calculate the diagonal elements of the
noise covariance matrix in pixel space; the correlations are beyond the scope of
this appendix.  Also, the noise calculated here should be added in quadrature
to the 0.2\% \WMAP\ calibration error.

For discussing beam smoothing, we use the same notation as 
Equation~(4) of \citet{hill/etal:2009}.  
\be
B_\ell = \Omega_B b_\ell = 2\pi \int^{1}_{-1} b(\theta) P_\ell(\cos \theta)\; d \cos \theta.
\ee
In this case, we use the beam to 
describe the additional smoothing that we apply to the map to bring the total
smoothing up to 1 degree FWHM.

The pixel temperature value, $T^{\rm convol}_p$, in a convolved map is a weighted sum 
of the nearby pixel values, 
\be
T^{\rm convol}_p = \sum_{i} w_{i,p}T_i,
\ee
where $w_{i,p}$ gives the weight that each original pixel with index $i$ gives
to convolved pixel $p$.  The weights $w_{i,p}$ define the beam used for smoothing.
From this formula and a noise estimate in the original pixels, we propagate errors directly,
assuming uncorrelated noise in the original pixels.
\be
\sigma^2\left( T^{\rm convol}_p \right) = \sum_{i} w_{i,p}^2 \sigma^2(T_i),
\ee
where $\sigma^2(T^{\rm convol}_p)$ is the noise variance in the convolved
pixel $p$ and $\sigma^2(T_i)$ is the noise variance in the original pixel $i$.

The noise in each convolved pixel can be rapidly computed by smoothing a map
of unsmoothed noise variance values, $\sigma_0^2/\Nobsi$.
However, the smoothing must be done using the squared weights, 
which requires determining the Legendre transform of the beam once it has been squared
in real space, $b(\theta)^2$.
\be
\Omega_b' b'_\ell = 2\pi \int_{-1}^1 b^2(\theta) P_\ell(\cos \theta)\; d\cos
\theta.
\ee
The values for the required beam smoothing, $\Omega_b' b'_\ell$, can be computed
numerically by calculating $b(\theta)$ on a one-dimensional finely spaced grid
in $\theta$, squaring it, and computing the above integral as a sum.

The above description of smoothed noise assumes it will be reported in a map
with a pixel size much smaller than the beam size.  In the opposite case, where
the final pixel size is much larger than the beam size, the noise can be
averaged down ignoring the beam, since the effect of the beam will be small.
However, there is an intermediate case where the pixel size and beam size are
comparable, such as with r6 maps of 1 degree smoothed data.  In this case, a
more careful treatment of the pixel window function could be useful.  Instead of
approximating the pixel window function as an azimuthally symmetric beam, we
take a more brute-force approach, outlined below.  

We have r9 maps of \Nobsi.  Suppose we want to know the noise properties
of the corresponding temperature map smoothed to 1 degree FWHM and then degraded
to r6.  To determine this, we calculate the real-space smoothing function
needed to bring the beam smoothing up to 1 degree; we call this $b(\theta)$.  
This will be a 1 degree FWHM beam $b_\ell^1$ divided by the \WMAP\ instrument beam
$b_\ell^\nu$ for that DA.
We approximate $b(\theta)$ numerically by finding the Legendre transform of the 
needed smoothing, $b_\ell = b_\ell^1 / b_\ell^\nu$, on a one-dimensional list of angles
$\theta$.
Then, for each r6 pixel,
we find all r9 pixels within 2 degrees of the r6 pixel center.  We
determine the weights $w_{i,p}$, where $i$ is an index over r9 pixels within
2 degrees of the r6 pixel center, and $p$ is an index over r9 pixels
inside the r6 pixel.  As before, we have
\be
w_{i,p} = b(\theta_{i,p})
\ee
where $\theta_{i,p}$ is the angle between the centers of pixels $i$ and $p$, and
the weights have been rescaled so that $\sum_i w_{i,p} = 1$.
The radius of two degrees was chosen so that noise outside of that circle would
be negligibly averaged into the r6 pixel, given our beam smoothing size.

Since the noise for the r9 pixels of the smoothed map is averaged into
an r6 pixel, we must account for this in our error propagation.  
We assume flat weighting for the degrade from r9 to r6, in the following description.
There are 64 r9 pixels in an r6 pixel.
The temperatures (pixels with index $p$) are averaged into an r6 pixel (with index $q$) as 
\be
T^{\rm degraded}_q = \frac{1}{64}\sum_p \sum_i w_{i,p} T_i.
\ee
The formula for propagation of errors is
\be
\sigma^2(T^{\rm degraded}_q) = \sum_i \left(\frac{\partial T_q}{\partial T_i}\right)^2 \sigma^2(T_i),
\ee
which then becomes
\be
\sigma^2(T^{\rm degraded}_q) = \sum_i \left(\frac{1}{64} \sum_p w_{i,p}\right)^2 \frac{\sigma_0^2}{\Nobsi}.
\ee
Alternatively, we can quote an effective $N_{{\rm obs},q}^{\rm eff}$ 
value for a r6 pixel as
\be
\frac{1}{N_{{\rm obs},q}^{\rm eff}} \equiv
\sum_i \left(\frac{1}{64} \sum_p w_{i,p}\right)^2 \frac{1}{\Nobsi}.
\ee
Since this is the number more commonly reported in our data files, we use this.

There appear to be artifacts in these $N_{{\rm obs},q}^{\rm eff}$ maps. 
This is most readily visible when a simple binned version of $N_{{\rm obs},q}$
which ignores the effects of smoothing is divided out.  In this case, the above
noise propagation predicts what appears to be suppressed noise levels (greater
$N_{\rm obs}$) near the edges of the base tiles in the polar cap regions of the
HEALPix pixelization.

These results can be verified by creating white noise realizations at r9,
smoothing them, binning them to r6, and then checking the variance of the
noise in each pixel.  When this comparison is done, some of these artifacts
remain in these simulations as well, so it appears the pixelization
(slightly varying pixel shapes) is causing
a real effect in the smoothed noise.  The fluctuations that appear to be due to
the HEALPix pixelization are on order of 10\% in $N_{{\rm obs},q}$ in all bands.

The median values of $N_{{\rm obs},q}$ over the whole sky 
for the two approaches (white noise sims
vs. the above propagation of errors) differ by about 5\%
at K-band (where the additional smoothing is smallest), and roughly 1\% in other bands.
The above propagation of errors appears to underestimate the noise slightly
(overestimate $N_{{\rm obs},q}$). 

\section{Bandpass Integration}

\label{sec:bandpass_integration}
In this section we first discuss the full integration over the bandpass based on
data from \citet{jarosik/etal:2003}, and then we discuss a useful approximation
to that integration based on three frequencies in each band.  This is the
approximation used for foreground fitting in Section~\ref{sec:six_band_fits}.

The full integration of different foreground spectra over the \WMAP\ bandpasses 
can be done as follows, based on the description of the radiometers in
\citet{jarosik/etal:2003}.
After computing $r_{\rm avg}(\nu_i)$ from Equation~(46) of that paper
using the discretized bandpass measurements, we
combine the measurements as if we were doing an unweighted average of the maps
in thermodynamic temperature, as follows.
First, we normalize the bandpass for each radiometer so that
\be
\sum_i r_{\rm avg}(\nu_i) = 1
\ee
We note the small shift in bandpass that we describe in Appendix~\ref{sec:band_center_frequencies}.
Then, we interpolate the foreground spectrum onto the specific frequencies at which
the \WMAP\ bands were measured, $\nu_i$, average the frequency over the
spectrum, and convert from antenna to thermodynamic temperature.  The
measured foreground thermodynamic temperature response to a foreground spectrum
$f(\nu)$ given in antenna temperature, averaged over all the radiometers in one
\WMAP\ band, is
\be
T_{\rm band}[f(\nu)] = \frac{1}{N_{\rm radiometers}} 
\sum_{j=1}^{N_{\rm radiometers}} \sum_i \frac{r_{{\rm avg},j}(\nu_i)}{w'(\nu_i)} f(\nu_i)
\ee
where $w'(\nu)$ is as defined in \citet{jarosik/etal:2003}: it is the 
derivative of the single-polarization Planck spectrum with respect to
temperature, divided by $k_{\rm B}$ to make it unitless.  It depends 
on both CMB temperature and frequency, but the derivative is taken with
respect to CMB temperature.
\be
w(\nu) \equiv \frac{h\nu}{e^x - 1} \qquad x \equiv \frac{h\nu}{k_B T}
\ee
\be
w'(\nu) \equiv \left|\frac{1}{k_B} \frac{dw(\nu)}{dT}\right|_{T=T_{\rm CMB}}
= \frac{x^2 e^x}{(e^x-1)^2}
\ee
Note that this assumes an unweighted average of the maps.  If we were to do an
optimal weighted average, the total bandpass would have some small spatial
dependence with pixel, as the number of observations varies between DAs.

In practice, it is the complexity and shape of the foregrounds that limits
the foreground fitting.  The detailed bandpass discussion above is more accurate, but
fast approximations are useful.  \citet{jarosik/etal:2003} provides a useful
approximation given by Equation~(50) of his paper for spectra that are power laws
in antenna temperature.  This allows one to determine the effective frequency of
the bandpass and therefore rapidly calculate the measured antenna temperature from the
power law.  However, power laws are always concave upward on a plot
of antenna temperature as a function of frequency with both axes linear.  Since
we also want to fit a spinning dust spectrum which is concave downward, we
invent another approximation.

Instead of doing the full integration discussed above for each band, 
this approximation only requires a weighted average of the antenna temperature 
at three frequencies.  
The thermodynamic temperature measured by \WMAP\ in a specific band is 
approximated as
\be
T = \frac{\Delta T}{\Delta T_A} \sum_{i=1}^3 w_i T_A(\nu_i)
\ee
where $T_A(\nu_i)$ is the antenna temperature foreground spectrum measured
at frequencies $\nu_i$, and $\Delta T / \Delta T_A$ is the conversion from
antenna to thermodynamic temperature.
The frequencies and weights used are in Table~\ref{tab:interp1}.  
The weights are chosen so that any spectrum that is a second
order polynomial in antenna temperature will have its integral evaluated
exactly (to the accuracy with which the bandpasses were measured).  
These weights are therefore including information about the full shape of the 
bandpass.
We do not expect to have spectra that are second order polynomials;
most of the antenna temperature spectra are either power laws 
(rarely with powers of precisely 0, 1, or 2) or special fitting functions, but they
can typically be approximated well as a smooth quadratic over the width of the \WMAP\
bandpasses.  
The fitting frequencies are somewhat arbitrary.  They were chosen by taking a canonical
center frequency for each band and two frequencies about 9\% higher and lower.
Then they were adjusted by hand so that the weights
were roughly equal and so the frequencies were multiples of 0.1 GHz.  
Further adjustment could be done, but the current numbers appear to work well.
Because of this arbitrariness of the frequencies in Table~\ref{tab:interp1},
they should not be taken to be a meaningful representation of the center or
width of the bandpass.  

The error in this approximation is typically less than the \WMAP\
calibration error of 0.2\%, for smooth spectra such as power laws.  
In Q band, for low frequency scale factors, the error in the spinning dust
spectrum can be on order of 1\%.  However, it is not clear that we know the
shape of the spinning dust spectrum to that accuracy.
This is intended to be a rapid and reasonably accurate way of integrating over
the \WMAP\ bands.  If more accurate methods are needed, such as for very steep
spectra or for spectra with emission lines, then a full integration over the
bandpass should be done.

\begin{deluxetable}{cccccccc}
  \tablewidth{0pt}
  \tablecolumns{8}
  \tablecaption{Interpolation data\tablenotemark{a} for $T = (\Delta T / \Delta T_A) \sum_{i=1}^3 w_i T_A(\nu_i)$ \label{tab:interp1}}
  \tablehead{
    \colhead{Band} &
    \colhead{$\nu_1$\tablenotemark{b}} &
    \colhead{$\nu_2$\tablenotemark{b}} &
    \colhead{$\nu_3$\tablenotemark{b}} &
    \colhead{$w_1$} &
    \colhead{$w_2$} &
    \colhead{$w_3$} &
    \colhead{$\Delta T / \Delta T_A$\tablenotemark{c}} 
    }
\startdata
K  &   20.6 &   22.8 &   24.9 &   0.332906 &   0.374325 &   0.292768 &   1.013438 \\
Ka &   30.4 &   33.0 &   35.6 &   0.322425 &   0.387532 &   0.290043 &   1.028413 \\
Q  &   37.8 &   40.7 &   43.8 &   0.353635 &   0.342752 &   0.303613 &   1.043500 \\
V  &   55.7 &   60.7 &   66.2 &   0.337805 &   0.370797 &   0.291399 &   1.098986 \\
W  &   87.0 &   93.5 &  100.8 &   0.337633 &   0.367513 &   0.294854 &   1.247521 \\
\enddata
\tablenotetext{a}{As stated in the text, the frequencies shown here have an
arbitrariness that prevents them from being a meaningful representation of the
center frequency or width of the \WMAP\ bandpasses.  The weights $w_i$ account for
this arbitrariness; they make the overall approximation accurate.
The weights and conversion factors are given to a precision of about 6
significant figures.  Our approximation is not that accurate; we provide this
precision to allow people to more easily reproduce our results and to make round-off
error negligible.
}
\tablenotetext{b}{Frequencies are given in GHz.}
\tablenotetext{c}{This is the antenna to thermodynamic conversion for an
unweighted average of radiometers, which should be used for this approximation.}
\end{deluxetable}

\bibliography{wmap}
\end{document}